\documentclass[notitlepage]{report}
\pdfoutput=1 
\usepackage{amsmath, amsthm, amssymb}
\usepackage{graphicx}
\usepackage{caption}
\usepackage{float}

\newcommand{\beq}{\begin{equation}}
\newcommand{\eeq}{\end{equation}}
\newcommand{\eqna}{\begin{eqnarray}}
\newcommand{\eqne}{\end{eqnarray}}
\newcommand{\eqnaa}{\begin{eqnarray*}}
\newcommand{\eqnae}{\end{eqnarray*}}
\newcommand{\dia}{\begin{displaymath}}
\newcommand{\die}{\end{displaymath}}
\newcommand{\bes}{\begin{split}}
\newcommand{\ees}{\end{split}}
\def\Sch{Schr\"odinger }
\def\vE{\mbox{\bf E}}
\def\vB{\mbox{\bf B}}
\def\vr{\mbox{\bf r}}
\def\vecr{\mbox{\bf r}}
\def\vk{\mbox{\bf k}}
\def\vA{\mbox{\bf A}}
\def\vC{\mbox{\bf C}}
\def\vD{\mbox{\bf D}}
\def\vL{\mbox{\bf L}}
\def\vS{\mbox{\bf S}}
\def\vJ{\mbox{\bf J}}
\def\vQ{\mbox{\bf Q}}
\def\vP{\mbox{\bf P}}
\def\cE{\mathcal{E}}

\def\cP{\mathcal{P}}
\def\cA{\mathcal{A}}
\def\vM{\mbox{\bf M}}
\def\vF{\mbox{\bf F}}

\def\vj{\mbox{\bf j}}
\def\vl{\mbox{\bf l}}
\def\ve{\mbox{\bf e}}
\def\vd{\mbox{\bf d}}
\def\vm{\mbox{\bf m}}
\def\vn{\mbox{\bf n}}
\def\vs{\mbox{\bf s}}
\def\vp{\mbox{\bf p}}
\def\vv{\mbox{\bf v}}
\def\va{\mbox{\bf a}}
\def\vb{\mbox{\bf b}}
\def\vv{\mbox{\bf v}}
\def\vq{\mbox{\bf q}}
\def\vlambda{\mbox{\boldmath$\lambda$}}
\def\vmu{\mbox {\boldmath $\mu$}}
\def\vphi{\mbox {\boldmath $\phi$}}
\def\valpha{\mbox {\boldmath $\alpha$}}
\def\hpsi{\hat{\psi}}
\def\hpsis{\hat{\psi}^{+}}
\def\psis{\hat{\psi}_\sigma}
\def\psiss{\hat{\psi}_\sigma^{+}}
\def\pd{\partial}
\def\arg{\mbox{{\bf r},t}}
\def\e/c{\frac{e}{c}}
\def\ha{\hat{a}}
\def\defint{\int_{-\infty}^{\infty}}

\begin{document}
\title{\bf{Quantum Mechanics Lecture Notes.} \\ Selected Chapters}
\author{ S. Levit \\  Department of Condensed Matter Physics \\  Weizmann Institute of Science, Rehovot, Israel \\
shimon.levit@weizmann.ac.il}
\maketitle
\begin{abstract}

These are extended lecture notes of the quantum mechanics course which I am teaching in the Weizmann Institute of Science physics program. They cover the topics listed below.  The first four chapters are posted here.   Their content is detailed on the next page.  The other chapters are planned to be added in the coming months.   

1. Motion in External Electromagnetic Field.  Gauge Fields in Quantum Mechanics. 

2. Quantum Mechanics  of Electromagnetic Field
 
3. Photon-Matter Interactions

4.  Quantization of the Schr\"odinger Field (The Second Quantization)

5.  Open Systems.  The Density Matrix

6.  Adiabatic Theory.  The Berry Phase.  The Born - Oppenheimer Approximation

7. Mean Field Approaches for Many Body Systems - Fermions and Bosons

\end{abstract}
\tableofcontents
\chapter{Motion in External Electromagnetic Field. Gauge Fields in  Quantum Mechanics}
Electromagnetic potentials $\vA(\vr,t)$ and $A_0(\vr,t)$ appear in classical physics as
auxiliary quantities which are introduced  in order to simplify the form and solutions
of the Maxwell equations, cf., Chapter 10 in Ref. \cite{Grif}. The fact that they are not
uniquely  defined  and can be changed without affecting  any physical  results by a
transformation bearing a strange name of "gauge"    seems to be rather an annoying  nuisance
than a fundamental symmetry of nature. 

This state of affairs undergoes drastic revision
when quantum mechanical description  is attempted. We do not know how to formulate  such a
description in the presence of the electromagnetic  field without making an essential use
of the  electromagnetic potentials.  Moreover the invariance under the gauge transformations
becomes  a profound symmetry  of our world which lies at the origin of  all  known
interactions. Because of this the fields   which carry these interactions  are
termed  {\em gauge fields}.

The problem of  the quantum mechanical motion in an external
electromagnetic field provides  the simplest setup  in which one
encounters some of the strange and beautiful phenomena appearing
as a result of   the symbiosis of  gauge fields and  quantum
mechanics.

{\bf{Note.  I have changed from CGS to SI units in Sections 1-8. The rest is in CGS. }}

\section{Electromagnetic Potentials. The Hamiltonian } 

\subsection{Electromagnetic potentials in classical physics \label{EMpot} }  

 Let us
begin by briefly recalling how the electromagnetic potentials  are introduced. Classical
 electromagnetic    field   is described by two vector fields $\vE (\arg)$ and
$\vB (\arg)$. In the present chapter these fields
 will be considered as external, i.e. produced by sources (electric charges and
 currents) which dynamically are not a part  of the physical system under consideration and
are not effected by it. This means that the back reaction of the system on the sources  of
the field  is negligible.  Although in such circumstances
$\vE$ and  $\vB$  should be regarded as   controlled  externally by  charge and current distributions $\rho(\vr, t) $ and $\vj(\vr, t) $ of  the sources via
$$
\nabla\cdot \vE = \frac{\rho} {\epsilon_0} \;\;\; , \;\;\; \nabla \times \vB = \mu_0 \vj + \frac{1}{c^2} \frac{\partial \vE} {\partial t}
$$
they can not be taken as
completely arbitrary.  Indeed irrespective  of the configuration of $\rho$ and $\vj$ 
these fields  must satisfy the  homogeneous pair of Maxwell equations
\beq
\nabla \cdot \vB = 0 \;\;\;  , \;\;\; \nabla \times \vE =
  -\frac{\pd \vB}{\pd t}
\eeq at every  point in space and time. In order to have these equations automatically
satisfied the  familiar vector and scalar potentials $\vA(\arg)$ and $A_0(\arg)$  are
introduced\footnote{Although we use "relativistic" notation for $A_0$ we use  "non relativistic" terminology and  call it a scalar potential}. This is done   by noticing that the first of the equations  above  means that
$\vB$ must be a curl of a  vector field  $\vA(\vr,t)$. Using this in the second equation gives
$$
\nabla \times (\vE +\frac{\pd \vA}{ \pd t})=0
$$
restricting the combination $\vE +\pd \vA /\pd t$ to be a gradient of a scalar
field. One has therefore
\eqna
\vE & = &-\frac{\pd \vA}{\pd t} - \nabla A_0 \; , \nonumber \\
\vB & = &\nabla \times \vA \; \label{empot}
\eqne 
Unlike the field strengths $\vE$ and $\vB$, the electromagnetic potentials can be
regarded as unrestricted so that  any $\vA(\vr,t)$  and $A_0(\vr,t)$ can be realised  by
the poper choice of the external charge and current distributions.

 The  use  of the
electromagnetic potentials however presents another problem. They are not unique since the
gauge transformation
\eqna
\vA^{\prime} (\arg) & = &\vA (\arg) + \nabla \chi (\arg)  \nonumber  \\ A_0^{\; \prime}
(\arg)  & = &  A_0 (\arg) - \frac{\pd \chi(\arg)}{\pd t}   \label{gt}
\eqne with an arbitrary function $\chi (\arg)$   leaves $\vE$ and $\vB$ invariant. As
was already mentioned above this  invariance, called the gauge invariance,  has profound
consequences in quantum mechanical systems and will be discussed at length below. At the
moment we just notice that because of it only three
 among the four functions $\vA$ and  $A_0$  are independent.  In general one combination of
the four functions can be eliminated by a suitably chosen  gauge transformation. For instance
choosing 
$$
\chi(\arg) = \int_{t_0}^{t}  \, A_0(\vr, t')\, dt'
$$ 
 (with arbitrary $t_0$) eliminates  $A_0$ and leaves
$\vA(\arg)$ as the only independent degrees of freedom of the electromagnetic field.

 \subsection{Classical Hamiltonian and equations of motion} 

 Classical  non relativistic  equation of motion for a    particle with electric charge
$q$ and mass $m$ in a given electromagnetic field is obtained by using the Lorenz force in
the Newton law
\beq m\frac{d^2 \vr}{dt^2} = q\vE + q \left ( \frac{d\vr} {dt} \times \vB \right)
\; \label{lzeqmo} .
\eeq
 In order to obtain the quantum mechanical description one can follow either the canonical or
the  path integral quantization procedures. We will start with the former. We first
determine  the classical canonical variables and the classical Hamiltonian function of the problem. 

The above equation is in terms of coordinates $\vr(t)$ and velocities $\vv(t) = d \vr/dt$ so it is is most convenient to
start by determining the Lagrangian of the system. This is 
\beq 
L(\vr,\vv, t) =\frac{1}{2}m\vv^2 +
q\vA\cdot\vv - qA_0(\vr)\; .   \label{cllgma}
\eeq
Indeed have
$$
\frac{d} {dt} \frac{\partial L } {\partial \vv} = m\frac{d \vv}{d t}  + q\frac{d\vA}{dt}  = m\frac{d \vv}{d t} + q\frac{\pd \vA}{\pd r_j}  \frac{dr_j}{dt} + q\frac{\pd \vA}{\pd t}
$$
and
$$
\frac{\pd L} {\pd \vr} = q \frac{\pd } {\pd \vr} (\vA\cdot\vv) - q  \frac{\pd A_0 } {\pd \vr} 
$$
In components
$$
m\frac{d v_i}{d t} +  q\frac{\pd A_i}{\pd r_j}  v_j  + q\frac{\pd A_i}{\pd t} = q  \frac{\pd A_j} {\pd r_i}  v_j  - q  \frac{\pd A_0 } {\pd r_i} 
$$
So have
$$
m\frac{d v_i}{d t}  =  q\left(- \frac{\pd A_i}{\pd t} -   \frac{\pd A_0 } {\pd r_i} \right) + q  \left( \frac{\pd A_j} {\pd r_i}    -  \frac{\pd A_i}{\pd r_j}   \right) v_j
$$
which is the Newton equation (\ref{lzeqmo}).  Indeed recalling  Eq.(\ref{empot})  one sees that the fist term is $q\vE$, while the last term can be transformed as
$$
\epsilon_{ijk} v_j B_k  = (\vv\times\vB)_i
$$
where we used the  antisymmetric symbol
$\epsilon_{ijk} $\footnote[1]{The Levi--Civita symbol
$\epsilon_{ijk}$ is defined by
$\epsilon_{123} = 1$ and the antisymmetry property under interchange of any indices,
$\epsilon_{ijk} = - \epsilon_{jik} = -\epsilon_{ikj},$ etc.  $\epsilon_{ijk}$ does not change
under cyclic permutations $\epsilon_{ijk} = \epsilon_{kij} = ...$.  .}
 to write vector products, e.g
\dia(\vC \times \vD)_i = \epsilon_{ijk}C_jD_k  \;\;,\;\;   C_iD_j - C_jD_i = \epsilon _{ijk}
(\vC\times\vD)_k \;.
\die
These two equalities are related by a useful identity 
$$
\epsilon_{ijk}\epsilon_{ij'k'} = \delta_{jj'}\delta_{kk'} -
\delta_{jk'}\delta_{j'k} \; .
$$

The above calculations show that the canonical momentum is
\beq
\vp =  \frac{\partial L } {\partial \vv} = m\frac {d\vr}{dt} + q \vA(\vr) \; . \label{cncjmo}
\eeq 
which expresses perhaps the most unusual aspect of the motion in the EM field - the fact that $\vp \ne m\vv$.  In the literature one often meets  the term "kinetic momentum" referring to the familiar $m \vv$.

Expressing $ \vv (\vp,\vr) = (\vp - q \vA(\vr))/m$ and using
$$
H= \vp \cdot \vv - L 
$$
with the above $\vv(\vp,\vr)$ we find the Hamiltonian function
\beq
 H(\vp,\vr) = \frac{1}{2m}\left(\vp - q \vA(\arg)\right)^2 + qA_0(\arg) \; ,
\eeq  
It is not difficult (and not surprising) to show that with this $H(\vp,\vr)$ the equation of motion (\ref{lzeqmo}) is equivalent to the two Hamilton equations
$$
  \frac{d\vr}{dt} = \frac{\pd H }{\pd \vp}   \;\;\; , \;\;\; \frac{d \vp}{ dt} = -\frac{\pd H}{\pd \vr}
 $$ 
\section{Quantization}

\subsection{The orbital part}

 Having established the form of H we follow the canonical quantization procedure   and consider the \Sch equation with the Hamiltonian
operator which is obtained by  replacing   $\vr$ and $\vp$ in $H$ by the operators
$\vr_{op} = \vr$  and  $\vp_{op} = -i\hbar \nabla$,
\beq  \label{hmwosp}
H_{op} = \frac{1}{2m}\left[-i\hbar\nabla - q\vA(\vr,t)\right]^2 + qA_0(\vr,t) \;
.\eeq 

\subsection{The spin magnetic moment}
Experimental  evidence shows that  this Hamiltonian  is capable of
describing   only   particles which do not carry spin. It must  be modified when the spin
degrees of freedom are present. This should not be too surprising since  already in
classical physics the energy of a  spinning {\em charged} particle  receives an additional
contribution  apart from the orbital motion. This  contribution arises from  the interaction
with the magnetic field $\vB$ of a localized  distribution of  electric current  $\bf j(r)$ which
a spinning charge creates. For a "point like" particle, i.e. a particle the size of 
which is much smaller than the scale over which $\vB(\vr)$ changes, the corresponding energy is
$$
E_{\rm spinning \; charge} = -\vmu\cdot\vB(\vr)
$$
 where
$\vmu$ is the   magnetic moment  of the current, cf.,  Chapter 5 of the  Ref. \cite{Grif},
$$ 
\vmu = \frac{1}{2}\int d^3 r\;\vr\times  \bf j(r) \; . 
$$
For composite particles the total current is a sum over internal components
 $$
 {\bf j(r)} =\sum_a q_a \vv_a \delta(\vr-\vr_a)
 $$
 each with its dynamics  so the calculation of $\vmu$ is in general 
not an easy task. However if all the components have an  equal charge to mass ratio
$q_1/m_1 = q_2/m_2 = . . . = q/m$ the magnetic moment  can be written as
\beq  \vmu  =\frac{1}{2}\sum_a q_a (\vr_a\times \vv_a) =
\frac{q}{2m}\sum_a \vr_a\times m_a\vv_a = \frac{q}{2m}\bf L \;. \eeq
  Experimental data as well as theoretical considerations (cf.,  Section \ref{gfac} below)
indicate that for elementary particles like electrons this classical linear relation
between $\vmu$ and the angular momentum of a system  holds also between the corresponding
quantum mechanical operators  of the magnetic moment
$\vmu_{op}$ and  the spin  $\vs_{op}$. However the proportionality coefficient  in general
does not coincide with the classical value. To emphasize this difference it is conventional (for charged particles) 
to write the relation between the operators  $\vmu$ and  $\vs$  as
\beq \label{mus}
\vmu_{op} = g\;\frac{q}{2m}\; \vs_{op} 
\eeq
with $q$ - the particle charge and $g$ - dimensionless   coefficient called the gyromagnetic  factor or for short
the g-factor.  Theoretical methods which allow to determine  $g$ and examples of
their applications  are considered in Section \ref{gfac}.

\subsection{The \Sch equaion}
Adding the term $-\vmu_{op}\cdot\vB$    to the Hamiltonian operator (\ref{hmwosp})  one
can write the Hamiltonian for an elementary  particle with a spin in an external EM field as 
\beq 
H_{op} = \frac{1}{2m}\left(-i\hbar\nabla -
 q\vA\right)^2 + qA_0 -g\frac{q}{2m} \;\;\vs_{op}\cdot\vB \; .
\label{schem0}
\eeq  
and the corresponding \Sch equation 
\beq
\label{schem}
i\hbar \frac{\pd \psi}{\pd t} = \left[\frac{1}{2m}\left(-i\hbar\nabla -
 q\vA\right)^2 + qA_0 -g\frac{q}{2m} \;\;\vs_{op}\cdot\vB \right]\psi
 \eeq
 where 
 $$
 \psi = \psi(\vr, \sigma; t)
 $$
 is a function of space  and spin variables $\vr$ and $\sigma$.
 
In writing out the square in this equation one should not forget that the operator
$\vp_{op} = -i\hbar\nabla$  in general does not commute with the vector $\vA$ which is a
function of coordinates. Since $\vp_{op}\cdot\vA - \vA\cdot\vp_{op} = -i\hbar\nabla\cdot\vA$,
one can write
$$
\frac{1}{2m}\left(-i\hbar\nabla -
 q\vA\right)^2 = -\frac {\hbar^2}{2m}\nabla^2 + \frac{i\hbar q }{2m}(\nabla\cdot\vA +
 2\vA\cdot\nabla)  + \frac{q^2}{2m}\vA^2 \;.
 $$  
 The operators $\vp_{op}$ and $\vA$ commute
if $\nabla\cdot\vA = 0$.  This happens e.g., for $\vA = (\vB \times \vr )/2$ which is a
possible choice of $\vA$ in a particular case of a uniform magnetic field\footnote{
Verifying \eqna
(\nabla \times \vA)_i  &=& \frac{1}{2}\epsilon_{ijk} \nabla_j \epsilon_{klm} B_l x_m =\frac{1}{2}(\delta_{il}\delta_{jm} - \delta_{im}\delta_{jl})\delta_{jm} B_l = \frac{1}{2}(3B_i - B_i) =  B_i   \nonumber \\
\nabla \cdot \vA &=& \nabla_i A_i =  \nabla_i \epsilon_{ijk} B_j x_k = \epsilon_{ijk} B_j \delta_{ik} =0
\eqne 
 }.

 In the following  sections we will examine various properties of the equation
(\ref{schem})  and will present its solutions for some particular simple choices of the
electric and magnetic fields.

\section{Gauge Invariance}

\subsection{Gauge transformations in quantum mechanics}

The electromagnetic field enters the classical and quantum equations
(\ref{lzeqmo}) and (\ref{schem}) via very  different sets of  variables. The
classical equation  depends on  the physically measurable variables  $\vE$
and $\vB$  of the field whereas in the \Sch equation   the field enters via non uniquely defined and
seemingly auxiliary objects $\vA$ and $A_0$.   This is not an accident.  At present no
formulation of quantum mechanics exists which does not explicitly use the electromagnetic
potentials.  \Sch and Heisenberg pictures require  the Hamiltonian while the path integral
quantization uses the  Lagrangian (cf., below, Section \ref{piexem}) 
and both objects can not be written without $\vA$ and $A_0$.  Since the
potentials are not uniquely defined and can be changed by  a gauge transformation one must
address the question of how  unambiguous  physical results are obtained in such a situation.

 Unlike in classical mechanics where  gauge  transformations do not change the equations of
motion the \Sch equation (\ref{schem}) and therefore also its solutions   $\psi (\arg)$
are transformed  in a non trivial way\footnote{
In this and many of the following sections the dependence of $\psi(\vr,\sigma;t)$ on the spin variable $\sigma$ 
will not be of interest and will be suppressed for brevity.
}.  It is not difficult to find how the transformation of
$\psi(\arg)$ is related to the transformation of the potentials. For this we notice that
$\vA$ and $A_0$ enter the equation only in the combinations
$$
(-i\hbar\nabla - q \vA)  \;\;\; {\rm and} \;\;\;\;
(i\hbar\frac{\pd}{\pd t}  - q A_0)\; .
$$
 Thus if
$\psi(\arg)$ is a  solution for a particular choice of $\vA$ and $A_0$ then
\beq
\psi(\arg) = \exp\left[-i\frac{q}{\hbar }\chi(\arg)\right]\psi^{\;\prime}(\arg)
\equiv S(\arg)\psi^{\;\prime}(\arg) \label{gt1}
\eeq  satisfies
\eqna -i\hbar\vD\psi & \equiv &
\left(-i\hbar\nabla - q \vA\right)\psi = S(\arg)\left(-i\hbar\nabla - q
\vA^{\prime}
\right)\psi^{\prime} = -i\hbar S(\arg) \vD^{\prime}\psi^{\prime}\;,\nonumber
\\ i\hbar D_0 \psi & \equiv &
\left(i\hbar\frac{\pd}{\pd t } - q A_0\right)\psi = S(\arg)\left(i\hbar\frac{\pd}{\pd t } - q
A^{\;\prime}_0\right)\psi\  ' = i\hbar S(\arg) D^{\;\prime}_0\psi^{\prime}\label{cvder}
\eqne  
 and therefore solves the \Sch equation for the transformed potentials  (\ref{gt}) 
 (please note the primed $\vD^{\prime}$ and  $D^{\;\prime}_0$ on the right hand side of the expressions above). 

We see that the classical concept of  the gauge  transformation undergoes a generalization in
quantum mechanics. Now not only the potentials which describe the electromagnetic field but
also the wave functions describing the material particles must change  simultaneously
according to  the rules (\ref{gt}) and  (\ref{gt1}).  This change is {\em local}, i.e. it
is different for different points in  space and time. One often emphasizes this aspect
by calling the transformation given by Eqs. (\ref{gt}), (\ref{gt1}) a {\em local gauge
transformation} to distinguish it from a  {\em global  transformation} in which the
wave function is multiplied by a constant phase factor.

  It is seen that  the combinations
$\vD\psi$ and $D_0\psi$ defined in (\ref{cvder}) transform
under a local gauge transformation in a particularly simple way -- i.e. as if it were
a global transformation. These combinations  are  called gauge covariant derivatives in
theories with gauge fields. The way to introduce the electromagnetic field in the dynamical
equations by replacing the ordinary derivatives
$\pd/\pd t$ and $\pd/\pd \vr$ by the gauge covariant combinations $D_0, \vD$ is known as {\em minimal coupling}.

 \subsection{Gauge symmetry vs gauge invariance}
 
 We may now ask a question as to whether  the  classical gauge invariance   also holds  in
quantum mechanics,  namely whether the  result of any measurement is invariant under  gauge
transformations which now include also the local transformation (\ref{gt1}) of the wave
function. It is an empirical fact that the answer to this question is positive.
Moreover it is also  clear that this invariance  known as {\em the local
gauge invariance} is a  profound  fundamental symmetry of the quantum mechanical description
in the presence of gauge fields. 

It is important to note that  this symmetry does not mean
that the wave functions must be  invariant. Like with other fundamental symmetries, e.g. the
invariance with respect to translations and rotations, the gauge symmetry means that the
wave functions
 transform in a particular  way given by   Eq.  (\ref{gt1}), i.e they  form a representation
of the corresponding  group of transformations. 

Here, however, the similarity ends. Unlike 
other symmetries the gauge symmetry demands that the  {\em observable quantities}  must not
be effected by the gauge transformations and therefore must be  "gauge scalars", i.e. depend
on gauge invariant combinations of  $\psi,\vA$ and $A_0$. No "gauge vectors", "gauge
tensors", etc, are ever observed. The origin of this difference can only be understood
when the  full quantum dynamics of the electromagnetic field and its coupling to matter are
discussed.

We conclude this section by noting  that explicit
appearance of the electromagnetic potentials in the equations of quantum mechanics   makes
the gauge invariance  a  very subtle symmetry.  Its consequences and
generalizations  are important aspects of the modern physics.  We will make a
special point in this chapter to illustrate  some of the related physical ideas and results.

\subsection{The Gauge Principle -- symmetry  dictates \protect \\interactions \label{ginv}}

In the previous section  we started with the known transformation properties of the potentials
and then on the basis of the special manner in which they  entered the \Sch equations --
i.e. in combinations ${\bf D}$ and
$D_0$, -- derived the required transformation properties of the wave functions which were
necessary in order to keep the
\Sch equation form-invariant.

Imagine now that we reverse this derivation in the following
manner. Let us begin by considering  the  free
\Sch equation 
$$
i\hbar\pd_t\psi = -\hbar^2\nabla^2\psi/2m \;\;. 
$$  
This equation is obviously
invariant under the {\em global gauge transformations} i.e. the  transformations (\ref{gt1})
with a constant
$\chi$ independent of ($\arg$). This global gauge  invariance is a fundamental feature  of
the \Sch equation. One of its notable consequences  is  the conservation  of the integral
$\int d\vr \;\psi^*(\arg)\psi(\arg)$. This integral is the total probability or, when
multiplied by $e$,  the total electric charge.
 The relation of its conservation to the global gauge invariance is not  intuitively
obvious  but can be rigorously derived by the applications of arguments of the Noether
theorem to the \Sch field.

 Now let us see what  happens if one  demands that the nature should be invariant not only
under the global but also under {\em local gauge transformations}, i.e. with  the
($\arg$)-dependent phase  $\chi$ in Eq. (\ref{gt1}). It is obvious that the free \Sch
equation will not satisfy this demand since its derivatives will act on the local  phase
producing  additional terms with $\nabla\chi$ and $\pd \chi/\pd t$. With the hindsight of
the previous section  we can  however   write a more general  \Sch equation which will be
locally gauge invariant.  

In order to compensate for the  derivatives $\nabla\chi$ and $\pd
\chi/\pd t$   and eliminate them from the transformed  \Sch equation  we must 

 (a)  "postulate" the existence of a field described by  the potentials $\vA$ and
$A_0$, 

 (b)  replace  the ordinary derivatives $\pd/\pd t$ and $\nabla$ in the equation by
the gauge covariant combinations $D_0$,
${\bf D}$   and

 (c)  require that the potentials transform according to  Eq. (\ref{gt})
simultaneously with the transformation (\ref{gt1}) of the wave functions.  

The demand of the {\it local gauge invariance} is  thus turned into a  powerful {\em heuristic} principle --
{\em The Gauge Principle}, which, had we not known about the electromagnetic field,  led us
to "discover" its existence and the way it must appear  in the \Sch  equation.

 Of course the last, spin-dependent term in (\ref{schem}) would not be deduced in such a
procedure and should be justified separately.  The need for this separate discussion of the
spin interaction with the electromagnetic field  disappears when a fully relativistic theory
of {\em  elementary particles} is considered,  cf. Section 8.1 in Ref. \cite{Che} or
Chapter 3 in Ref.\,\cite{Ryd}. Moreover it can be shown  that the entire Maxwell electrodynamics
is fully consistent with the  The Gauge Principle supplemented by very general requirements
of the time-space translational invariance  and the Lorenz invariance.

 It also turns out that
the fields responsible for all other known interactions, i.e.  weak, strong and
gravitational are consistent with The  Gauge Principle  in a similar way.  
Namely for every known interaction there exist a
 a global symmetry of a non interacting theory which  becomes a local  symmetry after the
interaction is introduced. The potentials describing the interaction are  the compensating
gauge  potentials which are necessary to introduce in order to satisfy this demand are the
fields of the fundamental interactions. Thus The Gauge Principle essentially  means that
{\em  Symmetry Dictates Interactions}. The Gauge Principle for general relativity  for example
means  that  the theory is invariant under  {\em local}\  Lorenz transformations. In Section
\ref{nonab} below we consider an example of how a so called non abelian gauge field  appears
as a result of the demand that the
\Sch equation  is invariant under  local non abelian transformations.

\section{Electric  Current Density.}
\subsection{The orbital part}
 Let us  derive the quantum mechanical expression for the current density of charged
particles.  We will start by considering the continuity equation for the   charge density
$\rho(\arg)  = q\psi^*(\arg)\psi(\arg)$. Multiplying the \Sch equation (\ref{schem}) on the
left by
$\psi^*$ and its complex conjugate by $\psi$ and subtracting one obtains in a standard way
that  $\pd \rho /\pd t +\nabla\cdot {\bf j} = 0$ with the current density

\beq
 {\bf j}_{\rm orbital}(\vr) = \frac{q}{2m}\left[\psi(\vr)(i\hbar\nabla -
q\vA(\vr))\psi^*(\vr) +
 \psi^*(\vr)(-i\hbar\nabla - q\vA(\vr))\psi(\vr)\right]\; .
\eeq 

This expression is the expectation value $<\psi|{\bf j}_{op}(\vr)|\psi>$ of the operator
\beq 
{\bf j}_{op}(\vr) = \frac{1}{2}\left[q\vv_{op}\delta(\vr - \vr_{op}) +
\delta(\vr -
\vr_{op})\, q\vv_{op}\right]
\eeq of the  current density due to orbital motion with the velocity 
 $$
 \vv_{op} = [\vp_{op}
- q\vA(\vr_{op})]/m\;.
$$ 
This operator is just what is obtained from  the classical
expression $\rho(\vr,t)\vv(t) =q \delta(\vr - \vr(t))\vv(t)$ for a point particle  by
replacing the classical quantities $ \vr(t)$ and $\vv(t)$ with the corresponding operators
$\vr_{op}$ and $\vp_{op}$  and symmetrizing the final
expression  in order to make it hermitian.

\subsection{The spin contribution}
The  missing feature in the above  expression for the current is the absence of the
contribution from the spin of the particle. This is the reason we have added to it the  index
$orbital$. As we have already discussed  a  spinning charged particle creates a local
distribution of electric current at its location and one  should expect to find an
appropriate term in the current density in addition to the contribution  of the orbital
motion. We have missed this term because as we will see in a moment it is in the form of a
rotor of a vector (a so called solenoidal term) and therefore can not be seen in  the
continuity equation which  depends only  upon the divergence of the current.

 In order to  correct  our result let us consider a physical system of charges $\{q_a\}$
placed in positions $\{\vr_a\}$ and put it under the influence of an external electric field
$\vE(\vr)$. We start classically and consider a time interval $dt$ during which these charges
move distances $d\vr_a = \vv_a dt$. As a result  their total energy is changed by
$$
dW =
\sum_a q_a\vE(\vr_a)\cdot d\vr_a = dt
\int d\vr [\sum_a q_a \frac{d\vr_a}{dt}\delta(\vr - \vr_a)]\cdot\vE(\vr) 
 $$
 The expression in the square
brackets here is the total current density flowing in the system, so that
\beq 
\frac{dW}{dt} = \int d\vr \;{\bf j}(\vr)\cdot\vE(\vr)
\; .\label{vara}
\eeq

We assume that this relation holds also for quantum mechanical expectation values.
Considering  for simplicity  one particle and let us form the expectation value of the Hamiltonian
 (\ref{schem0})
\eqna 
W & = &\int d\vr \psi^*H\psi \nonumber \\ & = &\int d\vr\psi^*
\left[\frac{1}{2m}\left(-i\hbar\nabla -
 q \vA\right)^2  + qA_0 -g\frac{q}{2m} \;\;\vs_{op}\cdot\vB\right] \psi \; .
\eqne  We also have
\beq
\frac{dW}{dt} = <\psi|\frac{\pd H}{\pd t}|\psi> \;\; .
\eeq 
This relation (sometimes called the Feynman--Hellmann theorem) is valid since the term
$<\pd\psi/\pd t|H|\psi> + <\psi|H|\pd\psi/\pd t>$ vanishes on account of the \Sch equation
$i\hbar\:\pd\psi/\pd t = H\psi$.

In order to find the time derivative of the Hamiltonian  we note that it depends on time only
via the time dependence of the potentials $\vA, A_0$. Part of this time dependence is not
physical and is related to the time dependent gauge transformations of $\vA$ and
$A_0$. In order to avoid this fake time dependence we fix the gauge by choosing $A_0 = 0$.
This choice does not fix the potentials completely but the only freedom left is {\em time
independent} gauge transformations, i.e. Eq.(\ref{gt}) with time independent
 $\chi (\vr)$. With this choice we have that
\beq
\frac{\pd H}{\pd t} = \int d\vr \;\frac{\delta H}{\delta \vA (\arg)}\;\frac{\pd
\vA(\arg)}{\pd t} \; .
\eeq  Using (\ref{empot}) with $A_0 =  0$ and Eq. (\ref{vara}) we obtain the general relation
for the electric current
\beq {\bf j}(\arg) = -<\psi|\frac{\delta H}{\delta \vA (\arg)}|\psi>\; .
\eeq Varying H with respect to $\vA$ and using  $\vB = \nabla\times\vA$  we obtain
\eqna 
<\psi|\delta H|\psi> & = & \int d\vr \left\{ \psi^*\left[ \frac{iq\hbar}{2m}(\nabla
\cdot\delta\vA +
\delta\vA\cdot\nabla) + \frac{q^2}{m}\delta\vA\cdot\vA\right]\psi \right.\\ &  &
\;\;\;\;\;\; \;\;\;\;\; - \left.
g\frac{q}{2m}(\psi^*\vs_{op}\psi)\cdot\left(\nabla\times\delta\vA\right)\right\}
\nonumber
 \eqne
Integrating by parts in the first term, using the identity 
$$
\va\cdot\nabla\times\vb
 = - \nabla\cdot(\va\times\vb) +\vb\cdot(\nabla\times\va)
 $$
  for the last term
 in this expression and assuming that  the surface terms vanish we obtain  the following
expression for the current
\beq \label{elcurr}
 {\bf j}(\vr) = \frac{iq\hbar}{2m}(\psi\nabla\psi^* - \psi^*\nabla\psi)
-\frac{q^2}{m}\vA\psi\psi^* +  g\frac{q}{2m}\nabla\times(\psi^*\vs_{op}\psi) 
\eeq 
The first two terms are just  the "orbital" current already  obtained earlier from the
continuity equation. The last, "solenoidal" term is the spin contribution which has the
appearance of the classical relation between the current and the magnetic moment  
$$
{\bf
j}_{spin}(\vr) = \nabla \times (gq/2m)\vs = \nabla \times
\vmu \;.
$$

\subsection{Convective, diamagnetic and spin parts of the current}

The first term in the expression (\ref{elcurr}) for the current ${\bf j}$ is called the convection current and
coincides with the usual expression for the current density in the absence of
 the electromagnetic field. It is not  gauge invariant without the second term which is
called the   diamagnetic current. The third, spin term in ${\bf j}$ is obviously gauge invariant by
itself.

In elementary quantum mechanics  one develops certain intuition about   currents  associated
with given wave functions. In particular one is used to the fact that non vanishing
current density  does not appear  if the wave function is real, that
 the current is related to the local complex phase of $\psi$, etc.  This intuition is
founded  entirely on the first term in Eq. (\ref{elcurr}) and  could be 
misleading in the presence of electromagnetic field. In this case   one finds for instance
a non vanishing orbital current density 
  $$
  {\bf{j}}_{\rm orbital}(\arg) = (q^2/m)\vA\psi^2(\vr) \;\; ({\rm for\;\; \psi - real} )
  $$
for a real wave function. Of course the freedom of local gauge transformations (\ref{gt1})
makes the    phase of $\psi$ and the difference between  real and complex  wave functions
into something which depends on the choice the gauge and therefore unphysical.

\section{Motion in a Uniform Electric Field}

Already such a simple problem as the motion of a charged particle in a constant uniform
electric field $\vE$ exhibits peculiarities of gauge fields in quantum mechanics. Classically
everything is simple. The particle moves with the constant acceleration
$q\mid\vE\mid/m$ in the direction of the field and has a constant, determined by initial
conditions  velocity perpendicular to this
 direction.  In quantum mechanics one may have  differently looking descriptions
 depending on which of the many (i.e. continuous number of) possible choices of
$\vA$ and $A_0$ is made leading to  the same constant $\vE$ and $\vB=0$. Of course  the gauge
invariance  will assure  that all  physical quantities are independent of the gauge choice
but  in actual calculations  it may require some efforts to see the connections.

\subsection{Static gauge}
We will explore in some detail two gauge choices,  the simplest  and most familiar  gauge
$\vA = 0$,
$A_0 = -\vE\cdot\vr$ and another,  time-dependent  gauge $\vA = -\vE t$, $A_0 = 0.$ In
the former case the time and the coordinate variables are separable in the
\Sch equation
\beq  i\hbar\frac{\pd \psi(\vr,t)}{\pd t} = \left(-\frac{\hbar^2}{2m}\nabla ^2 -
q\vE\cdot\vr\right)\psi(\vr,t)  \; , \label{elsch}
\eeq and moreover also separable are the coordinates parallel and perpendicular to
$\vE.$ Choosing  the
 $x$ axis parallel to $\vE$ and denoting by  subscript $\perp$ vectors which  are
perpendicular to $\vE$ one  can write the stationary solution as
\beq
\psi(\vr,t) = \phi_{\varepsilon }(x)
\exp\left(i{\bf k}_{\perp} \cdot\vr_{\perp}\right)
\exp\left[-\frac{it}{\hbar}\left(\varepsilon  +
\frac{\hbar^2 {\bf k}_{\perp}^2}{2m}\right)\right] \; , \label{gsol}
\eeq where $\varepsilon $ and
$\phi_{\varepsilon }(x)$  are the   eigenenergies and the corresponding  eigenfunctions  of the
motion parallel to x. They satisfy  the one dimensional \Sch equation
\beq
\left(-\;\frac{\hbar^2}{2m}\;\;\frac{d^2 }{dx^2} -   Fx\right)\phi_{\varepsilon } =
\varepsilon \phi_{\varepsilon } \label{fiep}
\eeq where we denoted $F = q\mid\vE\mid$.

In the  equation for $\phi_{\varepsilon }(x)$ the behavior of the potential
$-Fx$ at infinite values  of x is such that the energy levels $\varepsilon $ form a continuous
spectrum of values from $-\infty$ to $+\infty$. They  should correspond to motion which is
bounded from $x = -\infty$ but unbounded in the direction $x \to +\infty$ . The wave
functions must vanish in the region of large and negative x and  therefore the energy levels
are non degenerate. Indeed if there were two solutions
$\phi_1(x)$ and $\phi_2(x)$ for the same $\varepsilon $ then
\beq
\frac{1}{\phi_1}\frac{d^2\phi_1}{dx^2} = \frac{2m}{\hbar^2}\left(\varepsilon  + Fx\right) =
\frac{1}{\phi_2}\frac{d^2\phi_2}{dx^2}
\eeq so that the Wronskian $w = \phi_1(d\phi_2/dx)  - \phi_2(d\phi_1/dx) = const$. The
condition that wave functions vanish at $x = -\infty$ means that $w = 0$   leading to
$\phi_1 = const \;\phi_2$ i.e. the two solutions would in fact coincide.

\subsection{Linear potential - the Airy function}
\begin{figure}
\centering \includegraphics[width=0.8\textwidth]{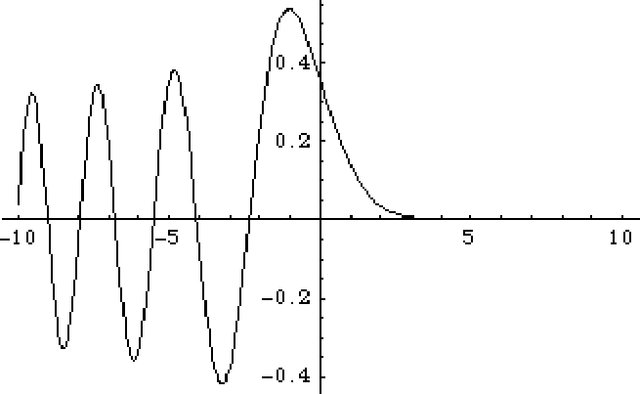}
\caption{Airy function $Ai(x)$ }
 \label{fig:Airy}
\end{figure}

The simplest way to solve the  equation for $\phi_{\varepsilon }$  is to consider it in the
momentum representation. Inserting the expansion
\beq
\phi_{\varepsilon }(x) =
\int_{-\infty}^{\infty}\frac{dp}{\sqrt{ 2\pi\hbar}}\; a_{\varepsilon }(p)
\;e^{ipx/\hbar}
\eeq in the equation for $\phi_{\varepsilon }(x)$ we easily obtain
\beq
\left(\frac{p^2}{2m} - i\hbar F \frac{\pd}{\pd p}\right) a_{\varepsilon }(p) = \varepsilon 
a_{\varepsilon }(p) \; .
\eeq    Integrating this  first order equation  we find
\beq a_{\varepsilon }(p) = const\;\;\exp\left[ \frac{i}{\hbar F}\left(\varepsilon  p - \frac{p^3}{6m}
\right)\right]  \; .
\eeq 
The constant in front of this expression must be determined by normalization.
 Choosing e.g., to normalize $a_{\varepsilon }(p)$ on the delta function in $\varepsilon $
\beq
\int_{-\infty}^{\infty}dp\; a^*_{\varepsilon '}(p)\;a_{\varepsilon }(p) = |C|^2 \defint dp\;
\exp\left[\frac{i}{\hbar F}\left(\varepsilon  - \varepsilon '\right)p \right] = \delta(\varepsilon  -
\varepsilon ')
\eeq we obtain $C = 1/\sqrt{2\pi\hbar F }$. The wave functions in the position representation
are
\beq
\phi_{\varepsilon }(x)  = \int_{-\infty}^{\infty}\frac{dp}{2\pi\hbar\sqrt{F}}\;\;
\exp\frac{i}{\hbar}\left[px + \frac{1}{F}\left(\varepsilon  p -
\frac{p^3}{6m}\right)\right]
  =  \frac{\alpha}{\pi\,\sqrt{F}} \;\;Ai\left[-\alpha(x +\varepsilon  /F)\right] \; ,
\label{pcoor}
\eeq 
where we denoted $\alpha = (2mF/\hbar^2)^{1/3}$ and introduced the notation
\beq 
Ai(\xi) = \frac{1}{\sqrt{4\pi}}\defint du \exp \left[i\left(\frac{u^3}{3} + \xi
u\right)\right] \;.\label{intair}
\eeq
The function $Ai(\xi)$ defined by this integral  is called the Airy function.  We will
explore some of its properties below, cf., also Ref. \cite{Air}. The graph of the Airy function 
is shown in Fig.\,\ref{fig:Airy}.

Eq. (\ref{pcoor}) together
with (\ref{gsol}) furnish the general solution of the
\Sch equation (\ref{elsch}).  

It is instructive to explore the asymptotic behavior of
the wave function
$\phi_{\varepsilon }(x)$ for $x \to \pm \infty$.  This can be found
 by using the saddle point approximation
 \begin{quotation}
cf., https://atmos.washington.edu/~breth/classes/AM568/lect/lect22.pdf
\end{quotation}
in order to evaluate the integral
in  (\ref{pcoor}). Differentiating  the exponent in the integrand we obtain that the
stationary value $p_0$ of p must satisfy
\beq
 \varepsilon  = \frac{p_0^2}{2m} - Fx \; .\eeq
This is  the classical -- energy momentum relation in the  potential $-Fx$.  It is an
example of  a   typical "cleverness" of the saddle point method -- when a phase of a rapidly
oscillating integral depends on  external physical parameters  (the coordinate x in
(\ref{pcoor}))    the saddle point condition frequently has a transparent
physically significance.   The
difference with the classical physics is that
$p_0$ does not have to be real. Only for
$x> - \varepsilon /F$, i.e. in the region where  the classical motion is allowed, $p_0$ is real,
but it is pure imaginary in the classically forbidden region $x < - \varepsilon /F$. In both
cases there are two solutions corresponding to the two signs in the square root
\beq p_0(x) = \sqrt{2m(\varepsilon  + Fx)} \; .
\eeq  According to the rules of the saddle point approximation both
saddle point solutions should be retained  in the real case while only the saddle point
with decaying exponential should be admitted in the imaginary case.  We thus find  for
(\ref{pcoor})
\eqna
\phi_{\varepsilon }(x) & \approx & \sqrt{\frac{2m}{\pi\hbar p_0
}}\cos\left(\frac{p_0\,^3(x)}{3\hbar Fm} -
\frac{\pi}{4}\right)\;\;,\;\;\;\; {\rm for }\;\; x \to \infty \nonumber \\
\phi_{\varepsilon }(x) & \approx &
\frac{1}{2}\sqrt{\frac{2m}{\pi\hbar |p_0|}}\exp\left(-\frac{|p_0(x)|^3}{3\hbar Fm}\right)\;\;
,\;\; {\rm for}\;\;x \to -\infty \; .
\eqne We will see in the section devoted to the semiclassical limit that these expressions
correspond to the semiclassical approximation for wave functions. As required the wave
function decays exponentially in the classically forbidden region $E<-Fx$.  In the
classically allowed region the positive and negative momenta
$p_0$ with equal amplitudes coexist for a stationary quantum mechanical state producing the
interference cosine with the argument which can be written as
\beq
\frac{p_0^3(x)}{3\hbar Fm} \equiv \frac{1}{\hbar}W(x) = \frac{1}{\hbar}\int_{-E/F}^x
p_0(x')\;dx'
\eeq 
in terms of the classical action $W$.  The classical momentum determines the {\em local
wave length}
$\lambda = 2\pi\hbar (dW/dx)^{-1}$ which decreases with increasing x in accordance with
the uniform classical acceleration in the direction of the field and the de Broglie
relation.

\subsection{Time dependent gauge}

 Let us  now examine how this problem looks in
another gauge
$\vA = -\vE t$,
$A_0 = 0$. We use the gauge transformation (\ref{gt1}) with $\chi = -\vE\cdot\vr t$ in the
equation (\ref{elsch}) and obtain
\eqna \label{tdep_gauge}
\psi(\arg) & = &\exp(i q \vE\cdot\vr t/\hbar)\psi'(\arg) \; , \\ \nonumber 
i\hbar\frac{\pd \psi'(\arg)}{\pd t} & = & \frac{1}{2m}(-i\hbar\nabla + q\vE t)^2
\psi'(\arg)\;,
\eqne where we denoted by $\psi'(\arg)$ the transformed wave function. 

A simple solution of
this equation is a plane wave and we obtain for $\psi'(\arg)$
\beq \label{wf_in_tdep_gauge}
\psi'_{\vk}(\arg) = A_{\vk}(t) \exp[i \vk\cdot\vr] 
\eeq 
with the time dependent amplitude  $A_{\vk}(t)$ satisfying
\beq
 i\hbar\frac{d A_{\vk}(t)}{d t} = \frac{1}{2m}(\hbar\vk + q\vE t)^2 A_{\vk}(t)\;.
\eeq 
Integrating we find
\beq A_{\vk}(t) =C_0\exp\left\{-\frac{i}{\hbar}\left[\frac{\hbar^2\vk_{\perp}^2}{2m}\;t\;\; +
\;\;\frac{1}{6mF}(p + Ft)^3\right]\right\}\; ,
\eeq
 where $C_0$ is an arbitrary constant, $p = \hbar k_x$ and we used the same notation for
$\vk_{\perp}$ and $F$ as in the previous section.

We note that with this solution the wave function before the gauge transformation (\ref{tdep_gauge}) is 
\beq \label{td_wf_in_static_gauge}
\psi_{\vk}(\arg) = A_{\vk}(t) \exp[i(\vk + q\vE t/\hbar)\cdot\vr] 
\eeq 
which of course is a solution of the \Sch equation (\ref{elsch}) in the static gauge. The set of these 
time dependent solutions with different $\vk$'s is identical to the
set (\ref{gsol}) of stationary solutions as far as the motion in $\vr_\perp$ is concerned.  However in
the direction of the field the sets look quite different  and the  point to note here is that
in different gauges the same problem may have a very different appearance.  

Of course mathematically both sets are equivalent and one can easily show that  each can be expressed
as a linear combination of the other. The time
dependent solution is  closer to the classical intuition of the accelerated motion under a
constant force.

It is instructive in this simple problem to compare the calculations of the currents for two solutions $\psi_{\vk}(\arg)$, Eq. (\ref{td_wf_in_static_gauge}), and the corresponding transformed one $\psi'_{\vk}(\arg)$, Eq. (\ref{wf_in_tdep_gauge}). 
One will get different results with the two solutions
$$
q \frac{\hbar \vk + q\vE t}{m} |C_0|^2 \;\;\;\; {\rm and} \;\;\;\;\;  q\frac{\hbar \vk}{m} |C_0|^2
$$
 for the  convective part  of the current given by the fist part of Eq.(\ref{elcurr}). 
But this difference is "counterbalanced" by the different second diamagnetic term $-(q^2/m) \vA|\psi|^2$
in the current expression. It is zero in the static gauge  but is
$$
q^2\frac{\vE t}{m} |C_0|^2
$$
 in the time dependent gauge.   The end results is of course the same expression as it must be for the gauge invariant quantity.  

\subsection{Translations in uniform $\vE$. Symmetries in the presence of gauge fields } 
Physics in a constant electric field must be invariant under translations of coordinates
$\vr \rightarrow \vr + {\bf a}$ with a constant vector ${\bf a}$.  Applying this
transformation in the \Sch equation (\ref{elsch}) one at first finds that it is not
invariant -- the term $-\vE\cdot\va$ is added to the Hamiltonian. This term however can
be removed if  one simultaneously performs a gauge transformation of the wave function
\beq
\psi(\arg) \rightarrow \exp\left(-iq\vE\cdot{\bf a} t/\hbar\right)\psi(\vr + {\bf a},t) =
\exp\left[i{\bf a}\cdot(-i\hbar\nabla - q\vE t)/\hbar\right]\psi(\arg)
\eeq   
The \Sch equation is invariant under this combined transformation which must be
therefore adopted as {\em the definition} of the translation in the present case. One can
call it "electric translation" in analogy with the modified "magnetic translations" in
a uniform magnetic field, cf.  Section \ref{trinv} below.

 This feature of modification of the standard symmetry transformations by  additional gauge
transformations is quite typical for theories with gauge fields.
 It accounts for the fact that changing the coordinate system  may also effect the gauge
choice and  care must be taken to return to the original gauge.  The generator of the
infinitesimal translations in the expression above is obviously
\beq 
{\bf g}_{op} = \vp_{op} - q\vE t \label{gmom}\;.
\eeq
 The symmetry means that it must be conserved and indeed one finds that
  $$
  d{\bf g}_{op}/dt =
\pd{\bf g}_{op}/\pd t + (i/\hbar)[H\,,\,{\bf g}_{op}] = 0
$$
 with $H$ as it appears in the right hand  side of (\ref{elsch}).

We will find  below  two  other  examples of the  gauge  field modifications  of the
symmetries -- translational  and  rotational   invariance  in a  uniform   magnetic  field
(Section \ref{trinv})   and  rotational   invariance  in the field of a magnetic   monopole
(Section \ref{rtinv}).

\section{Motion in a Uniform Magnetic Field}

 We will now consider the quantum mechanical motion of a charged particle  in a uniform
external magnetic field  $\vB$ which is constant in magnitude and direction over the  entire space. 
For convenience we present the discussion for electrons, i.e. we take the value of the charge 
$$
q=-e \; .
$$

\subsection{Classical motion.  The guiding centers}
It is instructive to recall first the classical solutions of the problem.  The classical equation of motion is
$$
m\,d \vv /dt = -e(\vv \times \vB)\;.
$$  
Let us choose the direction of the $z$ axis
parallel to $\vB$. Then the  motion along $z$ is free,
 $$
 m\, dv_z/dt = 0 \;\; , \;\; z = v_z t + z_0
 $$ 
 with constant $v_z$ and  $z_0$ determined by initial conditions. 
 
 The equations for the $x$ and $y$ components are
\eqna \label{eq:eqs_in_uniform_B}
m \frac{dv_x}{dt} & = &- eBv_y \; , \nonumber \\ m \frac{dv_y}{dt} & = &
eB v_x \; .
\eqne
 An important observation to be made  here is that these Newton equations for the velocities
of motion in a plane perpendicular to $\vB$ have the same formal appearance as the  Hamilton equations of a {\em one
dimensional oscillator} with  $v_x$ and  $v_{y}$  formally proportional to the  respective
coordinate and momentum of the oscillator.   The solution of these  equations is "harmonic motion" in the
"velocity space" with the  frequency 
$$
\omega_c = eB/m
$$ 
called the cyclotron frequency and 
\beq v_x = v \cos (\omega_c t + \alpha)\;\; ,\;\; v_y = v \sin (\omega_c t + \alpha)\;\; ,
\eeq  
so that the trajectory in the (x,y) plane is
\eqna  \label{mag_circles}
x & = & \frac{v}{\omega_c} \sin (\omega_c t + \alpha) + x_0  = \frac{v_y}{\omega_c} +
x_0 \;\;,\nonumber  \\ y & =  & - \frac{v}{\omega_c} \cos (\omega_c t +
\alpha) + y_0  =
-\frac{v_x}{\omega_c} + y_0
\eqne  
Here  $v, \alpha, x_0$ and $y_0$ are constants of the motion the values of which are fixed by the initial conditions  $x(t_0), y(t_0), v_x(t_0), v_y(t_0)$ at some initial time $t_0$. 

The physical meaning of these constants is the following. The above solution
describes a circle with the radius $v/\omega_c$. The  position of the centre of the circle is given by the coordinates $x_0$ and $y_0$ which are therefore conventionally called the coordinates of the {\em guiding center}, cf., Fig.\,\ref{fig:smrd_orbits}  below. The value of $v$ also determines the energy
$$
m(v_x^2 + v_y^2)/2 = mv^2/2
$$ 
of the motion in the $(x,y)$ plane\footnote{The conservation of this quantity is trivially "discovered'' by multiplying the two equations (\ref{eq:eqs_in_uniform_B}) respectively by $v_x$ and $v_y$ and adding}. This energy is independent
of where on the $x,y$ plane the orbit is situated, i.e. is independent of the values of $x_0$
and $y_0\,$. 

Using the terminology of quantum mechanics we can say that the above circular motion is degenerate - all circles with the same radius $v/\omega_c$ have the same energy. This degeneracy is characterized by different values of the guiding centre coordinates  $x_0$ and $y_0$ so one can say that the classical motion is $\infty^2$ degenerate. As we will see in the next section in quantum mechanics the motion is "only" $\infty$ degenerate. It is important to observe that the expressions of the guiding centre coordinates
as given by resolving (\ref{mag_circles})
\beq \label{gc}
x_0=x -\frac{v_y}{\omega_c} \;\; , \;\; y_0 = y + \frac{v_x}{\omega_c} 
\eeq
are constants of the motion
$$
\frac{d x_0}{d t} = \frac{d y_0}{dt} = 0
$$
 This fact, which is trivial in classical mechanics will play a very important role in the quantum mechanical treatment of the problem.

\subsection{Landau levels}
The quantum mechanics of this problem was first worked out by Landau and the corresponding solution is known as Landau levels. 
\subsubsection{The eigenenergies}
The quantum Hamiltonian  of a particle without spin in this case is
\beq  \label{hb}
H_{op} = \frac{1}{2m}\left(-i\hbar \nabla  + e A(\vr)\right)^2 = \frac{m \vv_{op}^2}{2}
\eeq
 where the vector potential must be chosen such that $\vB = \nabla\times\vA$ is a constant
vector parallel to $z$. With  simple choices of $\vA$ it is possible to find explicit
solutions of the corresponding \Sch equation as we will discuss in detail    below.  At the
moment however we prefer  to proceed in a more general manner and show that many  features of
the solution can be anticipated on the basis of simple  considerations which  are useful to
follow in order to gain a better understanding of the physics of the problem.

 We start by  considering the commutators of the components
 $$
 \hat{v}_i = (-i\hbar\nabla_i + e A_i)/m
 $$
  of the velocity operators  which enter the Hamiltonian (\ref{hb}). They  are
easily calculated,
\beq 
 \left[\hat{v}_j,\hat{v}_k\right] = -\frac{ie\hbar}{m^2}\left\{\left[\nabla_j, A_k\right] + \left[A_j, \nabla_k\right] \right\}=  -\frac{ie\hbar}{m^2 }\left\{\frac{\pd A_k}{\pd x_j}-  \frac{\pd A_j}{\pd x_k}\right\}=-i\frac{e\hbar}{m^2 }\epsilon_{jkl} B_l \; .
\eeq  
These non vanishing commutators  show that for a general magnetic field $\vB(\vr)$ one
can not have definite values simultaneously for all 3 components of  the velocity. 

In our particular case of a constant $\vB$ along the $z$ axis only the commutator
$[\hat{v}_x,\hat{v}_y]$ is not zero. This  means that   $\hat{v}_z$ commutes with $H_{op}$, Eq. (\ref{hb}).
Since moreover one  can choose
$A_z = 0$ and $A_x$ and $A_y$ to be functions of only $(x,y)$  one has
$$
H_{op}= \frac{m}{2}\left(\hat{v}_x^2 + \hat{v}_y^2\right)  + \frac{\hat{p}_z^2}{2m} \; \; ,\;\; \hat{p}_z = -i\hbar \frac{\pd}{\pd z}
$$
so that  the parts of $H_{op}$ depending on $(x,y)$  and on $z$ are separable. 
The $z$--dependent  part of the wave function must be
a plane wave $\exp(i k_z z)$ with $k_z = mv_z/\hbar$ describing quantum free motion in accordance
with the classical case.

The part of $H_{op}$  describing the motion in the  $(x,y)$ plane,
\beq
h_{op}\equiv \frac{m}{2}\left(\hat{v}_x^2 + \hat{v}_y^2\right)
\eeq
 is proportional to the sum of squares of the operators $\hat{v}_x$ and $\hat{v}_y$  with a {\em constant commutator}
$$
[\hat{v}_x\,, \,\hat{v}_y] = -i(e\hbar B/m^2 )\, .
$$ 
This suggests to define rescaled variables  
$$ 
\hat{p}_{\xi} = m\hat{v}_x \;\;, \;\;  \hat{\xi} = (m/eB) \hat{v}_y
$$  
with the canonical commutator 
$$
[\hat{p}_{\xi} \, , \, \hat{\xi}] = - i \hbar
$$   in
terms  of which the operator $h_{op}$ takes the form
$$
h_{op}= \frac{\hat{p}_{\xi}^2}{2m}+ \frac{(eB)^2}{2m}\hat{\xi}^2 = \frac{\hat{p}_{\xi}^2}{2m} +\frac{m \omega_c^2}{2} \hat{\xi}^2
$$
 of the Hamiltonian of  a
one dimensional oscillator with mass $m$ and frequency $\omega_c$ in accordance with the
character of the corresponding classical motion. The spectrum of the oscillator is  well known and
adding it  to the free motion eigenvalues of the $\hat{p}_z^2/2m$  term  we obtain  the eigenvalues
of $H_{op}$  as
\beq E(n,k_z) = \hbar \omega_c \left(n+\frac{1}{2}\right) + \frac{\hbar^2 k_z^2}{2m}\;,\;\; n
= 0,1,2,.... \label{enmg}
\eeq

We have  succeeded to obtain the  eigenvalues  of the Hamiltonian  (\ref{hb}) on the basis of the
commutation relations  without solving the \Sch equation.  There  however remains a problem.
The eigenvalues   depend only on two quantum numbers
$n$ and $k_z$  whereas  dealing with  three degrees of   freedom one must   find three quantum
numbers which characterise the  eigenfunctions of $H_{op}$.  

\subsubsection{Degeneracy of the Landau levels. Quantum guiding centers} 
The independence of $E(n,k_z)$ on the third quantum number means that the energy levels of the problem 
 are degenerate and we will presently determine the reason and the nature of this degeneracy.  For
this purpose let us consider the quantum mechanical operators corresponding to the  guiding
center coordinates, Eq. (\ref{gc}),
\beq \hat{x}_0   =  x - \frac{\hat{v}_y}{\omega_c} \;\;\;\; , \;\; \;\; \hat{y}_0  =  y +
\frac{\hat{v}_x}{\omega_c}\; . \label{gcnt}
\eeq  We easily  find  that they both commute with all the components of the  velocity
operators,
\beq \label{com_of_v}
[\hat{x}_0,\hat{v}_i] = [\hat{y}_0,\hat{v}_i]  = 0
\eeq
Indeed, e.g.
$$
[\hat{x}_0,\hat{v}_x] =  [x, \hat{v}_x ] - \frac{1}{\omega_c} [\hat{v}_y, \hat{v_x}]  = 
\frac{1}{m} [x, \hat{p}_x]  -i  \frac{\hbar }{m}  =0  \;\; ;\;\;  [\hat{x}_0 , \hat{v}_y]  =  [x, \hat{v}_y ] =0 \;\; , \;\; {\rm etc }
$$
 Therefore $\hat{x}_0$ and $\hat{y}_0$ commute with $H_{op}$,  i.e. are conserved as in the classical case.   As in the
classical treatment the energy of the motion is independent of these quantities.  However we
find  that their commutator is not zero. Indeed using (\ref{com_of_v})
\beq 
[\hat{x}_0,\hat{y}_0] =  [\hat{x}_0,y] = -[\hat{v}_y,y] /\omega_c = -[\hat{p}_y, y]/m\omega_c = i(\hbar/m\omega_c)
\eeq
This relation is commonly written as 
\beq \label{eq:comm_of_guid_c}
[\hat{x}_0,\hat{y}_0] =  i\ell^2
\eeq
where the  constant
 $$
 \ell = (\hbar / eB)^{1/2}
 $$ 
 is called the magnetic length.
 
 The non vanishing  commutator between $\hat{x}_0$ and $\hat{y}_0$ means that they  can not both  have
simultaneously definite values and moreover the constant value of the commutator  shows that
like the velocity operators above, their properties are  similar to a canonical
coordinate--momentum pair.  Only one of the two   can be specified and since it is conserved
its  eigenvalues should provide the missing   quantum number which we are looking  for  in
order to  characterize the degenerate eigenfunctions belonging to the  same eigenenergy
$E(n,k_z)$. In fact the existence of the pair of {\em non commuting conserved}  operators is
the cause of the degeneracy of $E(n,k_z)$. If we choose  the states of the system  to be
eigenfunctions of, say,  $\hat{x}_0$ operator, acting on one of them  with $\hat{y}_0$ will produce a
different state with the same  energy. As we will show below there is a deep relation between
the properties of the operators $\hat{x}_0$ and $\hat{y}_0$ and the basic symmetry of the system -- the
translational invariance.

\subsubsection{The eigenfunctions}
From the commutation relations Eq.\,(\ref{eq:comm_of_guid_c}) it follows  that for the eigenstates with definite   $x_0$ the values of
$y_0$ are  completely  undetermined so that  the position of the center of the quantized
cyclotron orbit will have equal probability to be found at any point along the line with the
given $x_0$. To see this explicitly we now turn to the  solutions of the  \Sch equation
which have definite values  of $x_0$.  We  need to choose first the gauge for the vector
potential $\vA$.  The explicit forms  of $\hat{x}_0$,  
$$
\hat{x}_0\equiv x - \frac{\hat{v}_y}{\omega_c} = x-(1/eB)\hat{p}_y -A_y/B
$$
 and of 
 $$
 m\hat{v}_x = \hat{p}_x +
e A_x
$$
 suggest the following convenient choice
\beq 
A_x = 0\;,\;A_y = Bx \;,\;A_z = 0 \;\; \to \;\; \vB = (0, 0 , B)
\eeq 
for which 
$$
\hat{x}_0= i\ell^2\pd/\pd y \;\;, \;\;  m\hat{v}_x = -i\hbar \pd/\pd x
$$ 
and
 the Hamiltonian
\beq 
H_{op} = \frac{\hat{p}_x^2}{2m}  + \frac{1}{2m}\left(\hat{p}_y +  eB  x\right)^2 +
\frac{\hat{p}_z^2}{2m}  \; .
 \eeq
 The eigenfunctions of $\hat{x}_0$ and $\hat{p}_z$   have the form
\beq \label{eq:sol_for_LLs}
 \psi(\vr) = const\; \phi(x) \; e^{-i x_0 y/\ell^2}
e^{i k_z z}  \;\; ,  \;\; const = \frac{1}{\sqrt{L_y}}\;
\;\frac{1}{\sqrt{L_z}}
 \eeq  
 with yet undetermined  $\phi(x)$.  For convenience we have assumed that the motion in the
$y$ and the $z$  directions is limited by  large but finite intervals $L_y$ and $L_z$ with periodic boundary conditions.

 Inserting in the \Sch equation $H_{op}\psi = E\psi$  and separating the variables we  obtain
\beq \label{eq:wf_in_x_for_LL}
 \left[\frac{\hat{p}_x^2}{2m}+ \frac{m\omega_c^2}{2}\left(x - x_0\right)^2 \right]\phi(x) =  \varepsilon \phi(x) \; ,
\eeq
 where we denoted $\varepsilon = E - (\hbar k_z)^2/2m$.   This is the equation  of a
harmonic oscillator centered around the eigennvalue of $x_0$. As anticipated the
eigenenergies are  given by (\ref{enmg}) and are independent of $x_0$. The
eigenfunctions are
\beq \label{llwf}
\phi_{n,x_0}(x) =  \chi_{n} (x - x_0) \; , 
 \eeq
where $\chi_n(x)$ are the normalized eigenfunctions of harmonic oscillator
\beq
\chi_{n}(x)  =  \left(\frac{1}{\pi\ell^2}\right)^{\frac{1}{4}}\frac{1}{\sqrt{2^n n!}}
\exp\left[-x^2/\ell^2\;\right]\; H_n\left[x/\ell \;\right] \eeq
 with $H_n(x)=(-1)^n\; e^{x^2} (d^n/dx^n)e^{-x^2}$ -- the  Hermite polynomials.  The first
few functions  $\chi_n(x)$ are
\eqna \chi_0(x) &=&
\left(\frac{1}{\pi\ell^2}\right)^{\frac{1}{4}}\exp\left(-\frac{x^2}{2\ell^2}\right)\;\;,\;\;\chi_1(x)
=\left(\frac{1}{\pi\ell^2}\right)^{\frac{1}{4}}
\frac{\sqrt{2}x}{\ell}\exp\left(-\frac{x^2}{2\ell^2}\right)
\; ,\\ \chi_2(x) &=&\left(\frac{1}{4\pi\ell^2}\right)^{\frac{1}{4}}\left(\frac{x^2}{\ell^2}
-1\right)\exp\left(-\frac{x^2}{2\ell^2}\right)
\nonumber
\;,\;\;  {\rm etc}\; . \eqne

Imposing periodic boundary conditions in the  $y$ direction we  find that $x_0$ in Eq.(\ref{eq:sol_for_LLs})  
takes  discrete values separated by distances $\Delta x_0 = 2\pi\ell^2/L_y$. 
We thus have one state per area $L_y\Delta x_0 = 2\pi\ell^2$ in the x-y plane. 
The dependence of the wave functions  on y via the plane wave
 phase means that the probability to find a particle is independent of this coordinate. 
 It also seems to suggest that like in the $z$ - direction there is a free motion also in the y-direction. This however
 is not corrects as it is based on the experience in situations in which there was no gauge field present. 
 In this case  the wavefunction's  phase is gauge dependent so to evaluate what motions it describes one must form gauge invariant 
 observables. We will do this below by calculating the current density components with physically interesting results. 
 
In the $x$ direction the state is centered around the value
$x_0$.  Its extension can be  determined, using  e.g.,   the equipartition property of the
oscillator  meaning that the average potential energy  is  one half of the total energy,
$m\omega_c^2\langle(x - x_0)^2\rangle/2 = \hbar\omega_c (n+1/2)/2$. This gives
 $\sqrt{\langle(x - x_0)^2\rangle} =  \ell\sqrt{n + 1/2}$. Each degenerate energy level  can
thus pictorially  be viewed as a two dimensional plane filled  with overlapping (for $L_y \gg
2\pi\ell/\sqrt{n+1/2}$) "strips" occupied by individual quantum states parallel to the y axis
representing  quantized cyclotron orbits uniformly "smeared" along every strip. This picture
repeats itself for every n and $k_z$ with  the radius of the orbits, i.e. the thickness of
the strips growing as $\ell\sqrt{n+1/2}$.  The smearing of the orbits is the result of the
Heisenberg--like uncertainty relation between the guiding center coordinates $x_0$ and $y_0$.

The degenerate energy levels  which we have just described  are called Landau levels.
Choosing $y_0$ to have defined values will lead to the same picture of  Landau levels  but with the strips
parallel to the x axis. It is amusing to consider what happens if more complicated functions
of $x_0$ and $y_0$ are chosen to have defined values. Suppose we fix $x_0^2 + y_0^2$. Then the strips in the picture
above will have the shape of concentric circles around the origin. Choosing fixed $x_0^2/a^2
+ y_0^2/b^2$ with some constants a and b will lead to strips of elliptic shapes, while fixing
the function $(x_0 y_0 + y_0 x_0)/2$ (symmetrized to make the corresponding operator  hermitian)  
will result in a hyperbolic shape of the strips, etc. 

\begin{figure}[H]
\centering \includegraphics[width=0.9\textwidth]{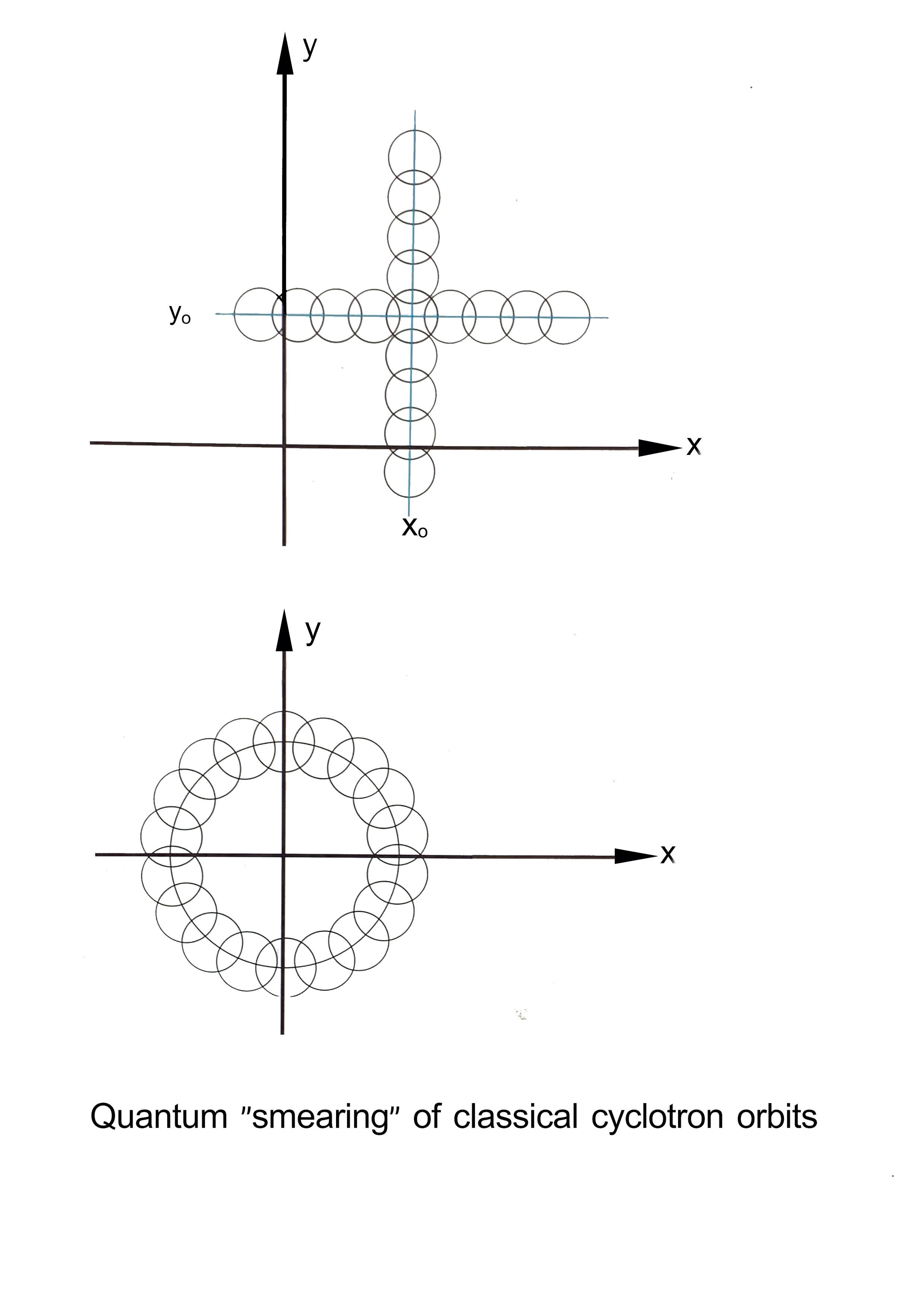}
\caption{Schematic illustration of  a single  classical cyclotron orbit and how it gets "smeared" in quantum mechanical description. The upper figure shows  the cases of $x_0$ fixed (an orbit is smeared in the y-direction)  or $y_0$ fixed (orbit is smeared in the x-direction).  In the lower figure $x_0^2 + y_0^2$ is fixed - an orbit is smeared along the corresponding circle}
 \label{fig:smrd_orbits}
\end{figure}

In Fig.\ref{fig:smrd_orbits} we illustrate some of these cases.  Of course all the choices above are equivalent as long as the degeneracy remains but some may be  singled out if a perturbation removing  this degeneracy is added to the Hamiltonian.

When choosing the eigenvalues of the operator $\hat{y}_0$ instead of $\hat{x}_0$ for the characterization of the degenerate
wave functions it should be more convenient to choose the  gauge $A_x = -By$, $A_y = A_z = 0$
in which $\hat{y}_0$ is just $-i\ell^2\pd/\pd x$. For a combination $\hat{x}_0^2 + \hat{y}_0^2$ the symmetric
choice $A_x = -By/2$, $A_y = Bx/2$, $A_z = 0$ is the most  appropriate. In the literature it
may sometimes seem that the choice of the gauge determines which combination of $\hat{x}_0$ and $\hat{y}_0$
will be diagonal. Our remark here is meant to   clarify  the correct order of choices .

The  square of the magnetic length $\ell$ appearing in the commutator of the guiding  center
coordinates plays the role of the "Planck constant" for these variables. Therefore the  analogue of
uncertainty relation $\Delta x_0 \Delta y_0 \geq \ell^2/2$ must hold. We also recall from the statistical 
 physics that in the semiclassical picture quantum states "occupy" phase space volume $\Delta p \Delta q =2\pi \hbar$.
  Here we may expect an analogous situation that
a single state of a degenerate Landau level "occupies" an area in the plane of
$(x_0,y_0)$ which is  $2\pi\ell^2$. And indeed we have seen this  in the
particular case of the solutions (\ref{eq:sol_for_LLs}).  The physical meaning  of this minimal area is
simple and profound -- the magnetic flux  through this area  is ratio of universal world constants
$$
B\cdot2\pi\ell^2 = 2 \pi B \frac{\hbar}{eB} = \frac{2\pi \hbar}{e} = \frac{h}{e}
$$
Such magnetic flux has a special notation 
\beq \label{mag_flux_quantum}
\Phi_0=  \frac{2\pi \hbar}{e} = \frac{h}{e}
\eeq
and a special name - magnetic flux quantum.  We will meet this quantity a number of  times in these notes (cf., below).  Let us 
stress that its name doesn't mean that the magnetic flux in such problems is quantized. Rather, as we see here and will be seen 
below certain physical features get repeated with $\Phi_0$ as a period. 

As can be seen from the above discussion the density of single states  in a degenerate 
Landau level is  the inverse of $2\pi\ell^2$ which is independent of n and of the way we
choose to classify the degeneracy.  The mnemonic rule of "one state per one flux quantum" is
something which is encountered in many quantum mechanical problems in the presence of
magnetic field and is therefore well worth remembering.

\subsubsection{Currents  and  edge  currents}
 Individual states in a Landau
level carry a non vanishing {\em current density}.  Apart from an obvious contribution from
the free motion in the z--direction one also finds current distribution in the x-y plane.
Qualitatively one expects that in this plane the quantum mechanically smeared cyclotron
orbits  with one fixed guiding center coordinate should combine to give opposite currents
parallel to  and concentrated on  the edges of the strip occupied by the state and have  zero
current on the midline of the strip.  

We easily find for the states  Eq.\,(\ref{eq:sol_for_LLs}) using Eq.\,(\ref{elcurr}) for the current
\beq \label{eq:current_for_LL}
j_x = 0\;,\; j_y(x) = e\omega_c (x_0 - x)\rho(x)\;,\; j_z(x) = e(\hbar k_z/m)\rho(x)\; 
\eeq 
where we denoted the particle density
$$
\rho(x) = \phi^2(x)/L_y\,L_z
$$ 
The appearance of the lengths $L_y$ and $L_z$ is related to  the (standard) normalization of the wavefunction (\ref{eq:sol_for_LLs}) to one particle. 

The wave function $\phi^2(x)$ as given by any of the solutions Eq.  (\ref{llwf})  is concentrated 
in a symmetric strip around $x_0$ which means that the current density $j_y(x)$  has
 an antisymmetric profile with respect to $x = x_0$.  Because of this antisymmetry the {\em total current} 
 \beq
 I_y = \int j_y(x) \, dx
 \eeq
 flowing  in the y-directions, i.e. along the state $\psi_{n, x_0}$ in the x-y plane is zero for these states.
 
  If one adds a  constant electric field parallel to the $x$--axis one can still find exact wave
functions (cf., homework problems or tutorial). The current density profile of these
wave functions will  change from antisymmetric to asymmetric and the total current {\em in
the $y$ direction} will not be zero. 

\begin{figure}[H]
\centering \includegraphics[width=0.4\textwidth]{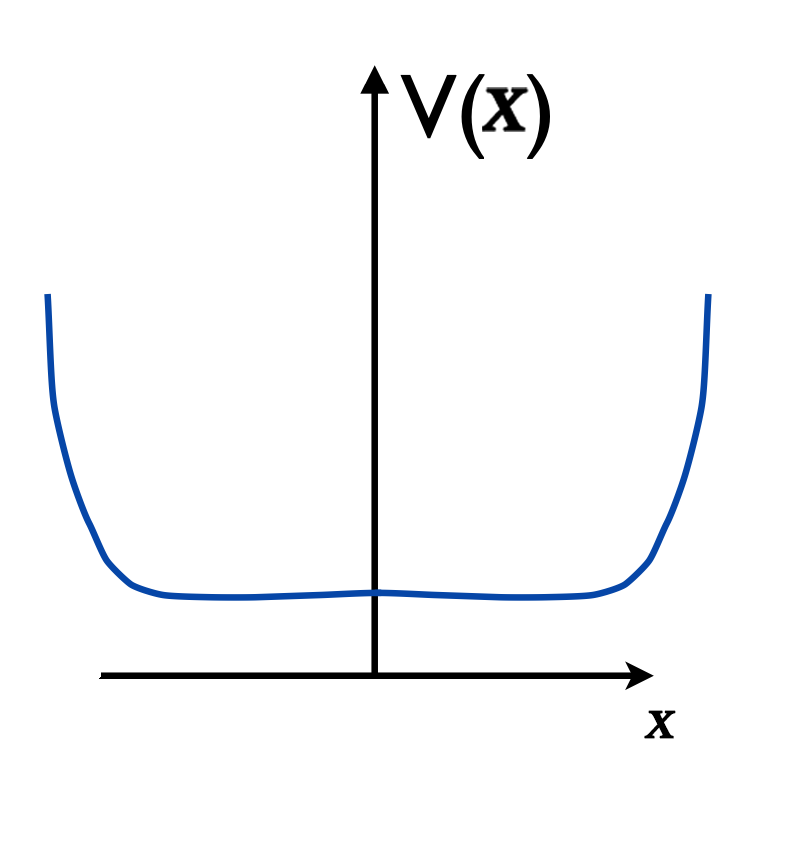}
\caption{Potential simulating edges  in the x-y plane}
 \label{fig:edge_poten}
\end{figure}

Another interesting non zero current carrying Landau states 
appear at the edges of the x-y plane.  Let us assume that an additional potential $V(x)$ with the shape
shown in Fig.\,\ref{fig:edge_poten} is added to the Eq.\,(\ref{eq:wf_in_x_for_LL})
\beq \label{eq:eq_with_U}
 \left[\frac{\hat{p}_x^2}{2m}+ \frac{m\omega_c^2}{2}\left(x - x_0\right)^2 + V(x) \right]\phi(x) =  \varepsilon \phi(x) \; ,
\eeq
This potential simulates the edges of the sample in the x direction. 

It is instructive to  examine how the combined potential 
$$
U(x) = \frac{m\omega_c^2}{2}\left(x - x_0\right)^2  + V(x)
$$
changes when plotted for different guiding center coordinate values $x_0$ relative to the positions 
of the  potential walls  representing the edges. The shape of $U(x)$ getting more narrow 
for the values of $x_0$  near and "inside" the edges 
indicates that eigenenergies $\epsilon_n$ will break the degeneracy of the Landau levels  in such a way 
that they become rising functions $\epsilon_n(x_0)$ for such values of $x_0$.  Recalling the width $\ell\sqrt{n + 1/2}$ which the unperturbed Landau levels occupy we can approximate for  low $n$ values and the potential $V(x)$ slowly varying on the magnetic length $\ell$  scale as
\beq \label{eq:en_vs_x0}
\bes
V(x) & \approx V(x_0)  \\ 
\epsilon_n(x_0) &\approx \hbar\omega_c(n+\frac{1}{2}) + V(x_0)
\end{split}
\eeq
In this approximation the modified Landau levels similar to the unperturbed ones form an "equidistant ladder" with each step having the shape of $V(x_0)$. 

The asymmetric shape of the combined $U(x)$ for $x_0$ near the edges means that the resulting eigenfunctions $\phi_{n,x_0}(x)$  
will not depend on $x_0$ via $x-x_0$ as in Eq.\,(\ref{llwf}) and will not have the harmonic oscillator symmetry around $x_0$ as in the unperturbed Landau states. This in turn means that the current density $j_y(x)$ along the edge will not have an antisymmetric profile with respect to $x = x_0$ and therefore the {\em total current}  flowing in the x-y plane for such states near the edges will not be zero.  Such currents are called "edge currents".  They correspond to the skipping  classical orbits near potential walls, move in opposite directions on the opposite edges  and  play important role in explaining the Quantum Hall Effect, cf., Ref.\,\cite{Hal}. 

Let us express the energy of a given state $\phi_{n, x_0}$ using
 $$
h_{op}(x_0)\phi_{n, x_0} = \epsilon_n(x_0) \phi_{n, x_0} \;\;\;  \to \;\; \epsilon_n(x_0) = \langle \phi_{n, x_0} | h_{op}(x_0) | \phi_{n, x_0} \rangle 
$$
 with $h_{op}(x_0)$ denoting the Hamiltonian operator in the left hand side of Eq.\,(\ref{eq:eq_with_U}). Using the Feynman-Hellman theorem\footnote{The theorem relates the derivative of the eigenenergy with respect to a parameter to the expectation value of the derivative of the Hamiltonian with respect to that parameter. The proof is straightforward
 \beq
 \bes
\frac{\partial \epsilon_n(x_0)} {\partial x_0} &= \frac{\partial } {\partial x_0}  \langle \phi_{n, x_0} | h_{op}(x_0) | \phi_{n, x_0} \rangle = \nonumber \\
 &= \langle  \frac{\partial \phi_{n, x_0}} {\partial x_0} | h_{op}(x_0) | \phi_{n, x_0} \rangle +   \langle \phi_{n, x_0} | h_{op}(x_0) | \frac{\partial \phi_{n, x_0}} {\partial x_0} \rangle + \langle \phi_{n, x_0} |  \frac{\partial h_{op}} {\partial x_0} | \phi_{n, x_0} \rangle =  \nonumber \\
  &= \epsilon_n(x_0) \left[\langle  \frac{\partial \phi_{n, x_0}} {\partial x_0} | \phi_{n, x_0} \rangle +   \langle \phi_{n, x_0} | \frac{\partial \phi_{n, x_0}} {\partial x_0} \rangle\right] + \langle \phi_{n, x_0} |  \frac{\partial h_{op}} {\partial x_0} | \phi_{n, x_0} \rangle = \langle \phi_{n, x_0} |  \frac{\partial h_{op}} {\partial x_0} | \phi_{n, x_0} \rangle
 \end{split}
  \eeq
  where it was used that 
  $$
   \frac{\partial } {\partial x_0} \langle \phi_{n, x_0}  | \phi_{n, x_0} \rangle = 0 = 
\langle  \frac{\partial \phi_{n, x_0}} {\partial x_0} | \phi_{n, x_0} \rangle + 
\langle \phi_{n, x_0} | \frac{\partial \phi_{n, x_0}} {\partial x_0} \rangle
 $$  
  }
one obtains
\beq
\bes
  \frac{\partial \epsilon_n(x_0)} {\partial x_0} &= \langle \phi_{n, x_0} |  \frac{\partial h_{op}} {\partial x_0} | \phi_{n, x_0} \rangle = \\
 &= m\omega_c^2 \langle \phi_{n, x_0}| (x-x_0) | \phi_{n, x_0} \rangle = m\omega_c^2 \int dx  (x-x_0) \, \phi_{n, x_0}^2 (x) 
 \end{split}
 \eeq
 Comparing with the expression for the current density $j_y(x)$ in Eq.\,(\ref{eq:current_for_LL})  and ignoring for convenience the $z$ direction we find the relation 
 \beq
 \frac{\partial \epsilon_n(x_0)} {\partial x_0} =  \frac{m\omega_c L_y }{e} \; I_y (n,x_0) 
 \eeq
 where $I_y(n,x_0)$ is the total current of a single particle in the $\psi_{n,x_0}$ state.  Referring to Eq.\,(\ref{eq:en_vs_x0}) with $V(x)$ as shown in  Fig.\,\ref{fig:edge_poten} one sees clearly where the edge currents are expected, their magnitude and direction.

 \subsection{Degeneracy of  Landau  levels and space symmetries \label{trinv}}
Conservation laws are always results of  symmetries and the  existence of the conserved
operators $\hat{x}_0$, $\hat{y}_0$ and $\hat{v}_z$ is not an exception. They are related to the basic symmetry of  the motion in
a uniform field --  invariance under translations. This  invariance  is however not explicit
in the Hamiltonian (\ref{hb})  which changes under the translation
$\vr$   to  $\vr + {\mbox{\bf a}}$  with an arbitrary constant vector $\va$.  We have already
encountered a similar phenomenon in the simpler case of a uniform electric field. Also here
the the  Hamiltonian (\ref{hb}) remains invariant if   simultaneously with the proper
translation one performs a suitably chosen gauge transformation. The conserved quantities
should be  the appropriate generators of these combined transformations.  

To  see this in
detail we observe that after a proper translation  the \Sch equation with the Hamiltonian
(\ref{hb})   has the same form but with  the different vector potential $\vA(\vr + {\mbox{\bf a}})$. {\em
For a constant} $\vB$ however the difference   $\vA(\vr + {\mbox{\bf a}}) - \vA(\vr)$ is a
gauge transformation, i.e. it is a gradient of a scalar function. It will be  sufficient  to
show this for an  infinitesimal ${\mbox{\bf a}}$ for which we have
$\vA(\vr + {\mbox{\bf a}}) \approx \vA(\vr) + ({\mbox{\bf a}}\cdot\nabla)\vA(\vr)$. The
 last term is
 $$
 a_i\pd_i A_j =  a_i (\pd_i A_j - \pd_j A_i ) + a_i\pd_j A_i = a_i(\epsilon_{ijk}B_k + \pd_jA_i)
 $$ 
 and for  a constant  $\vB$ it is indeed a gradient
\beq
\pd_j\alpha \;\;\; {\rm with} \; \;\;  \alpha= a_i(\epsilon_{ijk}x_j B_k + A_i)= {\mbox{\bf a}}\cdot[\vr\times\vB + \vA(\vr)] 
\eeq
 It can be removed
from $H_{op}$  by a gauge transformation of the wave function in addition to the proper translation.
The symmetry transformation is therefore
\beq
\psi(\vr) \rightarrow \left\{1 +\frac{ie}{\hbar }{\mbox{\bf a}}\cdot[\vr\times\vB +
\vA(\vr)] \right\} (1 + i{\mbox{\bf a}}\cdot\vp_{op}/\hbar)\psi(\vr)\;\;\;
 (\rm infinitesimal \;\;\;{\mbox{\bf a}})\;, \label{trin}
\eeq where we have used the proper translation operator $\exp(i{\mbox{\bf a}}\cdot\vp_{op}/\hbar)$
for infinitesimal $\mbox{\bf a}$ to write $\psi(\vr + {\mbox{\bf a}})$ in terms of
$\psi(\vr)$.

The combined transformation (\ref{trin}) is what should be called translation in the presence
of  a magnetic field (the term  "magnetic translation" is sometimes used). The generators of
this transformation are read off the  linear term in
${\bf a}$ found after multiplying the brackets in Eq.(\ref{trin}). They are
\beq
\vp_{op} +e[\vA(\vr) +\vr \times \vB] = m\vv_{op} + eB (\vr\times {\mbox{\bf e}}_z) \label{trop}\;.
\eeq For translations along the z axis this is just $mv_z$ whereas for the translations along
the x and y axes we obtain respectively $ eB\, \hat{y}_0$ and $-eB \, \hat{x}_0$ in terms of the
operators of the guiding center coordinates.

 It should now become intuitively clear why these
operators do not commute. We expect that the result of translating  the wave function
parallel to x and then parallel to y should not be the same as  translating it in the
opposite order. The difference should be related to the  Aharonov-Bohm phase 
(see Section \ref{AB} below  for its definition) induced by the
flux of the magnetic field through the rectangle  obtained in the course of these  reversed
order translations. Let us see how it happens. Transporting a wave function by an
infinitesimal $\Delta x$ followed by $\Delta y$ and then by $-\Delta x$ and $-\Delta y$
respectively one indeed obtains keeping the terms up to a 2nd order in the translations $\Delta x$ and $\Delta y$
\eqna 
&& (1 - i\Delta y K_y - \frac{1}{2}(\Delta y)^2 K_y^2)(1 - i\Delta x K_x  -
\frac{1}{2}(\Delta x)^2 K_x^2 ) \times \nonumber \\ 
&& \times (1 + i\Delta y K_y -
\frac{1}{2}(\Delta y)^2 K_y^2)
 (1 + i\Delta x K_x - \frac{1}{2}(\Delta x)^2 K_x^2)\psi(\vr) = \nonumber
\\
 && = (1 + \Delta x\Delta y[K_x\, , \, K_y])\psi(\vr) = \left[1 + 2\pi i(\Delta
\Phi/\Phi_0)\right]\psi(\vr) 
\eqne 
where we denoted by $\hbar K_x$ and $\hbar K_y$ the
corresponding vector components of  the operator of translations (\ref{trop})  and
$\Delta\Phi = B\Delta x\Delta y$ -- is the flux through the rectangle. Since the first non
vanishing term in the expression above apart of unity was  quadratic and  proportional to
$\Delta x\Delta y$  it was necessary
to keep the quadratic terms in the operators of each translation.

A similar discussion concerning the generalization of  transformations and their generators
can be worked out  for another  symmetry  of the problems  -- the rotational symmetry around
the direction  of  the magnetic field $\vB$.  We will leave this for homework  or tutorials.

\section{The Aharonov - Bohm Effect  \label{AB}}

\subsection{Local and non local gauge invariant quantities} 

 We have emphasized in Section \ref{ginv} that all observable quantities in a theory with a
gauge field are  gauge invariant. Perhaps the simplest such quantities are the electric and
magnetic fields and the particle density
$\rho(\arg) = |\psi(\vr,t)|^2$. In the expression for the electric current  density considered in the previous
section we encountered another set involving the derivatives of $\psi$ -- the  combinations
$\psi^*(\arg)\vD\psi(\arg)$ to which we can  also add their  time dependent partner
$\psi^*(\arg)D_0\psi(\arg)$. These  combinations are  gauge invariant due to the  simple
transformation properties of the  gauge covariant  derivatives (\ref{cvder}).  

A distinct feature of all these invariants   is that  they depend on
$\psi, \vA$ and $A_0$ and their first derivatives at the {\em same  space--time point}, i.e.
they are {\em local}. Consider however a circulation integral $\oint
\vA\cdot d\vr$ taken around some closed contour drawn in space. By the
Stockes theorem this integral is equal to the flux
of $\vB$ through the contour and is therefore  gauge invariant.
This is an example of a {\em non local} gauge invariant quantity. In the following sections we will
discuss  situations  in which the non trivial dependence  on $\oint
\vA\cdot d\vr$  leads to
unexpected  quantum mechanical effects  which are collectively known as the
Aharonov--Bohm effect,  Ref.\,\cite{Aha}.  The  sensitivity of the quantum theory to non local gauge
invariants can be traced to essential non locality  of the quantum mechanical description
---  eigenvalues  and   expectation values of various physical quantities such as energy,
angular momentum, etc., depend on  what happens with the wave function in the entire
configuration space of the system.

Concluding this section we  mention that in addition to the circulation of the vector
potential another type of non local gauge invariants appears in certain physical
applications. These are   bi-local quantities of the type
$$
\psi^*(\vr')\exp[i(e/\hbar )\int_{\vr}^{\vr'}\vA(\vr'')\cdot d\vr'']\psi(\vr,t)\; .
$$
 Under gauge transformations the
exponential and the wave functions produce phase factors which cancel each other.  Quantities
like this are often met in the field theoretical context  and recently in certain many body
problems.

\subsection{Quantum mechanics "feels" non zero $\oint_C \vA\cdot d\vr$ even if $\vE=\vB=0$ on and near the contour C} 

\begin{figure}[H]
\centering \includegraphics[width=0.8\textwidth]{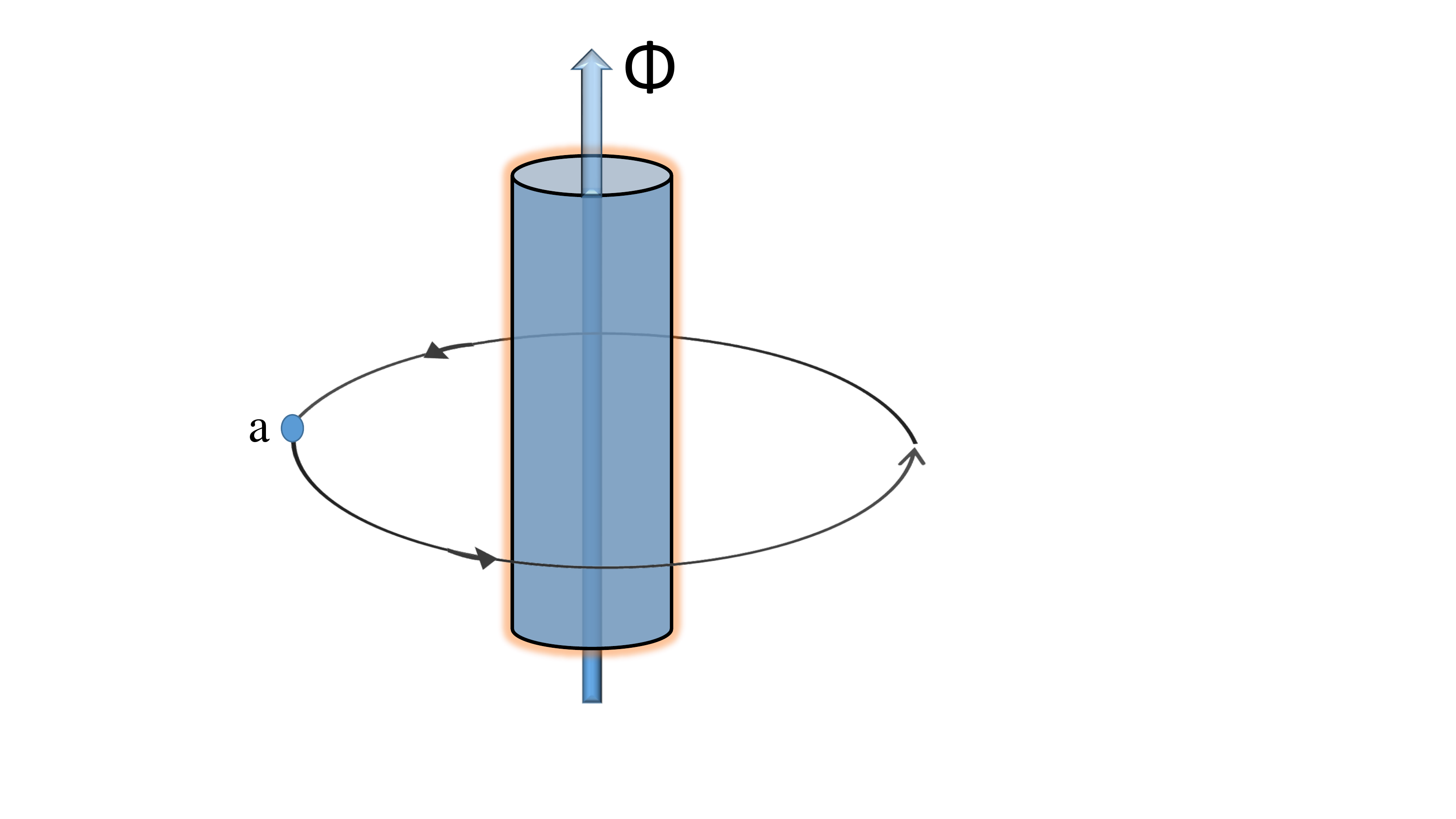}
\caption{Example of the Aharonov-Bohm flux $\Phi$ in an impenetrable tube and a closed contour encircling it in the region with zero $\vE$ and $\vB$. The non zero circulation  $\oint_C \vA\cdot d\vr = \Phi$ around such contours has no effect in classical description of charged particle motion (trajectories) in this region but produces observable effects in its quantum mechanics (wave functions) }
 \label{fig:ABflux}
\end{figure}

Let us consider  a region of space in which  local invariant quantities
$\vE$ and $\vB$ are zero  but in which contours  can be found for which  $\oint
\vA\cdot d\vr$ does not vanish.  A simple example   is a region {\em outside} of  a long thin
tube with impenetrable walls and  a non zero magnetic field concentrated inside and running
parallel  to the tube, Fig.\ref{fig:ABflux} 

Non zero circulation integrals
$\oint\vA\cdot d\vr$   are obtained for the integration  contours which  wind around the
 tube.  Since by assumption $\vB=\nabla\times\vA=0$ outside the tube the details of a particular  contour 
 are of no importance except for  the number of
times it winds around the tube and the direction of this winding. Denoting this number by $n$ one can write 
\beq
\oint_{C}\vA\cdot d\vr = n \Phi \;\;, \;\; n=0, \pm1, ...
\eeq
 Here $\Phi$ denotes the  magnitude of the total  flux of the magnetic field in the tube. The circulation integrals outside
  the tube  depend only on $\Phi$ and not  to the details of the magnetic field distribution.
 One conventionally refers to such a tube as  a solenoid  and to such an isolated magnetic
flux $\Phi$ as the Aharonov--Bohm flux (AB flux for brevity).

\subsection{"Gauging out" the AB flux. Periodic dependence on its value}

Classically the {\em free motion} of a particle in the outside region is not
influenced by the presence of the field inside the tube.  At first sight one may reach a similar conclusion
in  the quantum mechanical description.  Indeed to write the \Sch equation one needs to determine
 first the electromagnetic potentials.  Since $\vB=\nabla\times\vA = 0$ in the outside region one must have
 that $\vA$ must be a gradient of some scalar, 
 \beq \label{eq:B_in_AB}
 \vA = \nabla\xi(\vr)
 \eeq
 With such $\vA$ (and $\vE=0$ ) it may appear 
 that  in the corresponding \Sch equation
 $$
i\hbar \frac{\pd \psi}{\pd t} = \frac{1}{2m}\left(-i\hbar\nabla +  e \vA(\vr) \right)^2  \psi
 $$
 one could remove the $e\vA$ term by a gauge transformation 
 $$
 \psi(\vr,t) = \psi'(\vr,t) \exp[- i\frac{e}{\hbar} \xi(\vr)]
 $$
 with
 $$
 \xi(\vr) = \xi_0 + \int_{\vr_0}^{\vr} \vA(\vr') \cdot d\vr' =  \xi_0+ \int_{\vr_0}^{\vr}\nabla\xi(\vr') \cdot d\vr'
 $$
 and $\xi_0$ some constant.   
 
 The problem however with this elimination of $\vA$  from the \Sch equation is that in the presence of the AB flux  $\Phi$ the scalar function $\xi(\vr)$ in Eq.\,(\ref{eq:B_in_AB})  is not single valued. It is a  {\em multivalued} function as can be seen  in the following way.  To have the required  value  of the AB flux  the function $\xi(\vr)$ must change by  $\Phi$ when  "taken (followed) continuously" along a contour $C$ around the solenoid in the positive direction
 \beq \label{eq:circ_int_of_A}
\Phi= \oint_C\vA\cdot d\vr =  \oint_C \nabla \xi (\vr)\cdot d\vr = \int_{\vr_i}^{\vr_f}\nabla\xi\cdot d\vr = \xi(\vr_f) - \xi(\vr_i) \;\;\; {\rm with}  \;\;\vr_f=\vr_i \;.
 \eeq
 Thus at every $\vr$ in the region outside the solenoid the function $\xi(\vr)$ has many (infinity) of values differing by $n\Phi$ with  (positive or negative)  integer $n$.
 
Given this the transformed $\psi'(\vr,t)$,
 $$
  \psi'(\vr,t) = \psi(\vr,t) \exp[ i\frac{e}{\hbar} \xi(\vr)]
  $$
  will also be multivalued - its phase will change  by 
  \beq \label{eq:mult_val}
  \Delta \xi = \frac{e}{\hbar}\Phi = 2\pi \frac{\Phi}{\Phi_0}
  \eeq
  when  "taken continuously" around the solenoid.  
  
  To understand what the demand of such a particular non single valuedness of the wave function produces  let us consider a specific example of the angular momentum.  Assuming the $z$-axis along the solenoid and the $z$ 
  component  $\hat{L}_z$ we have for its eigenfunctions
  $$
  \hat{L}_z \psi(\phi) =  \hbar\nu \psi(\phi) \;\; \to \;\; \psi(\phi) = const \, e^{i\nu\phi}
  $$
  In the usual case, i.e. in the absence of the AB flux one applies the condition $\psi(\phi+2n\pi) =\psi(\phi)$, i.e. the condition of single
   valuedness of $\psi(\phi)$ which leads to the usual integer quantization 
   $$
   \nu = M \;\;, \;\; M=0, \pm 1. \pm 2, ...
   $$ 
   For the multivalued function condition Eq.\,(\ref{eq:mult_val}) we have 
  \beq \label{eq:L_via_mult_val}
  \begin{split}
  e^{i\nu (\phi + 2n\pi)} =& e^{i\nu \phi} e^{i 2\pi n\Phi/\Phi_0}  \to 2n\pi(\nu - \Phi/\Phi_0)  = 2n\pi M \to \\
   &\to \; \nu = M+\Phi/\Phi_0
  \end{split}
  \eeq
  This shows  that despite our "gauging out" of the vector potential $\vA = \nabla \xi$  its gauge invariant content, i.e. the AB flux $\Phi$ in Eq.\,(\ref{eq:circ_int_of_A}), if not zero modifies the physics via the resulting multivalued wave function condition Eq.\,(\ref{eq:mult_val}).   We will see below that this modifications is (not surprisingly) identical to the straightforward solution with such a vector potential. 
  
 As an important additional observation we note that when $\Phi = n \Phi_0$ there is no effect!  The transformed $\psi'$ remains single valued and such AB flux is non observable "from outside". This observation is probably one of the advantages of the multivalued wave function formulation. It also means that the Aharonov-Bohm effects have periodic dependence on the magnitude $\Phi$ of the AB flux with the period of the flux quantum $\Phi_0$.  One can see this in the dependence of the eigenvalues $\nu$ on $\Phi$, Eq.\,(\ref{eq:L_via_mult_val}). They change from integer to non integer with the period  $\Phi_0$.  We will also see this periodicity in the examples considered in the next section and provide a more general point of view  in Section \ref{sec:Hom_period}.   
 
 Another important observation is the following. The view of the Aharonov--Bohm effect   as a modification of the condition that the wave
function repeats itself as it is taken around a solenoid  stresses that
in order to "feel" this modification  the wave function must extend all around the solenoid.
Otherwise there will be no observable consequences of the Aharonov--Bohm flux.  Below  we
will consider  an example of a ring pierced by the Aharonov--Bohm flux with  a particle
localized on a finite sector of the ring.  There is no Aharonov--Bohm effect in this
case.

 \subsection{Example of the AB flux}
 
Assume that the solenoid with the AB flux $\Phi$ is placed along the $z$-axis.  
 A possible simple choice for the vector potential  outside such a solenoid is
\beq
\vA = \nabla\xi(\vr)\;,\;\; {\rm with}  \;\; \xi(\vr) = \frac{\Phi}{2\pi}\arctan(y/x) \equiv  \frac{\Phi}{2\pi} \phi \; 
\label{abfl} 
\eeq 
where $\phi$ is the azimuthal
angle.  Recalling the expression of the gradient in cylindrical coordinates $r,\phi,z$ 
$$
\nabla =   {\bf e}_r \frac{\partial}{\partial r}  +{\bf e}_{\phi} \frac{1}{r}\frac{\partial}{\partial \phi} + {\bf e}_z  \frac{\partial}{\partial z}  
$$
one finds 
\beq \label{vec_pot_forAB}
 A_\phi = \frac{\Phi}{2\pi r} \;\; ; \;\;  A_r=A_z=0
 \eeq
 and therefore  the circulation integral outside the solenoid along a circular contour in a plane perpendicular to the solenoid
 $$
 \oint\vA \cdot d\vr = \int_0^{2\pi} A_\phi \; rd\phi = \Phi
 $$ 
 Since $\nabla\times \vA=\vB=0$ outside the solenoid one can deform the above circular contour without changing the integral as long  as the new contour  has "the same topology" - i.e. encircles the flux once in the same direction.  One can also change the particular
 $\vA$ in (\ref{abfl})  by
adding a {\em single valued function} to $\xi$   without influencing  $\vB = 0$ or
circulation integrals $\oint\vA \cdot d\vr$ outside the solenoid.  We observe that the dependence of the outside  vector  potential
 on the magnetic field is via the flux $\Phi$ irrespective of a particular radial dependence of  $\vB$ inside the solenoid. 

To have a convenient example of the AB flux one can think of $\vB = B(r){\bf e}_z$  with a  constant B inside and
zero outside.   With this magnetic field one can write  for all $r$'s,
\eqna
\vA  &=& (Br/2)\;{\bf e}_{\phi}\; \; \; \;\;\;\;{\rm inside\;\;\; the\;\;\; solenoid}
\;,\nonumber \\
\vA &=& (B r_0^2)/2r\; {\bf e}_{\phi} = \frac{\Phi}{2\pi r}  {\bf e}_{\phi} \; \; \; {\rm outside
\;\;\;the\;\;\; solenoid} \; , \eqne
 where $r_0$ is the radius of the solenoid. Since $\Phi = B(\pi r_0^2)$,  this expression for
the outside region  is the same as  (\ref{abfl}).

\subsubsection{The Hamiltonian and the spectrum}

The Hamiltonian with the vector potential (\ref{vec_pot_forAB}) outside the AB flux 
has a simple form in cylindrical
coordinates. Using $\vp = p_r{\bf e}_r + p_{\phi}{\bf e}_{\phi} + p_z {\bf e}_z$ and
$\vA = A_{\phi}{\bf e}_{\phi}$ we have
\beq 
H = \frac{p_r^2}{2m} +\frac{1}{2m}\left(p_{\phi} +  e A_{\phi}\right)^2  +
\frac{p_z^2}{2m} + U(r) \;, \label{hmba}
\eeq   
where we disregarded the spin degrees of freedom and added U(r) - the  potential
which should account for the impenetrable walls of the solenoid. 
Note that in our notation here $p_{\phi}$ is a projection of
$\vp$ on ${\bf e}_{\phi}$ and is related to  the z-projection of the angular momentum
as 
$$
(\vr \times \vp)_z = r p_{\phi} = L_z
$$
Compared to the situation without the magnetic flux the Hamiltonian (\ref{hmba}) is
modified by the presence of the potential
$A_{\phi}$ in the centrifugal term  which depends on the combination
 $$
 p_{\phi} +  e A_{\phi} = \frac{1}{r} (L_z + e B r_0^2 /2 ) = \frac{1}{r} (L_z + e\Phi/2\pi )
$$  
  In classical mechanics one could  absorb the
constant $e\Phi/2\pi $ into $L_z$ and
 completely eliminate $A_{\phi}$ from the equations of motion. However in quantum mechanics
this freedom does nor exist since $L_z $ becomes an operator $L_z = -i\hbar\:\pd/\pd\phi$  which has discrete
eigenvalues $\hbar M$ (M -- integer).  The eigenvalues'  selection follows from the
requirement that the wave function is single valued which  imposes the periodic boundary
conditions
 $$
 \psi(r,\phi,z) = \psi(r,\phi + 2\pi,z) .
 $$
 The eigenvalues of the angular part $L_z + e\Phi/2\pi$ in the expression for $ p_{\phi} +  e A_{\phi}$  are 
 therefore  
$$
\hbar( M +e\Phi/2\pi \hbar) \equiv \hbar (M + \Phi/\Phi_0) \;\;, \;\; \Phi_0 = h/e
$$  
which is identical with what was obtained in Eq.\,(\ref{eq:L_via_mult_val})  of our discussion of the effect
  of the multivalued wave function condition obtained after "gauging out" the vector potential. 

In the \Sch equation $H_{op} \psi =E\psi$ one can separate the z-part and use 
$$
\psi(r,\phi) = R(r) \frac{1}{\sqrt{2\pi}}e^{iM\phi} 
$$
to write the radial part of the equation as
$$
\left[ \frac{\hat{p}_r^2}{2m} + \frac{\hbar^2(M + \Phi/\Phi_0)^2 }{2mr^2} + U(r) \right] R(r) =\varepsilon  \,R(r)
$$
with $\varepsilon $ the corresponding part of the total energy  $E$.

The above change of the  spectrum   of the centrifugal part of the potential is  a  manifestation of the Aharonov-Bohm effect in this example. A
classically unobservable  magnetic flux  inside  an impenetrable solenoid
causes an observable effect in the outside region when the problem is treated quantum mechanically. The dependence on the magnitude $\Phi$ of the flux exhibits periodicity with magnetic flux quantum $\Phi_0$ as a period. 

\subsubsection{Thin ring solution}

 Let us see how this happens in a  simple model of a thin ring.  To construct this model we
add  to the Hamiltonian (\ref{hmba}) a potential $V(r,z)$  constraining the motion in the $r$
and $z$ directions to a  very narrow ring region. It is the simplest  to choose
$V(r,z)$ as  zero for $|r - a|\le b$, $|z|\le b$ and infinite otherwise. This gives   a ring
of thickness $b$ with radius $a$ lying in the $z=0$ plane.  For a very small b the radial
coordinate in the second term in (\ref{hmba}) can be set to the fixed radius $a$ and the
motion in the azimuthal direction becomes decoupled from $r$. The Hamiltonian of this motion
is just
\beq
    H_{\phi} =  \frac{1}{2ma^2}\left(-i\hbar\frac{\pd}{\pd\phi} + \frac{e\Phi}{2\pi
}\right)^2 \label{hrng}
\eeq 
with eigenfunctions
$$
\psi_M(\phi) = \frac{1}{\sqrt{2\pi}}\; \exp(iM\phi) \;\;, \;\; M = 0, \pm 1 , \pm 2 , ... 
$$ 
and the corresponding eigenvalues
\beq
 E_M = \frac{\hbar^2}{2ma^2}\left(M + \frac{\Phi}{\Phi_0}\right)^2\;. \label{enrn}
\eeq  
The energies of the motion in the $r$ and the $z$ directions in this approximation are
independent of $\Phi$ and we will not be concerned with them.

\begin{figure}[H]
\centering \includegraphics[width=0.8\textwidth]{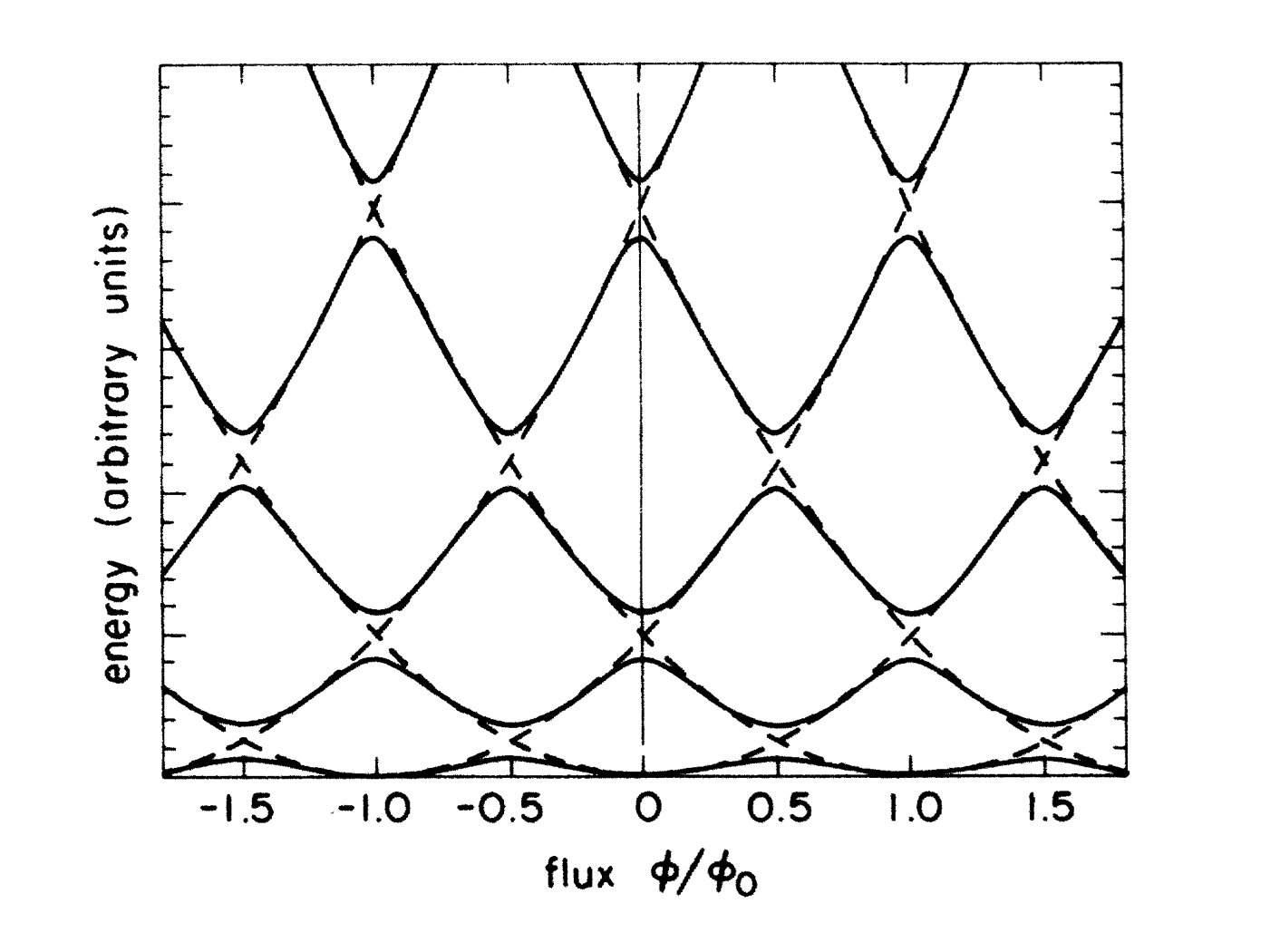}
\caption{Schematic diagram of the electron energy levels as a function of the flux $\Phi/\Phi_0$ 
 in a one-dimensional ring encircling the flux, Ref.\cite{Ho}. Solid and dashed curves, respectively, are 
 for the ring with and without weak  additional   potential
$V(\phi)$ acting on a particle on the ring, Ref.\cite{Ho}. }
 \label{fig:AB_ring}
\end{figure}

In Fig.\,\ref{fig:AB_ring}
 we plot the dependence of the energy levels on the magnetic flux which shows the
$\Phi_0$ periodicity of the Aharonov--Bohm effect.  An analysis  which we do not reproduce
here shows that if there is a weak  additional   potential
$V(\phi)$ acting on a particle on the ring the behavior of the levels will   follow the
pattern of the solid lines in Fig.\,\ref{fig:AB_ring}  which retain the same 
periodicity, Ref.\cite{Ho}.

Consider now a  case of a strong potential  $V(\phi)$,  so strong that  the particle is
localized in a finite sector of the ring  as opposed to the free motion around the
entire circumference of the ring  as in (\ref{hrng}).  A simple such
$V(\phi)$ is a potential  "well" $V(\phi) = 0$ for $0 < \phi < \phi_0 <2\pi$  and infinite
outside this interval. The  eigenfunctions in this case are zero  except in the
interval with zero potential where they are easily found to be
\beq
\psi_n (\phi) =\sqrt{
\frac{2}{\phi_0}}\exp\left[i\frac{\Phi}{\Phi_0}\;\phi\right] \sin\left(\frac{\pi
n\phi}{\phi_0}\right)\;\;\; , \;\;\; n=1,2,3,...
\eeq
 The dependence on the flux enters in the phase of these functions but the corresponding
eigenenergies do not depend on it at all,
\beq 
 E_n = \frac{\hbar^2\pi^2 n^2}{2ma^2\phi_0^2}\;. 
 \eeq 
  In Fig.\,\ref{fig:AB_ring} they would be
represented by horizontal straight lines  giving   a trivial limiting case of the general
periodic dependence on $\Phi$ referred to above. Here we have an example in which the
localization of the eigenfunctions  on a part of  the ring leads to the disappearance  of
the the Aharonov--Bohm effect (the
$\Phi$ dependent phase is the same for all solutions and is therefore not observable in this
case). As we have already stressed, in order to have   a sensitivity to the Aharonov -- Bohm
flux  the wave function must have   a "tail"  extending   all around the flux.

\subsubsection{AB effect in quantum interference and scattering off the AB flux}

Here we will briefly consider two additional manifestations of the AB effect.

{\bf 2-slit with AB flux}

\begin{figure}[H]
\centering \includegraphics[width=0.5\textwidth]{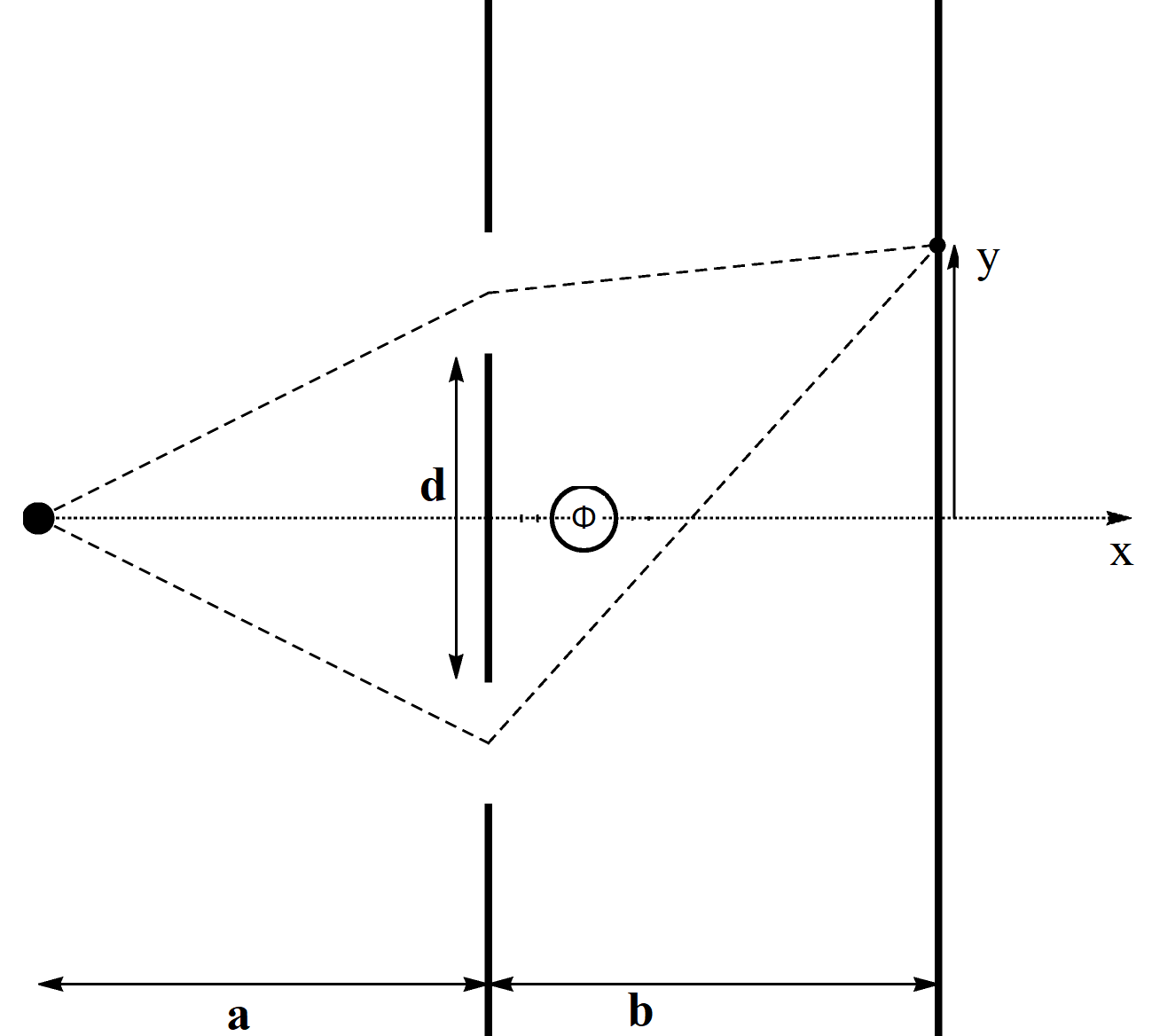}
\caption{Double slit interference in the presence of the Aharonov-Bohm flux $\Phi$}
 \label{fig:AB_inter}
\end{figure}

The understanding  that Aharonov--Bohm flux modifies  the phase of the wave function leads to
an intuitive way of describing  the Aharonov -- Bohm effect as the change of the interference
of the quantum mechanical waves as they propagate on each side of the solenoid. Let us
consider the classic 2-slit  experiment  as depicted in Fig.\ref{fig:AB_inter}. 
 Electrons pass
from a point source through a wall with 2 narrow slits and fall on a screen behind it, cf.,
Ref. \cite{Fey}.
As long as it is  not detected which slit the electrons pass through,  they
produce an interference pattern according to the phase difference for paths going via each
of the slits. 

If the Aharonov--Bohm solenoid is placed behind the wall between the slits
this phase difference will change by the amount
\beq
 \Delta \beta  = \frac{e}{\hbar }\left(\int_1\vA\cdot d\vr  -
\int_2\vA\cdot d\vr\right) = \frac{e}{\hbar }\oint \vA\cdot d\vr =
\frac{2\pi \Phi}{\Phi_0}
\eeq 
where as before $\Phi$ is the flux through the solenoid and the subscripts 1 and 2
denote integrals along the two trajectories in Fig.\,\ref{fig:AB_inter}. For a position $y$ on the screen
(measured from its centre)  the phase difference between waves from the two slits in the absence 
of the solenoid is $\beta = k\Delta L$ where $k = \sqrt{2mE}/\hbar$  is the wave number  and $\Delta L$ --  the difference
in the paths the waves travel from the slits to the screen. 

For a distance $b$ from the slits to the screen and for $y, d << b$ one can approximate
$\Delta L = (y/b)d$ where d is the distance between the slits. Therefore a given phase difference $\beta$ 
will be found at $y = (\beta/kd)b$. The
additional phase difference  $\Delta \beta$
 due to the Aharonov -- Bohm flux will result in a shift in the interference pattern by the amount
\beq
\Delta y = \frac{\Delta \beta}{kd}b = \frac{2\pi b}{kd}\frac{\Phi}{\Phi_0}
\eeq
\newpage
{\bf Scattering off the AB flux}

Yet another way to see the phase difference between the waves which pass on different sides
of the solenoid is to consider a scattering of a plane wave from it. This was discussed in
the original paper by Aharonov and Bohm,  Ref.\,\cite{Aha}. For the vanishing magnetic flux
one finds a standard picture of a cylindrical  wave scattered from the solenoid with the
amplitude which falls like $r^{-1}$ superimposed on the initial  plane  wave. For non zero
$\Phi$  the  wavefronts  in the region "down stream" and far away from  the solenoid form a
pattern of two flat fronts shifted with respect to each other in an   abrupt, almost
discontinuous fashion along   the line stretching from the solenoid in the direction of the
propagation of the original plane wave. The magnitude of the shift is given by the phase
difference $ 2\pi\Phi/\Phi_0$ divided by the wave number $k$. We refer reader to Ref.\,\cite{Aha} for the details of this discussion.

\subsection{Multiply  connected  regions.  Homotopy \label{sec:Hom_period}} 

Certain features of the results obtained
in  examples above are quite  general in nature. In any region with zero $\vE$ and $\vB$
Eqs.\,(\ref{empot}) imply that the vector  potential must be the gradient of  a
(time-independent) function, $\vA(\vr) = \nabla\xi(\vr)$   and that $A_0 = const$. The
integral
$\int\vA \cdot d\vr = \int\nabla\xi\cdot d\vr$  is equal to the difference between the values of
$\xi$ at the initial and the final points of the contour of the integration so that it must
be zero for a closed contour unless (a) the function $\xi$  is {\em not single valued} and
(b) the contour  takes $\xi$ from one of its branches to another. This can not happen in
simply connected regions, i.e. such in which  {\em all closed
 contours   are contractable to a point}.  Since  continuous deformations of the   contour in
the integral $\oint\nabla\xi\cdot d\vr$ can not change its value  all such integrals  will be
zero for contractable  contours. Equivalently stated, a regular  function like $\xi(\vr)$
must be single valued in a simply connected region.

Consider however  multiply   connected regions. These are regions where one can find closed
contours which can not be contracted to a point without crossing the boundaries.  The
impenetrable solenoid and the ring  discussed above are examples of such regions.  The
contours around the "tube" of the solenoid or   around the "hole" of the ring  can not be
contracted to zero.  Non zero values for the circulation  integral $\int\vA \cdot d\vr$  are
to be expected for such contours   and  actually occur when  there is a magnetic flux
through the excluded regions. We note in passing that not every shape of excluded region
 will lead to the existence of non contractable contours.  Excluded cavity of a spherical
shape for instance will not. Its existence  will create uncontractable closed {\em surfaces
but not curves} and will be relevant for considerations of e.g. non vanishing surface
integrals of a vector field with zero divergence.

 All possible closed curves in a multiply connected region can be divided into classes of
curves which can be contracted into each other. Such classes are called homotopy classes of
curves, cf., \cite{hom}. Among all homotopy classes one can define a complete set of
elementary classes of curves
$C_k$ out of which every other non elementary class can be obtained by multiple traverses of
curves belonging to the elementary classes. For a solenoid there is one elementary class
of    curves encircling the solenoid {\rm once} in, say, a clockwise  direction and
one in the counter clockwise  direction.   Clearly the changes of
$\xi$ on  closed curves within each  elementary  class must be the same since the curves can
be continuously deformed into each other. For different elementary classes however they in
general will be different reflecting possible different values of the Aharonov--Bohm fluxes
through different excluded regions or their different signs.

Let us apply these considerations to a general system of charged particles in a multiply
connected field free region, cf. Ref.\,\cite{Blo}. Their \Sch equation  is
\beq
 H[\vp_{a} - q_a\vA(\vr_a),\;\vr_a]\psi(\{\vr_a\}) = E \psi(\{\vr_a\})
\eeq 
where we assumed a  general Hamiltonian depending on the momenta $\vp_{a} =
-i\hbar\nabla_a$  and coordinates $\vr_a$ of the particles with charges $q_a, \; a =
1,2,...,N$. Since in the field free  region the vector potential $\vA$ is "pure gauge",  $\vA
= \nabla \xi$,   we can apply a gauge transformation
\beq  \label{shbl}
\psi(\{\vr_a\}) =  \exp[ i \sum_{a=1}^N q_a \xi(\vr_a)/\hbar ] \psi'(\{\vr_a\})
\eeq and find that $\psi'$ satisfies
\beq H[\vp_{a},\vr_a]\psi'(\{\vr_a\}) = E \psi'(\{\vr_a\})
\eeq with the Hamiltonian in which the potential $\vA$ was "gauged out".  Since $\psi$ is
single valued  and since going around any elementary non contractable contour $C_k$
increases  $\xi$ by the corresponding Aharonov--Bohm flux $ \Phi_k = \oint_{C_k}
\vA \cdot d\vr $ ,  we must demand that $\psi'(\{\vr_a\})$ is multiplied by the factor
$\exp(-iq_a \Phi_k/\hbar )$ when the particle $a$ is brought around $C_k$.  Thus the
boundary conditions are different from the case of zero fluxes and one should expect that the
energy levels will depend on the values of  $\Phi_k$.  Since charges of all particles are
multiples of the elementary electronic charge
$e$ the  change in the boundary conditions is  the same for  the fluxes $\Phi_k$ which differ
by multiples of the  flux quantum $\Phi_0$.  This periodicity should occur in the solutions
$\psi'$ and therefore in the set of energy levels obtained from Eq.(\ref{shbl}). Physical
quantities which are determined by this set must therefore exhibit this periodicity. This
conclusion as well as the entire set of the  preceding arguments are very general and based
solely on the fundamental principles of gauge invariance, requirement of single valued wave
functions and the elementary nature of the electric charge $e$.

\section{Magnetic Moments \label{gfac}}
\subsection{The  g--factors} 
We now return to the relation (\ref{mus})
between the magnetic moment and the spin operators.  It is customary to
quote the numerical value of  the magnetic moment of a particle as equal to the maximum
value of its projection, i.e. the value
$\mu_z =  g(q/2mc) s_z$   for $s_z = s$.  In this Section we will discuss the dimensionless gyromagnetic ratio $g$ in this relation 
called the g-factor.

For {\em elementary particles} g   is
determined by relativistic quantum mechanical wave equations. E.g. for the electron the
Dirac equation gives  $g = 2$,   i.e. {\em twice the classical value}.  Unlike  orbital
angular momentum or the  spin of a composite particle the spin of an  elementary particle
has a fixed value and therefore its magnetic moment is {\em fixed}   and must be regarded as
one of  the characteristics of the particle like its charge, mass,  etc. The electron magnetic moment (spin 1/2)  is to a
good approximation given by the Dirac value\footnote{Small deviations from this value are
very accurately described in Quantum Electrodynamics by the effects of the interaction with
the surrounding cloud of virtual photons and electron--positron pairs.}
\beq
\mu_0 = \frac{\mid e \mid \hbar}{2mc} \; .\label{bmag}
\eeq This quantity is called the {\em Bohr magneton} and serves as a convenient unit in which
magnetic moments are measured in atomic physics. Its numerical value is
$5.79\cdot 10^{-9}$ eV/Gauss. 

 In nuclear physics a more appropriate unit is the {\em nuclear
magneton} defined as in (\ref{bmag}) but with the mass  of the proton used for $m$.
Experimentally measured value for the magnetic moment of the proton is 2.793 nuclear
magnetons meaning that the g-factor is 5.586. For neutrons the values are --1.913 and --3.826
respectively. The deviation of these g-factors from the corresponding Dirac values
$g = 2$ and $g = 0$  was among the first  experimental indications that protons  and neutrons
are not elementary but rather composite particles. In general  the calculation of the
g-factors  for composite particles requires the knowledge of the intrinsic dynamics, i.e. the
wave function of the elementary constituents, their spins, etc. We will consider  examples of
such   calculations below.

\subsection{Atoms in a magnetic field}

\subsubsection{The Hamiltonian} 
Consider an atom placed in a uniform magnetic field.  Assuming fixed heavy nucleus atomic  electrons are  described by the Hamiltonian\footnote{In this and the following Sections we use CGS units}
\beq H = \frac{1}{2m}\sum_a\left(\vp_a + \e/c\vA(\vr_a)\right)^2 + U({\vr_a}) +
\frac{e}{mc}\vB\cdot\sum_a\vs_a\; ,
\eeq  where we denoted by $\vr_a$ , $\vp_a$ and $\vs_a$ the coordinates, momenta and spin
operators of the electrons and included in $U({\vr_a})$ the interaction of the electrons with
the atomic nucleus as well as their Coulomb interaction  with each other.  We used $q=-e$ for electrons and for simplicity disregarded the nuclear spin. 

Choosing the vector potential in the form $\vA = (\vr \times
\vB)/2$ for which $\nabla\cdot\vA = 0$ we can write the Hamiltonian in the form
\eqna H  & =  & H_0 + \frac{e}{2mc}\vB \cdot\sum_a(\vr_a\times\vp_a) +
\frac{e^2}{8mc^2}\sum_a(\vB\times\vr_a)^2  + \frac{e}{mc}\vB\cdot\sum_a\vs_a \\  &  = & H_0 +
\mu_0 (\vL + 2\vS)\cdot\vB + \frac{e^2}{8mc^2}\sum_a(\vB\times\vr_a)^2 \;,
\label{hmzm} \eqne where $H_0$ is the Hamiltonian in the absence of the magnetic
field, $\mu _0$ is the Bohr magneton  and we used the expressions
$\vL = \sum_a(\vr_a\times\vp_a)$ and $\vS = \sum_a\vs_a$ for the total orbital angular
momentum and  spin. The terms in $H$ which depend on the magnetic field can be written as
$-\vmu\cdot\vB$ with the operator of the magnetic moment
\beq
\vmu = -\mu_0 (\vL + 2\vS) + \frac{e^2}{8mc^2}\sum_a[\vr_a^2\vB -\vr_a(\vr_a\cdot\vB)]\;.
\eeq 

The first term in this expression is independent of $\vB$ and can be considered as the
operator of the intrinsic  magnetic moment of the atom which exists in the absence of the
field. It is a sum of the orbital and the spin contributions in which  the latter
enters with  twice as large coefficient.  It is crucial to observe that because of this non classical
 Dirac value of the spin g--factor the intrinsic magnetic moment is not parallel to the system total
angular momentum $\vJ = \vL + \vS$. As we will presently see this is the main
reason why  in general the atomic   g--factors do not have the universal classical value
$g=1$ but depend on the state of the atom.

 The second term in $\vmu$ depends on $\vB$ and must be regarded as the  operator of the
magnetic moment which is {\em induced} by the magnetic field. Its magnitude
$-(e^2/8m^2c^2)\sum_j I_{ij}B_j$ is  proportional to the moment of inertia
$I_{ij} = \sum_a m(r_{a,i}r_{a,j} - \delta_{ij}\vr_a^2)$ which of is one of the
manifestations of the Larmor theorem.

\subsubsection{Treating the $\vB$ dependent terms perturbatively. LS and jj couplings}

Exact diagonalization of the Hamiltonian (\ref{hmzm}) is not feasible even when the solutions
in the absence of the magnetic field are known. The standard way of treating this problem is
to use the perturbation theory with respect to the B--dependent terms. Let us start with the
linear term in (\ref{hmzm}).  Because of the rotational symmetry the  states of the atom are
characterized by the eigenvalues $J(J + 1)$ of $\vJ^2$ and for non zero $J$ are multiplets of
degenerate states  which  can be  labeled  by one of  the projections of $\vJ$.  One must
therefore use degenerate perturbation theory and  to lowest order diagonalize the
perturbation $H_1 = \mu_0(\vL + 2\vS)\cdot\vB$ in the subspace of each multiplet.    The
magnetic field breaks the rotational symmetry and removes the multiplet degeneracies.  The
remaining axial symmetry of  rotations around the direction of $\vB$  indicates that  within
each multiplet  of the degenerate states the correct combinations  which  diagonalize $H_1$
are the eigenstates of  the projection of $\vJ$ on $\vB$. The energy shift of these states
with respect to the unperturbed value is simply the expectation value of  $H_1$,
\beq
\Delta E = \mu_0 B <\alpha;J,M|L_z + 2S_z|\alpha;J,M> = \mu_0 B <\alpha;J,M|J_z +
S_z|\alpha;J,M> \label{splz} \; ,
\eeq  where we have chosen the z-axis along the direction of $\vB$ and denoted by $\alpha$
 the additional quantum numbers apart of $J$ and its projection  $M$ which are needed in
order to specify an atomic  state. 

According to the Wigner--Eckart theorem, cf. Ref.\,\cite{Sak}, 
the matrix element of a component of any  vector operator between states of a
multiplet with a given J is proportional to the same matrix element of the same component of
the operator
$\vJ$ with the proportionality constant which is independent of M. We can therefore write
\beq
\Delta E = \mu_0 g_{\alpha,J}B <\alpha;J,M|J_z|\alpha;J,M> = \mu_0 g_{\alpha,J}BM \; ,
\label{spl1}
\eeq where the yet undetermined proportionality constant $g_{\alpha,J}$  obviously represents
the g--factor of the atomic state. Finding explicit  expression for $g_{\alpha,J}$ requires
further  information  about the structure of
$|\alpha,J,M>$ and can only be made in certain limiting cases.

 If the  interactions in atoms
were the ordinary Coulomb forces the total orbital and  spin angular momenta and their
projections $M_L$ and $M_S$ would be separately conserved and in this case
\beq
\Delta E = \mu_0 B (M_L + M_S) \; .
 \eeq 
 In reality, however relativistic effects are important
and produce the so called {\em fine structure}  of atomic levels. The main relativistic
effect turns out to be  the presence in the atomic Hamiltonian $H_0$ of the spin--orbit term
$\sum_a V_{so}(|\vr_a|)\;\vl_a\cdot\vs_a$  with
$V_{so}(r)$ proportional to $r^{-1}$ times the derivative with respect to r of the atomic
potential $-Z e^2/r$.  When this term is relatively weak (as happens for most atomic states)
one can treat it as a perturbation and diagonalize it separately within each degenerate
multiplet of  $(2L + 1)(2S + 1)$ states with given $L$ and $S$. 

The  result is what is called the    {\em
fine  splitting}  of the multiplet  into  closely lying states which  have definite values
of  $J$.  In this zero order of the perturbation treatment  they are linear combinations of the unperturbed wave functions with same
values of L and S but different $M_L$ and $M_S$. Formally these {\em zeroth order} atomic states are
written  as 
$$
|n; L, S; J, M> =
\sum_{M_L+M_S=M}<L,M_L;S,M_S|L,S;J,M>|n,L,M_L>|S,M_S>
$$
 and are referred to as states of the
"LS -- coupling" scheme.  By $n$ we denoted here the remaining quantum numbers  for the
orbital motion  and the coefficients in the sum  are the standard Clebsh-Gordan coefficients
for coupling of two angular momenta.  

In the opposite extreme case of the strong spin--orbit
interaction one can not talk about separate conservation of the orbital and spin angular
momenta. Individual electrons must be characterized by their total angular momenta j which
must be combined to produce the total J. Such a scheme of constructing the zeroth order wave
functions is called the "jj -- coupling". This extreme limit is rarely  found in atoms but
plays a central role in nuclear spectroscopy.

\subsubsection{Lande formula}

For states with LS -- coupling a general expression for the  g-factors called the Lande
formula can be derived,
\beq g = 1 + \frac{J(J + 1) - L(L + 1) + S(S + 1)}{2J(J + 1)} \label{lnde}\; .
\eeq  This is found as follows. As was already mentioned the  Wigner -- Eckart  theorem gives
\beq <\vS> = const\cdot<\vJ>
\eeq                 where we use the angular brackets to denote averages with respect to the
state
$|n; L, S; J, M>$ .  Since the operator $\vJ$ commutes with $\vL$ and $\vS$ it does not
change the quantum numbers of this state so we can write
$$
<\vS\cdot\vJ> = const<\vJ\cdot\vJ> 
$$
with the same constant. Using $<\vJ\cdot\vJ> = J(J+1)$ have
\beq 
<S_z> = const \; M = M \frac{<\vS\cdot\vJ>}{J(J +1)} 
\eeq 
Using $\vL\cdot\vL = (\vJ - \vS)^2 = \vJ\cdot\vJ + \vS\cdot\vS - 2\vJ\cdot\vS$ and the
properties of the LS -- coupling  state we find that
\beq 
<\vS\cdot\vJ> = \frac{1}{2}[J(J +1) - L(L + 1) + S(S + 1)] \; 
\eeq 
Collecting the
results in Eq. (\ref{splz}) we obtain that $\Delta E $ is in the form (\ref{spl1}) with
$g_{\alpha,J}$ given by the Lande expression (\ref{lnde}). As usual with the results of the
perturbation theory this formula  is valid when $\Delta E$ are small as compared to the
intervals between the unperturbed atomic energy levels. In the present case these  are the
intervals due to the fine structure splitting.

\subsection{The Zeeman effect }

The general phenomenon of the energy splitting of atomic levels in magnetic field is called
the Zeeman effect. The Lande formula gives the classical value $g = 1$ in the case $S = 0$
and the Dirac value $g = 2$ when $L = 0$. Historically   the measured deviations  of $g$
from the classical value 1 were termed the anomalous Zeeman effect. In the case when the
magnetic field is so intense that $\mu_0 B$ is larger than the intervals of the fine
structure   the energy splittings
$\Delta E$ deviate from the predictions of the Lande formula. This is called the Pashen --
Back effect.  We will not discuss the details of it. 

Let us now turn to the   last  term in the Hamiltonian (\ref{hmzm}) which is quadratic and
describes as we already mentioned the interaction of induced magnetic moment with the field
$\vB$. This interaction is sometimes called  diamagnetic to distinguish it from the  linear
term which is called the paramagnetic interaction . The   relative magnitude of these two
terms  can be estimated as
$(e/\hbar c)\; r^2 B \sim 4\cdot 10^{6} (r/{\rm cm})^2 B/Gauss$ and one finds that for
typical magnetic fields in laboratory  the diamagnetic term is  negligible if r has atomic
dimensions.      However  when an atomic state has zero spin and orbital angular momentum ($L
= S = 0$), the linear term  does not effect the  energy levels in any order of the
perturbation since it has vanishing matrix elements. In this case the entire effect  is
determined by  the quadratic term. In first order of the perturbation theory the
corresponding energy shift is
\beq
\Delta E  = \frac{e^2}{8mc^2}\sum_a<(\vr_a\times \vB)^2>\; ,
\eeq 
where the average is with respect to a (non degenerate) state with L = S = 0.
 Since  $<(\vr_a\times\vB)^2> = B^2 <r_a^2 \sin^2\theta_a>$ and since the wave function of a
state with L = S = 0 is spherically symmetric one can average first over the angle and
obtain  $ <(\vr_a\times\vB)^2> = 2 B^2<r_a^2>/3$ where we used
$$
<\sin^2\theta_a> = \int \sin^2\theta_a  \; 2\pi \cos \theta_a\; d\theta_a/4\pi = 2/3
$$
Therefore
\beq
\Delta E = \frac{e^2}{12mc^2}B^2 \sum_a <r_a^2>
\eeq
 Having in mind the general expression $-\vmu\cdot\vB$ we see that the change of
  the induced magnetic moment with the field in this case is negative which
means that such a state is   diamagnetic.

\section{Time Reversal in  Magnetic Field. Kramers  Degeneracy}

In the absence of  magnetic field and for spinless particles the \Sch equation  with a time
independent Hamiltonian is invariant under  the substitution $t \rightarrow -t$ provided one
also changes $\psi \rightarrow \psi^*$.  One adopts
\beq
\psi(\arg) \to T\psi(\arg) \equiv \psi^*(\vr, -t)\label{ortirv}
\eeq as a definition of the time  reversal transformation in this  case.
Magnetic field   and  the particle spin require modifications of
this definition.  Since  magnetic field acts also on the spin variables it is natural to
discuss them together.

Even time independent magnetic field  breaks the time reversal symmetry.  This is
already  known in classical physics. The equation  of motion (\ref{lzeqmo}) is time
reversal invariant for any static
$\vE$ if $\vB = 0$. For non vanishing  $\vB(\vr)$ this symmetry is lost but one
observes that the  equation   retains its form  if together with the sign change of
$t$ one changes  the sign of the magnetic field. Thus  all solutions
$\vr (t)$ found in a given $\vE(\vr)$ and $\vB(\vr)$ must have "partners" in the form $\vr
(-t)$ in a related problem with  $\vE(\vr)$ and $-
\vB(\vr)$. Of course one must take care in defining properly matched initial conditions for
related solutions, i.e. impose time reversed initial
velocities. One easily understands why the sign of $\vB$ must be reversed -- this is
consistent with   Maxwell equations which  relate $\vB$ to  external currents which change
their direction under time reversal. Similar arguments  make it clear why $\vE$ should stay
the same.

Let us now turn to quantum mechanics in a static electromagnetic field. We first
notice  that changing the sign of $t$  and of $\vB$ without changing $\vE$ simply means that
$\vA \rightarrow - \vA$ together with $t \rightarrow -t$.  Transforming also $\psi (\vr,t)
\rightarrow \psi^*(\vr,-t)$  in the \Sch equation   (\ref{schem}) we see that such a combined
transformation  leaves  invariant all the terms in the equation except  for  the last,
spin dependent term which  becomes  $ge\vB \cdot \vs^* \psi^*/2mc$ rather than $-ge\vB \cdot
\vs \psi^*/2mc$. By analogy with the orbital angular momentum one needs the reversal of the
sign  of the spin  operators and the   complex conjugation does not
accomplish this.   Indeed recalling  the standard representation  of the spin operators
in terms of the Pauli matrices,
$$
s_x = \frac{\hbar}{2}\left( \begin{array}{cc} 0  &  1 \\ 1  &  0 \end{array}\right)\;\;\;
\;\; s_y = \frac{\hbar}{2}\left( \begin{array}{cc} 0  &  -i \\ i  &  0
\end{array}\right)\;\;\;
\;\;
s_z = \frac{\hbar}{2}\left( \begin{array}{cc} 1  &  0 \\ 0  &  -1 \end{array}\right)\;
$$
 one sees that ({\em in this particular representation}, in which  $s_z$ is diagonal) the
complex conjugation causes only
$s_x^* = s_x$,  $s_y^* = - s_y$ and
$s_z = s_z^*$.   Hence one must
together with the  complex conjugation also change the sign of
$s_x$ and $s_z$ without changing $s_y$. This can be accomplished  by the rotation by the
angle  $\pi$ around the y--axis in the  "space" of the spin variables. Simce $\vs$ is
the operator of infinitesimal rotations in this space such a rotation  is achieved by
the operator $\exp{[\,i\pi s_y/\hbar]}$.  Accordingly,  we generalize the  time reversal
transformation of the wave function for particles with spin as
\beq
\psi(\arg)  \to T\psi(\arg) \equiv \exp{[\,i\pi s_y/\hbar]} \psi^*(\vr,-t) \;,
 \label{gntirv}
\eeq which must be supplemented with the sign change of  $\vA$ and $\vB$ in the presence of
the magnetic field.   Now of course all the terms in Eq.(\ref{schem}) will transform
correctly.   We note that the transformation (\ref{gntirv}) (as well as the incomplete
(\ref{ortirv})) is antilinear, i.e.
$T(\alpha\psi_1 + \beta\psi_2) =
\alpha^* T\psi_1 + \beta^* T\psi_2$ and antiunitary, i.e. $<T\psi|T\phi> = <\psi|\phi>^*$.

The transformation properties of any (possibly time dependent)  quantum mechanical  operator
$O_{op}$ under time reversal are  determined by considering
$$T(O_{op}\psi) =
\exp{[\,i\pi s_y/\hbar]}\; O_{op}^*(-t) \psi^*(\vr,-t)$$ and comparing with $T(O_{op}\psi)
=    (T\;O_{op}\;T^{-1})(T\psi)$. We thus find
\beq
 T\:\vr\:T^{-1}  = \vr\;\;, \;\; T\:\vp\:T^{-1} = - \vp\;\;,\;\;T\:\vs\:T^{-1} = -\vs\;,
\eeq in line with the intuition.

It is important to remember that the explicit form of the time reversal operator as  given
above was derived  in the particular representation,  i.e. in the coordinate representation
and diagonal  spin projection $s_z$.  It is  in general  not valid in other representations,
but can be derived following the rules of transformations between representations.  E. g., a
plane wave
$\exp(i\vk\cdot\vr)$ in  the coordinate representation becomes $\delta(\vp - \hbar\vk)$ in
the momentum representation for which  the complex conjugation is obviously not producing
what expected under the time reversal -- the change of the sign of $\vk$. Using the relation
between the coordinate and momentum  representations we obtain
\beq  <\vp\;|\;T\psi> = \int d\vr<\vp\;|\;\vr><\vr\;|\psi(-t)>^* = <-\vp\;|\;\psi(-t)>^*\;,
\eeq     where we ignored the spin and used $<\vp\;|\;\vr> = <-\vp\;|\;\vr>^*$.  It is seen
that the time reversal in momentum representation is  a combined action of the complex
conjugation and the change of  sign of the momenta ---  not surprising.

 Let us return to  physical systems without  external magnetic field. Their   Hamiltonians
are symmetric under time reversal, $[H,T] = 0.$
 For an eigenstate $\psi_n$ of H this gives  $HT\psi_n = TH\psi_n = E_n\psi_n$ which means
that $\psi_n$ and $T\psi_n$ have the same energy. There are two strong results which follow
from this fact :
\begin{itemize}
\item for spinless particles  non degenerate eigenstates of H can always be chosen to be
real and
\item  eigenstates with {\em half--integer total spin} are always at least doubly
degenerate. This degeneracy is called {\em the Kramers degeneracy}.
\end{itemize}

To prove the first result we note that  since for spinless
 particles $\;T\psi_n(\vr) =\psi_n^*(\vr)$ and since by  the assumption $E_n$ is not
degenerate the function
$\psi_n(\vr)$ must coincide with $\psi_n^*(\vr)$ up to a constant independent of $\vr$.
For normalized wave functions this is at most a phase factor which is inessential   for
any physical results and can be disregarded.

In order to prove the second result consider an eigenstate wave function $\psi_n =
\psi_{\alpha jm}$ and its time reversed partner
$T\psi_{\alpha jm}$, where we denoted by $jm$ the total spin of the system and its projection
and by $\alpha$ all  other quantum numbers.  Should these functions represent the same
state as in the spinless case they would be related as
$T\psi_{\alpha jm} =  C\psi_{\alpha jm}$ with some complex constant C. Applying T once again
we would get  $T^2\psi_{\alpha jm} = |C|^2\psi_{\alpha jm}$. But on the other hand   $\;T^2 =
\exp{[\,2i\pi s_y/\hbar]}$ which gives $(-1)^{2j}$ when applied to $\psi_{\alpha jm}$, cf.,
  Problem \ref{sptr}. This can not be equal to a positive
$|C|^2$ for half--integer  j. We are therefore led to conclude that $\psi_{\alpha jm}$ and
$T\psi_{\alpha jm}$ must correspond to different states for half--integer j which means that
the corresponding eigenvalue $E_n$ is at least doubly degenerate.  This Kramers
degeneracy  means, for instance,  that for a system with odd number of electrons the energy
levels will always be at least twofold degenerate even if it is placed in  any,
however complicated, electric field

\section{Path Integrals with  the External Electromagnetic Field \label{piexem}} The
derivation of the path integral quantization of a particle in the presence of the
electromagnetic field follows the standard route.  The
propagator
$$
K(\vr_f,t_f;\vr_i,t_i) \equiv \langle \vr_f | e^{-i H_{op}(t_f - t_i)}  | r_i\rangle
$$
 is represented as a multiple integral
 \eqna K(\vr_f,t_f;\vr_i,t_i) & = &\lim_{N\to\infty}\int d^3 r_N \int d^3 r_{N-1}\;
.\; .\; . \int d^3 r_1\;K(\vr_f,t_f; \vr_N,t_N)\times \nonumber
\\ & &\times\;\; . \;\; . \;\; . \times K(\vr_1,t_1;\vr_i,t_i)\; ,
\eqne
over infinitesimal propagators which should be calculated for the Hamiltonian operator
given by Eq.\,(\ref{hmwosp}) (we do not consider  the
spin dependent term  -- such terms require special treatment in the path integral
formulation). Based on  the  experience with path integrals one
should expect  that the   propagator $K(\vr,t+\epsilon;\vr',t)$ over an infinitesimal  time
interval $\epsilon$ is expressed as $$\left(\frac{m}{2\pi i
\hbar\epsilon}\right)^{3/2}\exp\{\frac{i}{\hbar}\epsilon L[(\vr +\vr')/2,(\vr-
\vr')/\epsilon]\}
$$  in terms of  the classical Lagrangian $L(\vr,\vv)$  given by  Eq. (\ref{cllgma}). An
explicit calculation indeed shows that
\beq 
\psi(\vr, t+\epsilon) = \int d\vr' K(\vr,t+\epsilon;\vr',t) \;\psi(\vr',t) 
\eeq
reproduces the  \Sch equation with  the infinitesimal propagator given by
\eqna \label{inprmg}
K(\vr,t+\epsilon;\vr',t) &=& \left(\frac{m}{2\pi i
\hbar\epsilon}\right)^{3/2}\exp\left\{\frac{i\epsilon}{\hbar}\left[\frac{m}{2}
\left(\frac{\vr- \vr'}{\epsilon}\right)^2 - eA_0\left(\frac{\vr +
\vr'}{2}\right)\right]\;\; + \right. \nonumber\\          &  &
\;\;\;\; \left. + \;\;\frac{ie}{\hbar c}(\vr - \vr')\cdot\vA\left(\frac{\vr +
\vr'}{2}\right)\right\} 
\eqne
 The details of this calculation are rather cumbersome and will not be
reproduced here.  They can be found in  Ref.\,\cite{Sch}.

Using the expression for the infinitesimal propagator in the multiple integral for
$K(\vr_f,t_f;\vr_i,t_i)$ we find  after combining the product of the exponentials into a
exponential of a sum and using the continuous notation
\beq K(\vr_f,t_f;\vr_i,t_i) = \int\limits_{\vr(t_i) = \vr_i }^{\vr(t_f)=\vr_f}
D[\vr(t)]\exp\left\{\frac{i}{\hbar}\int\limits_{t_1}^{t_2} dt\left[\frac{m\vv^2}{2}
- eA_0(\vr) + \e/c\vA(\vr)\cdot\vv\right] \right\} \label{pimgfl}\eeq
where as usual the definition of $D[\vr(t)]$  includes the product of $N$ 
$d^3 r_i$'s  each multiplied by the pre-exponential factors from Eq.\,(\ref{inprmg}).

 The first two terms in Eq. (\ref{pimgfl}) are the usual kinetic and potential energies but
the  last term  is a new feature of this path integral.  It mixes coordinates and
velocities  but its linear dependence on
$\vv$ is special. Making  the replacement $\vv\cdot dt=d\vr$ the contribution of this term
  for every path in the path integration  can be written as
$$\exp\left[\frac{i e}{\hbar c}\int\limits_{\vr_i}^{\vr_f}\vA[\vr(t)]\cdot d\vr\right]\;.$$
This phase factor depends  on the path but not on the velocity of propagation along it.
If one considers a  difference of these phases  between two arbitrary paths one can
write it  as the circulation of
$\vA$ along a closed path which is obtained by traversing from $\vr_i$ to $\vr_f$ along
one path and then back to $\vr_i$ along the other.  Using the Stokes theorem
$\oint\vA\cdot d\vr = \int \nabla\times\vA\cdot d\vS=\int \vB\cdot d\vS$  one can write this
phase difference as
 \beq \exp\left[2\pi\,i\,\frac{\Phi}{\Phi_0}\right]\;, \label{exflf0}\eeq
where  $\Phi$ is the flux of the magnetic field through the closed contour defined by
the two paths and $\Phi_0 = hc/e$ is the magnetic flux quantum  already familiar from our discussions
 of the Aharonov-Bohm effect.

We would like  to point out  an important  subtlety related to the appearance of terms like
 $\vA(\vr)\cdot\vv$  in the path integration. One will get different answers depending on
 whether  $\vA(\vr)$ is evaluated  at $(\vr +\vr')/2$,  at
$\vr$,  at $\vr'$ or somewhere in between  in the infinitesimal propagator
(\ref{inprmg}). This  ambiguity  is known as the  {Ito ambiguity} and is discussed in detail
in  Ref. \cite{Sch}.   It is
shown there that the correct prescription is to take $\vA$  as it is written in
Eq.(\ref{inprmg}), i.e.  {\em at a midpoint}. This is sometimes referred to as
the {\em  mid--point rule}. Only with this rule the correct \Sch
equation is reproduced.  The mid--point rule  is important
for a term $A\cdot\vv$ and not  for the conventional potential term $eA_0(\vr)$. This is 
 because of  {\em the different
powers of $\epsilon$}, i.e. $\epsilon^{0}$ and $\epsilon^{1}$  which multiply
the discretized version of $\vA\cdot\vv$ and
$e A_0$  respectively in the expression for the infinitesimal propagator.   As is shown
in standard discussions of the path integrals  the typical distances between propagation points
obey the estimate  $|\vr-\vr'|
\sim \sqrt{\epsilon}$. Changes  of this  order of magnitude  in the
argument of $eA_0[(\vr+\vr')/2]$ combined with $\epsilon^{1}$ in front of it will contribute
a negligible difference of the order $\sim
\epsilon^{3/2}$.    The same change in $\vA[(\vr+\vr')/2]$ combined with the term
$\vr-\vr'\sim\epsilon^{1/2}$ which multiplies it  contributes
$O(\epsilon^{1})$ which can not be neglected.

 Let us now examine how the gauge transformations effect the path integral (\ref{pimgfl}).
Performing  a gauge transformation (\ref{gt}) of the potentials  $\vA$ and
$A_0$ adds  in the action the term proportional to
\beq \int_{t_i}^{t_f}dt\left[\frac{d\vr}{dt}\cdot\nabla\chi(\arg) +
\frac{\pd\chi(\arg)}{\pd t}\right] =
\int_{t_i}^{t_f} \frac{d\chi(\arg)}{dt}dt = \chi(\vr_f,t_f) - \chi(\vr_i,t_i)\;,
\eeq where the last equality holds because of the mid--point rule of the discretization of
the integral and gives the result which is the same for all paths. Using this we find that
under the gauge transformation the propagator  changes as
\beq K\,'(\vr_f,t_f;\vr_i,t_i) = \exp[ie\chi(\vr_f,t_f)/\hbar
c]K(\vr_f,t_f;\vr_i,t_i)\exp[-ie\chi(\vr_i,t_i)/\hbar c]\; .\eeq This of course is of the
same origin  as the change of the phase of the wave function (\ref{gt1}).  The
phase change  of $K(\vr_f,t_f;\vr_i,t_i)$ depends only on the initial and the final
coordinates.  The phase difference between different paths is strictly gauge invariant.

In a {\em uniform} electric and magnetic field the Lagrangian (\ref{cllgma}) is a
quadratic function of the coordinates and velocities and the path integral in this case is
of the Gaussian type and   can be evaluated exactly (cf., Problem \ref{prprmg} at the end of
the chapter).

\section{Dirac Magnetic Monopoles}

\subsection{Multivalued wave functions. Non integrable phases \label{nswf}}

An instructive discussion related in a surprising way to the general issue of  the gauge
transformations arises when one examines in depth the requirement  that the solutions of the
\Sch equation must be single valued. This requirement  is usually imposed as natural and  is
the main reason for finding the standard quantized values of physical quantities such as the
energy, the angular momentum, etc.  Following the  discussion by Dirac\,\footnote{In this
section we draw freely on the original paper of P.A.M. Dirac, Ref. \cite{Dir}.} let
us try  to see what happens if this requirement is removed. 

Of course one still must obtain
unambiguous results for   quantities which have direct physical meaning. This certainly
means that the amplitude  of the wave function must be single valued since its square is a
physical density  function. The phase of the wave function on the other hand  does not have
to have  a unique value at a particular point so in general the wave function can be written
as
$\psi(\arg) = \phi(\arg) \exp[i\beta]$ with $\phi(\arg)$ the ordinary single valued complex
function  and all multivaluedness residing in the properties of the  phase
$\beta$. A useful way to characterize this multivaluedness is  to consider how $\beta$
changes when one goes along some curve connecting two points in space--time. Since
$\psi(\arg)$ satisfies the \Sch equation it must be continuous and  therefore   it is natural
to assume that $\beta$ must have a definite derivative almost at
 every point $(\arg)$. We will discuss later the points where this does not happen. 
 
 The change of $\beta$ along a curve which does not pass through such singular points can be
expressed by the integral $\sum_\mu \int \kappa_\mu dx_{\mu}$  taken along this curve  with
$\kappa_i = \pd\beta(\arg)/\pd x_i$ and $\kappa_0 = \pd\beta(\arg)/\pd t$. Since
$\kappa_\mu$ in general do not satisfy the conditions of integrability
$\pd \kappa_\mu/\pd x_\nu = \pd \kappa_\nu/\pd x_\mu$ the value of this integral depends on
the curve and in particular the total change in the phase $\beta$ need not vanish when the
integral is calculated round a closed curve.  The values of such circulation integrals for
all imaginable closed curves completely characterize the multivalued properties of the
non--integrable phase $\beta$. 

We now show that in order to have   unambiguous  results for
physical quantities {\em any such circulation integral must be the same for all the wave
functions}.  Indeed probabilities to measure physical quantities  are given by squares of
moduli of overlap integrals $\int
\psi_m^* \psi_n d^3 r$  with different wave functions $\psi_m$ and $\psi_n$. In order that any
such integral will have a definite modulus the integrand, although it need not have a
definite phase at each point, must have a definite phase difference between any two points.
Thus the change of phase of $\psi_m^* \psi_n$ round a closed curve must vanish. This requires
that the change in phase in $\psi_n$ round a closed curve shall be equal to that in $\psi_m$
and since $\psi_m$ is arbitrary  it must be a universal value for a given curve for all wave
functions.

This result means that without loss of generality the possible non integrable phase factor
$\exp(i\beta)$ in the wave function may be taken as  {\em universal} for all wave  functions.
Let us now consider the \Sch equation for $\psi$. Since
\beq -i\hbar\frac{\pd}{\pd x}\psi = e^{i\beta}\left(-i\hbar\frac{\pd}{\pd x} +
\hbar\kappa_x\right)\phi
\eeq with similar relations for the $y$, $z$ and $t$ derivatives one obtains that the single
valued part $\phi$ of the general wave function $\psi$  satisfies the \Sch  equation with
{\em gauge potentials} which are proportional to the derivatives of the non integrable phase
$\beta$.  In the most common case these would have to be identified with the electromagnetic
potentials
\beq
\vA = (\hbar c /e) {\mbox {\boldmath $\kappa$}}\;\;, \;\; A_0 = -(\hbar/e)\kappa_0\; .
\eeq 
We therefore conclude  that multivalued wave
functions need not be considered in quantum mechanical description since they are equivalent
to single valued  wave functions in the presence of an external gauge field.

Although this conclusion is certainly valid there are two   ambiguities  which remain  in the
above discussion. The first is related to the Aharonov--Bohm effect and can occur  in
multiply connected regions such as the inside of a  ring  as was already discussed in Section
\ref{AB} above.  In this case even  for a vanishing  electromagnetic field {\em inside the
region} one can not in general  assume that the wave function must be single valued. If one
can find non contractable closed curves in the region one must first classify these curves
according to different homotopy classes as in Section \ref{AB}.  One may then assign an
arbitrary but fixed phase factor
$\exp(i\beta_k)$ for every  elementary homotopy class $C_k$ and demand that only  solutions
of the \Sch equation  which change their phase by these assigned factors are allowed.

Intuitively one can interpret this situation by thinking  about a multiply connected region
as a region with "holes". Even when  the electromagnetic field vanishes inside the region one
can still have arbitrary magnetic fluxes "in the holes".  These fluxes will give rise to
Aharonov - Bohm phases for closed curves surrounding the "holes" provided these curves can
not be continuously deformed to a point.  Hence assigning different sets of phase factors
$\exp(i\beta_k)$ for elementary classes of curves corresponds to assuming different
distributions of Aharonov-Bohm fluxes $\exp(2\pi i \Phi_k/\Phi_0)$ in the "holes".

There exists another important  ambiguity in the  discussion of possible appearance of non
integrable phases in quantum mechanics.
 This was first observed by Dirac and  is related to the fact that although {\em in the
absence of the electromagnetic field in a singly connected region} the factor $\exp(i\beta)$
can be taken as single valued the phase $\beta$ itself may change by an arbitrary integer
multiple of $2\pi$. Allowing for such changes requires a reconsideration of the connection
between the derivatives
$\kappa$ of the non integrable phase $\beta$  and the electromagnetic potentials and leads to
a new physical phenomenon -- a possible existence of {\em magnetic monopoles with quantized
charges}. We will now discuss this fascinating subject.

\subsection{Magnetic  monopoles}

The Maxwell equation $\nabla\cdot\vB = 0$ means that  there are no sources of the magnetic
field, i.e. that the  magnetic charges  do not exist in nature. However
 nothing conceptually wrong should occur in  the classical theory  if one assumes a non zero
$\nabla\cdot\vB = 4\pi\,  \sigma$ with $\sigma$ -- the  density of  magnetic charges. In
fact  the theory would be  more symmetrical in this case  since a  symmetry under the so
called duality transformation $\vE \to
\vB$, $\vB \to -\vE$  would then exist  if one simultaneously exchanges the magnetic and the
electric charges. The non zero $\nabla\cdot\vB$ poses however a problem     in quantum
theory  where the canonical or path integral quantization  in the presence  of a magnetic
field require an explicit introduction of the vector potential  via $\vB =
\nabla\times\vA$. Without  this relation   one is not able to define the  Hamiltonian or the
Lagrangian of the theory  but  it is valid only  for divergenceless $\vB$.  Let us analyze
this problem  more closely  and  consider a hypothetical point--like particle, called
magnetic monopole,  which carries a magnetic charge $g$.  In its presence
\beq
\nabla\cdot\vB = 4\pi g \delta(\vr - \vr_0) \;\;\; , \;\;\; \vB = g\frac{\vr - \vr_0}{|\vr -
\vr_0|^3}
\; ,  \label{mgmn}
\eeq where $\vr_0$ denotes the position of the monopole and $g$ is its magnetic charge.

For all  points  in space apart from  an infinitesimal  vicinity of  $\vr_0$   we have a
divergence-less $\vB$ and can  write $\vB(\vr) =
\nabla\times\vA(\vr)$. Although correct {\em locally} the function    $\vA(\vr)$ is not
single valued.  This is seen  by considering  the integral form of the relation
$\vB = \nabla\times\vA$,  i.e.  the  Stokes theorem,  $$\int_S\vB\cdot d\vS = \oint_C\vA\cdot
d\vr$$  where $C$ is some closed curve    in space and the  integral on the left hand side is
over  an arbitrary surface $S$ with $C$ as its boundary. Such an integral -- the flux of
$\vB$ --  is  however {\em not unique}  in the present case.  It does not change for all
surfaces which can be continuously deformed into each other without crossing the position
 of the monopole but once the surface crosses $\vr_0$ the flux changes.  The difference
between the fluxes for surfaces "on both sides" of the monopole is equal to the total flux
through the closed surface which these two surfaces form.  Integrating (\ref{mgmn}) over the
volume inside this surface and using the Gauss theorem one finds that this flux is equal to
$4\pi \, g$.  The non zero $\nabla\cdot\vB$  thus effects the definition of $\vA$ globally
and not  just near  $\vr_0$.  Using the Stokes theorem  with  continuously changed  contour
$C$ as a way of continuous definition of the relation between the functions $\vA(\vr)$ and
$\vB(\vr)$ we will find two "branches" of this relation depending on "which side" of $\vr_0$
we choose the surface
$S$.

There is a number of ways of overcoming  this  difficulty. Historically the first was
suggested by P. M. Dirac,  Ref.\,\cite{Dir}.  We will
follow a  more modern way of presenting this approach.   The idea is somewhat similar to what
is done in the theory of multivalued analytic functions, i.e.  to introduce a branch cut
extending from a branch point.  Viewing the position of the magnetic monopole as analogous to
such a branch point one can avoid the ambiguity in  the use of the Stokes theorem  for
determining the relation between $\vA(\vr)$ and
$\vB(\vr)$  if together with $\vr_0$ a thin tube extending from it  to infinity (or to
another, oppositely charged monopole)  is  excluded from the space. By excluding we mean that
the surface $S$ for the contour
$C$  can never be chosen such that it is pierced by the tube.  This  uniquely  defines "the
side" of the monopole which one should choose to draw the surface in the Stokes formula. One
can thus  assure the   single valuedness  of the $\vB(\vr) \to \vA(\vr)$ relation   everywhere
in space apart from the inside of the excluded tube. We can make the tube as thin as we like
and send it in any direction.

Let us illustrate this discussion and consider  an example of a monopole placed at the origin
and  let us choose the excluded tube along a positive $z$ axis. It is easy to verify
 that the vector potential  the curl of which gives the magnetic field (\ref{mgmn})
everywhere  except on the positive $z$   can be chosen as
\beq A_r = A_{\theta} = 0 \;\;, \;\;A_\phi = -\frac{g}{r}\frac{(1 + \cos\theta)}{\sin\theta}
\label{vp1}
\eeq where $A_r, A_{\theta}$ and $A_\phi$ are spherical components of $\vA$ and
$\theta$ and $\phi$ are the conventional polar and azimuthal angles.  On the $z$ axis this
potential does not reproduce the field (\ref{mgmn}) of the monopole but rather gives a
singular magnetic field directed towards the monopole and carrying  a flux $4\pi\, g$. The
total effective magnetic field represented by the curl of (\ref{vp1}) is  therefore
\beq \label{eq:monop_mag_fld}
\vB_{eff} = g\frac{\vr}{r^3} - g\;\;\delta(x)\;\delta(y)\;{\mbox{\bf e}}_z \; .
\eeq
where $\theta(z)$ denote the step function.  
  
  We could choose another  vector potential
\beq 
A^{\prime}_r = A^{\prime}_{\theta} = 0\; \;,\;\; A^{\prime}_\phi = \frac{g}{r}\frac{(1 -
\cos\theta)}{\sin\theta} \label{vp2}
\eeq  
which  also gives the required  magnetic  field (\ref{mgmn})  but with the excluded
tube along the negative $z$ axis.  The corresponding "effective" field again has a singular
component along this tube  in the direction of the monopole.  The flux along the tube is
equal to the total flux of the first component of $\vB_{eff}$, i.e. the flux of the monopole
field. 

It is easy to understand  now the logic behind the excluded tube construction. The
magnetic flux  along the  tube
 "feeds"  the radially directed field of the monopole  so that the resulting "effective"
field is divergenceless, $\nabla\cdot\vB_{eff} = 0$ and can be represented as a curl of a
vector potential like the examples (\ref{vp1}) and (\ref{vp2}) above.

\begin{figure}
\centering \includegraphics[width=0.4\textwidth]{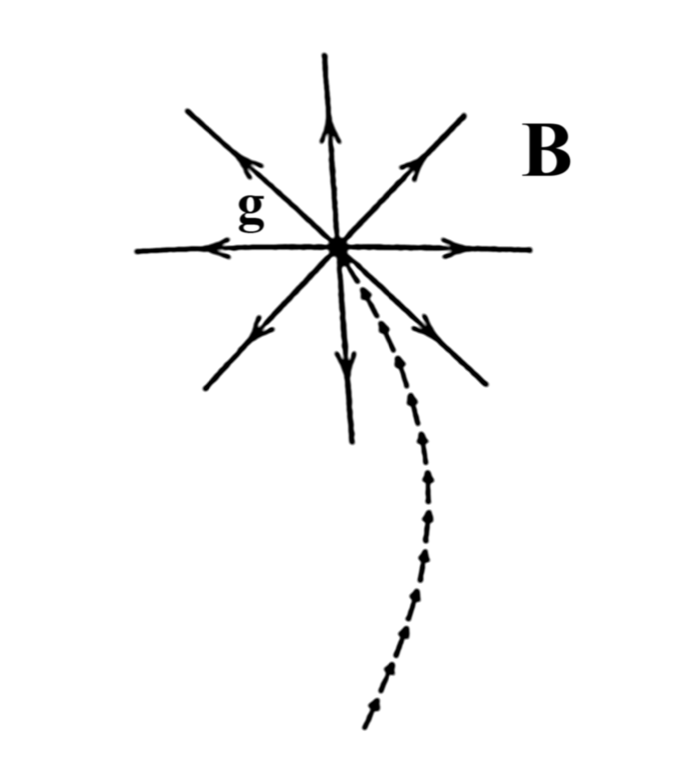}
\caption{Magnetic field of Dirac monopole including the singular string. Note that the string is plotted as curved, which is allowed, but in the examples in the text it was chosen to be a straight line along the positive $z$ axis for simplicity, Eq. (\ref{eq:monop_mag_fld}). }
 \label{fig:Dirac_monopole}
\end{figure}

At this point a crucial question arises. We have replaced the desired magnetic field  of the
monopole  by the effective field with the flux tube.  How does one make sure that this
modification has not changed the physics of the problem? Since the entire construction was
invented for quantum mechanics we must worry if  the presence of the flux tube  added to the
field of the monopole   influences  the solutions of the \Sch equation.  In fact we know that
such a flux tube does have a {\em global} influence in the form of the Aharonov--Bohm effect.
It is also clear   how to avoid  this effect and make the flux tube {\em unobservable} at
large distances. One must demand that the flux carried by the  tube is equal to an integer
multiple of the magnetic flux quanta,
$$
4\pi\, g = n\Phi_0 = 2\pi n \hbar c/e
$$ 
This imposes  a {\em quantization condition} on the
possible values  of $g$,
\beq 
 \label{drc}
eg = \frac{1}{2}n\hbar c \; , n = \pm 1, \pm 2, \dots 
\eeq
  This relation is called {\em the Dirac quantization condition}. The {\em unobservable flux
tube} carrying  integer number of magnetic flux quanta is called the {\em Dirac string}. The
entire construction which we just described is called the {\em Dirac monopole}, cf., Fig. \ref{fig:Dirac_monopole}

The quantization condition (\ref{drc}) for the monopole charge  implies  that  if there exists 
a magnetic monopole
anywhere in the universe all electric charges will  be quantized:
$e = n (\hbar c / 2 g)$. Note that this quantization condition has an explicit dependence on
the Plank constant and therefore on the quantum theory. 

Experimental search for the presence
of the  magnetic monopoles in nature has so far given negative results. We note that since $g
= (137/2)\;e$ the force between two monopoles is $(137/2)^2 \cong 4692$ larger than between
two electrons. This may mean perhaps that all the monopoles in nature are tightly bound in
pairs of opposite sign. In order to decide whether this is true one needs to know the masses
of the monopoles about which the  theory above gives no information. 

In recent years
another  theory of magnetic monopoles was suggested by t\  'Hooft and Polyakov, Ref.\,\cite{tHo}, 
in the context of the so called non abelian gauge theories with broken symmetry.
This theory predicts that the mass of the monopoles should be very large. Viewed from large
distances both Dirac and non abelian monopoles should look exactly the same and our
discussion of quantum mechanics in the field of monopoles is expected to remain valid at
such distances.

There is another way to introduce magnetic monopoles in quantum mechanics which avoids  the
appearance of the singular Dirac string. It  was proposed by  Wu and Yang, Ref.\,\cite{Wu},
and adopts an approach of sections similar to what is done by mapmakers when they  map the
spherical surface of the earth onto a plane map. A single map would obviously have a
singularity at one point. Indeed imagine a rubber sheet with rectangular coordinate grid on
it and try to wrap it around the globe. In order to avoid the singularity of a single  map
two maps are introduced, one for say a northern hemisphere and one for the southern.  The
two maps together form a singularity--free mapping of the earth. In order to be able to pass
smoothly from one map to another one should let each to cover more than its own hemisphere
so that an overlap is created in the region of the equator.  In this overlapping region the
coordinates of both maps must be in one to one correspondence for identical points of the
globe surface. In a similar way singularity-free vector potential can be found for a
magnetic monopole. 

The emerging formulation is in essence  the so called {\em
fiber bundle} formulation of gauge fields in quantum mechanics.  We will not go into this
here referring the reader to the literature.

\subsection{Angular momentum  and  rotational   symmetry in  the presence  of  a
monopole \label{rtinv}}
 Although the magnetic field of the monopole is spherically symmetric it should  be
intuitively clear that    the Lorenz force acting perpendicular to the velocity of a moving
particle will not conserve the ordinary expression
$\vr\times m\vv$ for the angular momentum relative to the origin. Take, e.g., a particle
which starts along a planar circular orbit around the monopole. The magnetic field will
deflect it away from the plane changing the initial
$\vr\times m\vv$. Using the equation of motion  one can calculate the rate of change of this
expression
\beq
\frac{d}{dt}(\vr\times m\vv) = \vr\times m \frac{d}{dt}\vv = \frac{eg}{c
r^3}(\vr\times(\vv\times\vr)) = \frac{d}{dt}\left(\frac{eg}{c}\frac{\vr}{r}\right) \; .
\eeq This suggests that we can define the total angular momentum as
\beq {\mbox{\bf J}} = \vr\times m\vv -
\left(\frac{eg}{c}\frac{\vr}{r}\right) = \vr\times\vp - \frac{e}{c}\left[\vr\times\vA +
g\frac{\vr}{r}\right] \;,\label{lmag}
\eeq so that it is conserved.

The appearance  of unusual terms  in the expressions of conserved quantities in the presence
of electromagnetic field should  already be familiar from the expressions of momenta in
uniform electric and magnetic fields, Eqs. (\ref{gmom}) and (\ref{trop}). In addition to the
generators of the symmetry  one must include the generator of the gauge transformation which
is needed to keep the formulation in the "same" gauge. One  can  see this explicitly  by
considering infinitesimal rotation
$\vr \to \vr' = \vr +\delta\vphi\times\vr$ and correspondingly
$$ \vA(\vr) \to \vA'(\vr) = \vA(\vr - \delta\vphi\times\vr) + \delta\vphi\times\vA(\vr)
\to \vA(\vr) + \delta\vphi\times\vA(\vr) -
\left[(\delta\vphi\times\vr)\cdot\nabla\right]\vA(\vr)$$ For $\vA(\vr)$ which corresponds to
the spherically symmetric magnetic field of the magnetic monopole  the  last term in this
expression must be a gauge transformation, i.e. equal to a gradient of a scalar function,
$\nabla\xi(\vr)$.  One finds
$$ \xi(\vr) = - \delta \vphi\cdot\left(\vr\times\vA(\vr) + g\frac{\vr}{r}\right)\;, $$
 The transformation of the wave function under rotation is therefore
$$[1 + \frac{i}{\hbar}\delta\vphi\cdot(\vr\times\vp)]\;[1 + i \frac{e}{\hbar c}
\xi(\vr)]\;\psi(\vr)\; . $$ In the brackets of the expression for
$\xi$ one finds just the two terms which must be added (after multiplication by
$e/c$) to the canonical $\vr\times\vp$ in order to obtain the conserved Eq.(\ref{lmag}).

 There exists another, quite different interpretation of the   last term in the expression
(\ref{lmag}). It is   the angular momentum contained in the electromagnetic field which
exists in the space surrounding the moving particle and the fixed monopole. Using the
expression for $\vB$ of the monopole and
$ \vE = e(\vr - \vr_0)/4\pi|\vr - \vr_0|^3$ for the electric field of the particle at
$\vr_0$ one  indeed finds
\beq {\mbox{\bf L}}_{em}  =  \int d^3 r\; \vr \times (\vE \times \vB) =
 \frac{egr_0}{4\pi}
\int d^3 r\; \frac{r^2(\hat{\vr}\cos\theta -\hat{\vr}_0)}{r^3(r^2 + r^2_0 -
2\,r\,r_0\,\cos\theta)^{3/2}}
\eeq where $\theta$ is the angle between $\vr$ and $\vr_0$. Straightforward evaluation of the
integral gives $-eg\,\vr_0/|\vr_0|\;,$ which for $eg = \hbar c/2$ gives the last term in
(\ref{lmag}).  The physical picture behind this interpretation of the additional term in
$\vL$ is obscure to the present author.

\section{ Non Abelian Gauge Fields \label{nonab}}
   In Section \ref{ginv}  we discussed how the existence of the electromagnetic field could
be predicted by demanding that  a global symmetry of the free \Sch equation becomes local,
i.e. by "gauging" this symmetry.  We give  now an example of what happens when a more
complicated {\em non abelian} symmetry is gauged leading to a concept of a  non abelian gauge
field, Ref.\,\cite{Yan}. Let us assume that  particles in our theory in addition to spin
carry another {\em discrete intrinsic dynamical} variable
$\tau$ which we will tentatively call pseudospin and which may take two values,
$\tau = 1,2$.  In analogy with the spin variables  the wave functions will now carry an
additional index $\tau$ so that $\mid \psi_{\tau} \mid ^2$ integrated  and summed over all
other variables (
$\vr$,  spin, etc.)  gives the probability to measure this particular value of  $\tau$. We
also have to introduce  operators which act on the variables  $\tau$ and in terms of which we
shall represent all observable quantities  involving this variable. These operators  must be
hermitian $2\times 2$ matrices. One can write any such operator
$O_{\alpha\beta}$ as a linear combination of a unit and Pauli matrices with real coefficients
 \beq O = O_0\delta_{\alpha\beta} + \sum_{a = 1}^3 O_a\tau^{(a)}_{\alpha\beta}\; ,\label{lc}
\eeq
\dia
\tau^{(1)} = \left( \begin{array} {cc} 0 & 1 \\ 1 & 0  \end{array}\right)\; ,
\tau^{(2)} =
\left( \begin{array} {cc} 0 & -i \\ i & 0  \end{array}\right)\; ,
\tau^{(3)} = \left( \begin{array} {cc} 1 & 0 \\ 0 & -1  \end{array}\right)\; ,
\die since the Pauli matrices together with a unit matrix represent a complete set for
expanding any  $2\times 2$ matrix cf., Density Matrix  chapter in the notes of this course.

Now let us consider a free
\Sch equation $$i\hbar\frac{\pd}{\pd t}\psi_{\tau}(\arg) =
-\frac{\hbar^2}{2m}\nabla^2\psi_{\tau}(\arg)$$ for  a particle with the pseudospin. This
equation is obviously invariant under a linear transformation
$\psi\ '_{\tau} =
\sum_{\tau '}S_{\tau\tau '}\psi_{\tau '}$ with an arbitrary matrix $S$.  In order to preserve
the normalization of
$\psi$ the matrix $S$ must be unitary.  Its general form therefore must be an imaginary
exponential of an arbitrary  $2\times 2$ hermitian matrix and therefore can be  written in
terms of the Pauli matrices  as
\beq S = \exp \left(i\varphi_0 + i \sum_a \varphi_a\tau^{(a)}\right) \; . \label{su}
\eeq with arbitrary {\em real} $\varphi$\  's. For simplicity we will limit ourselves to the
transformations with
$\varphi_0 = 0$ which is equivalent to imposing the  condition $\det S = 1$ on the allowed
matrices $S$. We say that  the
\Sch equation is {\em invariant under the global SU(2) transformations} , i.e. under the
transformations which belong to the group SU(2) of all unitary $2\times 2$ matrices with
unit determinant.  This group is non abelian -- two arbitrary SU(2) matrices
 in general do not commute.  Let us now employ {\em The Gauge Principle} of   Section
\ref{ginv} for this SU(2) symmetry and demand that our theory must be not only globally but
also {\em locally} invariant under the above SU(2)  transformations. This means that
transformations with matrices   $S$ having their parameters $\phi_a$ as arbitrary  functions
of $\vr$ and $t$ should leave the \Sch equation invariant. The way  to   achieve the
invariance  under such {\em local gauge } transformations is to introduce the  gauge field
potentials which will compensate for the derivatives of $S$ when the transformed
\beq\psi\  '(\arg) = S(\arg)\psi(\arg) \label{nagt} \eeq is inserted in the \Sch equation.
Since the derivatives of the matrix
$S$ are obviously  also matrices the compensating potentials should  be matrices. Since there
are four derivatives $\pd/\pd t$ and $\nabla = (\pd /\pd x, \pd/\pd y , \pd/\pd z)$ in the
\Sch equation one must introduce four such matrix compensating potentials
$A^{(\mu)},\; \mu = 0, 1, 2, 3, 4$. They can be  represented as linear combinations
\beq A^{(\mu)}(\arg) = \sum_{a=1}^3 A^{(\mu)}_a(\arg) \frac{\tau^{(a)}}{2} =
A^{(\mu)}\cdot\frac{\tau}{2} \; ,
\eeq where we employed obvious short hand notation for the sum of products of arbitrary real
functions $A^{(\mu)}_a(\arg)$ and Pauli matrices $\tau^{(a)}$ and introduced the factor $1/2$
to follow the conventional definitions in this field. In the fixed basis of
$\tau^{(a)}$'s   to represent a matrix $A^{(\mu)}$ is equivalent to giving three functions
$A^{(\mu)}_a$.

In analogy with electromagnetism we introduce now gauge covariant derivatives
\eqna D_0 & = & \frac{\pd}{\pd t} + \frac{ig}{\hbar}A^{(0)}\cdot\frac{\tau}{2} \; ,
\nonumber \\
\vD & = &   \nabla - \frac{ig}{\hbar c}\vA\cdot\frac{\tau}{2} \; ,
\eqne  and use them in the \Sch equation in place of the ordinary derivatives ,
\beq i\hbar D_0\psi(\arg)  =  - \frac{\hbar^2}{2m}\vD^2 \psi(\arg) \; . \label{nasc}
\eeq The constant $g$ introduced here is analogous to the electric charge $e$ in the
electromagnetism. It determines the strength of the coupling of the particle described by
this equation to the non abelian gauge fields $A^{(\mu)}_a$.  In order to achieve the
invariance of the equation under the local gauge transformations (\ref{nagt}) we demand that
$D_0\psi$ and
$\vD\psi$ have the same transformation properties as $\psi$ itself, i.e.
\eqna D_0\  '\psi\  ' \equiv \left(\frac{\pd}{\pd t} + \frac{ig}{\hbar} A^{(0)\
\prime}\cdot\frac{\tau}{2}\right)\psi\  \  ' & = & SD_0\psi \equiv S\left(\frac{\pd}{\pd t} +
\frac{ig}{\hbar}A^{(0)}\cdot\frac{\tau}{2}\right)\psi \\
\vD\  \  ' \psi\  ' \equiv \left(\nabla - \frac{ig}{\hbar c}\vA\
'\cdot\frac{\tau}{2}\right)\psi\  '
 & = & S\vD\psi \equiv  S\left( \nabla - \frac{ig}{\hbar c}\vA\cdot\frac{\tau}{2}\right)\psi
\;.
\eqne This is obviously a sufficient condition for the invariance of Eq.(\ref{nasc}) and
determines the transformation properties of the gauge potentials
\beq  A^{(\mu)\prime}(\arg)\cdot\frac{\tau}{2} =
S(\arg)A^{(\mu)}(\arg)\cdot\frac{\tau}{2}S^{-1}(\arg) - \frac{i}{g}\left[\frac{\pd
S(\arg)}{\pd x_{(\mu)}}S^{-1}(\arg)\right] \; . \label{gt2}
\eeq  We see that  under a gauge transformation each  matrix gauge potential is locally
"rotated" at every space--time point by the gauge transformation matrix $S(\arg)$ and at the
same time it is shifted by an amount which depends on the corresponding  derivative of
$S(\arg)$.  This expression as well as the  relations above are valid for any unitary group
SU(N) with the  appropriate generalization of the transformation matrix $S$ and the Pauli
matrices $\tau^{(a)}$. For the abelian group U(1) we will obviously recover the known \Sch
equation and the gauge potentials of the electromagnetic field. In general there will be
$d\times(N^2 - 1 )$ gauge potentials with $d = 4$ -- the dimensionality of the space--time
and $(N^2 - 1)$  -- the number of the independent generators of the group SU(N).  The gauge
freedom expressed by (\ref{gt2}) means that in general only
$(d-1)\times(N^2 - 1)$  combinations of  the gauge potentials are independent. We finally
remark that among the gauge fields known in nature the unified electromagnetic and the weak
interactions (often called electroweak) are described by  $U(1)\times SU(2)$
 and the strong interactions by SU(3) non abelian gauge potentials. The intrinsic quantum
numbers for these interactions (analog of what we called pseudospin) are the standard
electric charge and the so called weak isospin and the color respectively. Since as already
mentioned the gravitational field is also a gauge field we have all four basic interactions
described by the  gauge  fields.

\vspace{1cm}
{\Large \bf Problems}
\begin{enumerate}
\item \label{sptr} Consider the spin  part of the time reversal operator
$U=\exp[i2\pi\, s_y/\hbar]$.
\begin{itemize}
\item Show that U = -1  when applied to the wave function of a spin 1/2 particle.
\item Components $\psi_{jm}$ of a spin wave function with a general spin j can be considered
as  far as their transformation properties are concerned as suitably chosen components of the
wave function of a system of 2j spin 1/2 particles. Use this to prove that $U \psi_{jm} =
(-1)^{2j}\psi_{jm}$.
\end{itemize}
\item Find the propagator in a uniform electric and magnetic fields by evaluating the
appropriate Gaussian  path integrals, cf., \cite{Kle} \label{prprmg}
\item  Electrons are confined to move in a plane $(x,y)$ and are placed in a uniform
magnetic  field  perpendicular to the plane.
\begin{enumerate}
\item Consider two different gauges choices  {\bf a)}  $A_x = - B y \ , A_y = A_z = 0$ and \
\ {\bf b)}   $A_x = -\frac{1}{2} By \ ,  Ay = \frac {1}{2} Bx \ , A_z =
0$.  electron  eigenfunctions  calculated in Which combinations of the guiding center
coordinates do they diagonalize?  How are these two sets of eigenfunctions related?
Calculate and explain the behaviour  of the current density in each
of the above cases.  What is the total current?  How the current
will change in the case  {\bf a)} above if one adds a uniform electric
field along the $x$ direction?  along the $y$ direction?  What is the
total current now?  In which direction does it flow?  In case {\bf b)}
above assume that a very thin solenoid with magnetic flux $\Phi$
is added along the $z$ axis (at $x = y = 0)$.  What and how will
it influence?  Consider your answer for various values of the
$\Phi$ and see if there are special values of $\phi$.
\end{enumerate}
In addition they are su
a potential $U = \frac{1}{2} \alpha x^2.$

{\bf a)}  How the Landau levels are changed by this potential?  What
is the current density in a single state of a Landau level as
compared to the case with $U=0?$

{\bf b)}  Now repeat for $U = \frac{1}{2} \alpha (x^2+y^2).$\\

2.  Show that the operators $x_o$ and $y_o$ of the guiding
center coordinates are generators of the translations in the
presence of the uniform magnetic field.

3.  Consider wave functions of Landau levels with definite
values of $x_o$.  How are  they  related to the similar wave
functions with $y_o$?  Consider now wave functions with definite
$x^2_o+y^2_0$.    Assume that a very thin solenoid with magnetic
flux $\phi  ?$

\item  Consider  quantum mechanics in a strong magnetic field.   

a.The projection on a  lowest
Landau level (LLL).  Show that the  eigenstates  of $x_0$ and $y_0$ are exactly equivalent to
eigenstates of p and x in a one dimensional quantum mechanics. 

 b.Derive  semiclassical
approximation in the limit of $\ell\to 0$. 

  c. Find semiclassical  energy levels of two
interacting electrons in 2 dimensions in a strong magnetic field.  Discuss also the  case of
oppositely charged particles (say an electron - positron  system   or electron-hole system in
a solid state).
\end{enumerate}
``

\chapter{Quantum Mechanics of  Electromagnetic Field}

In this Chapter we will show how the quantum mechanical description of one or several particles is extended to the quantization of electromagnetic field. In contrast to particles which are described by the  coordinates of their positions $\vecr_a$, $a=1, ..., N$ (N- the number of particles) the electromagnetic field is described by the configuration of the electric and magnetic fields $\vE(\vecr)$ and $\vB(\vecr)$. In order  to learn how such extended systems are treated in quantum mechanics we shall start with a much simpler system - that of a one dimensional string.

\section{Simple System First - Quantum Mechanics of a Guitar String}
\subsection{Classical string}
\begin{figure}[H]
\centering \includegraphics[width=0.8\textwidth]{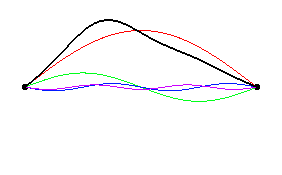}
\caption{Configurations of a guitar string. Denoting the abscissa of the figure (the equilibrium position of the string)  by x and the ordinate (the string deviations from the equilibrium) by $\phi$ the string configurations are described by a function $\phi(x)$. }
 \label{fig:string}
\end{figure}

We consider a string depicted in Fig. \ref{fig:string}
Classically its general configuration is  conveniently described  by a function $\phi(x)$ which  determines  the deviations $\phi$ of the string  from the equilibrium position $\phi=0$ at every point of the axis $x$. For simplicity we shall assume that the ends of the string are fixed at $x=0$ and $x=L$
\beq  \label{eq:bcforstring}
\phi(0)=0 \;\; , \;\; \phi(L)=0\;\; .
\eeq
In the following section we will extend our discussion to a more relevant example of a string with periodic boundary conditions - the so called closed string.

We will assume that classically the string is described by a simple linear wave equation
\beq \label{eq:stringeq}
\frac{\partial^2 \phi(x,t)}{\partial t^2}=v^2\frac{\partial^2 \phi(x,t)}{\partial x^2}
\eeq
where $v$ has dimensionality of velocity and is actually the phase (as well as group) velocity of the waves
$$
\phi(x,t)=A\sin[kx\pm\omega(k) t+\alpha]\;.
$$
These are solutions of the wave equation as can be easily verified by direct substitution. Here $A$ and $\alpha$ are arbitrary constant amplitude  and phase and the dispersion relation is
\beq \label{eq:stringdisprelation}
\omega(k)= v k \;.
\eeq
It will be very useful to  view the function $\phi(x)$ as a collection of the coordinates describing the "position" of the string. To emphasize this one might think of $\phi(x)$ as a set $\{\phi_x\}$ indicating  that $x$ is  actually an index numbering different coordinates. To make it even more precise the $x$  variable can be discretized and $\phi(x)$ reduced to  $N+1$ variables as follows
$$
\phi_x\equiv\phi(x=n\Delta x)\;\;, \;\; \Delta x =L/N \;\; , \;\; n=0,1, ..., N
$$
Formally one should at the end let $N\to \infty$, $\Delta x \to 0$ but in practice it is enough to have $\Delta x$ much smaller than  the smallest wave length $\Lambda$ of the waves which one intends to consider. The physical reasons behind the cutoff $\Lambda$   may actually be  the requirement  that it is much larger than the microscopic length scales related to say the distances between the constituents  of which the string  is built.

\subsection{Quantum description of the string}
\subsubsection{The wave functional}
Our goal is to quantize the classical string  as described above. We shall use the the straightforward generalization of the canonical quantization procedure for system with few degrees of freedom like one or several particles. This means that instead of having a definite $\phi(x)$ describing the string configuration we must assume that for each $\phi(x)$ there is a probability amplitude $\Psi[\phi(x)]$ which contains all the (quantum) information  about the string. In particular $|\Psi[\phi(x)]|^2$ gives the probability density to find a particular configuration $\phi(x)$.

Mathematically $\Psi[\phi(x)]$ represents a correspondence between the set of all functions $\phi(x)$ subject to the conditions 
Eq.\,(\ref{eq:bcforstring}) and a set of complex numbers $\Psi$.  Such a correspondence is  called a functional. Examples of functionals  should be familiar to the reader already from classical mechanics where the classical action $S[q(t)]$ is a functional of the trajectories $q(t)$.

The functional $\Psi[\phi(x)]$ is called the wave functional.  Using the approximate discretized form $\phi_x$ of the functions $\phi(x)$ the functional  $\Psi[\phi(x)]$ can actually be viewed  as a function of  $N-1$ variables $\{\phi_{x=n\Delta x}\}$.  The variables $\phi_0$ and $\phi_{x=N\Delta x}$ are fixed to $0$ to comply with Eq.\,(\ref{eq:bcforstring}).

Classically string dynamics is described by the time dependence $\phi(x,t)$ as governed by the equation (\ref{eq:stringeq}). Quantum mechanical time evolution should be described by the time dependence of the wave functional $\Psi[\phi(x),t]$. What governs this time evolution? Continuing the analogy with the few degrees of freedom system this should be the \Sch equation
\beq \label{eq:stringScheq}
i\hbar\frac{\partial \Psi[\phi(x),t]}{\partial t}=H_{\rm op}\Psi[\phi(x),t]
\eeq
with   $H_{\rm op}$ the Hamiltonian operator of the string. We will now determine this operator following the standard route.

\subsubsection{The string Hamiltonian}
We will start  by finding the classical Hamiltonian function of the string. For this we shall  rewrite the string equation (\ref{eq:stringeq}) in the Hamiltonian form. It is useful to  note that this equation represents a set of coupled  Newton equations for the string coordinates $\phi_x$. This can be seen by rewriting it in a discretized form
\beq \label{eq:strinNewtoneq}
\frac{d^2 \phi_x}{d t^2}=\frac{v^2}{\Delta x^2}\left( \phi_{x+\Delta x}-2\phi_x+\phi_{x-\Delta x}\right)
\eeq
where we used the  discretized form of the second derivative
$$
\frac{\partial^2 \phi(x)}{\partial x^2} \to \frac{1}{\Delta x}\left(\frac{\phi_{x+\Delta x}-\phi_x}{\Delta x}-\frac{\phi_x-\phi_{x-\Delta x}}{\Delta x}\right)
$$

We will rewrite the second order in time wave (Newton) equation of the string  as a pair of first order equations
\beq \label{eq:pairforstring}
\frac{\partial \phi(x,t)}{\partial t}=\pi(x,t) \;\;\;, \;\;\; \frac{\partial \pi(x,t)}{\partial t}=v^2 \frac{\partial^2 \phi(x,t)}{\partial x^2}
\eeq
where (as usual) the first equation is actually the definition  of the momenta. As a next step let us introduce the following functional
\beq \label{eq:stringHamiltonian1}
  H[\pi(x), \phi(x)]=\int_0^L dx\left[\frac{1}{2}\pi^2(x)+\frac{v^2}{2}\left(\frac{\partial \phi(x)}{\partial x}\right)^2\right]
\eeq
Using it we can write the pair (\ref{eq:pairforstring}) as
\beq \label{eq:stringHameqs}
\frac{\partial \phi(x,t)}{\partial t}=\frac{\delta H[\pi(x,t), \phi(x,t)]}{\delta \pi(x,t)} \;\;\; , \;\;\; \frac{\partial \pi(x,t)}{\partial t}=-\frac{\delta H[\pi(x,t), \phi(x,t)]}{\delta \phi(x,t)}\;\;.
\eeq
Here the notation $\delta/\delta \pi(x,t)$ and $\delta/\delta \phi(x,t)$ stands for the variational derivative (see below) with respect to $\pi(x,t)$ and $\phi(x,t)$ respectively.  We now show that the above equations are indeed equivalent to the pair (\ref{eq:pairforstring}) recalling in passing how the variational derivatives are defined and calculated.

We let   the functions $\pi(x)$ and $\phi(x)$ in the functional (\ref{eq:stringHamiltonian1}) to have  infinitesimal variations $\delta\pi(x)$ and $\delta \phi(x)$. The corresponding   variation    $\delta H$ due to this is
\eqna
\delta H&\equiv&  H[\pi(x)+\delta \pi(x), \phi(x)+\delta \phi(x)]- H[\pi(x), \phi(x)] = \nonumber \\
&=&\int_0^L dx \left[\pi(x)\delta \pi(x) +v^2 \frac{\partial \phi(x)}{\partial  x}\frac{\partial \delta \phi(x)}{\partial x}\right]\;\;  + \;\; {\rm higher\;\; order\;\; terms} \nonumber \\
&=&\int_0^L dx \left[\pi(x)\delta \pi(x) - v^2 \frac{\partial^2 \phi(x)}{\partial  x^2}\delta \phi(x)\right] \;\; + \;\; {\rm higher\;\; order\;\; terms}\nonumber
\eqne
where we  used integration by parts in the second term.

In analogy with the relation of the differential $dF$ of a function of many variables $F(q_1,q_2, ..., q_N)$ and its partial derivatives
$$
dF=\sum_{n=1}^N\frac{\partial F}{\partial q_n} dq_n
$$
the functional derivatives of $H[\pi(x,t), \phi(x,t)]$ are by definition the functions which multiply $\delta \pi(x,t)$ and $\delta \phi(x,t)$ respectively in the expression for the variation $\delta H$,
\beq
\frac{\delta H[\pi(x), \phi(x)]}{\delta \pi(x)}=\pi(x) \;\;\; , \frac{\delta H[\pi(x), \phi(x)]}{\delta \phi(x)}=
-v^2\frac{\partial^2 \phi(x)}{\partial  x^2}
\eeq
Inserting these relations into Eq.\,(\ref{eq:stringHameqs}) we see that they indeed reproduce Eq.\,(\ref{eq:pairforstring}).

Equations (\ref{eq:stringHameqs}) have the Hamiltonian form with $H[\pi(x), \phi(x)]$ as the Hamiltonian and $\pi(x)$, $\phi(x)$ as the momenta and coordinates. It should perhaps be  more clear if for a moment we use the  notation $\pi_x$ and $\phi_x$ instead of $\pi(x)$ and $\phi(x)$  treating $x$ as a label. The equations (\ref{eq:stringHameqs}) in these notations are
$$
\frac{\partial \phi_x(t)}{\partial t}=\frac{\partial H[\pi_x(t), \phi_x(t)]}{\partial \pi_x(t)} \;\;\; , \;\;\; \frac{\partial \pi_x(t)}{\partial t}=-\frac{\delta H[\pi_x(t), \phi_x(t)]}{\partial \phi_x(t)}\;\;.
$$

\subsubsection{Basic quantum operators for the string}
We shall now proceed to define the quantum mechanical operator $H_{\rm op}$. We will do this by first determining what are the operators corresponding to $\pi(x)$ and $\phi(x)$ and  then replacing with them the latter in the expression (\ref{eq:stringHamiltonian1}) for the classical Hamiltonian.

Since in our formulation $\phi(x)$ are the coordinates of the string the corresponding operator $\phi_{\rm op}(x)$ should be  just the operator of multiplication by $\phi(x)$, i.e. its action on an arbitrary wave functional is
\beq 
\label{eq:string_op_phi}
\phi_{\rm op}(x)\Psi[\phi(x')]=\phi(x)\Psi[\phi(x')]
\eeq
To avoid confusion we use different arguments of $\phi$'s in the operator and in $\Psi$. This would perhaps be easier to understand if (again momentarily) we shall switch to the notation $\phi_{x'}$ instead of $\phi(x')$. Then the functional $\Psi[\phi(x')]$ is just a function $\Psi(\{\phi_{x'}\})$ of the set of  all $\phi_{x'}$ variables. The action of the operator $\hat{\phi}_x$, i.e. the operator of the $x-{\rm th}$ component the coordinates of the string is just a multiplication by $\phi_x$ with this particular $x$.   Note that in order  to avoid the double subscript  we here used $\hat{\phi}$ to denote the operator.

 In the same way we can determine the operator corresponding to the momentum $\pi(x)$. In the "simplified" notations it should
  be $\hat{\pi}_x=-i\hbar\partial/\partial \phi_x$ which means that in terms of the functional derivatives it is
 \beq
 \pi_{\rm op}(x)\Psi[\phi(x')]=-i\hbar\frac{\delta}{\delta \phi(x)}\Psi[\phi(x')]
 \eeq
 We note that the commutator of the basic operators is
 \beq \label{eq:paiphicommutator}
 [\phi_{\rm op}(x), \pi_{\rm op}(y)]=i\hbar \delta(x-y)
 \eeq
 This can verified by acting with the commutator  on an arbitrary wave functional
 $$
 [\phi_{\rm op}(x), \pi_{\rm op}(y)]\Psi[\phi(x')]=-i\hbar\left(\phi(x)\frac{\delta}{\delta\phi(y)} \Psi[\phi(x')]-
 \frac{\delta}{\delta\phi(y)}\phi(x)\Psi[\phi(x')]\right)=
 $$
 $$
 =-i\hbar\left(\phi(x)\frac{\delta}{\delta\phi(y)} \Psi[\phi(x')]-\frac{\delta\phi(x)}{\delta\phi(y)}\Psi[\phi(x')]
 -\phi(x)\frac{\delta}{\delta\phi(y)}\Psi[\phi(x')]\right)= $$
 $$
 =i\hbar \delta(x-y)\Psi[\phi(x')]
 $$
 Substituting the operators $\phi_{\rm op}(x)$  and $\pi_{\rm op}(x)$ in the Hamiltonian (\ref{eq:stringHamiltonian1}) we obtain
 \beq \label{eq:stringHamiltonian2}
  H_{\rm op}=\int_0^L dx\left[\frac{1}{2}\pi_{\rm op}^2(x)+\frac{v^2}{2}\left(\frac{\partial \phi_{\rm op}(x)}{\partial x}\right)^2\right]
\eeq

Armed with this explicit form of the Hamiltonian operator of our system we can proceed to solve the \Sch equation (\ref{eq:stringScheq}).  Since the Hamiltonian is time independent it will be sufficient to solve the stationary equation
\beq
H_{\rm op}\Psi=E\Psi
\eeq
Knowing all its solutions will allow to find the most general solution of (\ref{eq:stringScheq}).

The string Hamiltonian operator (\ref{eq:stringHamiltonian1}) may look formidable but is actually quite simple because  of its  quadratic dependence on the coordinates and momenta. This of course  is a direct consequence of the linearity of the string equation (\ref{eq:stringeq}). As is seen from the discretized form (\ref{eq:strinNewtoneq}) such equations describe coupled harmonic oscillators. The standard way of solving such problems is to make a transformation to normal modes.

 \subsection{Reminder - normal modes of vibrations}
 Let us recall how the transformation to normal modes is done in the general context represented by the set of N coupled equations
\beq \label{eq:coupledosc1}
m_l\ddot{q}_l=-\sum_{n=1}^N k_{ln}q_n \;\;, l = 1, ..., N
\eeq
with masses $m_l$ and $N$ by $N$ symmetric matrix of elastic constants $k_{ln} =k_{nl}$. For simplicity we shall assume in the following that  all the masses are equal $m_1=...=m_N=m$. The Hamiltonian of this problem is the standard sum of the kinetic and potential energies
\beq \label{eq:Haminexample}
H=\frac{1}{2m}\left[\sum_{l=1}^N p^2_l+ \sum_{l,n=1}^N mk_{ln}q_lq_n\right]
\eeq
Let us try the following  solution of the equations (\ref{eq:coupledosc1})
$$
q_l=Re(C_l e^{i\omega t}), \;\; l=1, ..., N
$$
where $Re$  stands for real part and  $C_l$'s are constants. This form assumes that all the degrees of freedom vibrate with the same frequency. Inserting this into the equations (\ref{eq:coupledosc1}) we obtain
\beq \label{eq:coupledosc2}
\sum_{n=1}^N (k_{ln} - m\omega^2 \delta_{ln})C_n = 0
\eeq
where we remind that we set for simplicity all $m_i=m$.
  To have a non trivial solution one must demand that
\beq \label{eq:coupledosc3}
\det(k_{ln}-m\omega^2\delta_{ln}) = 0
\eeq
which shows that $\omega^2$ is an eigenvalue of the matrix $k_{ln}/m$ which in turns means that in general one will have $N$ such solutions which will have  $\omega^2>0$ provided $k_{ln}$ is positive definite.

 Let us denote by $\omega_{\nu}$ and $\{C_n^{\nu}\}$ the set of N solutions of Eq.(\ref{eq:coupledosc2}). The symmetry of $k_{ij}$ assures orthogonality of the eigenvectors $\{C_n^{\nu}\}$'s  with different eigenvalues $\omega_\nu$. For a degenerate case i.e. if  some $\omega_\mu = \omega_\nu$ one has a freedom to choose $\{C_n^{\mu}\}$ and    $\{C_n^{\nu}\}$ to ensure that orthogonality holds also in this case.  We also note that since the equations (\ref{eq:coupledosc2}) are homogeneous at least one of the components in a given vector $\{C_n^{\nu}\}$ is arbitrary and can be used to set normalization of $\{C_n^{\nu}\}$'s to unity. We thus have orthonormality
$$
\sum_{n=1}^N C_n^{\mu} C_n^{\nu}=\delta_{\mu\nu}
$$
The N vectors $\{C_n^{\nu}\}$ each with N components form an $N \;\times \;N$ matrix. The orthonormality conditions (together with completeness which we do not discuss) mean that this matrix is orthogonal (unitary for complex $C_n$'s). Let  us use it to make the transformation to new coordinates
\beq \label{eq:transtonorm}
q_n=\sum_{\nu=1}^N C_n^{\nu} Q_\nu
\eeq
Inserting this in Eqs. (\ref{eq:coupledosc1}) (with $m_i=m$) and using Eqs. (\ref{eq:coupledosc2}) one obtains
\beq
\sum_{\nu=1}^N [m C_l^{\nu} \ddot{Q_\nu}+\sum_{n=1}^N k_{ln} C_n^{\nu} Q_\nu]=\sum_{\nu=1}^N m C_l^{\nu}[ \ddot{Q_\nu}+ \omega_\nu^2 Q_\nu]=0
\eeq
Due to orthogonality of $C$'s one finds that the equations for $Q_{\nu}$'s are decoupled. Indeed multiplying by $C_l^{\mu}$ and summing over $l$ one obtains
\beq
\ddot{Q_\mu}+ \omega_\mu^2 Q_\mu=0 \;\; , \;\; \mu=1,...,N \; .
\eeq
The transformation (\ref{eq:transtonorm}) from the original coordinates $q_l$ to the new $Q_\nu$ is called the transformation to normal modes and $Q_\nu$ -- the normal mode coordinates.

How does the Hamiltonian look in the new coordinates? To answer this we need to add to (\ref{eq:transtonorm}) also the transformation to the corresponding normal modes momenta.  Since in our case $p_l=m\dot{q}_l$ it is clear that  $p$'s transform like $q$'s
\beq
p_n=\sum_{\nu=1}^N C_n^{\nu} P_\nu
\eeq
Inserting this and (\ref{eq:transtonorm}) into  the Hamiltonian (\ref{eq:Haminexample}) we obtain  using the orthonormality of $C_n^{\mu}$'s and equations (\ref{eq:coupledosc2})
\beq
H=\frac{1}{2 }\sum_{\mu=1}^\infty[ P_\mu^2 + \omega_\mu^2 Q_\mu^2]
\eeq
where for simplicity we have set $m=1$. In normal mode variables the Hamiltonian is just a collection of independent oscillators.

\subsection{String as a collection of decoupled oscillators}

\subsubsection{Normal modes of the guitar string}
We will now use the technique described in the previous subsection to transform the string Hamiltonian to a collection of independent oscillators.  We are looking for the analog of the transformation (\ref{eq:transtonorm}) from the string coordinates $\phi(x)$ to the normal modes coordinates. Since $x$ here plays the role of the index $n$ in $q_n$ the analog of the matrix $C_n^\nu$ of the transformation  should be functions of $x$ defined on the interval $0\le x \le L$. We denote the set of these  functions by $u_n(x)$ and write
\beq
\phi(x)=\sum_\nu u_\nu (x) Q_\nu \;\; , \;\; u_\nu (0)=u_\nu (L)=0
\eeq
where we  indicated that $u_\nu (x)$ should vanish at the end points of the string to assure the  boundary conditions (\ref{eq:bcforstring}).

 By comparing the string equation (\ref{eq:stringeq}) and the coupled oscillators equation (\ref{eq:coupledosc1}) we see that the role of the coupling matrix $k_{ij}$ is played by $-v^2\partial^2/\partial x^2$ so the functions $u_n(x)$ must satisfy (cf., Eq. (\ref{eq:coupledosc2}))
\beq \label{eq:stringnormalmodeseq}
-v^2\frac{\partial^2}{\partial x^2}u_\nu(x)=\omega^2 u_\nu(x)\;\; , \;\; u_\nu (0)=u_\nu (L)=0
\eeq
which has orthonormal eigenfunction solutions
\beq \label{eq:stringnormmodes}
u_\nu(x)=\sqrt{\frac{2}{L}}\sin k_\nu x  \;\; ,\;\;\ k_\nu =\frac{\pi \nu}{L}\;\;,\;\; \nu=1, 2, ....
\eeq
with eigenvalues
\beq \label{eq:stringdisprel}
\omega_\nu=vk_\nu
\eeq
We have fixed the coefficients in $u_\nu(x)$'s  so that these functions are normalized.

To conclude - the functions (\ref{eq:stringnormmodes}) represent the configurations of the string normal modes in which all the points of the string oscillate with the same frequency $\omega_\nu$ which depends on the wave number $k$ of the mode, Eq. (\ref{eq:stringdisprel}).  We remind that in general the  relation $\omega=\omega(k)$ of the frequency upon the wave vector is called the dispersion relation. It is the most important characteristic of linear waves. 

\subsubsection{Quantum mechanics of string normal modes}

Using $u_\nu(x)$'s  we can transform the Hamiltonian operator (\ref{eq:stringHamiltonian2}) of the string to a sum of independent oscillators.  We view  the operators $\phi_{\rm op}(x)$ and $\pi_{\rm op}(x)$ as functions of $x$ and expand
\beq \label{eq:expansionofphiandpai}
 \phi_{\rm op}(x)=\sum_{\nu=1}^\infty \sqrt{\frac{2}{L}}\sin k_\nu x \; \hat{Q}_\nu \;\;,\;\;\pi_{\rm op}(x)=
 \sum_{\nu=1}^\infty \sqrt{\frac{2}{L}}\sin k_\nu x \; \hat{P}_\nu
 \eeq
Since we expand operator valued functions the coefficients $\hat{Q}_\nu$ and $\hat{P}_\nu$ here are operators which we denoted by hats above to avoid double subscripts.

There are important relations which these operators must satisfy in order to preserve the canonical commutation relations (\ref{eq:paiphicommutator}) between $\phi_{\rm op}(x)$ and $\pi_{\rm op}(x)$. These operators  must themselves be canonical, i.e. they must obey
\beq \label{eq:stringcommutatorsforPandQ}
[\hat{Q}_\mu,\hat{P}_\nu]=i\hbar\delta_{\mu\nu} \;\;, \;\; [\hat{Q}_\mu,\hat{Q}_\nu]=[\hat{P}_\mu,\hat{P}_\nu]=0
\eeq
This can be verified in one of the two ways. We can insert the expansions (\ref{eq:expansionofphiandpai}) in
$[\phi_{\rm op}(x), \pi_{\rm op}(y)]$.  Using the first commutator above and the completeness relation $\sum_\nu u_\nu(x) u_\nu(y)=\delta(x-y)$ we will obtain that (\ref{eq:paiphicommutator}) is indeed satisfied. The other two commutators simply assure that $\phi_{\rm op}(x)$ and $\pi_{\rm op}(x)$ commute at different points.   The other way is to "invert" (\ref{eq:expansionofphiandpai})
\beq
\label{eq:invert_phi_pi_to_Q_P}
\hat{Q}_\nu=\int_0^L \phi_{\rm op}(x)\sqrt{\frac{2}{L}}\sin k_\nu x\; dx \;\;\;, \;\;\; \hat{P}_\nu = \int_0^L \pi_{\rm op}(x)\sqrt{\frac{2}{L}}\sin k_\nu x \;dx
\eeq
and calculate the needed commutators. Incidentally the above relations also demonstrate how the operators $\hat{Q}_\nu$ and $\hat{P}_\nu$ should act on wave functionals $\Psi[\phi(x)]$.

 The commutation relations (\ref{eq:stringcommutatorsforPandQ}) mean that $\hat{Q}_\mu$ and $\hat{P}_\mu$ are respectively coordinate and momentum operators of the normal modes of the string. As can be seen from (\ref{eq:expansionofphiandpai}) classically they are coordinates  and momenta representing the amplitudes and their velocities  of all the harmonic  standing waves which the string can support.
 
Inserting the above expansions in (\ref{eq:stringHamiltonian2}) and using the orthonormality property of the set (\ref{eq:stringnormmodes}) we obtain
  \beq  \label{eq:normalmodes_H}
   H_{\rm op}=\frac{1}{2}\sum_{\nu=1}^\infty(\hat{P}_\nu^2 + \omega_\nu^2 \hat{Q}_\nu^2)
  \eeq
 The Hamiltonian operator is reduced to a sum of terms each representing simple Harmonic oscillator  with unit mass and frequency $\omega_\nu$.  It is important to observe that the underlying waves on the elastic string  can only be seen in the dependence of $\omega$ of the oscillators on the corresponding wave vectors encoded in the dispersion relation (\ref{eq:stringdisprel}).
 
Classically the simple form (\ref{eq:normalmodes_H}) of the Hamiltonian in terms of the normal modes' dynamical variables suggests to switch the string description from $\phi(x,t), \pi(x,t)$ to the set $\{Q_\nu(t), P_\nu(t)\}$. Quantum mechanically we note that the relation (\ref{eq:string_op_phi}) and the first of (\ref{eq:invert_phi_pi_to_Q_P}) implies that the operators $\hat{Q}_\nu$ are simple multiplication operators 
$$
\hat{Q}_\nu  \Psi[\phi(x)] = Q_\nu \Psi[\phi(x)]
$$
Following the commutation relations (\ref{eq:stringcommutatorsforPandQ}) the  canonically conjugate operators $\hat{P}_\nu$ can be taken as
$$
\hat{P}_\nu  \Psi[\phi(x)] = -i\hbar \frac{\partial}{\partial Q_\nu} \Psi[\phi(x)]
$$
This suggests to switch to the description in which wave functionals $\Psi[\phi(x)]$ are viewed as functions of (formally infinite number of) the variables $Q_\nu$
$$
\Psi[\phi(x)]  \to \Psi(\{Q_ \nu\}) 
$$

\begin{figure} [H]
\centering \includegraphics[width=0.8\textwidth]{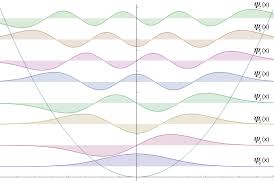}
\caption{Energy levels and corresponding wave functions of harmonic oscillator. The energy levels are "equidistant", separated by equal energy intervals}
 \label{fig:harm_osc}
\end{figure}

We now note that the terms in the sum representing $H_{\rm op}$, Eq. (\ref{eq:normalmodes_H}) commute  between themselves on account of the last pair of commutators in  Eq. (\ref{eq:stringcommutatorsforPandQ}). This means that the eigenfunctions of $H_{\rm op}$ are products of the eigenfunctions  of all individual terms in the sum  and the corresponding eigenvalues are  sums of individual eigenvalues.
 \beq \label{eq:eigenfn_of_string}
 \bes
 E_{\{N_\nu\}}=\sum_{\nu=1}^\infty \hbar\omega_\nu(N_\nu + \frac{1}{2}) \; = \;  & E_{\rm ground \; state}    + \sum_{\nu=1}^\infty \hbar\omega_\nu N_\nu  \;, \;N_\nu=0, 1, 2, . . .   \\
 E_{\rm ground \; state} = \sum_{\nu=1}^\infty \frac{\hbar\omega_\nu}{2}  \;  , \;  \Psi_{\{N_\nu\}}(\{Q_ \nu\}) & =\prod_{\nu=1}^\infty \psi_{N_\nu}(\beta_\nu Q_\nu)\; ,  \; \beta_\nu =\sqrt{\omega_\nu/\hbar} 
 \end{split}
 \eeq
 The  eigenfunctions $\psi_{N}$ are  the well known harmonic oscillator eigenfunctions
 \beq \label{eq:harm_osc_wfs}
 \psi_N(y)=\frac{1}{\sqrt{2^N N!}}\left(\frac{\omega}{\pi \hbar} \right)^{1/4} e^{- y^2/2}H_N (y) 
 \eeq
 where $H_N(y)$ denotes N-th order Hermite polynomial. In Fig. \ref{fig:harm_osc} graphs of several of these functions are shown. 

The  energies (\ref{eq:eigenfn_of_string}) exhibit the most important result of our discussion of string quantization -  that it can be viewed as a collection of independent quanta with energies $\hbar \omega_\nu=\hbar v k_\nu$. This is a consequence of two  general features - string is a linear dynamical system and therefore a collection of normal modes oscillators while the quantum energy levels of oscillators are "equidistant", i.e. separated by equal energy intervals $\hbar\omega$. 

 In the next Section we will consider a closed string which can support not only normal modes in a form of standing waves like the present fixed end string but also traveling waves.  We will show that the corresponding energy quanta of such modes carry mechanical momentum and could therefore be considered as particles.     
 
\subsubsection{String ground state. The Casimir effect \label{sec:string_Cas_effect}} 
The ground state energy in (\ref{eq:eigenfn_of_string}) is formally a sum of infinitely many "zero point motion" terms.  This is an "ultraviolet" infinity related to the formal possibility to have waves with $k_\nu \to \infty$, i.e. vanishingly small wavelengths $\lambda_\nu= 2\pi/k_\nu$. In practice of course the simple description given by Eq. (\ref{eq:stringeq}) ceases to be valid at atomic scales and should be replaced by a more elaborate model. As a (much more practical) alternative one could introduce a cutoff $k_{cutoff}$ for "allowed" normal modes in the model and limit the validity of the model (\ref{eq:stringeq}) to scales $\gg \lambda_{cutoff} =2\pi/k_{cutoff}$. 

Denoting by $\nu_c$ the largest integer $\nu$ corresponding to the $ k_{cutoff} =\pi\nu_c/L$ the ground state wavefunction is a product of $\nu_c$ Gaussians 
\beq \label{gr_st_of_string}
\bes
 \Psi_{\rm ground \; state }(\{Q_ \nu\}) &=\prod_{\nu=1}^{\nu_c} \psi_0(\beta_\nu Q_\nu) = \\
 &=\prod_{\nu=1}^{\nu_c} \left(\frac{\omega_\nu}{\pi \hbar} \right)^{1/4}  \exp\left(-\sum_{\nu=1}^{\nu_c}\omega_\nu Q_\nu^2/2\hbar \right)
 \end{split} 
 \eeq
which express the "zero point" fluctuations of the quantum string which is not at rest even in its lowest energy state.   

As we will see in the forthcoming sections the ground state of the EM field is expected to exhibit similar zero point fluctuations of the fields in 
its ground state which is the vacuum of the theory. Is it possible to observe these vacuum fluctuations?  In a 1948 famous paper Ref.\,\cite{Cas} Casimir proposed a way to do this using what has become known as a Casimir effect. We will now explain its principle idea in the simple example of the guitar string ground state.

\begin{figure}[H]
\centering \includegraphics[width=0.5\textwidth]{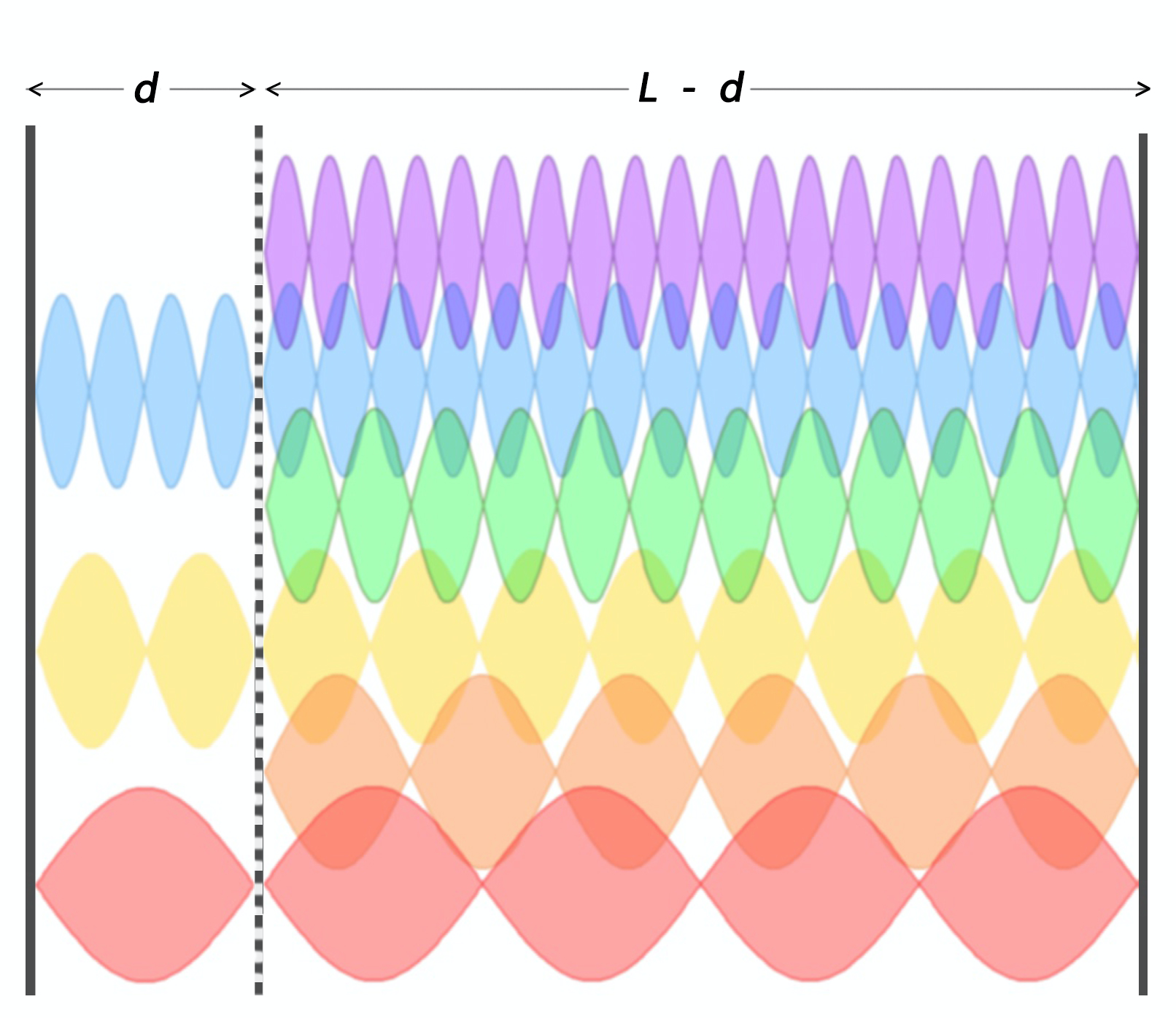}
\caption{Guitar string fretted at $x=d$ and its (schematically drawn) normal modes}
 \label{fig:Casimir_in1D}
\end{figure}

Let us consider what will happen with the quantum guitar string if we "fret" it, i.e. press with an imaginary finger hard at some position $x=d$ so that the string will not vibrate at this point, Fig.\,\ref{fig:Casimir_in1D}.  Obviously this changes the normal modes of the string by excluding the modes which do not vanish at $x=d$.  This means that the ground  state  energy density will change.  In fact the new normal modes will consist of two families with $\omega^{\prime}_\nu = vk_\nu = v\pi \nu / d$ and $\omega^{\prime\prime}_\nu = v \pi \nu / (L-d)$. The corresponding ground state energy will correspondingly consist of two parts 
\beq
E_0(d)= \sum_{\nu = 1}^{\nu_c}  \frac{\hbar v\pi }{2} \left[\frac{\nu}{d} + \frac{\nu}{L-d}\right]
\eeq
It is clear that for $d=L/2$ both parts are equal while for $d < L/2$ ($d > L/2$) the first term, i.e. the energy of the narrower (wider) part is smaller (larger) than the second term.   

The finite cutoff frequency $\sim \nu_c$ in the above expression "regularizes"  the (ultraviolet) divergence of the sum $\sum_\nu \nu$.
To eliminate $\nu_c$ from the final result one must "renormalize" it which can be done, cf., Ref.\,\cite{Boy}, by calculating $E_0(d)$ relative to the symmetric configuration at $d=L/2$ with the result\footnote{The common way of calculating is to use a "soft" cutoff, i.e. to replace  e.g. $\sum_{\nu=1}^{\nu_c} \nu$ by $\sum_{\nu=1}^\infty \nu e^{-\nu/\nu_c}$, calculate the last sum for 
$\nu_c \to \infty$ using
$$
\sum_{\nu=1}^\infty \nu e^{-\alpha\nu} = - \frac{\partial}{ \partial \alpha} \sum_{\nu=1}^\infty e^{-\alpha\nu}  =  \frac{\partial}{ \partial \alpha}
\frac{1}{1-e^\alpha} \underset{\alpha\to 0} {\longrightarrow} \frac{1}{\alpha^2} -\frac{1}{12} + ...
$$
with $\alpha = 1/\nu_c $}
$$
\Delta E_0(d) \equiv E_0(d) - E_0(L/2) = -   \frac{\pi \hbar v }{24}\left( \frac{1}{d}  + \frac{1}{L-d} - \frac{4}{L}  \right)  
$$
 It is seen that $\Delta E_0(d)$ is symmetric with respect to $d \to L-d$ and decreases monotonically as $d \to 0$ and $d \to L$
 as
 \beq
 \Delta E_0(d)|_{d\ll L/2}  \to  -   \frac{\pi \hbar v }{24}\frac{1}{d}  \;\;  , \;\;  \Delta E_0(d)|_{L-d\ll L/2}  \to  -   \frac{\pi \hbar v }{24}\frac{1}{L-d}
\eeq
 The dependence on $d$ means that the function $E_0(d)$ can be considered as the potential energy of the separation point of the string at $x=d$ and that there is a force 
$$
F(d) = -\frac{\partial E_0(d)}{\partial d} 
$$
acting on what causes the separation between the two parts of the string (the imaginary fretting finger). This force "tries to drive" the separation towards the end points of the string.  A simple physical intuition behind this force is the imbalance of the ground state fluctuations radiation pressure on both sides of the separation point $x=d$ when $d\ne L/2$.  One must be aware however that  things are more delicate as the sign of the force depends on the type of boundary conditions assumed at $x=d$.  For details cf., Ref. \cite{Boy}.
 
The force $F(d)$ is called the Casimir force and its appearance is a manifestation of a Casimir effect. We will return to this effect below in the context of the vacuum fluctuations of the quantized EM field. 

\section{Quantization of Traveling Waves. Closed String}
The normal modes of a string were described in the previous section are standing waves as can be most clearly seen
by considering the  expansions  (\ref{eq:expansionofphiandpai}). Classically $\hat{Q}_\nu$'s and $\hat{P}_\nu$ are functions $Q_\nu(t),
 P_\nu(t)$  each depending harmonically on time with frequency $\omega_\nu$. Thus each term in (\ref{eq:expansionofphiandpai}) is a standing wave $\sim Q_\nu(t)\sin k_\nu x$ and $\sim P_\nu(t)\sin k_\nu x$.

 \subsection{Expansion in traveling waves}
 \subsubsection{Periodic boundary conditions}
 The standing wave solution of the equations (\ref{eq:stringnormalmodeseq}) defining the normal modes is a consequence of the fixed ends boundary conditions (\ref{eq:bcforstring}) for the guitar string.  These were reflected in the boundary conditions $u_\nu(0)=u_\nu(L)=0$ in  the normal modes equations Eqs. (\ref{eq:stringnormalmodeseq}). In this section we will explore a more interesting and practical situation  when the normal modes are traveling waves.  This is realized if one assumes periodic boundary conditions for a string, i.e. for every $x$
\beq
\phi(x,t)=\phi(x+L,t)
\eeq
This effectively means that such string does not have ends,  i.e. it is closed  and equivalent to a circle.
Note in passing that  differentiating the periodicity condition with respect to $x$ shows that also the derivatives $\pd \phi/\pd x$, $\pd^2 \phi/\pd x^2$, etc of $\phi$  are periodic. According to the string equation (\ref{eq:stringeq}) this means that so are the time derivatives.

The periodic boundary conditions for $\phi(x)$ are translated into conditions
$$
u_\nu(x)=u_\nu(x+L)
$$
 replacing the fixed ends conditions in the normal modes equation (\ref{eq:stringnormalmodeseq}). The solutions are now
coming as an infinite set of degenerate pairs each with the same frequency
\beq
\sqrt{\frac{2}{L}}\sin k_\nu x \;\; , \;\; \sqrt{\frac{2}{L}}\cos k_\nu x \;\;, \;\;  k_\nu=\frac{2\pi \nu}{L} \;\; , \;\; \omega_\nu=v k_\nu\;\; \; , \;\; \nu= 1, 2, ...
\eeq
This degeneracy is "compensated" by the the values of $k_\nu$ being at twice larger intervals $\Delta k = 2\pi/L$  than in the standing wave solutions (\ref{eq:stringnormmodes}) with $\Delta k = \pi/L$.   
In real space  this means that the normal modes of the closed string  have integer  number 
$L/\lambda_\nu =\nu$ of the wavelengths $\lambda_\nu =2\pi/k_\nu =L/\nu$ over the string length $L$ rather than integer number of half wavelengths $\lambda_\nu= 2L/\nu$ as it was in the fixed ends string case. 

Solutions belonging to different frequencies are automatically orthogonal and we chose them to be orthogonal  also within each degenerate pair.  Here is a helpful integral 
\eqnaa
&&\int_0^L \sin k_\nu x \cos k_\mu x dx =\int_0^L \frac{1}{2}\left[\sin(k_\nu+k_\mu)x + \sin(k_\nu-k_\mu)x\right] dx = \\
 && \;\;\;\;\;\;\;\;\;\;\;\;\;\;\; = -\frac{1}{2}\left[\frac{1}{k_\nu+k_\mu}\cos(k_\nu+k_\mu)x \Big\vert_0^L  +  \frac{1}{k_\nu-k_\mu}\cos(k_\nu-k_\mu)x\Big\vert_0^L\right] = 0
\eqnae
We also normalize them as in the fixed ends case. 

As always with degeneracies the choice above is of course not unique.  Another useful possibility  is
 \beq
 \frac{1}{\sqrt{L}}e^{ik_\nu x}\;\; , \;\; k_\nu=\frac{2\pi \nu}{L} \;\; , \;\;\;  \omega_\nu=v |k_\nu| \;\;\; , \;\;\; \nu= \pm 1, \pm 2, ...
 \eeq
 again with degenerate in frequency orthogonal pairs. We will see the results of such a choice below, cf., 
 Eq.(\ref{eq:expansion_in_a_a_dagger}).  In the following sections we will use the freedom in specifying the degeneracy of normal modes in a closed string to find the traveling waves expansion.

Let us also note that a non vibrating constant solution  $u_0(x) = {\rm const}$ exists with $\omega=0$. This means that the string configuration in this mode is constant independent of $x$ and has linear time dependence $\phi(x,t)=at+b$.  It describes a uniformly moving string  and plays an important role in the so called super string theory.   It will not be of interest to us and will not be included in our considerations.

 \subsubsection{Trying the simplest expansion}

 The most natural expansion using the above degenerate modes would be
 \beq \label{eq:stringexpinstandwaves}
 \bes
 \phi(x,t)&=\sqrt{\frac{2}{L}}\sum_{\nu=1}^\infty \left[ Q_{1,\nu}(t) \sin k_\nu x  + Q_{2,\nu}(t) \cos k_\nu x \right] \\
  \pi(x,t)&=\sqrt{\frac{2}{L}}\sum_{\nu=1}^\infty \left[ P_{1,\nu}(t)  \sin k_\nu x + P_{2,\nu}(t) \cos k_\nu x \right]  
\end{split}
 \eeq
with two independent sets of amplitudes $\{Q_{1,\nu}(t) , P_{1,\nu}(t) \}$ and  $\{Q_{2,\nu}(t) , P_{2,\nu}(t) \}$  for the two degenerate modes. Inserting this in the string equation and projecting each mode leads to the decoupled equations 
\beq
\ddot{Q}_{i,\nu}+\omega_\nu^2 Q_{i,\nu}=0\;\;\; ; \;\;  i=1,2 \;\; ; \;\; \nu= 1, 2, ... 
\eeq
and the corresponding Hamiltonian 
$$
H=\sum_{i=1,2}\sum_{\nu=1}^\infty H_{i,\nu} \;\; , \;\;  H_{i,\nu} = \frac{1}{2}\left(P_{i,\nu}^2 +\omega_\nu Q_{i,\nu}^2\right)
$$
The general solution
\beq \label{eq:solution_for_Q}
\bes
Q_{i,\nu}(t)&=Q_{i,\nu}(0)\cos\omega_\nu t + P_{i,\nu}(0)/\omega_\nu\sin\omega_\nu t   \\
P_{i,\nu} (t) &= -\omega_\nu Q_\nu(0)\sin\omega_\nu t + P_\nu(0)\cos\omega_\nu t 
\end{split}
\eeq
with arbitrary initial  conditions $Q_{i,\nu}(0)$ and $P_{i,\nu}(0) = \dot{Q}_{i,\nu}(0)$ shows however that the expansion above will be in terms of standing or traveling waves depending on these conditions.  

\subsubsection{Transforming to  new normal modes variables}

To obtain an expansion in traveling waves let us use the degeneracy of the two modes at every $k_\nu$ and do the following transformation
\beq \label{eq:fromQi_to_Qpm}
\bes
Q_{1,\nu} &= \frac{1}{\sqrt{2}}(Q_{k} -Q_{-k}) \;\;\; , \;\;\;  P_{1,\nu} = \frac{1}{\sqrt{2}} (P_k -P_{-k})  \\
Q_{2,\nu} &= \frac{1}{\omega_k\sqrt{2}}(P_{k}  + P_{-k}) \;\;\; , \;\;\;  P_{2,\nu} = -\frac{\omega_k}{\sqrt{2}} (Q_k + Q_{-k}) 
\end{split}
\eeq
to new variables $Q_{\pm k} , P_{\pm k}$. Note that to simplify notations we dropped the subscript $\nu$ in the right hand side and denoted accordingly  $\omega_k = v|k|$. We note that the above  transformation mixes coordinates and momenta.  In  the Appendix we show that this transformation is canonical.  

Substituting (\ref{eq:fromQi_to_Qpm}) in the expansion (\ref{eq:stringexpinstandwaves}) we obtain 
\beq \label{eq:stringexpintravelwaves2}
\bes
\phi(x,t)&=\sqrt{\frac{1}{L}}\sum_k\left[Q_k(t) \sin k  x  + \frac{P_k(t)} {v|k|} \cos k x \right]   \\
\pi(x,t) &= \sqrt{\frac{1}{L}}\sum_k\left[ P_k(t) \sin k  x -  v|k| Q_k(t) \cos k x \right] \;, \;  k=\frac{2\pi \nu}{L} \;,\; \nu = \pm 1, \pm 2 , ... 
\end{split}
\eeq
where we have combined together the sums over  $Q_k, P_k$ and $Q_{-k}, P_{-k}$  modes by extending the sums to include the negative values of $k$.  We show in Appendix that this is the desired expansion in traveling waves  - waves with positive and negative $k$'s  moving in opposite directions. The hamiltonian in the new variables has the sum of independent oscillators form 
\beq
H = \frac{1}{2}\sum_k  \left(P^2_k  + \omega_k^2 Q^2_k \right) 
\eeq
with the sum extending over both positive and negarive $k$'s.  The key point to note is that  compared to (\ref{eq:stringexpinstandwaves}) this is achieved in the  expansion (\ref{eq:stringexpintravelwaves2}) by making the amplitudes of the second degenerate mode  not  independent  but  proportional to the canonical conjugate of the amplitudes of the first mode and extending the the sum to the negative k's. 

\subsubsection{Inverting the transformation}

We remark that inverting (\ref{eq:stringexpintravelwaves2}) requires  some care.  The simplest  is to recall that $\sqrt{2/L} \sin k  x$ and $\sqrt{2/L} \cos k x$ form orthonormal set for positive $k>0$ and use the expansion (\ref{eq:stringexpinstandwaves}) together with the relations (\ref{eq:fromQi_to_Qpm}).   This means that it is the following combinations of $Q_k$ and $P_k$ which are simple projections
\beq
\bes 
\frac{1}{\sqrt{2}}(Q_{k} -Q_{-k}) &= \sqrt{\frac{2}{L}} \int_0^L dx \; \phi(x) \sin kx   \\ 
\frac{1}{\sqrt{2}} (P_k -P_{-k}) & =  \sqrt{\frac{2}{L}} \int_0^L dx \; \pi(x) \sin kx  \\
\frac{1}{\omega_k\sqrt{2}}(P_k  + P_{-k}) &= \sqrt{\frac{2}{L}} \int_0^L dx \; \phi(x) \cos kx  \\
-\frac{\omega_k}{\sqrt{2}} (Q_k + Q_{-k}) &= \sqrt{\frac{2}{L}} \int_0^L dx \;\pi(x) \cos kx 
\end{split}
\eeq
from which the expressions for each of the $Q_{k}$ and $P_{k}$ follow by a simple calculation. 
\beq \label{eq:P_k_Q_k_vs_phi_and_pi}
\bes 
Q_{k} &=  \frac{1}{\sqrt{L}}  \int_0^L dx [ \phi(x) \sin kx  -  \frac{1}{\omega_k} \pi(x) \cos kx ]     \\
P_k &= \frac{1}{\sqrt{L}} \int_0^L dx [ \pi(x) \sin kx + \omega_k \phi(x) \cos kx ]
\end{split}
\eeq

\subsubsection{The physics of the new variables}
Let us write the expressions for the terms in the expansions (\ref{eq:stringexpintravelwaves2}) as a single function. For this purpose let us transform
\beq \label{eq:amp_phase_of_mode}
Q_k = C_k \cos\alpha_k \;\; , \;\; P_k= -\omega_k C_k \sin\alpha_k   \;\;, \;\; 
C_k  =\frac{1} {\omega_k} \sqrt{ P_k^2 +\omega_k^2 Q_k^2 } 
 \eeq
This gives
\beq
\bes
& Q_k \sin k  x  + \frac{P_k} {\omega_k} \cos k x =  C_k \sin(kx - \alpha_k)  \\ 
 & P_k \sin k  x -  \omega_k Q_k  \cos k x  =  -  \omega_k C_k \cos(kx  - \alpha_k) 
\end{split}
\eeq
Since the amplitude $C_k$ is proportional to the square root of the energy of the mode  it is a constant of the motion for the mode time oscillations. It is not difficult to show that the  phase $\alpha_k(t)$ is  just
$$
\alpha_k(t) = \omega_k t + \alpha_k(0)
$$
Indeed writing 
$$ 
C_k = \sqrt{ 2 I_k /\omega_k}   
$$
one finds that Eq.(\ref{eq:amp_phase_of_mode}) is essentially a canonical transformation from $P_k,Q_k$ to the action-angle variables $I_k,\alpha_k$ for a harmonic oscillator, cf., Ref.\cite{Act}, with 
\beq
\bes
I_k & =\frac{1} {2\omega_k} (P_k^2 +\omega_k^2 Q_k^2 ) \;\; , \;\; \tan\alpha = - P_k/\omega_k Q_k \\
&\dot{I_k} = 0 \;\;, \;\; \dot{\alpha}_k= \omega_k
\end{split}
\eeq

 \subsubsection{Mechanical momentum of the string}
Apart of the Hamiltonian an important quantity  describing the physics of a string is its mechanical momentum (please do not confuse this $\mathbb{P}$ with the canonical $P_k$'s)
\beq \label{eq:string_momentum}
\mathbb{P}=-\int_0^L \pi(x,t)\frac{\pd}{\pd x}\phi(x,t) dx 
\eeq
It is conserved by the string equations of motion (\ref{eq:stringeq}) as can be seen from the following calculation. Defining the density of $P$
$$
{\mathcal{P}}(x,t)=-\pi(x,t)\frac{\pd}{\pd x}\phi(x,t)  
$$
we have
$$
\frac{\pd{\mathcal{P}}(x,t)}{\pd t}=-\frac{\pd\pi(x,t)}{\pd t}\frac{\pd \phi(x,t)}{\pd x}-\pi(x,t)\frac {\pd^2\phi(x,t)}{\pd t\pd x} $$
$$ =-v^2\frac{\pd^2\phi(x,t)}{\pd x^2} \frac{\pd \phi(x,t)}{\pd x}-\pi(x,t)\frac{\pd\pi(x,t)}{\pd x} $$
$$ = -\frac{1}{2}\frac{\pd}{\pd x}\left[v^2 \left(\frac{\pd \phi(x,t)}{\pd x}\right)^2 + \pi^2(x,t)\right] $$
This is one dimensional analogue of the continuity equation which connects the time derivative of ${\mathcal{P}}(x,t)$ and  the space derivative of the density of the Hamiltonian
$$
{\mathcal{H}}(x,t)=\frac{1}{2}\left[\pi^2(x,t)+v^2 \left(\frac{\pd \phi(x,t)}{\pd x}\right)^2\right]
$$
Integrating this and using the boundary conditions (must use periodic not fixed ends) we obtain the conservation law
\beq
\frac{\pd \mathbb{P}}{\pd t}  = \frac{\pd}{\pd t}\int_0^L {\mathcal{P}}(x,t) dx =0
\eeq

\subsection{Quantum mechanics of the traveling waves}
\subsubsection{The basic operators. String Hamiltonian and momentum} 

As in the case of the fixed ends string the closed string is quantized by introducing wave functionals $\Psi[\phi(x),t]$ for the string coordinates and the operators $\phi_{op}(x) =\phi(x)$ and $\pi_{op}(x)=-i\hbar \delta/\delta\phi(x)$. The Hamiltonian operator is the same given by Eq.(\ref{eq:stringHamiltonian2}) since the string equations are the same.  What is different are the boundary conditions which led to modified (degenerate) normal modes and the corresponding expansions (\ref{eq:stringexpintravelwaves2}).  

Using these  expansion for the operators
\eqna \label{eq:expintravelwaves}
\phi_{op}(x)&=& \sqrt{\frac{1}{L}}\sum_k\left[\sin k  x \;\hat{Q}_k + \frac{1}{v|k|}\cos k x \;\hat{P}_k \right]  \nonumber \\
\pi_{op}(x)&=& \sqrt{\frac{1}{L}}\sum_k\left[\sin k  x \;\hat{P}_k  - v|k|\cos k x \;\hat{Q}_k \right]
\eqne
we are led to the canonical commutators for  the traveling waves amplitudes
\beq 
[\hat{Q}_k, \hat{P}_{k'} ]=i\hbar \delta_{k k'} \;\; , \;\; [\hat{Q}_k, \hat{Q}_{k'} ]=[\hat{P}_k, \hat{P}_{k'} ] = 0
\eeq
These of course follow from the basic commutators (\ref{eq:paiphicommutator}) and the expressions (\ref{eq:P_k_Q_k_vs_phi_and_pi}). 
The Hamiltonian operator in terms of $\hat{Q}_k$'s  and  $\hat{P}_k$'s has the same form of decoupled oscillators
  \beq \label{eq:Ham_of_travel_waves}
  H_{op}=\frac{1}{2}\sum_k\left[\hat{P}^2_k +\omega^2_k \hat{Q}^2_k\right]
  \eeq
  
  Quantum mechanically  the string momentum $\mathbb{P}$, Eq. (\ref{eq:string_momentum}), becomes an operator 
  \beq
P_{\rm op}= -\frac{1}{2}\int_0^L dx\left\{\pi_{\rm op}(x)\frac{\pd}{\pd x}\phi_{\rm op}(x) dx + \left[\frac{\pd}{\pd x}\phi_{\rm op}(x)\right]\pi_{\rm op}(x)\right\}
\eeq
As usual with products of  non commuting  operators, here $\pi_{\rm op}(x)$ and  $\phi_{\rm op}(x)$,  one  must use a symmetrized expression. 

The operator $P_{\rm op}$ is the generator of translations
$$
\phi_{\rm op}(x)\; \to\; \phi_{\rm op}(x+a)\;\;\;, \;\;\; \pi_{\rm op}(x)\; \to\; \pi_{\rm op}(x+a)
$$
Indeed using the basic commutators (\ref{eq:paiphicommutator})   one can easily verify that
\beq
[P_{\rm op},\phi_{\rm op}(x)]=-i\hbar\frac{\pd}{\pd x} \phi_{\rm op}(x) \;\;\; , \;\;\; [P_{\rm op},\pi_{\rm op}(x)]=-i\hbar\frac{\pd}{\pd x} \pi_{\rm op}(x)
\eeq
as it should be for the generator of translations.

Inserting the expansions (\ref{eq:expintravelwaves}) in the momentum $P_{\rm op}$ we obtain
\beq \label{eq:Momntum_of_travel_waves}
\bes
P_{\rm op}&=-\frac{1}{L}\sum_{k k'}\int_0^L \left[\sin k  x \;\hat{P}_k  - v|k|\cos k x \;\hat{Q}_k \right]
\left[k' \cos k'  x \;\hat{Q}_{k'} - \frac{k'}{v|k'|}\sin k' x \;\hat{P}_{k'} \right]dx \nonumber \\
&=\frac{1}{2}\sum_k\left[\frac{k}{v|k|}\hat{P}^2_k + v k|k|\hat{Q}^2_k \right]
=\sum_k\frac{k}{\omega_k}\frac{1}{2}\left[\hat{P}^2_k + \omega_k^2\hat{Q}^2_k \right]
\end{split}
\eeq

 \subsubsection{The eigenstates. Energies and momenta of traveling waves quanta}
 The traveling waves Hamiltonian (\ref{eq:Ham_of_travel_waves}) has the same decouple normal modes oscillators form as the one for the standing waves (\ref{eq:normalmodes_H}) so formally its solutions have the same form as (\ref{eq:eigenfn_of_string}) 
 \beq 
 \bes 
 & E_{\{N_k\}} = E_{\rm ground \; state}  + \sum_{k}\hbar\omega_k  N_k  \;\;, \;\; N_k=0, 1, 2, .... \\
 & \Psi_{\{N_k\}}(\{Q_k\})  =\prod_k \psi_{N_k}(\beta_k Q_k) \; \; , \; \;  \beta_k=\sqrt{\omega_k/\hbar}
 \end{split}
 \eeq
with familiar harmonic oscillator eigenfunctions  $\psi_{N}(\beta Q)$, Eq. (\ref{eq:harm_osc_wfs}).  There are however important  differences.

  Since $\omega_k=v|k|$ the traveling waves energy quanta  $\epsilon_k=\hbar\omega_k$  are doubly degenerate with respect to the sign (direction) of k.  Even more profound is that these quanta also carry momentum.   Indeed comparing the expression (\ref{eq:Momntum_of_travel_waves}) for the string momentum $P_{op}$ with the Hamiltonian $H_{op}$ one observes that $P_{op}$ has the same eigenfunctions (not surprising) with the eigenvalues\footnote{The presence of $\pm \hbar k$ terms in this expression helps to cancel the $1/2$ "zero point motion" term present in the expression for the energy.}
\beq
P_{\{N_k\}} = \sum_k \hbar k N_k
\eeq
Each quantum has "mechanical" momentum $p_k=\hbar k$. So the closed string can be considered as a collection of traveling waves   "quasi"particles with energy momentum relation 
\beq
\epsilon_k \equiv \hbar\omega_k= \hbar v|k| =v|p_k|  \;\; \Rightarrow \;\;  \epsilon(p) = v|p|  
\eeq
It is useful to pay attention that this result can be viewed (obtained by a shortcut)  as a consequence of the three fundamental relations - 
two basic quantum mechanical relations  - the Plank-Einstein $\epsilon=\hbar\omega$ and the de Broglie  $p=\hbar k$ and the string dispersion relation $\omega=v|k|$. In a similar way we will find below that the quanta of the EM field will be particles (photons) with energy-momentum relation $\epsilon=\hbar \omega = \hbar ck = cp$ i.e. of massless relativistic particles.  The quanta of the Schr\"odinger field will have $\epsilon=\hbar \omega = \hbar^2k^2/2m = p^2/2m$,  i.e. the energy-momentum relation of non relativistic particles.

\subsubsection{Transformation to creation and annihilation operators}

In practice it is very convenient to introduce creation and annihilation operators in the standard way
\beq
\bes
 \hat{Q}_k=\sqrt{\hbar/ 2 \omega_k}\left(\hat{a}_k + \hat{a}^{\dagger}_k \right) \;\;\; &,\;\;\;  \hat{P}_{k}=i\sqrt{\hbar\omega_k/2} \left(\hat{a}^\dagger_{k} - \hat{a}_{k} \right) \\
  \ha_k  =  \sqrt{1/ 2\hbar\omega_k}  \left( i\hat{P}_k  + \omega_k \; \hat{Q}_{k}\right)   \;\; & , \;\;   \ha^\dagger_k  = \sqrt{1/2\hbar\omega_k} \left(-i\hat{P}_k  + \omega_k \; \hat{Q}_{k} \right)  \\
[\ha_{k},\ha^{\dagger}_{k'}]=\delta_{k k'}  \;\;\; & , \;\;\; [\ha_{k},\ha_{k'}]  =0 =[\ha^{\dagger}_{k},\ha^{\dagger}_{k'}]
\end{split}
\eeq
Using these operators we can write the Hamiltonian
\beq
\hat{H}_r=E_0+\sum_{k}\hbar\omega_k \ha^{\dagger}_{k}\ha_{k}
\eeq
and its eigenstates
\beq
|\{N_{k}\}>=\prod_{k}|N_{k}>=
\prod_{k}\frac{(\ha^{\dagger}_{k})^{N_{k}}}
{(N_{k}!)^{1/2}}|0>
\eeq
Great advantage of using $\ha$ and $\ha^{\dagger}$ operators rather than $\hat{P}$ and $\hat{Q}$ is the simplicity of the "action" of these operators on the "number states", i.e the states with a fixed numbers of quasi particles in each normal mode. Schematically
 $$
 \ha|n>=\sqrt{n}|n-1> \;\;\; , \;\;\; \ha^{\dagger}|n>=\sqrt{n+1}|n+1>
 $$
 In detailed notation
 $$
 \ha_{k}|\{N_{k}\}>=\sqrt{N_{k}}|N_{k}-1>\prod_{k'\ne k}
 |\{N_{k'}\}>
 $$
 $$
 \ha^{\dagger}_{k}|\{N_{k}\}>=\sqrt{N_{k}+1}|N_{k}+1>\prod_{k'\ne k}
 |\{N_{k'}\}>
 $$
 One says that the operators $\ha^{\dagger}_{k}$ and $\ha_{k}$ create and destroy quasi particles of energy $\hbar\omega$ and momentum $\hbar k$.

 It is useful to express the field operators in terms of $a$ and $a^{\dagger}$. Using (\ref{eq:expintravelwaves}) we obtain
 \beq \label{eq:expansion_in_a_a_dagger}
 \bes
 \phi_{op}(x)&=-i\sum_k  \sqrt{\frac{\hbar}{2\omega_k L}}  \left[ \hat{a}_{k} e^{i k x}  - \hat{a}^{\dagger}_k e^{-i k x} \right] =
 -i\sum_k  \sqrt{\frac{\hbar}{2\omega_k L}}\left[\hat{a}_k   - \hat{a}^{\dagger}_{-k}\right]e^{ik x} \\
 \pi_{op}(x) &= - \sum_k  \sqrt{\frac{\hbar\omega_k}{2 L}} \left[  \hat{a}_k e^{ikx} + \hat{a}^\dagger_{k} e^{-ikx} \right] = 
   - \sum_k  \sqrt{\frac{\hbar\omega_k}{2 L}} \left[ \hat{a}_{k} + \hat{a}^{\dagger}_{-k}\right] e^{ik x}  
\end{split}
\eeq
which are sums of terms which either create a quantum with momentum $\hbar k$ or annihilate one with the opposite momentum 
$-\hbar k$.

\section{Quantization of the EM Field}
The quantization of the electromagnetic field follows the same route as with the simple string above. The classical equations of the field  are the Maxwell equations. We will now cast them into Hamilton form and identify the Hamiltonian and the canonical coordinates and momenta of the field.  We  will then replace them by operators with canonical commutation relations acting on the appropriate wave functionals.

\subsection{Hamilton form of the Maxwell equations}

The Maxwell equations have the familiar form\footnote{In this Chapter we use the SI system of units.}
\eqna \label{eq:Maxwelleqs}
\nabla\cdot\vE(\vecr,t)=\frac{\rho(\vecr)}{\epsilon_0} \;\;\; &,& \;\;\; \nabla\cdot\vB(\vecr,t)=0 \\
\nabla\times\vE(\vecr,t)=-\frac{\pd \vB(\vecr,t)}{\pd t}\;\; &,& \;\;\; \nabla\times\vB(\vecr,t)=\frac{1}{c^2}\frac{\pd \vE(\vecr,t)}{\pd t} +\frac{ \vj(\vecr,t)}{c^2\epsilon_0} \nonumber
\eqne
Here $c$ is the light velocity and $\epsilon_0$ is a constant $\epsilon_0=8.85 \cdot 10^{-12} F m^{-1}$ called vacuum permittivity  which is related to our choice of the SI measurement  unit system.

The Maxwell equations describe the EM field configuration for a given distribution of the electric current $\vj(\vecr,t)$ and density $\rho(\vecr,t)$ of electric charges.  Assuming that we are dealing with a system of $N$ charges and denoting by $\vecr_a(t)$ , $\vv_a(t),  a=1, ..., N$ their positions and velocities we have
\beq \label{eq:denscurrent}
\rho(\vecr,t)=\sum_{a=1}^N q_a\delta(\vecr-\vecr_a(t)) \;\;, \;\; \vj(\vecr,t)=\sum_{a=1}^N q_a \vv_a\delta(\vecr-\vecr_a(t))
\eeq
These expressions must be supplemented by the mechanical equations of motion for the charges as they move in the given $\vE(\vecr,t)$ and $\vB(\vecr,t)$.   These equations  are just the Newton equations for the charges
\eqna \label{eq:NewtforNcharges}
m_a\frac{d\vv_a}{dt}&=&q_a\vE(\vecr_a,t)+ q_a\left(\vv_a\times \vB(\vecr_a,t)\right)\;\;, \;\;  \vv_a= \frac{d\vecr_a}{dt} \;\;\;, \;\;\; a=, ... , N
 \eqne
 The coupled equations
(\ref{eq:Maxwelleqs}), (\ref{eq:denscurrent}) and (\ref{eq:NewtforNcharges}) provide the complete system  which determines how the positions of the charges and their motion determine the EM field and how this field determines the motion  of the charges.  Our first goal will be to cast this system in the Hamiltonian form thereby determining its canonical variables and the Hamiltonian.

\subsubsection{Vector potential. The $A_0=0$ gauge}  

We start by noting that the first pair of Maxwell equations (\ref{eq:Maxwelleqs}) does not involve  time derivatives. They are in a sense constraints on the possible functional dependence of $\vE(\vecr)$ and $\vB(\vecr)$. Both constraints are easy to resolve. The condition $\nabla\cdot\vB=0$ means that there are no magnetic charges in nature and that $\vB$ can be represented as a curl of an arbitrary vector function
\beq \label{eq:BtoArelation}
\vB=\nabla\times \vA
\eeq
which is conventionally  called the vector potential.  In the chapter where we considered the motion in an external EM field we have seen that the quantum mechanical formulation was impossible without an explicit use of this function.  Also presently we will find that the quantization of  EM field can not avoid using $\vA$.

The second pair of the Maxwell equations consists of dynamical equations. Inserting Eq.\,(\ref{eq:BtoArelation}) in the  first of these equations we obtain
\beq
\nabla\times(\vE+\frac{\pd \vA}{\pd t})=0 \;\;\; \Rightarrow \;\;\; \vE=-\frac{\pd \vA}{\pd t}-\nabla A_0
\eeq
Here  $A_0$ is (in non relativistic parlance) the "scalar potential"  which together with $\vA$ completely determine the fields $\vE$ and $\vB$. The potentials  $\vA$ and $A_0$ are not uniquely defined. We can choose instead different function $A_0'(\vecr,t)$  and $\vA'(\vecr,t)$ related to $A_0$ and $\vA$ by the gauge transformation
\beq
A_0'=A_0 - \frac{\pd\chi}{\pd t} \;\; \; , \;\;\; \vA'=\vA+\nabla \chi
\eeq
with arbitrary function $\chi(\vecr,t)$.   We shall use this freedom and take  $A_0$ to be identically equal to  zero and write
 \beq \label{eq:Azeroeqtozero}
\frac{\pd \vA}{\pd t} = -  \vE
 \eeq
This choice is called "working in the $A_0=0$ gauge". Importantly this choice does not exhaust the full gauge freedom. We can still add to $\vA$ a gradient of a \emph{time independent} function $\chi(\vecr)$ without changing our $A_0=0$  assumption.

Inserting (\ref{eq:BtoArelation}) in the last of the Maxwell equations (\ref{eq:Maxwelleqs}) we obtain
\beq \label{eq:2nddynameqforA}
\frac{\pd \vE }{\pd t} =c^2 \nabla\times\nabla\times\vA  -\frac{ \vj}{\epsilon_0}
\eeq
Using (\ref{eq:Azeroeqtozero}) this equation becomes
\beq
\frac{\pd^2 \vA}{\pd t^2} =-c^2 \nabla\times\nabla\times\vA  +\frac{ \vj}{\epsilon_0}
\eeq
Regarding the 2nd time derivative on the left as acceleration of $\vA(\vecr, t)$ one can view the above equations as coupled Newton equations for the degrees of freedom $\vA(\vecr)$. In this view at every point in space there are three such degrees of freedom (field coordinates) which  can symbolically be represented as $A_{i,\vecr}$.  The 3 dimensional vector index is $i=1,..., 3$ and $\vecr$ is running over all points in the 3 dimensional space  in a way similar to $x$ running over  points of the $x$ axis in the example of a string. The coupling between different $A_{i,\vecr}$ is via complicated  combination of second order vector derivative $\nabla \times \nabla \times $  connecting different vector components of $\vA(\vecr)$ in neighboring points.

 The last term in (\ref{eq:2nddynameqforA}) is the "force" acting on the field coordinates on the part of the matter.   Ignoring this force for a moment (i.e. considering the EM field in an empty space region) we can view the coordinates $\vA(\vecr)$ as representing  coupled oscillators. This is  because the above equation  without the last term is linear. Although complicated from  the vector analysis  point of  view the derivatives combination $\nabla \times \nabla \times $  is a linear operation.
  
 \subsubsection{The Hamiltonian}

 Continuing with  the "mechanical" interpretation of the EM field dynamics we notice that the pair of the 1st order equations (\ref{eq:Azeroeqtozero}) and (\ref{eq:2nddynameqforA}) without the last term can be regarded together as Hamilton equations with the following field Hamiltonian
 \beq \label{eq:fieldHamiltonian}
 H_f=\frac{\epsilon_0}{2}\int d^3 r \left[ \vE^2(\vecr)+c^2(\nabla\times\vA(\vecr))^2\right]
 \eeq
and canonical variables $A_i(\vecr)$ as coordinates  and $-\epsilon_0 E_i(\vecr)$  as momenta. We will verify this in a moment  but first we note that perhaps the simplest way to guess the expression of the Hamiltonian  is to notice that on account of Eq. (\ref{eq:Azeroeqtozero}) the first term in it has the form of the kinetic energy. One can determine how it changes with time by  forming a scalar product of the left hand side of  (\ref{eq:2nddynameqforA})
with $\vE$. Multiplying  the first term on the right hand side (remember we still are ignoring the current term) with the equal quantity $-\pd \vA/\pd t$ we can  integrate both sides over $\vecr$. After simple manipulations\footnote{Here are the details
\eqna
\int d^3 r \;\vE\cdot\frac{\pd \vE }{\pd t} =-c^2 \int d^3 r  \frac{\pd \vA}{\pd t}\cdot\nabla\times\nabla\times\vA
&=& -c^2\int d^3 r\pd_t A_i\epsilon_{ijk}\pd_j\epsilon_{klm}\pd_l A_m  =\nonumber \\
=({\rm integrate \; by \; parts})= c^2 \int d^3 r \epsilon_{ijk}\pd_j \pd_t A_i\epsilon_{klm}\pd_l A_m
&=&-c^2\int d^3 r \left(\nabla\times \frac{\pd \vA}{\pd t}\right)\cdot\left( \nabla\times \vA\right)\;. \nonumber \\
{\rm Rewrite \; as} \;\; \frac{d}{d t}\;\frac{1}{2}\;\int d^3 r\; \vE^2 = -\frac{c^2}{2} \frac{d}{dt}\int d^3 r (\nabla\times\vA)^2 && \nonumber
\eqne }
one can show that the change in time of the kinetic energy   is equal to minus the change in time of the second term in $H_f$ which has the meaning  of the potential energy. This of course verifies that $H_f$ is conserved, $dH_f/dt=0$.

Returning to the Hamiltonian (\ref{eq:fieldHamiltonian}) we form  its first variation
\eqnaa
\delta H_f&=&\int d^3 r \left[ \epsilon_0 \vE\cdot\delta\vE + \epsilon_0c^2 (\nabla\times\vA)\cdot(\nabla\times\delta\vA)\right] \\
&=& \int d^3 \left[ \epsilon_0 \vE\cdot\delta\vE + \epsilon_0c^2 \epsilon_{ijk}\pd_j A_k \epsilon_{ilm}\pd_l \delta A_m\right] \\
&=&\int d^3 r \left[\epsilon_0 \vE\cdot\delta\vE - \epsilon_0 c^2 \delta A_m\epsilon_{ilm}\epsilon_{ijk}\pd_l \pd_j A_k  \right]\\
&=&\int d^3 r \left[ \epsilon_0 \vE\cdot\delta\vE + \epsilon_0c^2 \delta\vA\cdot \nabla\times\nabla\times\vA)\right]
\eqnae
where we performed integration by parts in the second term. From this it follows  that
\beq
\frac{\delta H_f}{\delta (-\epsilon_0\vE)}= -\vE \;\;\;, \;\;\;\frac{\delta H_f}{\delta\vA}=\epsilon_0c^2\nabla\times\nabla\times\vA
\eeq
showing that  the Hamilton equations with this  Hamiltonian and  canonical variables $\vA(\vecr)$ and $-\epsilon_0\vE(\vecr)$
\beq \label{eq:MaxwellasHamltneqs}
\frac{\pd \vA}{\pd t}= \frac{\delta H_f}{\delta (-\epsilon_0\vE)}=-\vE  \;\;\;, \;\;\; \frac{\pd (-\epsilon_0\vE)}{\pd t}=-\frac{\delta H_f}{\delta\vA}=-\epsilon_0c^2\nabla\times\nabla\times\vA
\eeq
indeed coincide with Eqs. (\ref{eq:Azeroeqtozero}) and (\ref{eq:2nddynameqforA}) without the current term.

Let us now show how  to account for the current term and the dynamical Newton equations  (\ref{eq:NewtforNcharges}) for the charges. Here we are guided by our knowledge of the Hamiltonian  of charges moving a given EM field  (cf. Chapter  Motion in External EM Field).   We simply add it to $H_f$ above and obtain
\beq \label{eq:fullHamofEMwithmatter1}
H=\frac{\epsilon_0}{2}\int d^3 r \left[ \vE^2(\vecr)+c^2(\nabla\times\vA(\vecr))^2\right] + \sum_{a=1}^N\frac{1}{2m_a}\left[\vp_a-q_a\vA(\vecr_a)\right]^2
\eeq
Using exactly the same calculation as in the chapter Motion in External EM Field we can show that the Hamilton equations
\beq
\frac{d\vecr_a}{dt}=\frac{\pd H}{\pd \vp_a} \;\;\;, \;\;\; \frac{d\vp_a}{dt}=-\frac{\pd H}{\pd \vecr_a}
\eeq
are equivalent to the  Newton equations  (\ref{eq:NewtforNcharges}). Let us now consider the first variation of the last term in (\ref{eq:fullHamofEMwithmatter1}) with respect to $\vA(\vecr)$. We obtain
\eqnaa
&&-\sum_{a=1}^N\frac{q_a}{m_a}\left[\vp_a-q_a\vA(\vecr_a)\right]\cdot \delta \vA(\vecr_a)=-\sum_{a=1}^N\frac{q_a}{m_a}\left[\vp_a-q_a\vA(\vecr_a)\right]\cdot\int d^3 r \delta(\vecr-\vecr_a)\delta\vA(\vecr) \\
&=&-\int d^3 r \sum_{a=1}^N\frac{q_a}{m_a}\left[\vp_a-q_a\vA(\vecr_a)\right]\delta(\vecr-\vecr_a)\cdot\delta\vA(\vecr)=-\int d^3 r \;\vj(\vecr)\cdot \delta\vA(\vecr)
\eqnae
where $\vj(\vecr)$ is the current as defined in  Eq. (\ref{eq:NewtforNcharges})  with
$$
\vv_a=\frac{1}{m_a}\left[\vp_a-q_a\vA(\vecr_a)\right]
$$
With this result the second equation of (\ref{eq:MaxwellasHamltneqs}) with $H_f$ replaced by the full $H$ (\ref{eq:fullHamofEMwithmatter1}) now reads
\beq
 \frac{\pd (-\epsilon_0\vE)}{\pd t}=-\frac{\delta H}{\delta\vA}=-\epsilon_0c^2\nabla\times\nabla\times\vA+\vj
 \eeq
reproducing the full equation (\ref{eq:2nddynameqforA}).  We also note that since the second term in the full $H$ does not depend on $\vE$ the first equation in (\ref{eq:MaxwellasHamltneqs})  remains unchanged when $H_f$ is replaced in it by $H$.

\subsection{Canonical quantization}
Having established the canonical structure of the theory we can now quantize it. Attentive reader should have noticed that we have not yet accounted for the first equation in the set (\ref{eq:Maxwelleqs}) expressing the Gauss law. We also seem to be missing from the Hamiltonian (\ref{eq:fullHamofEMwithmatter1}) the regular Coulomb interaction energy between the charges $\{q_a\}$.  We will address  these issues shortly but meanwhile let us proceed with the quantization.

Moving from classical to quantum description we recognise that the coordinate set of our system consists of the vector potential $\vA(\vecr)$ (i.e. 3 vector components in each point of the position space, i.e.  $3 \times \infty^3$ variables ) and 3N vectors $\{\vecr_a\}$  of the particles' positions. Accordingly we introduce the wave functional of the field $\vA(\vecr)$ which also depends (is a function of) the N particles' positions
and the
 \beq
\Psi=\Psi[\vA(\vecr),\vecr_1, ..., \vecr_N, t]
\eeq
This should be viewed as a correspondence between all field configurations and set of N particles' positions $\{\vecr_a\}$ and (in general) complex  probability amplitudes which in general change with time.

The physical operators are constructed from the corresponding classical quantities by the canonical substitution
\eqna \label{eq:canonicalops}
\vA(\vecr) \to \vA_{op}(\vecr)&=& \vA(\vecr) \;\; , \;\; \vE(\vecr) \to \vE_{op}(\vecr)= \frac{i\hbar}{\epsilon_0} \frac{\delta}{\delta \vA(\vecr) }\\
\vecr_a \;\;\to \;\; \hat{\vecr}_a &=&  {\vecr_a} \;\;\;, \;\;\;  \vp_a \;\;\to \;\; \hat{\vp}_a = -i\hbar \nabla_a \nonumber
\eqne
where we accounted for the fact that it is the combination  $-\epsilon_0\vE(\vecr)$ which is canonical to $\vA(\vecr) $ not just $\vE(\vecr)$.
Using the equality
$$
\frac{\delta A_j(\vecr') }{\delta A_i(\vecr)} = \delta_{ij}\delta(\vecr-\vecr').
$$
it follows that the field operators obey the commutation relations
\beq \label{eq:field_comm}
\left[\hat{E}_i(\vecr),\hat{A}_j(\vecr')\right] = \frac{i\hbar}{\epsilon_0}\delta_{ij}\delta(\vecr-\vecr')\;\; , \;\; \left[\hat{A}_i(\vecr),\hat{A}_j(\vecr')\right] =\left[\hat{E}_i(\vecr),\hat{E}_j(\vecr')\right]  =0
\eeq

The time dependence of the wave functional/function $\Psi[\vA(\vecr),\vecr_1, ..., \vecr_N, t]$ is governed by the Schr\"odinger equation
\beq
i\hbar \frac{\partial \Psi(t)}{\partial t}= H_{op}\Psi(t)
\eeq
 in which the Hamiltonian operator $H_{op}$ is obtained by replacing in the classical expression (\ref{eq:fullHamofEMwithmatter1}) the fields $\vA(\vecr)$ , $\vE(\vecr)$ and the particle variables $\{\vecr_a\}$, $\{\vp_a\}$ by the corresponding operators (\ref{eq:canonicalops}),
 \beq \label{eq:HamOperofEMwithmatter}
H_{op}=\frac{\epsilon_0}{2}\int d^3 r \left[ \vE_{op}^2(\vecr)+c^2(\nabla\times\vA_{op}(\vecr))^2\right] + \sum_{a=1}^N\frac{1}{2m_a}\left[-i\hbar\nabla_a-q_a\vA_{op}(\hat{\vecr}_a)\right]^2 \; .
\eeq

 The above Hamiltonian does not depend on time as we are dealing with a closed EM field + matter system. The energy is therefore conserved and quantum mechanically we can reduce in the standard way the solution of the above time dependent Schr\"{o}dinger equation to solving the  static equation
 \beq
  H_{op}\Psi=E \Psi
  \eeq
  This is a complicated equation for the coupled field-matter system. No exact solution is possible. We will discuss approximate solutions below. But before that we have to clarify several formal  but very important issues which will allow us to somewhat simplify the problem.

\subsection{ Gauge  invariance}
\subsubsection{The Gauss law} 
We will now show that the Hamiltonian formulation presented above neatly  accounts  for both the Gauss law and the Coulomb interaction between the charges. The key to this is to observe that the Hamiltonian is invariant under  the gauge transformation
\eqna \label{eq:residgaugetransf}
 \vA(\vecr,t)\to \vA(\vecr,t)+\nabla\chi(\vecr)\;\; \;\;\;\; &,& \;\;\,\vE(\vecr,t)\to\vE(\vecr,t) \\
 \;\;\;\vp_a(t) \to \vp_a(t)+q_a \nabla\chi(\vecr_a(t)) \;\; &,& \;\;\;\vecr_a(t)\to\vecr_a(t)  \nonumber
 \eqne
 with an arbitrary  time independent function $\chi(\vecr)$.  This is the residual gauge transformation  we have briefly mentioned after Eq.(\ref{eq:2nddynameqforA}).   
 
 The symmetry of $H$ under (\ref{eq:residgaugetransf}) is the result of the way the vector potential $\vA(\vecr)$ enters it, i.e. only via the combinations $\nabla\times\vA$ and $[\vp_a-q_a\vA(\vecr_a)]$.  It is a local symmetry meaning that it is characterised by parameters $\chi(\vecr)$ which depend on $\vecr$. Schematically  there are $\infty^3$ parameters corresponding  to the "number" of points in the 3D space of vectors $\vecr$.    As we will show below the generators of this symmetry are
 \beq \label{eq:genofgauge}
 g_{op}(\vecr)=-\epsilon_0\nabla\cdot\vE_{op}(\vecr)+\rho_{op}(\vecr)
 \eeq
 Their dependence on $\vecr$  means that there are  $\infty^3$ generators  corresponding to $\infty^3$ parameters $\chi(\vecr)$ in (\ref{eq:residgaugetransf}).

 Classically expressions corresponding to the generators of symmetries of  the Hamiltonian are conserved by the Hamilton equations. 
 Momentum and angular momentum are of course the classic examples of such conservations.  Accordingly let us show that equations (\ref{eq:Azeroeqtozero}), (\ref{eq:2nddynameqforA}) and (\ref{eq:NewtforNcharges}) conserve the above expression for the generator $g_{op}(\vecr,t)$ when it is taken as classical and when  $\vE$ and $\rho$ in it are allowed to evolve according  to these equations. We have
 \beq
 \frac{\pd}{\pd t}[-\epsilon_0\nabla\cdot\vE(\vecr)+\rho(\vecr)]=-\epsilon_0\nabla\cdot\frac{\pd \vE}{\pd t}+\frac{\pd \rho}{\pd t}=\nabla\cdot\vj+\frac{\pd \rho}{\pd t}=0
 \eeq
 where we used Eq.(\ref{eq:2nddynameqforA})  and the continuity equation for the charges.\footnote{Continuity equation is a general relation between $\rho(\vecr,t)$ and $\vj(\vecr,t)$ given by  (\ref{eq:denscurrent})
 \eqna  \frac{\partial\rho(\vecr,t)}{\partial t} = \frac{\partial}{\partial t} \sum_{a=1}^N q_a \delta(\vecr - \vecr_a(t)) = \sum_{a=1}^N q_a \nabla_{\vecr_a}\delta(\vecr-\vecr_a(t))
\cdot \frac{d \vecr_a}{ dt} = \nonumber \\= \left({\rm using} \;\; \nabla_{\vecr_a}\delta(\vecr-\vecr_a)=-\nabla_{\vecr}\delta(\vecr-\vecr_a)\;\right)=  - \nabla_{\vecr} \cdot \vj(\vecr,t) \nonumber
 \eqne}

 The vanishing of $\pd g(\vecr)/\pd t$ means that local quantities $-\epsilon_0\nabla\cdot\vE(\vecr)+\rho(\vecr)$ form a constant, time independent function of $\vecr$. It is natural to denote this function by $\rho_0(\vecr)$
  $$
  -\epsilon_0\nabla\cdot\vE(\vecr)+\rho(\vecr)=\rho_0(\vecr)
  $$
  and  interpret it as a density of fixed static electric charges.  We notice that these charges appear in addition to the dynamical charges $q_a$ described by  the equations  (\ref{eq:NewtforNcharges}).  Under normal circumstances  there are no such extra static charges. In fact there presence would violate such symmetries as translational, rotational, Lorenz.   So one should assume that $\rho_0=0$.  Using this in the above relation we recover the Gauss law.

 \subsubsection{Quantum mechanics of the gauge transformation}

 Let us work out the quantum mechanics of the gauge transformation (\ref{eq:residgaugetransf}) . What we want to show is that it is generated by the $\infty^3$ operators $g_{op}(\vecr)$, Eq. (\ref{eq:genofgauge}), i.e. that the following relations hold
 \beq \label{eq:unitarygaugetrans}
 e^{-(i/\hbar)\int d^3 r \chi(\vecr) g_{op}(\vecr)}
 \left(
\begin{array}{c}
  \vA_{op}(\vecr') \\
  \hat{\vp}_a
\end{array}
\right) e^{(i/\hbar)\int d^3 r \chi(\vecr) g_{op}(\vecr)}
=\left(\begin{array}{c}
  \vA_{op}(\vecr') +\nabla\chi(\vecr') \\
  \hat{\vp}_a + q_a\nabla\chi(\vecr_a)
\end{array}\right)
\eeq
As usual it is enough to consider the infinitesimal $\chi(\vecr)$ for which the left hand side reduces to 
\beq
-\frac{i}{\hbar} \int d^3 r \chi(\vecr)\left[g_{op}(\vecr), \left(
\begin{array}{c}
  \vA_{op}(\vecr') \\
  \hat{\vp}_a
\end{array}\right)\right] = -\frac{i}{\hbar} \int d^3 r \chi(\vecr)\left(
\begin{array}{c}
 - i\hbar\nabla_{\vecr}\delta(\vecr-\vecr') \\
 -i\hbar q_a\nabla_{\vecr}\delta(\vecr-\vecr_a)
\end{array}\right)
\eeq
Here  we omitted the identity term and used
\beq
\left[g_{op}(\vecr), \vA_{op}(\vecr')\right]=\left[-\epsilon_0\nabla\cdot\vE_{op}(\vecr),\vA_{op}(\vecr')\right] =-i\hbar\nabla_{\vecr}\delta(\vecr-\vecr')
\eeq
and
\beq
\left[g_{op}(\vecr), \hat{\vp}_a \right]=\left[\rho_{op}(\vecr), \hat{\vp}_a\right]=i\hbar\nabla_{\vecr_a}\delta(\vecr-\vecr_a)=-i\hbar \nabla_{\vecr}\delta(\vecr-\vecr_a)
\eeq
Now we do the integration by parts and use the delta function
\beq
-\frac{i}{\hbar} \int d^3 r \chi(\vecr)\left(
\begin{array}{c}
 - i\hbar\nabla_{\vecr}\delta(\vecr-\vecr') \\
 -i\hbar q_a\nabla_{\vecr}\delta(\vecr-\vecr_a)
\end{array}\right) = \int d^3 r \nabla\chi(\vecr) \left(
\begin{array}{c}
 \delta(\vecr-\vecr') \\
 q_a\delta(\vecr-\vecr_a)
\end{array}\right) = \left(
\begin{array}{c}
 \nabla\chi(\vecr') \\
 q_a\nabla\chi(\vecr_a)
\end{array}\right)
\eeq
obtaining exactly what is needed to get the $\chi(\vecr)$ dependent term in the right hand
side of Eq.(\ref{eq:unitarygaugetrans}).

Thus $\infty^3$ operators $g_{op}(\vecr)$ are indeed the generators of the gauge transformation. Since the Hamiltonian operator $H_{op}$ is invariant under this transformation one must have that
$H_{op}$ commutes with $g_{op}(\vecr)$
\beq
\left[H_{op}, g_{op}(\vecr) \right]= 0
\eeq
As in simpler quantum mechanical systems this means that $H_{op}$ and $g_{op}(\vecr)$ have common eigenfunctions. We write symbolically
\beq
H_{op}\Psi=E\Psi \;\;\; , \;\;\; g_{op}(\vecr)\Psi=\rho_0(\vecr)\Psi
\eeq
where we denoted by $\rho_0(\vecr)$ the eigenvalues of the $\infty^3$ operators $q_{op}(\vecr)$. As in the classical case the meaning of $\rho_0(\vecr)$ is the density of static ("background") electric charges. They are "background"  because  they are not a part of the dynamics. Just sit there as a part of initial conditions.  Their presence would violate basic symmetries (translational, rotational, Lorenz) so the physics dictates that one must select only the eigenfunctions which belong in the "sector" of the system Hilbert space  for which
\beq
g_{op}(\vecr)\Psi=0
\eeq
In other words - the \emph{gauge invariant sector}.

\subsubsection{Separating the longitudinal components of the fields} 

In the Hamiltonian formulation the electromagnetic field has $3\times\infty^3$ degrees of freedom which in our formulation are described by the coordinates $A_i(\vecr)$ and the corresponding momenta $-\epsilon_0 E_i(\vecr)$ , $i=1,2,3$.  Using the local gauge symmetry (\ref{eq:residgaugetransf}) one can eliminate one third of these  degrees of freedom. For this reason let us represent the functions $\vA(\vecr)$ and $\vE(\vecr)$ as sum of the so called transverse and longitudinal components
\beq \label{eq:sumoflongandtrans}
\vA(\vecr)=\vA_{T}(\vecr) + \vA_{L}(\vecr) \;\;, \;\;  \vE(\vecr)=\vE_{T}(\vecr) + \vE_{L}(\vecr)
\eeq
where $\vA_{T}$, $\vA_{L}$, $\vE_{T}$ and  $\vE_{L}$ satisfy
\beq
\nabla\cdot\vA_T(\vecr)=\nabla\cdot\vE_T(\vecr)=0 \;\;\; , \;\;\; \nabla\times\vA_L(\vecr)=\nabla\times\vE_L(\vecr)=0
\eeq
Such a representation is possible for any vector field. This can be shown (and the origin of the names longitudinal and transverse understood) using Fourier expansions. Let us take for example $\vA(\vecr)$  and expand
\beq
\vA(\vecr)=\sum_{\vk}\vA_{\vk} e^{i{\vk}\cdot\vecr}
\eeq
For convenience  in order to have discrete  values of $\vk$ we consider the fields in a large but finite volume.  The precise boundary conditions are not important for this discussion.

 The Fourier  amplitudes  $\vA_{\vk}$ are vectors. Their directions in principle bear no relation to the direction of the  corresponding wave vectors $\vk$. We can however represent each of them as a sum of two vectors which are parallel and perpendicular to "their" $\vk$
$$
\vA={\vA}_{\vk}^{(L)} +{\vA}_{\vk}^{(T)}
\;\;\;, \;\;\; {\rm with} \;\; {\vA}_{\vk}^{(L)}\times\vk=0\;\;,\;\; {\vA}_{\vk}^{(T)}\cdot\vk=0
$$
Using this we can write the Fourier expansion as a sum
\beq
\vA(\vecr)=\sum_{\vk}{\vA}_{\vk}^{(L)}e^{i{\vk}\cdot\vecr}+\sum_{\vk}{\vA}_{\vk}^{(T)}e^{i{\vk}\cdot\vecr}
\eeq
Using
$$
\nabla\cdot({\va} e^{i{\vk}\cdot\vecr}) = i{\vk}\cdot{\va} e^{i{\vk}\cdot\vecr} \;\;, \;\; \nabla \times({\va} e^{i{\vk}\cdot\vecr}) = i{\vk}\times{\va} e^{i{\vk}\cdot\vecr}
$$
we see that the two terms in the Fourier expansion of $\vA(\vecr)$ are respectively $\vA_L(\vecr)$ and $\vA_T(\vecr)$ as appear in (\ref{eq:sumoflongandtrans}).

We note also that longitudinal components of the vector fields can be written as a gradient of a scalar function. Therefore we can write
\beq
\vA(\vecr)=\vA_{T}(\vecr) + \nabla\xi(\vecr) \;\;\;\;, \;\;\;\;  \vE(\vecr)=\vE_{T}(\vecr) - \nabla\phi(\vecr)
\eeq
where two scalar functions $\xi(\vecr)$ and -$\phi(\vecr)$ fully determine the longitudinal components $\vA_L(\vecr)$ and $\vE_L(\vecr)$ respectively.  As will become clear in the next section $\phi(\vecr)$ is the scalar electric potential  so familiar from the Coulomb and other electrostatic problems. 

\subsubsection{Recovering the Coulomb interaction. Resulting Hamiltonian}

 Inserting the above expressions  for $\vA(\vecr)$ and $ \vE(\vecr)$ in the Hamiltonian (\ref{eq:fullHamofEMwithmatter1}) we obtain
\beq
H=\frac{\epsilon_0}{2}\int d^3 r \left[ \vE_T^2(\vecr)+(\nabla\phi)^2 + c^2(\nabla\times\vA_T(\vecr))^2\right] + \sum_{a=1}^N\frac{1}{2m_a}\left[\vp_a-q_a\vA_T(\vecr_a)\right]^2
\eeq
We have  transformed $\vp_a$ to $\vp_a +q_a\nabla\xi(\vecr_a)$. The mixed term containing $\vE_T\cdot\nabla\phi$ does not appear since it vanishes as can be seen after integrating it by parts
$$
\int d^3 r \;\vE_T\cdot\nabla \phi = - \int d^3 r \;(\nabla\cdot\vE_T) \phi \; =\; 0
$$
As a last step in transforming $H$ we note that the Gauss law allows to express $\phi(\vecr)$ in terms of $\rho(\vecr)$
\beq
\nabla\cdot\vE(\vecr)=-\nabla^2\phi(\vecr)=\frac{\rho(\vecr)}{\epsilon_0} \;\;\;\Rightarrow  \;\;\;
\phi(\vecr)=\frac{1}{4\pi\epsilon_0}\int d^3r' \frac{1}{|\vecr-\vecr'|}\rho(\vecr')
\eeq
We see that in this formulation the familiar scalar potential appears in the longitudinal component of the electric field $\vE_L(\vecr)$  and is completely determined by the density of the charge. This allows to express the term in $H$ containing $(\nabla \phi)^2$ as
$$
\frac{\epsilon_0}{2}\int d^3 r (\nabla \phi)^2=-\frac{\epsilon_0}{2}\int d^3 r \phi\nabla^2 \phi=\frac{1}{2}\int d^3 r \phi(\vecr)\rho(\vecr)=\frac{1}{8\pi\epsilon_0}\int d^3 r d^3r' \frac{\rho(\vecr)\rho(\vecr')}{|\vecr-\vecr'|}
$$
This is just the Coulomb interaction between the charges in  $\rho(\vecr)$  and can be written using the expression for $\rho(\vecr)$  given in (\ref{eq:NewtforNcharges})  as
\beq
\frac{\epsilon_0}{2}\int d^3 r (\nabla \phi)^2=\frac{1}{8\pi\epsilon_0}\sum_{a\ne b}^N\frac{q_a q_b}{|\vecr_a-\vecr_b|} + \sum_{a=1}^N\epsilon^a_{\rm self \;\; interaction}
\eeq
where $\epsilon^a_{\rm self \;\; interaction}$ are constants which express the Coulomb self-energy of each particle. They diverge for point particles. We will not deal with this in details but assuming  that particles have small but finite  sizes (cutoffs)  we will  simply disregard this constant term.

To conclude, the Hamiltonian has the form
\beq \label{eq:fullEM_matter_Hamilt1}
H = H_r  + \sum_{a=1}^N\frac{1}{2m_a}\left[\vp_a-q_a\vA_T(\vecr_a)\right]^2 + V_{\rm Coul}
\eeq
where we defined the radiation and the Coulomb interaction parts of the Hamiltonian
\eqna \label{radiation_H}
H_r\;\; &=& \frac{\epsilon_0}{2}\int d^3 r \left[ \vE_T^2(\vecr) + c^2(\nabla\times\vA_T(\vecr))^2\right]  \\
 V_{\rm Coul}&=& \frac{1}{8\pi\epsilon_0}\sum_{a\ne b}^N\frac{q_a q_b}{|\vecr_a-\vecr_b|} \nonumber
\eqne
It is often convenient to write this Hamiltonian as a sum of three parts
\beq
H=H_r+H_{\rm matter}+U_{\rm radiation-matter\;interaction}
\eeq
where $H_r$ is given by (\ref{radiation_H}) and
\eqna
H_{\rm matter}&=&\sum_{a=1}^N\frac{\vp_a^2}{2m_a} +  V_{\rm Coul} \\
U_{\rm radiation-matter\;interaction}&=& -\sum_{a=1}^N\frac{q_a}{m_a}\vp_a\cdot\vA_T(\vecr_a) + \\
& & + \sum_{a=1}^N \frac{q^2_a}{2 m_a} \vA_T(\vecr_a)\cdot \vA_T(\vecr_a) \nonumber
\eqne
Note that when switching to operators there will be no operator ordering ambiguity in the term $\hat{\vp}_a\cdot\vA_T(\vecr_a)$ since the difference, i.e. the commutator 
 $$
\hat{\vp}_a\cdot\vA_T(\vecr_a) -\vA_T(\vecr_a)\cdot \hat{\vp}_a  = -i\hbar \nabla_a\cdot \vA_T(\vecr_a) =0
$$

\subsubsection{An aside - separating transverse and longitudinal Maxwell equations}  
Let us now examine how the Hamilton (Maxwell) equations (\ref{eq:Azeroeqtozero}) and (\ref{eq:2nddynameqforA}) look in terms of the transverse fields $\vA_T(\vecr)$, $\vE_T(\vecr)$. One can see that each equation separates into two relating separately the transverse and longitudinal components
\eqna \label{eq:Maxwellfortransverse}
\frac{\pd \vA_T}{\pd t} &=& -  \vE_T \;\;\;\;\;\;\;\;\;\;\;\;\;\;\;\;\;\;\;\;\;\;\;\;\;\;,\;\; \frac{\pd \vA_L}{\pd t} = -  \vE_L \\
\frac{\pd \vE_T }{\pd t} &=&c^2 \nabla\times\nabla\times\vA_T  -\frac{ \vj_T}{\epsilon_0} \;\;\;, \;\; \frac{\pd \vE_L }{\pd t} = -\frac{ \vj_L}{\epsilon_0} \nonumber
\eqne
The last equation is equivalent to the continuity equation for the current
\beq
\frac{\partial \rho(\vecr,t)}{\partial t} = -\frac{\nabla\cdot\vj_L(\vecr,t)}{\epsilon_0} = -\frac{\nabla\cdot\vj(\vecr,t)}{\epsilon_0}
\eeq
as can be see by taking divergence of both parts and using the Gauss law for $\nabla\cdot \vE_L $. The equation  $\pd \vA_L/\pd t = -  \vE_L$,  shows how the longitudinal component of $\vA$ which does not enter the Hamiltonian (i.e. is the "cyclic" coordinate) develops in time for a given  $\vE_L$ which in turn is determined by the Gauss law via the charge density. One can see this as analogous to say the motion of the angular coordinates in a spherically symmetric problem as determined by the (conserved) angular momentum.

\section{Photons}
\subsection{Field oscillators} 
In the present and following sections we will disregard the radiation-matter interaction and will concentrate of the radiation part described by $H_r$.   Since this Hamiltonian is quadratic we will continue as in the case of a string. We will impose periodic boundary conditions and will expand $\vE_T(\vecr)$ and $\vA_T(\vecr)$ in terms of traveling waves. As we will see $H_r$ will become a sum of decoupled oscillators so the traveling waves are the normal modes of the radiation.

\subsubsection{Expansion in traveling waves}

Following the string example, cf., Eq.(\ref{eq:stringexpintravelwaves2}) and the footnote\footnote{For easy comparison we reproduce this expansion here 
\beq 
\bes
\phi(x,t)&=\sqrt{\frac{1}{L}}\sum_k\left[Q_k(t) \sin k  x  + \frac{P_k(t)} {v|k|} \cos k x \right]   \\
\pi(x,t) &= \sqrt{\frac{1}{L}}\sum_k\left[ P_k(t) \sin k  x -  v|k| Q_k(t) \cos k x \right] \;, \;  k=\frac{2\pi \nu}{L} \;,\; \nu = \pm 1, \pm 2 , ... \nonumber
\end{split}
\eeq
}
 below we expand the field canonical coordinates and momenta $\vA(\vecr)$ and $-\epsilon_0 \vE(\vecr)$, cf. Eq.\,(\ref{eq:MaxwellasHamltneqs}) in a large volume $\Omega$ 

\eqna \label{eq:expoffields1}
\vA(\vecr)&=&\frac{1}{\sqrt{\Omega\epsilon_0}}\sum_{\vk}\left(\vq_{\vk}\sin (\vk\cdot \vecr) +\frac{1}{ck}\vp_{\vk}\cos (\vk\cdot\vecr) \right)  \\
\vE(\vecr)&=& -\frac{1}{\sqrt{\Omega}\epsilon_0}\sum_{\vk}\left( \vp_{\vk}\sin (\vk\cdot\vecr)-ck\,\vq_{\vk}\cos (\vk\cdot \vecr)\right) \nonumber
\eqne
with vector expansion coefficients $\vq_{\vk}$ and $\vp_{\vk}$. To make the above expansions more symmetric with respects to the appearance of $\epsilon_0$ we changed our canonical variables to 
\[
\vA(\vecr) \to \vA(\vecr)/\sqrt{\epsilon_0} \;\; ,  \;\; -\epsilon_0 \vE(\vecr) \to  -\sqrt{\epsilon_0} \,\vE(\vecr) 
\]
 The periodic boundary conditions lead to discrete values of the wave vectors
\beq
\vk=\left\{(n_x, n_y, n_x)\frac{2\pi}{\Omega^{1/3}}\right\}\;\;, \;\; n_i=0, \pm 1, \pm 2 \, ...
\eeq
where we assumed the volume to be a cube, i.e. have the same  length, width and height each equal to $\Omega^{1/3}$.

The more conventional form  of the expansion Eq.\,(\ref{eq:expoffields1}(  found in the literature (cf., cf. Landau and Lifshitz, Classical Field Theory, Sec.52 or Ref.\,\cite{Coh} ) is written in terms of canonically transformed variables 
\beq
 \vq_{\vk}  \to - \frac{1}{\omega_k} \vp_{\vk} \;\;; \;\;  \vp_{\vk} \to  \omega_k  \vq_{\vk}  
 \eeq
 which for the transverse  components of the fields results in 
\eqna \label{eq:expoffields}
\vA_T(\vecr)&=&\frac{1}{\sqrt{\Omega\epsilon_0}}\sum_{\vk}\left(\vQ_{\vk} \cos (\vk\cdot\vecr)- \frac{1}{\omega_k} \vP_{\vk} \sin (\vk\cdot \vecr)  \right)  \\
\vE_T(\vecr)&=&-\frac{1}{\sqrt{\Omega\epsilon_0}}\sum_{\vk}\left( \vP_{\vk} \cos (\vk\cdot \vecr) +\omega_k  \vQ_{\vk} \sin (\vk\cdot\vecr)\right) \nonumber
\eqne
The transversality of $\vA_T(\vecr)$ and $\vE_T(\vecr)$ means that the vectors $\vQ_{\vk}$ and $\vP_{\vk}$ are always orthogonal to the corresponding $\vk$,

$$
\vk\cdot \vQ_{\vk}=0 \;\;\;,\;\;\; \vk\cdot\vP_{\vk}=0.
$$
It is convenient to use a pair of fixed unit polarization vectors $\vlambda_{\vk\alpha}$, $\alpha=1,2$ with
\beq
 \vlambda_{\vk 1}\cdot \vlambda_{\vk 2}=0 \;\;\;\;,\;\;\;\;  \vlambda_{\vk\alpha}\cdot\vk =0 \;\;\;  , \;\;\; \alpha=1,2
\eeq
so we can write
\beq \label{eq:expandinpolaris}
\vQ_{\vk}=\sum_{\alpha=1,2}Q_{\vk \alpha}\vlambda_{\vk\alpha} \;\;\;, \;\;\; \vP_{\vk}=\sum_{\alpha=1,2}P_{\vk \alpha}\vlambda_{\vk\alpha}
\eeq

We now insert expansions (\ref{eq:expoffields}) and (\ref{eq:expandinpolaris}) into the Maxwell equations for the transverse component, i.e. into the 1st and 3rd equation of the set (\ref{eq:Maxwellfortransverse}) in the absence of the current (recall we are discussing pure radiation). We obtain in the straightforward manner separate linear equations for $Q_{\vk \alpha}$ and $P_{\vk \alpha}$
\beq
\dot{Q}_{\vk \alpha}=P_{\vk \alpha} \;\;\; , \;\;\; \dot{P}_{\vk \alpha}= -\omega_k^2Q_{\vk \alpha} \;\;\; {\rm with} \;\; \omega_k=ck
\eeq
One clearly sees that these are Hamilton equations of  harmonic oscillators labeled by $\vk \alpha$ each with the Hamiltonian
 \beq
 H_{\vk\alpha}=\frac{1}{2}\left(P_{\vk\alpha}^2 +\omega_k^2 Q_{\vk\alpha}^2\right)
 \eeq
and $Q_{\vk\alpha}$ and $P_{\vk\alpha}$ being the generalized coordinates and momenta.
One could also obtain this by inserting expansions (\ref{eq:expoffields}) and (\ref{eq:expandinpolaris}) into $H_r$ to find
\beq \label{eq:decoplradiationmodes}
H_r=\sum_{\vk \alpha} H_{\vk\alpha}=\frac{1}{2}\sum_{\vk \alpha}\left(P_{\vk\alpha}^2 +\omega_k^2 Q_{\vk\alpha}^2\right) \;\;\;\; {\rm with} \;\;\; \omega_k=ck
\eeq
It is seen that indeed we represent $H_r$ as a sum of decoupled oscillators with frequencies given by the well know dispersion relation of the EM waves. There are two oscillators with different polarizations for each $\vk$. Since $\omega_k$ depends on the magnitude of $k$ all the oscillators with $|\vk|=k$ have the same frequency.

\subsubsection{Field wave functions and eigenstates. Photons appear}

We now turn to the quantum mechanics of the radiation. It is the easiest to do this in the decoupled eigenmodes of the field as encoded in the Hamiltonian (\ref{eq:decoplradiationmodes}).  Instead of classical time dependent variables $Q_{\vk\alpha}(t)$ and $P_{\vk\alpha}(t)$ we consider wave function $\Psi(\{Q_{\vk\alpha}\},t)$ which contains all the quantum information. This is "extracted" by using operators for every physical quantity which are build of two sets of basic operators of the "coordinate" and "momentum".
\eqna
Q_{\vk\alpha}\;\; \to \;\; \hat{Q}_{\vk\alpha} = Q_{\vk\alpha} \;\;\;\;; \;\;\;\; P_{\vk\alpha}\;\;\; \to \;\;\; \hat{P}_{\vk\alpha} = -i\hbar \frac{\pd}{\pd Q_{\vk\alpha}}  \\
\left[P_{\vk\alpha}, P_{\vk'\alpha'}\right]= \left[Q_{\vk\alpha}, Q_{\vk'\alpha'}\right]=0 \;\;\; ,
\left[Q_{\vk\alpha}, P_{\vk'\alpha'}\right]=i\hbar \delta_{\vk\vk'}\delta_{\alpha\alpha'} \nonumber
\eqne
All in the usual way as in quantum mechanics of mechanical systems.

The time dependence of  $\Psi(\{Q_{\vk\alpha}\},t)$ is governed by the Schr\"{o}dinger equation
 \beq
 i\hbar \frac{\pd \Psi(\{Q_{\vk\alpha}\},t)}{\pd t}=\hat{H}_r \Psi(\{Q_{\vk\alpha}\},t)
 \eeq
with the Hamiltonian operator $\hat{H}_r$ obtained from Eq.\,(\ref{eq:decoplradiationmodes}) by replacing in it the coordinates and momenta with the corresponding operators
\beq \label{eq:EM_H_sum_of_osc}
\hat{H}_r=\frac{1}{2}\sum_{\vk \alpha}\left(\hat{P}_{\vk\alpha}^2 +\omega_k^2 \hat{Q}_{\vk\alpha}^2\right)
\eeq
The most important solutions of the Schr\"{o}dinger equation are the stationary states which are the eigenstates of $\hat{H}_r$
\beq
\hat{H}_r \Psi(\{Q_{\vk\alpha}\})=E \Psi(\{Q_{\vk\alpha}\})
\eeq
Since $\hat{H}_r $ is a sum of independent terms each representing an oscillator the eigenvalues of $\hat{H}_r $ are sums of eigenenergies of independent oscillators
\beq
\bes
 \mathcal{E}_{\{N_{\vk\alpha}\}}&= \sum_{\vk \alpha}\hbar \omega_k \left(N_{\vk\alpha} +\frac{1}{2}\right) \equiv  
 \mathcal{E}+\sum_{\vk \alpha}\hbar \omega_k  N_{\vk\alpha} \\
 \mathcal{E}_0 & =  \sum_{\vk \alpha}\frac{ \hbar \omega_k}{2}  \;\;\; , \;\;\; N_{\vk\alpha}=0, 1, 2, 3, ... 
\end{split}
 \eeq
 The corresponding eigenfunctions are products
 \beq
 \Psi_{\{N_{\vk\alpha}\}}(\{Q_{\vk\alpha}\})=\prod_{\vk\alpha} \psi_{N_{\vk\alpha}}(Q_{\vk\alpha})   \;\;\; , \;\;\; N_{\vk\alpha}=0, 1, 2, 3, ... 
 \eeq
where $\psi_{N_{\vk\alpha}}(Q_{\vk\alpha})$ are the standard eigenfunctions of a harmonic oscillator, cf.,  Eq.(\ref{eq:harm_osc_wfs}),  with unit mass and frequency $\omega_k=ck$.

As in our discussion of the string quantization (and actually in the quantization of any linear dynamical system) we find that the EM field can be viewed as a collection of energy quanta
$$
\epsilon_{\vk\alpha} = \hbar\omega_k
$$
 in its normal modes.  In the following sections we will show that these quanta have all the characteristics of particles. They carry momentum, angular momentum and spin and have energy-momentum relation of massless particles moving with the speed of light, cf., Eq.(\ref{eq:phot_en_mom}). These quantum particles  are photons.
 
  Focusing on the details we note that  the EM modes are 3D vector waves. In our developments we have chosen them as plane waves with wave vectors $\vk$ and polarisations $\vlambda_{\vk\alpha}$.  But let us note that the eigenfrequencies of these modes $\omega_k=ck$ and as a result the energies $\epsilon_{\vk\alpha}$ of the quanta  depend only on the magnitude of $\vk$, i.e. on the wavelength and not 
  on  the direction of the vector $\vk$  and  the polarisation of the modes. 
  
  As with the ordinary matter particles with e.g. $\epsilon_{\vk}=\hbar^2 k^2/2m$ this means that there is a continuum degeneracy of the modes and therefore of the quantum mechanical states of the (free) photons. This degeneracy results in a freedom to change the basis states with a given energy $\epsilon$ from the  vector plane wave (like we did above)  to e.g. spherical (vector spherical!) or cylindrical (vector cylindrical!) etc waves.  The quantum numbers  $\vk\alpha$ will then be replaced by appropriately changed ones like $k, l, m$ replacing $k_x, k_y, k_z$ in the scalar waves. A recent reference to the vector spherical waves is e.g. Ref.\cite{Kri}. 
  
\subsubsection{The wave function of the vacuum. The Casimir effect}

The ground state  of the EM field is  the vacuum of the theory in the absence of matter and other quantum fields. Its wave function is the product of Gaussians familiar from our discussion of the guitar string, Eq. (\ref{gr_st_of_string}), 
\beq
\Psi_{\{N_{\vk\alpha} = 0\}}(\{Q_{\vk\alpha}\})=\prod_{\vk\alpha} \psi_{0}(Q_{\vk\alpha})  = \prod_{\vk\alpha} \left(\frac{\omega_k}{\pi \hbar} \right)^{1/4} 
 \exp \left(-\sum_{\vk\alpha} \omega_k Q_{\vk\alpha}^2/2\hbar \right)
\eeq
It provides perhaps the simplest example of quantum vacuum fluctuations of field degrees of freedom in a quantum field theory. 

\begin{figure}[H]
\centering \includegraphics[width=0.6\textwidth]{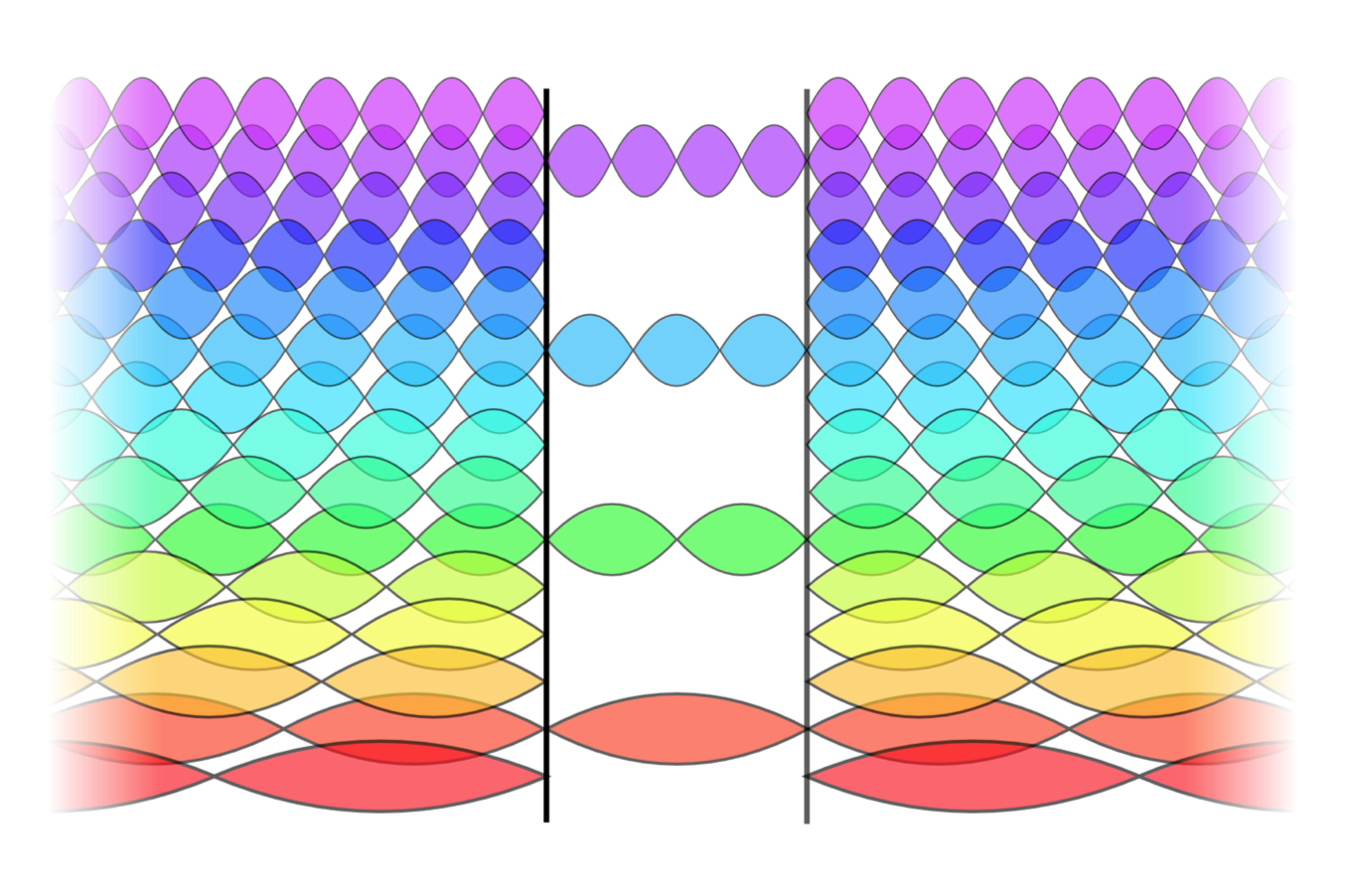}
\caption{Schematic depiction of the normal modes of the field vibrations in the presence of two plates   which enforce vanishing of the field  at the plates positions (Fig.1 from Ref.\cite{Kin}). This 
leads to the difference in the frequency densities of the field normal modes between the plates as compared to the outside free space. This difference depends on the distance between the plates and leads to the Casimir effect of plate attraction}
 \label{fig:Casimir_in3D}
\end{figure}

Can one observe these fluctuations?  In a ground breaking paper, Ref.\cite{Cas}, Casimir addressed this issue. He suggested that such fluctuations induce "attraction between two perfectly conducting plates".  On such plates the parallel to the plates components of the electric field must vanish so that the field normal modes for which this doesn't happen will be excluded from the field degrees of freedom and consequently from the vacuum fluctuations.  This is schematically illustrated in Fig.\ref{fig:Casimir_in3D}. The density of the normal modes frequencies between the plates will be smaller than in the free space outside. 

We have considered a one dimensional version of this effect in the context of the quantum guitar string, Sec.\ref{sec:string_Cas_effect}. We have shown there that it leads to an attractive force between the (analogue of) the plates with the more narrow spacing than that of the other part of the string.  The same happens in the realistic 3D case with quantized EM field. Casimir calculations predicted that an attractive force per unit area (pressure) at plate separation $a$ is given by 
 $$
 P= - \frac{\hbar \pi^2 c} {240 a^4}
 $$
 Note that the inverse quartic dependence on the distance is most unusual in physics. It is also worth mentioning that for certain special combinations of the plates materials, the Casimir force can be repulsive. The results obtained by Casimir were later extended to various geometries of the plates and his predictions were confirmed experimentally, cf., Ref.\cite{Sta}.

\subsubsection{Photon creation and annihilation operators. Field operators}

In practice it is convenient to introduce creation and annihilation operators of  photons in the standard way
\beq
\bes
 \hat{Q}_{\vk\alpha} =\sqrt{\hbar/2\omega_k}\left(\ha^\dagger_{\vk\alpha} + \ha_{\vk\alpha}\right) \;\;\; &,\;\;\;  \hat{P}_{\vk\alpha}=i\sqrt{\hbar\omega_k/2}\left(\ha^\dagger_{\vk\alpha} - \ha_{\vk\alpha}\right) \\
  \ha_{\vk\alpha}  =  \sqrt{1/ 2\hbar\omega_k}  \left( i\hat{P}_{\vk\alpha}  + \omega_k \; \hat{Q}_{\vk\alpha}\right)   \;\; & , \;\;   \ha^\dagger_{\vk\alpha}  = \sqrt{1/2\hbar\omega_k} \left(-i\hat{P}_{\vk\alpha}  + \omega_k \; \hat{Q}_{\vk\alpha} \right)  \\
[\ha_{\vk\alpha},\ha^{\dagger}_{\vk'\alpha'}]=\delta_{\vk\vk'}\delta_{\alpha\alpha'} \;\;\;  &, \;\;\; [\ha_{\vk\alpha},\ha_{\vk'\alpha'}]=0=[\ha^{\dagger}_{\vk\alpha},\ha^{\dagger}_{\vk'\alpha'}]
 \end{split}
\eeq
Using these operators we can write the Hamiltonian
\beq \label{eq:Hr_in_a_adagger}
\hat{H}_r=E_0+\sum_{\vk\alpha}\hbar\omega_k \ha^{\dagger}_{\vk\alpha} \ha_{\vk\alpha}
\eeq
and its eigenstates
\beq
|\{N_{\vk\alpha}\}>=\prod_{\vk\alpha}|N_{\vk\alpha}>=
\prod_{\vk\alpha}\frac{(\ha^{\dagger}_{\vk\alpha})^{N_{\vk\alpha}}}
{(N_{\vk\alpha}!)^{1/2}}|0>
\eeq
Great advantage of using $\ha$ and $\ha^{\dagger}$ operators rather than $\hat{P}$ and $\hat{Q}$ in dealing with photons is the simplicity of the "action" of these operators on the "number states", i.e the states with a fixed photon numbers in each normal mode. Schematically
 $$
 \ha|n>=\sqrt{n}|n-1> \;\;\; , \;\;\;  \ha^{\dagger}|n>=\sqrt{n+1}|n+1>
 $$
 In "full glory"
 $$
 \ha_{\vk\alpha}|\{N_{\vk\alpha}\}>=\sqrt{N_{\vk\alpha}}|N_{\vk\alpha}-1>\prod_{\vk'\ne\vk,\alpha'\ne\alpha}
 |\{N_{\vk'\alpha'}\}>
 $$
 $$
 a^{\dagger}_{\vk\alpha}|\{N_{\vk\alpha}\}>=\sqrt{N_{\vk\alpha}+1}|N_{\vk\alpha}+1>\prod_{\vk'\ne\vk,\alpha'\ne\alpha}
 |\{N_{\vk'\alpha'}\}>
 $$
 One says that the operators $\ha^{\dagger}_{\vk\alpha}$ and $\ha_{\vk\alpha}$ create and destroy(annihilate) photons.
 
It is useful to express the operators of the fields (\ref{eq:expoffields}) as well as $\hat{\vB}(\vecr)$
in terms of the creation and annihilation operators
 \beq\label{EM_fld_via_a}
 \bes
 \hat{\vA}_T(\vecr)&=\sum_{\vk \alpha}\left(\frac{\hbar }{2\epsilon_0\omega_k\Omega}\right)^{1/2}\left[\ha_{\vk\alpha}\vlambda_{\vk\alpha}e^{i\vk\cdot\vecr}+\ha^{\dagger}_{\vk\alpha}\vlambda_{\vk\alpha}e^{-i\vk\cdot\vecr} \right]=  \\
 &=\sum_{\vk \alpha}\left(\frac{\hbar }{2\epsilon_0\omega_k\Omega}\right)^{1/2}\vlambda_{\vk\alpha}e^{i\vk\cdot\vecr}(\ha_{\vk\alpha}+\ha^{\dagger}_{-\vk\alpha})  \\
 \hat{\vE}_T(\vecr) &=\sum_{\vk \alpha}i\left(\frac{\hbar\omega_k }{2\epsilon_0\Omega}\right)^{1/2}\left[\ha_{\vk\alpha}\vlambda_{\vk\alpha}e^{i\vk\cdot\vecr}-\ha^{\dagger}_{\vk\alpha}\vlambda_{\vk\alpha}e^{-i\vk\cdot\vecr} \right] \\
 \hat{\vB}(\vecr) &=\sum_{\vk \alpha}i\left(\frac{\hbar }{2\epsilon_0 \omega_k\Omega}\right)^{1/2}\left[\ha_{\vk\alpha}(\vk\times\vlambda_{\vk\alpha})e^{i\vk\cdot\vecr}-\ha^{\dagger}_{\vk\alpha}(\vk\times\vlambda_{\vk\alpha})e^{-i\vk\cdot\vecr} \right]
 \end{split}
 \eeq
  Once the longitudinal components of the fields $\vA_{op}(\vecr)$ and $\vE_{op}(\vecr)$ were separated the remaining transverse parts  $\hat{\vA}_T(\vecr)$ and $\hat{\vE}_T(\vecr)$ do not obey the canonical commutations (\ref{eq:field_comm}).  Rather the delta function there gets replaced by the so called "transverse" delta function cf., Ref.\cite{Coh}, Ch.\.III.A.1.  
 
 \subsection{Photon momentum}
 In this subsection we will consider the operator of the momentum of the EM field $\vP_{\rm field}$. Using it we will be able to show that photons are not just "portions" of energy of the EM field but that they also carry a "corresponding" portion of its momentum. Moreover the relation between the energies and  momenta of these portions are as of {\em massless} particles traveling with the speed of light.

 \subsubsection{Generators of translations in the matter-field system}
  Quantum mechanically it is probably the easiest to guess the expression of the momentum by recalling that it is the generator of infinitesimal translations.  For the interacting system
 of particles (matter) and EM field described by the Hamiltonian (\ref{eq:HamOperofEMwithmatter}) the operation of infinitesimal translation is the transformation
  \eqna\label{eq:translations}
e^{(i/\hbar)\va\cdot\hat{\vP}}{\vA}_{op}(\vecr) e^{(-i/\hbar)\va\cdot\hat{\vP}}&=& {\vA}_{op}(\vecr+\va)\approx{\vA}_{op}(\vecr)+(\va\cdot\nabla) {\vA}_{op}(\vecr)   \nonumber \\
e^{(i/\hbar)\va\cdot\hat{\vP}}{\vE}_{op}(\vecr) e^{(-i/\hbar)\va\cdot\hat{\vP}} &=& {\vE}_{op}(\vecr+\va) \approx {\vE}_{op}(\vecr)+(\va\cdot\nabla) {\vE}_{op}(\vecr) \\
 e^{(i/\hbar)\va\cdot\hat{\vP}}\hat{\vecr}_a e^{(-i/\hbar)\va\cdot\hat{\vP}} &=&  \hat{\vecr}_a +\va , \;\; a=1,...,N \nonumber
 \eqne
 Therefore the(vector) momentum operator $\hat{\vP}=\{\hat{P}_x, \hat{P}_y, \hat{P}_z\}$ of the system should be such that for each of its component $\hat{P}_k $,   the commutators hold
 \beq
  [\hat{P}_k, \hat{A}_j(\vecr)]  =  -i\hbar \partial_k \hat{A}_j(\vecr)  \;\;\;, \;\;
  [\hat{P}_k, \hat{E}_j(\vecr)]  =  -i\hbar \partial_k{E}_j(\vecr)
  \eeq
  \beq
   [\hat{P}_k, r_{j,a}]  =  -i\hbar\frac{ \partial r_{j,a}}{\partial r_{k,a}} =-i\hbar \delta_{kj}
  \eeq
It is actually very easy to guess what such $\hat{\vP}$ should be
\beq \label{eq:totalmom1}
\hat{\vP}= \hat{\vP}_{\rm matter} + \hat{\vP}_{\rm field} = \sum_{a=1}^N \hat{\vp}_a +
\frac{\epsilon_0}{2}\sum_{j=1}^3 \int d^3 r \left\{\hat{E}_j (\vecr)\nabla \hat{A}_j (\vecr) + h.c.\right\}
\eeq
 where $\hat{p}_a=-i\hbar \nabla_a \;, \; a=1, ... N$  and the "h.c." abbreviation stands for "hermitian conjugate".

  Indeed  the first term $\hat{\vP}_{\rm matter} $ has the required commutator with $\hat{\vecr}_a$ while commuting with ${\vE}_{op}(\vecr)$ and   ${\vA}_{op}(\vecr)$ and the second term $\hat{\vP}_{\rm field}$ commutes with  $\hat{\vecr}_a$ and satisfies
 \eqnaa
 [\hat{P}_{{\rm field},k} ,\hat{A}_j(\vecr)] &=&\frac{\epsilon_0}{2} \sum_{n=1}^3 \int d^3 r'\left\{ [\hat{E}_n (\vecr')\partial'_k \hat{A}_n (\vecr'),\hat{A}_j(\vecr)] + ... \right\}= \nonumber  \\ &=&-\epsilon_0 \sum_{n=1}^3 \int d^3 r' \frac{i\hbar \delta_{nj}}{\epsilon_0}\delta(\vecr'-\vecr))\partial'_k \hat{A}_n (\vecr') = -i\hbar\partial_k \hat{A}_j (\vecr)
 \eqnae
 \eqnaa
 [\hat{P}_{{\rm field},k} ,\hat{E}_j(\vecr)] &=&\frac{\epsilon_0}{2} \sum_{n=1}^3 \int d^3 r'\left\{ [\hat{E}_n (\vecr')\partial'_k \hat{A}_n (\vecr'),\hat{E}_j(\vecr)]  + ... \right\}= \nonumber  \\ &=&-\epsilon_0 \sum_{n=1}^3 \int d^3 r' \hat{E}_n (\vecr'))\partial'_k\frac{-i\hbar \delta_{nj}}{\epsilon_0}\delta(\vecr'-\vecr)) = -i\hbar\partial_k \hat{E}_j (\vecr)
  \eqnae
where in the last line we used the integration by parts.

Let us recall the classical expression for the conserved momentum in the presence of the EM field. On the basis of the Maxwell and Newton equations Eqs.(\ref{eq:Maxwelleqs} - \ref{eq:NewtforNcharges}) one finds that
 \beq \label{eq:TotalMomentum}
 \vP=\sum_{a=1}^N m_a \vv_a +\epsilon_0 \int d^3 r \, \vE(\vecr)\times \vB(\vecr)
 \eeq
 is the conserved total momentum of the field-matter system
 \beq
 \frac{d\vP }{dt}=0
 \eeq
 cf. p. 61, in Ref.\cite{Coh}   or a less formal text - Ref.\cite{Fey}.
 In Appendix\,\ref{sec:EM_momentum} below we show the equivalence of the expressions (\ref{eq:totalmom1}) and (\ref{eq:TotalMomentum}). 

\subsubsection{Momentum of the EM radiation}
  In the absence of  charged particles we can use  $\nabla\cdot \hat{\vE} =0$ and  replace $\hat{E}_j$ by the transversal $\hat{E}_{Tj}$ in the field part of the momentum in (\ref{eq:totalmom1}). The same can be done with $\hat{A}_j$ in it. Indeed the longitudinal part of $\hat{\vA}$ can be written as a gradient  $\vA_L=\nabla \hat{\xi}$.  Thus  it contributes (after the replacement $\hat{E}_j \to \hat{E}_{Tj}$) the term
  $$
  \frac{\epsilon_0}{2}\sum_{j=1}^3 \int d^3 r \left\{\hat{E}_{Tj} (\vecr)\nabla \hat{A}_{Lj} (\vecr) + h.c.\right\} =
   \frac{\epsilon_0}{2}\sum_{j=1}^3 \int d^3 r \left\{\hat{E}_{Tj} (\vecr)\nabla \partial_j \hat{\xi}(\vecr) + h.c.\right\}
 $$
 in $\hat{\vP}_{\rm field}$. This term is however zero as can be seen by integrating by parts in the right hand side and using $\partial_j \hat{E}_{Tj} =0$.

 Thus we can write the momentum of the "pure" radiation as
 \beq \label{radiation_mom}
\hat{\vP}_r  = \frac{\epsilon_0}{2} \int d^3 r \,\sum_{j=1}^3 \left[\hat{E}_{T,j}(\vecr) \nabla \hat{A}_{T,j} (\vecr) + h.c.\right]
\eeq
Inserting expressions for the operators $\vA_T(\vecr)$ and $\vE_T(\vecr)$ one obtains
\beq
 \hat{\vP}_r  =\sum_{\vk\alpha}\; \hbar\vk \; \ha^{\dagger}_{\vk\alpha} \ha_{\vk\alpha}
\eeq
 where we used $\sum_{\vk\alpha}\hbar\vk =0$ \footnote{There is a subtle point here - this sum diverges and must be regularized by, say, assuming a cutoff at some large $k_c$. }.
 As it should $ \hat{\vP}_r $ commutes with the Hamiltonian $\hat{H}_r$. Its eigenvalues are
 \beq
  \vP_{\{N_{\vk\alpha}\}} =  \sum_{\vk\alpha} \hbar\vk \; N_{\vk\alpha}\;\;, \;\;\; N_{\vk\alpha} = 0,1,2,3,...
  \eeq
 We can see that every state with $N_{\vk\alpha}$ quanta has momentum $\hbar\vk N_{\vk\alpha}$ so that
 every energy  quantum with  $\epsilon_k=\hbar\omega_k$ carries  momentum $\vp_k=\hbar\vk$.
 Using the dispersion relation $\omega_k=c|\vk|$ of the (classal)  light waves (EM normal modes) we find the energy-momentum relation of light quanta
 \beq \label{eq:phot_en_mom}
 \epsilon_k=c|\vp_k|
 \eeq
 i.e. of the massless particle moving with the light velocity.
 
  \subsection{Common states of light}
 \subsubsection{Number states}
 These are just the eigenstates $|\{N_{\vk\alpha}\}\rangle$ of the $\hat{H}_r$, cf., Eq. (\ref{eq:Hr_in_a_adagger}). Although most natural from the formal point of view they are highly nonclassical and in fact are extremely hard to produce "on demand"\footnote{E.g.
M. Oxborrow and A.G. Sinclair, Contemp. Phys. 46, 173 (2005).}.
Number states are states of well defined energy but not of the states of well defined EM field. As an example consider a single mode of the electric fields, i.e.  just one term with a given $\vk, \alpha$ in the expression for $\vE_T$ in (\ref{EM_fld_via_a}) and calculate 
\beq
 \langle N_{\vk\alpha}  |  \hat{\vE}(\vecr)  | N_{\vk\alpha}  \rangle =i\vlambda_{\vk\alpha}\left(\frac{\hbar\omega_k }{2\epsilon_0\Omega}\right)^{1/2} \langle N_{\vk\alpha}  | \left[\ha_{\vk\alpha}e^{i\vk\cdot\vecr}-\ha^{+}_{\vk\alpha}e^{-i\vk\cdot\vecr} \right]  | N_{\vk\alpha}  \rangle  = 0
 \eeq
 and 
 \beq
 \bes
 \Delta E &= \sqrt{ \langle N_{\vk\alpha}  |  \hat{\vE}(\vecr)\cdot  \hat{\vE}(\vecr)  | N_{\vk\alpha}  \rangle }= \\
 &=\left(\frac{\hbar\omega_k }{2\epsilon_0\Omega}\right)^{1/2}  \langle N_{\vk\alpha}  |   \ha_{\vk\alpha} \ha^{+}_{\vk\alpha} + \ha^{+}_{\vk\alpha}  \ha_{\vk\alpha}  | N_{\vk\alpha}  \rangle = \left(\frac{\hbar\omega_k }{2\epsilon_0\Omega}\right)^{1/2}(2N_{\vk\alpha} +1)^{1/2}
 \end{split}
 \eeq
 so that the everage value of $ \hat{\vE}$ is zero while the fluctuations grow with the number of photons. 
 
  \subsubsection{Quantum mechanics behind the classical EM field.  Coherent states of light}
 The correct description of the world is quantum mechanical while the classical physics is just an approximation. So it is natural to ask what is the quantum mechanical state behind the classical EM field?  Since the quantum mechanical operators of electric and magnetic components 
 $\hat{\vE}(\vecr)$ and $ \hat{\vA}(\vecr)$ of the field are non commuting there is no state in which they  both have definite values.   
 
 Under these restrictions the appropriate quantum state  $|\Psi (t)\rangle$ behind the classical EM field must be such that the averages, i.e. the expectation values of the field operators
 $$
 \vE(\vecr,t)\equiv \langle \Psi(t) | \hat{\vE}(\vecr)|\Psi(t)\rangle \;\;\; , \;\;\; \vA(\vecr,t) \equiv \langle \Psi(t) | \hat{\vA}(\vecr)|\Psi(t)\rangle
 $$
 will be developing in time as solutions of the classical Maxwell equation and be "classically large" i.e. much larger than the quantum uncertainties i.e. the standard deviations of these fields from the averages..

 It is not hard to find the state with the above properties for a free EM field. Since such a field can be represented as a collection of independents modes it is useful to start by considering a simple case of just a single mode with a given wave number $\vk$ and polarisation $\vlambda$.  Concentrating on the electric field we have the operator
\beq \label{eq:El_fld_single_mode}
\hat{\vE}(\vecr)=\frac{\vlambda_{\vk}}{\sqrt{\Omega\epsilon_0}}\left[ck \hat{Q}_{\vk}\cos (\vk\cdot\vecr) -\hat{P}_{\vk}\sin (\vk\cdot\vecr)\right] 
\eeq
 The quantum mechanics of $\hat{\vE}(\vecr)$ and its non non commutativity with $\hat{\vA}(\vecr)$  is "encoded" in the canonical non 
 commuting pair of the operators $\hat{P}_{\vk}, \hat{Q}_{\vk}$.  Their dynamics (for a free field) is simple - just that of  harmonic oscillator, cf., Eq. (\ref{eq:EM_H_sum_of_osc}).  
 
 So the task is to find a quantum state of harmonic oscillator, i.e.  solutions $|\psi(t)\rangle$ of the \Sch  equation
 \beq
 i\hbar \frac{\partial |\psi(t)\rangle }{\partial t} = \hat{h} |\psi(t)\rangle \;\;  {\rm with} \;\; \;\; \hat{h} = \frac{1}{2}\left(\hat{p}^2 + \omega^2 \hat{q}^2\right)
 \eeq
 for which the averages 
 $$
 q(t) = \langle \psi(t)| \hat{q} |\psi(t)\rangle \;\; \; , \;\;\; p(t) = \langle \psi(t)| \hat{p} |\psi(t)\rangle
 $$
 obey the classical equations of the harmonic oscillator
 \beq \label{eq:eqs_ho}
 \dot{q} = p \;\;, \;\; \dot{p} = -\omega^2 q
 \eeq
 and have smallest possible quantum uncertainties. 
 
 Such a state was first discussed by \Sch already in 1926 and has a name - coherent state.  Its common formal definition is that it is an eigenstate of the annihilation operator 
 \beq  \label{eq:coh_state}
 \ha |\alpha\rangle = \alpha |\alpha \rangle    \;\; 
 \Rightarrow  \;\;\; \left(\hbar \frac{\partial }{\partial q} + \omega \; q \right)  \psi_{\alpha} (q) = \sqrt{2\hbar\omega}  \alpha \psi_{\alpha} (q) 
 \eeq
 with eigenstates $|\alpha\rangle$ labeled by the eigenvalues $\alpha$ and where we used the coordinate representation of $\ha$
 $$
 \hat{a} =   \left( i\hat{p}  + \omega \; \hat{q} \right) /  \sqrt{2\hbar\omega}
 $$
 Note that since $\ha$  is non hermitian $\alpha$'s are in general complex valued\footnote{There are many unfamiliar features of $|\alpha\rangle$ states as a result of this.  Like non orthogonality at different $\alpha$'s or over completeness. This will be partly covered in the Appendix}.  Also note that for $\alpha = 0$ the coherent state is just a ground state of the harmonic oscillator 
 \beq \label{eq:gs_ho}
 \psi_0(q) = A e^{-\omega q^2/2\hbar} \;\; , \;\; A=(\omega/\pi\hbar)^{1/4}
 \eeq
 
\begin{figure}[H]
\centering \includegraphics[width=0.6\textwidth]{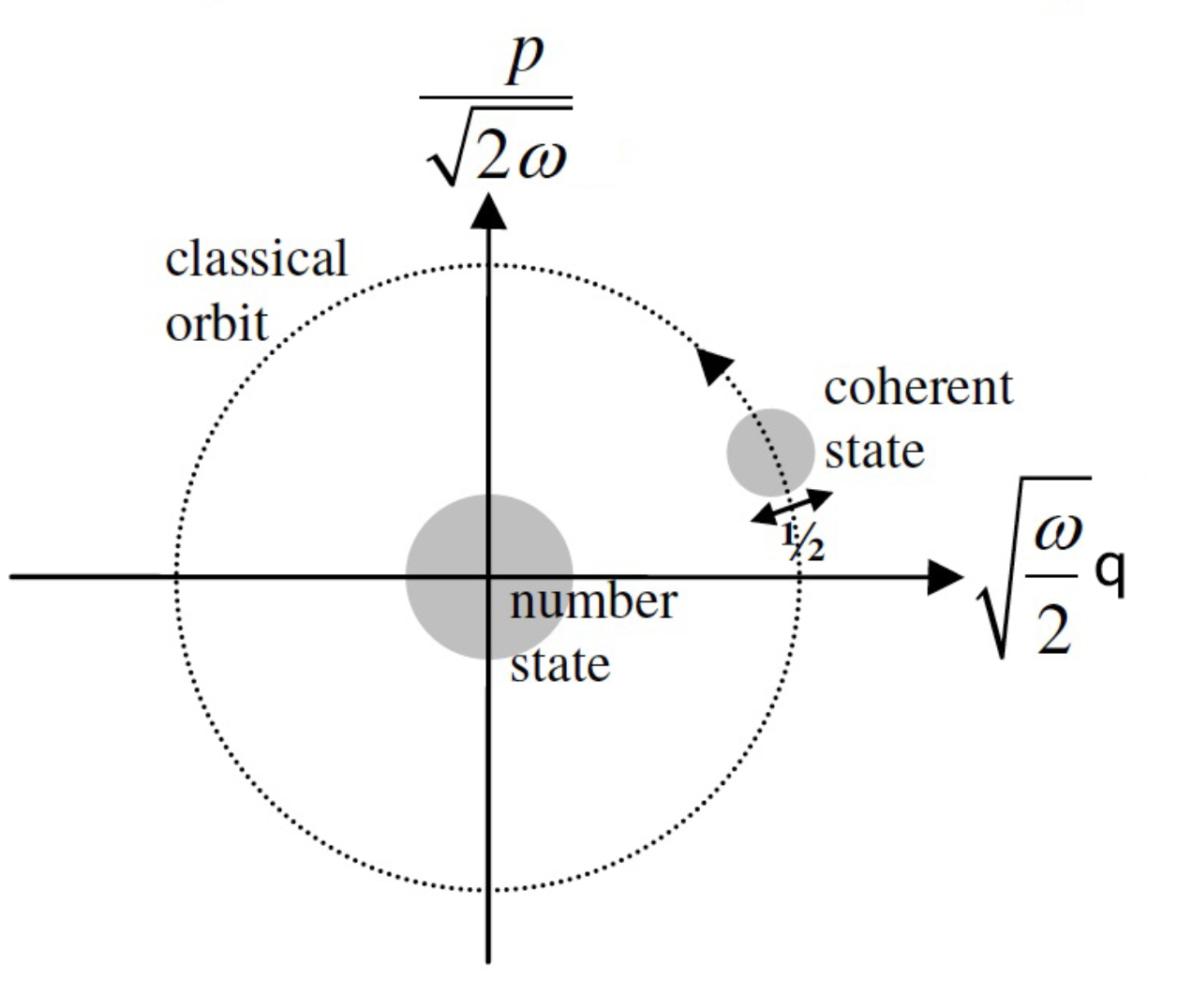}
\caption{Schematic representation of a coherent state and its motion as a smeared distribution (like e.g. Wigner distribution)  in a classical phase space, cf. Eq.(\ref{eq:alpha_via_p_q}). We use here $\hbar =1$ units. For comparison also a number state $\psi_n(q) =\langle q |n\rangle$ centered at the phase space origin ($\langle n |\hat{q}|n\rangle = \langle n |\hat{p}|n\rangle =0$)  is shown schematically.}
 \label{fig:coh_state}
\end{figure}

 The properties of the  coherent state are discussed in the Appendix of this chapter. It is shown there that coherent state is a wave packet  the  dynamics of which is such that the  averages 
 \beq
 q_0\equiv  \langle \alpha | \hat{q} | \alpha \rangle \;\; , \;\;  p_0 \equiv \langle \alpha | \hat{p} | \alpha \rangle
 \eeq
 move along the corresponding classical trajectories with uncertainties obeying the minimum uncertainties relation 
 \beq \label{eq:min_unc}
\Delta q \Delta p = \hbar/2
\eeq
 It is useful to schematically present this picture in the classical phase space as is explained in Fig. \ref{fig:coh_state}.

Let us briefly consider how the EM field "looks like" in a coherent state.  Consider a single mode  (\ref{eq:El_fld_single_mode}) written in terms of the photon and assume it is in a coherent state  $|\alpha_{\vk}(t) \rangle$.  Then 
\beq
\bes
 \langle \alpha_{\vk} (t) |  \hat{\vE}(\vecr) & | \alpha_{\vk} (t) \rangle =i\vlambda_{\vk}\left(\frac{\hbar\omega_k }{2\epsilon_0\Omega}\right)^{1/2}\left[\alpha_{\vk}(t)e^{i\vk\cdot\vecr}-\alpha^{*}_{\vk}(t)e^{-i\vk\cdot\vecr} \right] = \\
&= \vlambda_{\vk}|\alpha_{\vk}(0)|\left(\frac{2\hbar\omega_k }{\epsilon_0\Omega}\right)^{1/2}\sin(\omega_kt -\vk\cdot\vecr  -\phi_{\vk})
\end{split}
\eeq
where we used the results (\ref{t_dep_coh_st}, \ref{t_dep_coh_st_1}) from the Appendix.  This expression for the average $\hat{\vE}(\vecr)$ has the form of a classical field. Its amplitude is controlled by $|\alpha_{\vk}(0)|$, cf., the radius of the classical trajectory in Fig. \ref{fig:coh_state}.

Calculating 
\beq
\langle \alpha_{\vk} (t) |  \hat{\vE}(\vecr)\cdot  \hat{\vE}(\vecr) | \alpha_{\vk} (t) \rangle  = 
\frac{\hbar\omega_k}{2\epsilon_0\Omega}  \left[1 + 4|\alpha_{\vk}(0)|^2\sin^2(\omega_kt -\vk\cdot\vecr  -\phi_{\vk})\right]
\eeq
we obtain for the quantum fluctuations
\beq 
\Delta E \equiv \sqrt {\langle \alpha_{\vk} (t) |  \hat{\vE}(\vecr)\cdot  \hat{\vE}(\vecr) | \alpha_{\vk} (t) \rangle -  \langle \alpha_{\vk} (t) |  \hat{\vE}(\vecr) | \alpha_{\vk} (t) \rangle^2 } = \frac{\hbar\omega_k}{2\epsilon_0\Omega} 
\eeq
which is independent of the magnitude of the average, cf., again Fig. \ref{fig:coh_state}.  So for the electric field $\gg$ than the quantum scale of the fluctuations $\Delta E$ the field can be viewed as classical.

 \subsubsection{Thermal light}
 Thermal radiation is radiation in thermal equilibrium, which means (as is usual in quantum statistical  physics) that this radiation is described not by a wave function (or rather wave functional) but by the density matrix.  This density matrix is diagonal in the eigenenergy basis  
 \beq
 \rho = \sum_{\{N_{\vk\alpha}\}} w (\{N_{\vk\alpha}\}) | \{N_{\vk\alpha}\} \rangle \langle \{N_{\vk\alpha}\} |
 \eeq
 with the probabilities given by the Boltzmann factor
 $$
 w (\{N_{\vk\alpha}\}) = \frac{1}{Z(T)} \exp\left[-\frac{\mathrm{\mathcal{E}}(\{N_{\vk\alpha}\})}{T}\right]  \;\;\;, \;\;\;  Z(T) =  \sum_{\{N_{\vk\alpha}\}} \exp\left[-\frac{\mathcal{E}(\{N_{\vk\alpha}\})}{T}\right] 
 $$
 which of course is equivalent to saying that  the radiation power follows the Plank law. Just to remind - by using
 $$
 \mathcal{E}(\{N_{\vk\alpha}\}) = \sum_{\vk\alpha} \hbar \omega_k N_{\vk\alpha}  
 $$
 separating exponentials in $w (\{N_{\vk\alpha}\})$ and $Z(T)$ into products with different $\vk\alpha$ and summing over  $N_{\vk\alpha}$ 
 for each $\vk\alpha$  in $Z(T)$ one obtains
 \beq
  w (\{N_{\vk\alpha}\}) = \prod_{\vk\alpha} w (N_{\vk\alpha}) \;\;\; {\rm with} \;\;\;  w (N_{\vk\alpha}) =
\left( 1- e^{-\beta\hbar\omega_k}\right)  \exp{(-\beta\hbar\omega_k N_{\vk\alpha})}
  \eeq
 and $\beta=(k_BT)^{-1}$.   
 The average energy per mode is
 $$
 \langle \mathcal{E}\rangle  _{\vk\alpha} =  \sum_{N_{\vk\alpha}}\left( \hbar \omega_k N_{\vk\alpha}  \right) w (N_{\vk\alpha}) = \hbar \omega_k \langle N \rangle_{\vk\alpha} = \frac{ \hbar \omega_k}{e^{\beta\hbar\omega_k} -1}
 $$
 and the Plank spectral energy density 
 $$
 dn = \sum_\alpha \int_{\gamma\in 4\pi}  \langle \mathcal{E}\rangle  _{\vk\alpha} \frac {d^3 k}{(2\pi)^3} =  \frac{ \hbar \omega_k}{e^{\beta\hbar\omega_k} -1}
 \int_{\gamma\in 4\pi}  \frac { 2 k^2 dk d\gamma}{(2\pi)^3} 
= \frac{8\pi h \nu^3}{c^3}   \frac{ h \nu }{e^{\beta h\nu} -1} d\nu  
 $$
 with $h\nu= \hbar\omega$.
 
\begin{figure}[H]
\centering \includegraphics[width=0.6\textwidth]{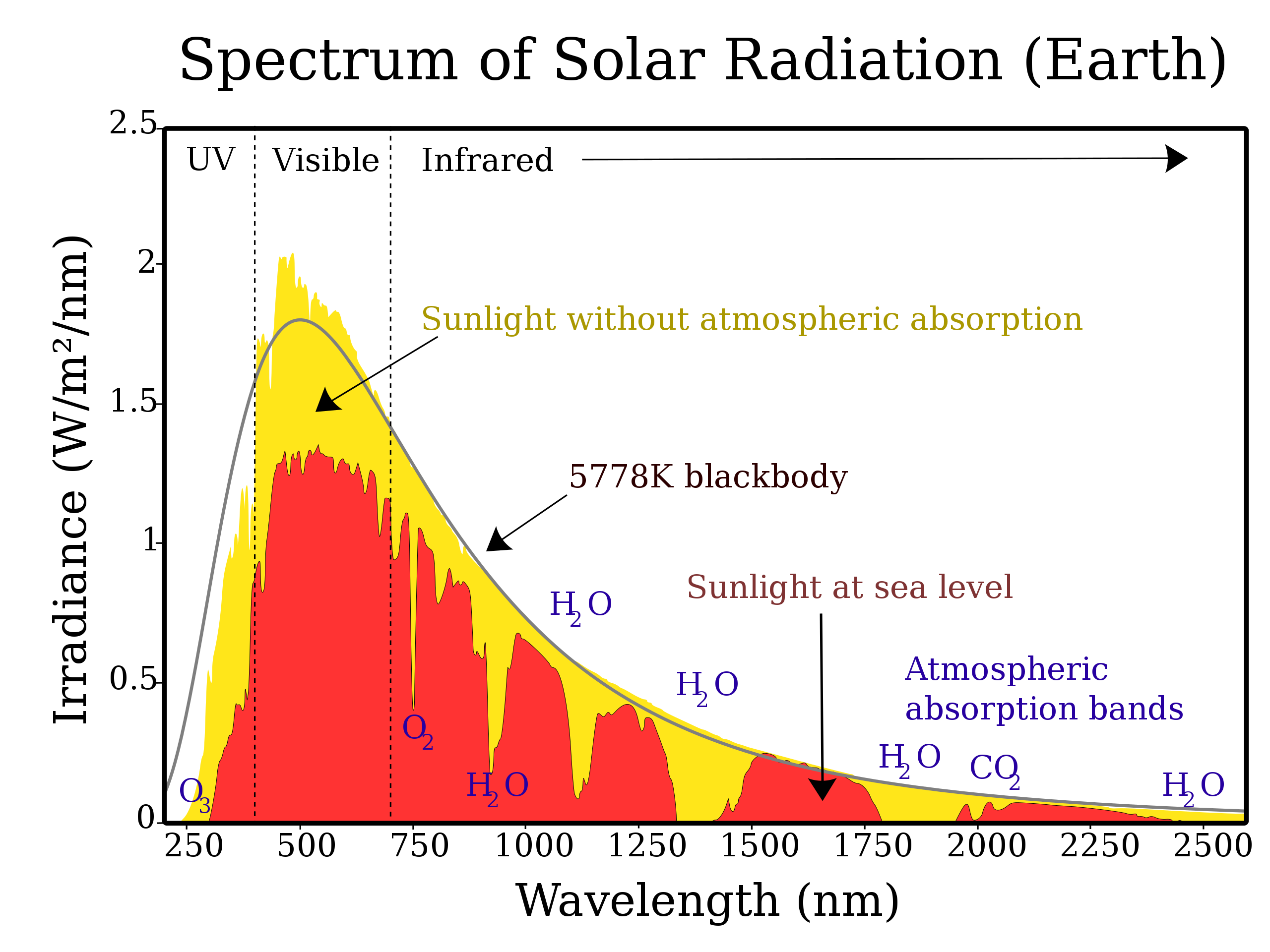}
\centering \includegraphics[width=0.6\textwidth]{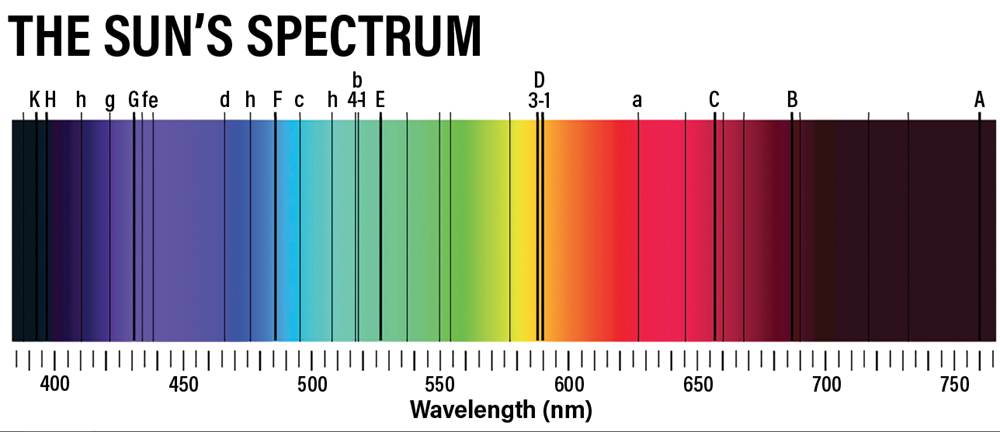}
\caption{Above - the Plank black body spectrum and its modifications in real world.  Below - discrete absorption lines on the background of the continuum solar light spectrum}
 \label{fig:Plank_rad}
\end{figure}

 Such a spectrum is an idealization of a radiation spectrum emitted by matter sources which by themselves are in a thermal equilibrium and  moreover the radiation which they emit "has enough time" inside the matter to reach equilibrium with it. The major factors "distorting" such spectra are layers of matter (like sun and earth atmospheres) between the equilibrated matter-radiation system and the observer. If such layers have different temperature and are too thin the light will "have no time" to re-equilibrate as it passes through them. The layers will just absorb some of the passing radiation at particular wave lengths depending on their chemical composition. This will produce the corresponding "absorption lines" in the radiation spectrum.   Hot excited atoms, molecules, etc, inside the layers will also emit and add non equilibrated light at particular wavelengths 
producing the "emission lines".  Example of the observed solar radiation spectrum, cf., Fig.\,\ref{fig:Plank_rad} provides a good illustration of these  features. 
 
 \subsection{Photon angular momentum and spin}

Using the Maxwell and Newton equations (\ref{eq:Maxwelleqs}),  (\ref{eq:NewtforNcharges}) together with the charge current and density  (\ref{eq:denscurrent}) one can show (cf., Ch 1A in Ref.\cite{Coh})  that the conserved angular momentum of the matter-field system is
\beq \label{eq:TotalAngMomentum}
 \vJ=\sum_{a=1}^N \vecr_a\times m_a \vv_a +\epsilon_0 \int d^3 r \,\vecr\times\left[ \vE(\vecr)\times \vB(\vecr)\right]
 \eeq
Comparing with the expression (\ref{eq:TotalMomentum}) for the matter-field momentum both terms have intuitively  clear meaning.

It is important to note that as is usual with the definition of  angular momentum the expression  (\ref{eq:TotalAngMomentum}) refers to a specific point - the origin of the chosen coordinate system -  with respect to which $\vJ$ is calculated. This of course can be  easily changed by replacing $ \vecr_a \to \vecr_a-\vecr_0$ and $\vecr \to \vecr -\vecr_0$ with an arbitrary vector $\vecr_0$ in both terms of $\vJ$ respectively.  This change leads to a straightforward  generalisation of the classical mechanics relation for such  transformations of angular momenta
$$
\vJ \;\; \Rightarrow \;\; \vJ'= \vJ -  \vecr_0 \times \vP
$$

\subsubsection{Generators of rotations in the matter-field system}

 Consider infinitesimal rotations of the coordinate system
 $$
 \vecr \to  \vecr'= \vecr +\delta\vecr= \vecr + \delta \valpha \times \vecr
 $$
where as usual the magnitude of  the vector $\delta \valpha$ is equal the rotation angle and it is directed along the axis of rotation (right hand rule).   We want to determine how the wave functional $\Psi[\vA(\vecr), \vecr_1, ..., \vecr_N]$ changes under this transformation.

Let us start by recalling that a scalar field change  obeys the intuitive rule
$$
\phi(\vecr) \to \phi'(\vecr') = \phi(\vecr)
$$
saying that the values of the rotated field $\phi'$  at rotated points $\vecr'$ are the same as non rotated field $\phi$  in original points
$\vecr$.  Using $\vecr = \vecr' -\delta \vecr$ and dropping the prime in $\vecr'$ on both sides can write
\eqna
\phi'(\vecr) &=& \phi(\vecr - \delta \vecr)  \approx  \phi(\vecr) -\delta \vecr \cdot \nabla  \phi(\vecr) = \nonumber  \\  &=&  \phi(\vecr) -(\delta \valpha \times \vecr) \cdot \nabla \phi(\vecr) =[1-\delta\valpha \cdot (\vecr\times\nabla)]\phi(\vecr) = \nonumber \\
&=& [1-\frac{i}{\hbar} \delta\valpha \cdot \hat{\vl} ]\phi(\vecr) \;\;\; {\rm with} \;\;\;  \hat{\vl} = -i\hbar[\vecr \times \nabla]
\eqne
 For a vector field one also has to rotate the field itself
 $$
 \vA(\vecr) \to \vA'(\vecr') =[1+ \delta\valpha \times] \vA(\vecr'- \delta \valpha \times \vecr)
 $$
 which gives (after dropping the prime on $\vecr'$)
 \eqna \label{rotating_vec_field}
\vA'(\vecr) &\approx& \vA(\vecr) + \delta \vA(\vecr) = \vA(\vecr) + \delta\valpha \times \vA(\vecr)  -  [(\delta \valpha \times \vecr)\cdot \nabla]  \vA(\vecr) = \nonumber \\
&=&  \vA(\vecr) + \delta\valpha \times \vA(\vecr)  -   [\delta\valpha \cdot (\vecr \times \nabla)] \vA(\vecr)
 \eqne
 As in the scalar field case the last term corresponds to the "orbital" rotation with $\hat{\vl} = -i\hbar \vecr\times \nabla$ while in the Appendix below we show that the additional second term is (not surprisingly) a rotation of the components of  the vector $\vA$  with spin one matrices.

 Let us now examine what happens to a wave functional $\Psi[\vA(\vecr)]$ when its argument is  transformed as in (\ref{rotating_vec_field}). To simplify things we leave out the particle coordinates $\{ \vecr_a \}$ aince the part of the rotation generator for them is obvious.  Have
 \eqna
 \Psi[\vA(\vecr)] &\to& \Psi[\vA(\vecr) + \delta \vA(\vecr) ] \approx \Psi[\vA(\vecr)]  + \int d^3 r \, \delta \vA(\vecr)\cdot \frac{\delta   \Psi[\vA(\vecr)] } {\delta \vA(\vecr) }   =  \nonumber \\
 &=& \left[ 1 +  \int d^3 r \, \delta \vA(\vecr) \cdot \frac{\delta } {\delta \vA(\vecr) } \right] \Psi[\vA(\vecr)] =   \nonumber \\
 &=&  \left[ 1 - \frac{i}{\hbar} \epsilon_0 \int d^3 r \, \delta \vA(\vecr) \cdot \vE_{op}(\vecr) \right] \Psi[\vA(\vecr)]  \nonumber
 \eqne
 where we used the expression for   $\vE_{op}(\vecr)$ as defined  in (\ref{eq:canonicalops}).  Using the explicit form of $\delta \vA$ from (\ref{rotating_vec_field}) we can write for the integral in the  second term
 \eqna
 \epsilon_0 \int d^3 r \, \delta \vA(\vecr) \cdot \vE_{op}(\vecr) &=&   \epsilon_0 \int d^3 r \, \left\{ \delta\valpha \times \vA(\vecr)  -   [\delta\valpha \cdot (\vecr \times \nabla)] \vA(\vecr) \right\} \cdot \vE_{op}(\vecr)  \nonumber \\
 &=& \epsilon_0 \delta \valpha \cdot \int d^3r \, \left\{ \vA \times\vE_{op} - \sum_i [(\vecr \times \nabla) A_i(\vecr)] \hat{E}_i \right\} \nonumber
 \eqne
 From this we can read off the generator of rotations for the field part. It can be written as a sum of two parts - spin and orbital
 \beq
 \hat{\vJ}_{field} = \hat{\vL}_{field} + \hat{\vS}_{field}
 \eeq
 with
 \eqna
 \hat{\vL} _{field} &=&     \epsilon_0 \int d^3r \, \sum_i \hat{E}_i  (\vecr \times \nabla)\hat{A}_i   \nonumber \\
 \hat{\vS}_{field}  &= & \epsilon_0 \int d^3r \, [\vE_{op} \times \vA_{op}  ]
 \eqne
 where we indicated that $\vA$ in this expression should be regarded as operator (although it is diagonal, $\vA_{op}=\vA$,  in the representation of $\Psi[\vA(\vecr)]$.  
 
 Note that in the expressions for $\hat{\vL} _{field}$ and $\hat{\vS}_{field}$  we were free to commute $ \hat{E}_i$ components to the left. 
 Indeed in $\hat{\vL} _{field}$ the commutator of $(\vecr \times \nabla) \hat{A}_i $ and $\hat{E}_i$ is proportional to the derivative of the delta function $\delta(\vecr-\vecr')$
$$
 \left[ (\vecr \times \nabla) \hat{A}_i(\vecr) ,  \hat{E}_i(\vecr')\right]_{\vecr=\vecr'} =  
 - \frac{i\hbar}{\epsilon_0} (\vecr \times \nabla) \delta(\vecr-\vecr') \big|_{\vecr=\vecr'} =0
 $$
 which vanishes at $\vecr = \vecr'$.  In $\hat{\vS}_{field}$ only different i.e. commuting components $\hat{A}_i$ and $\hat{E}_j$ with $i\ne j$ enter in their vector product.

 \subsection{Photon parity and photon  statistics}

The vector potential $\vA(\vecr)$ is a polar vector - it changes it sign under parity transformation
\beq
\vA(\vecr) \to -\vA(-\vecr)
\eeq
This property of $\vA(\vecr)$ is the basis of the statement that the photon, i.e. the quanta of the vibrations of $\vA(\vecr)$ have negative parity.  We note that at this stage this is a fairly cryptic statement which becomes clear when photon emission by matter system is studied (cf., later in the course).

  Photons are bosons! This too is a somewhat cryptic statement at this  stage. It will become clear when dealing with the second quantitation formalism of the Schr\"odinger field in relation to quantum many body systems. Here we only remark that one can have any number of photons in the same state, i.e. in the same mode characterised by $\vk,\alpha$ quantum numbers.

One can calculate the commutator of the operators of the electric and magnetic fields $\hat{\vE}_T(\vecr)$ and $\hat{\vB}(\vecr) = \nabla\times \hat{\vA}_T(\vecr)$ and find that they do not commute. This have all the usual quantum mechanical consequences. In fact in tutorials and home works we/you will deal with issues related to questions like "what is the electric/magnetic field of a photon?"

\section{Appendix}
\subsection{Details of the standing to traveling waves transformation} 

\subsubsection{What does the transformation  Eq.\,(\ref{eq:fromQi_to_Qpm}) achieve}

Let us start by noting  that the Hamiltonian for a given $k_\nu$ has the same form in the new variables
$$
H_\nu= \frac{1}{2}\sum_{i=1,2} ( P^2_{i,\nu} + \omega_\nu^2 Q^2_{i,\nu}) =  \frac{1}{2}[( P^2_{k} + \omega_k^2 Q^2_{k}) + ( P^2_{-k} + \omega_k^2 Q^2_{-k})] \;\;\; 
$$
 and due to their canonicity (cf., below) the dynamical equations for $Q_{\pm k} , P_{\pm k}$  are the same as for $Q_{i,\nu} , P_{i,\nu} , i=1,2$ so their solutions have the same form as in Eq.\,(\ref{eq:solution_for_Q}). Now both terms in this solution give traveling waves when inserted in the expansion (\ref{eq:stringexpintravelwaves2}).  Indeed  have for the first terms  in these solutions when inserted in the expansion for
 $\phi(x)$
$$
\sqrt{\frac{1}{L}}\sum_k Q_k(0)\left[ \sin k x \cos\omega t - \cos k x \sin\omega  t\right] \nonumber =\sqrt{\frac{1}{L}}\sum_k Q_k0)\sin(k x - \omega t) 
$$
and for the second terms 
$$
\sqrt{\frac{1}{L}}\sum_k \frac{P_k(0)}{\omega_\nu} \left[ \sin k x \sin\omega t +\cos k x \cos\omega t\right]\nonumber 
=\sqrt{\frac{1}{L}}\sum_k \frac{P_k(0)}{\omega} \cos(k x - \omega t)
$$
These traveling waves are  "running" in the positive or negative $x$-direction depending on the sign of $k$. 

 \subsubsection{Verifying canonicity} 
 Our transformation from the standing waves expansion (\ref{eq:stringexpinstandwaves}) to the traveling waves (\ref{eq:stringexpintravelwaves2}) amounted to transforming from  $Q_{i,\nu} , P_{i,\nu}$ phase space variables
to $Q_{\pm k} , P_{\pm k}$, Eq. (\ref{eq:fromQi_to_Qpm}). Let us now check the canonicity of this transformation. 

 Let us recall that  in a mechanical system described by a set of generalised coordinates and momenta $\{q,p\}$ the transformation to a canonically conjugate set $\{Q,P\}$ must satisfy
$$
\sum_i p_i d q_i = \sum_k P_k d Q_k + d F
$$
where $dF$ denote a complete differential. In our case  the set $\{q,p\}$ is $Q_{i,\nu} , P_{i,\nu}$ and we are transforming to $Q_{\pm k} , P_{\pm k}$.  We obtain
\eqnaa
\sum_i P_{i,\nu} d Q_{i,\nu} &=& \frac{1}{2} \left[ (P_k -P_{-k}) d  (Q_{k} -Q_{-k}) - (Q_k + Q_{-k}) d (P_{k} + P_{-k})\right]= \\         
&=&\frac{1}{2} [P_k  d  Q_{k} + P_{-k} d  Q_{-k}  - P_k d  Q_{-k} - P_{-k} d  Q_{k}    - \\ 
& & - \; Q_k d P_{k}  - Q_{-k} d P_{-k}  - Q_k  d P_{-k}  -  Q_{-k} d P_{k}] =  \\
&=& P_k d Q_{k} + P_{-k} d  Q_{-k} - \frac{1}{2} (P_k  d  Q_{k} + P_{-k} d  Q_{-k} +  Q_k d P_{k}  + Q_{-k} d P_{-k} ) - \\
& & - \;\frac{1}{2}(P_k d  Q_{-k} + P_{-k} d  Q_{k} + Q_k  d P_{-k}  +  Q_{-k} d P_{k})  = \\
& =& P_k  d  Q_{k} + P_{-k} d  Q_{-k} - \frac{1}{2}  d (P_k  Q_{k} + P_{-k}  Q_{-k} - P_k  Q_{-k}  - P_{-k} Q_{k})
\eqnae

It is instructive also to verify the canonicity of the general transformation (\ref{eq:stringexpintravelwaves2}). In this  case the set  $\{q,p\}$ is $\{\phi(x) , \pi(x)\}$, the sum over $i$ is integral over $x$ and we are transforming to $Q_k, P_k$.  So we have
\eqna
&&\ \int_0^L dx \pi(x,t)\frac{\partial \phi(x,t)}{\partial t} =  \nonumber \\
&=& \frac{1}{L}  \sum_{k k'}\int_0^L dx \left[\sin k'  x \;P_{k'}(t) - v|k'|\cos k' x \;Q_{k'}(t)\right]\times  \nonumber \\
&\times& \left[\sin k  x \;\dot{Q}_k(t) + \frac{1}{v|k|}\cos k x \;\dot{P}_k(t)\right] = \nonumber \\
&=&\sum_{k} \frac{1}{2} \left[P_k(t) \dot{Q}_k(t) - Q_k(t)\dot{P}_k(t)\right] =
\nonumber \\
&=& \sum_{k}  P_k(t) \dot{Q}_k(t)  -   \frac{1}{2} \sum_{k} \frac{d}{dt} [Q_k(t)P_k(t)] \nonumber
\eqne
which shows the canonicity of $P_k$ and $Q_k$.

\subsection{Details of the coherent states}
\subsubsection{Useful averages. Minimum uncertainty}
 It easy to find an explicit solution  of the equation (\ref{eq:coh_state}). But before doing that it is useful first to calculate the following averages  
 \beq
 \bes
q_0\equiv  \langle \alpha | \hat{q} | \alpha \rangle &=  \sqrt{\hbar/(2 \omega)}   \langle \alpha | (\ha + \ha^+)  | \alpha \rangle  = \sqrt{\hbar/ (2\omega)} (\alpha + \alpha^*)  =  \sqrt{2\hbar/ \omega} {\rm \; Re}\; \alpha \\
 p_0 \equiv \langle \alpha | \hat{p} | \alpha \rangle  &= i\sqrt{\hbar \omega/2}  \langle \alpha | (\ha^+ - \ha)  | \alpha \rangle =i \sqrt{\hbar \omega/2} (\alpha^* - \alpha)  =  \sqrt{2\hbar \omega} {\rm \; Im}\; \alpha 
 \end{split}
 \eeq
 which give
 \beq \label{eq:alpha_via_p_q}
 \alpha =  \sqrt{\omega/2\hbar} \; q_0 + i  \sqrt{1/(2\hbar \omega)} \; p_0
 \eeq
 Also have
 \beq
 \bes
  \langle \alpha | \hat{q}^2 | \alpha \rangle &=  (\hbar/2  \omega) \langle \alpha | \ha^2 +\ha\ha^+ + \ha^+\ha + (\ha^+)^2 | \alpha \rangle = \\
  &= (\hbar/2 \omega) \langle \alpha | \ha^2 +2\ha^+\ha+1  + (\ha^+)^2 | \alpha \rangle = \\
  &= (\hbar/2 \,\omega)[(\alpha +  \alpha^*)^2 +1)  = \langle \alpha | \hat{q} | \alpha \rangle ^2 + \hbar/(2 \omega) 
  \end{split}
  \eeq
  and
  \beq
  \bes
  \langle \alpha | p^2 | \alpha \rangle &=  - (\hbar \omega/2)  \langle \alpha | \ha^{+2}  - \ha^+\ha - \ha\ha^+ + \ha^2  | \alpha \rangle = \\
  &= - (\hbar \omega/2)[(\alpha - \alpha^*)^2 - 1] = \langle \alpha | \hat{p} | \alpha \rangle^2 + \hbar \omega/2
 \end{split}
 \eeq
which shows that the coordinate and momentum uncertainties in this state are independent of $\alpha$ 
\beq
\bes
\Delta q &\equiv \sqrt { \langle \alpha | \hat{q}^2 | \alpha \rangle -  \langle \alpha | \hat{q} | \alpha \rangle^2} = \sqrt{\hbar/(2 \omega)} \\
\Delta p &\equiv \sqrt{ \langle \alpha | \hat{p}^2 | \alpha \rangle -  \langle \alpha | \hat{p} | \alpha \rangle^2} = \sqrt{\hbar \omega/2}
\end{split}
\eeq
which in turns means that for large (classical) values of $q_0$ and $p_0$, i.e. for large $|\alpha|$, cf., Eq.\;(\ref{eq:alpha_via_p_q}), the quantum uncertainties are negligible.  The actual values of $\Delta q$ and $\Delta p$ show that $|\alpha\rangle$ is a minimum uncertainty state, Eq. (\ref{eq:min_unc}).

\subsubsection{Dynamics of coherent states}

Let is now consider the dynamics of a coherent state, i.e.  find
 $$
 |\alpha(t)\rangle \equiv e^{-i\hat{h}t/\hbar} |\alpha\rangle \;\; {\rm with}\;\; \hat{h} = \hbar \omega (\ha^+\ha +1/2)
 $$
 For this we use the Heisenberg representation $\ha(t) = e^{i\hat{h}t/\hbar} \; \ha \; e^{-i\hat{h}t/\hbar}$ of $\ha$ and the corresponding Heisenberg  equation which is easily solved 
 \beq 
 \bes
 i\hbar \frac{\partial \ha(t)}{\partial t} &=-\hat{h} \ha(t) + \ha(t)\hat{h}  = \hbar\omega[ - \ha^+(t)\ha(t) \ha(t) + \ha(t) \ha^+(t)\ha(t) ] = \\
 & =\hbar\omega \ha(t)  \;\; \Rightarrow \;\; \ha(t) = \ha e^{-i\omega t}
 \end{split} 
 \eeq
This gives
\beq \label{t_dep_coh_st}
 \ha  |\alpha(t)\rangle = \ha e^{-i\hat{h}t/\hbar} |\alpha\rangle =  e^{-i\hat{h}t/\hbar} \ha(t) |\alpha \rangle = e^{-i\hat{h}t/\hbar} \ha e^{-i\omega t} |\alpha \rangle =\alpha e^{-i\omega t}  |\alpha(t)\rangle 
 \eeq
 which shows that $ |\alpha(t)\rangle$ remains a coherent state with 
 \beq  \label{t_dep_coh_st_1}
 \alpha(t) = \alpha e^{-i\omega t} 
 \eeq
 In terms of the corresponding $q_0(t)$ and $p_0(t)$ 
 \beq
 \bes
 q_0(t) & =   \sqrt{2\hbar/ \omega} {\rm \; Re}\; \alpha (t) =  \sqrt{2\hbar/ \omega} [{\rm \; Re}\; \alpha \cos \omega t + {\rm \; Im}\; \alpha \sin\omega t ] =  q_0  \cos \omega t + (p_0/\omega) \sin\omega t \\
p_0(t) & =  
\sqrt{2\hbar \omega} {\rm \; Im}\; \alpha (t) = 
  \sqrt{2\hbar \omega} [ {\rm \; Im}\; \alpha \cos\omega t - 
 {\rm \; Re}\; \alpha \sin\omega t ] = p_0 \cos\omega t -  \omega q_0 \sin\omega t  \nonumber
\end{split}
 \eeq
 which coincide with the solution of the classical equations (\ref{eq:eqs_ho}).

\subsubsection{Explicit expressions. Ground state of a shifted harmonic oscillator}
Using the explicit expression  (\ref{eq:gs_ho}) for the coherent state at $\alpha=0$ it is easy to find solutions of Eq.\;(\ref{eq:coh_state}) for a general $\alpha$ by using the decomposition (\ref{eq:alpha_via_p_q}) in (\ref{eq:coh_state})
 \beq
  \left(\hbar \frac{\partial }{\partial q} + \omega \; q \right)  \psi_{\alpha} (q) =  [ \omega q_0 + i \; p_0] \psi_{\alpha} (q) \;
 \Rightarrow \; \left[ \left(\hbar \frac{\partial }{\partial q} - ip_0\right)  + \omega (q - q_0) \right]  \psi_{\alpha} (q) =0 \nonumber
\eeq
and noticing that this equation is similar to the one with $\alpha=0$, Eq. (\ref{eq:gs_ho}),  but with a shift $q\to q-q_0$ and a $p_0$ dependent phase
\beq
\psi_\alpha (q) = A \exp\{-[\omega (q-q_0)^2 +ip_0]/\hbar\}  \;\;, \;\;  A=(\omega/\pi\hbar)^{1/4}
\eeq
It clearly can be regarded as a ground state of a shifted harmonic oscillator, i.e. of 
\beq
\hat{h} = \frac{1}{2}\left[ (\hat{p} - p_0)^2 + \omega^2 (\hat{q}-q_o)^2\right]
\eeq
This observation is important for a qualitative discussion of the laser light. 

Let us note that the coherent state can also be written as an expansion in a complete set of number states, i.e. the harmonic oscillator eigenstates $|n\rangle$
$$
|\alpha\rangle = \sum_{n=0}^\infty c_n |n\rangle
$$
Acting with $\hat{a}$ we obtain
\beq 
\bes 
\hat{a}|\alpha\rangle &= \sum_{n=0} c_n \hat{a} |n\rangle =  \sum_{n=0}^\infty c_n\sqrt{n}  |n-1\rangle  
 = \alpha\sum_{n=0}^\infty c_n |n\rangle =\\  &= \alpha\sum_{k=1}^\infty c_{k-1} |k-1\rangle \;\; \Rightarrow 
  \;\sqrt{n}c_n=\alpha c_{n-1} \\
 &\Rightarrow c_n = \frac{\alpha^n}{\sqrt{n!} }  c_0  \;\;\; \Rightarrow  \;\;\; |\alpha\rangle = c_0\sum_{n=0}^\infty \frac{\alpha^n}{\sqrt{n!} } |n\rangle   \nonumber
\end{split}
\eeq
Find $c_0$ from normalization
$$
1=\langle \alpha | \alpha \rangle = |c_0|^2 \sum_{n=0}^ \infty \frac{|\alpha|^2}{n!} =  |c_0|^2 e^{|\alpha|^2} \;\; \;\; 
\Rightarrow  \;\;\;\; c_0= e^{-|\alpha|^2/2}
$$
so 
\beq \label{eq:expan_alpha}
|\alpha\rangle = e^{-|\alpha|^2/2} \sum_{n=0}^\infty \frac{\alpha^n}{\sqrt{n!} } |n\rangle 
\eeq
It is also easy to calculate the overlap
$$
\langle \alpha | \beta \rangle  = e^{-|\alpha|^2/2} e^{-|\beta||^2/2} e^{\alpha^* \beta}\;\;\;
 \Rightarrow  \;\;\;\; |\langle \alpha | \beta \rangle |^2 = e^{-|\alpha -\beta|^2}
$$
showing non orthogonality of different $|\alpha\rangle$ states. 
  The  set $|\alpha\rangle$ is over-complete but satisfies a useful resolution of  unity relation
  $$
  \int \frac{d^2\alpha}{\pi} | \alpha \rangle \langle \alpha | =\sum_{n=0}^\infty |n\rangle  \langle n | = \hat{1}\;\;\; {\rm with}\;\; d^2\alpha = d  {\rm \; Re} \, \alpha \; d  {\rm \; Im}\, \alpha
  $$
  which is easy to prove by using the expansion (\ref{eq:expan_alpha}) and changing to polar coordinates $\alpha = r e^{i\phi}, 
  d^2\alpha = r dr d\phi$  in the integral.

\subsection{More on the EM field momentum \label{sec:EM_momentum}}
\subsubsection{Relation to the classical expressions for the matter-field momentum}

  Classical expression (\ref{eq:TotalMomentum}) can be written  
  \beq \label{EMmomentum}
  \vP_f = \epsilon_0 \int d^3 r \,\vE(\vecr)\times \vB(\vecr) = \epsilon_0 \int d^3 r \,\vE(\vecr)\times \nabla\times \vA(\vecr)
  \eeq
 It is related to the integral of the Poynting vector, cf., the reference to the Feynman lectures given above for the physics discussion of this result.

 Let us write this expression in components (using the Levi-Civita tensor and the summation convention)
  \eqna
 ( \vE\times\nabla\times \vA)_i &=& \epsilon_{ijk}E_j \epsilon_{klm}\partial_l A_m =
 \epsilon_{kij}\epsilon_ {klm}E_j \partial_l A_m= \\
 &=& (\delta_{il}\delta_{jm}-\delta_{im}\delta_{lj}) E_j \partial_l A_m = E_j \partial_i A_j -
 E_j \partial_jA_i \nonumber
 \eqne
so that
\beq \label{EMmom2}
\epsilon_0 \int d^3 r \,( \vE\times\nabla\times \vA)_i = \epsilon_0 \int d^3 r \,(E_j \partial_i A_j -
 E_j \partial_jA_i )=\epsilon_0 \int d^3 r \,(E_j \partial_i A_j +
\partial_j E_j A_i )
 \eeq
where we integrated by parts in the last equality. Using the Gauss law $\partial_j E_j = \rho/\epsilon_0$ this gives
\beq
(\vP_{f})_i= \epsilon_0 \int d^3 r \,E_j(\vecr) \partial_i A_j (\vecr) +
\int d^3 r\rho(\vecr) A_i (\vecr) )= \epsilon_0 \int d^3 r \,E_j(\vecr) \partial_i A_j (\vecr) +
\sum_{a=1}^N q_a A_i(\vecr_a)
\eeq
 where we used $\rho(\vecr)=\sum_{a=1}^N q_a\delta(\vecr-\vecr_a)$ to integrate.

 Using this  in the expression for the total momentum (\ref{eq:TotalMomentum}) (and restoring for better clarity the summation symbol for the repeated index j) we obtain
 \beq \label{TotalMomentum2}
 \vP= \sum_{a=1}^N \vp_a + \epsilon_0 \int d^3 r \,\sum_{j=1}^3 E_j(\vecr) \nabla A_j (\vecr)
 \eeq
 with
 \beq
 \vp_a = m_a\vv_a+q_a \vA(\vecr_a)
 \eeq
This coincides with the expression (\ref{eq:totalmom1}).

\subsubsection{Field momentum in terms of the transverse components}
Both terms in the expression (\ref{eq:TotalMomentum}) are separately gauge invariant. However the two terms in the transformed expression (\ref{TotalMomentum2}) are not. Only their sum is. We can repair this if we repeat the calculation (\ref{EMmom2}) but first replacing $\vA$ by $\vA_T$ in the starting left hand side. This will lead to the same expression as (\ref{TotalMomentum2}) but with $\vA_T$ replacing $\vA$ in it
$$
\vP= \sum_{a=1}^N (m_a\vv_a+q_a \vA_T(\vecr_a))
+ \epsilon_0 \int d^3 r \,\sum_{j=1}^3 E_j(\vecr) \nabla A_{T,j} (\vecr)
$$
Now both terms are gauge invariant. We can moreover in the second term replace $E$ by $E_T$. Indeed writing
$$
E_j= E_{T,j} + E_{L,j} =  E_{T,j} -\partial_j\phi
$$
and using
$$
\int d^3 r \sum_{j=1}^3 \partial_j\phi (\vecr) \nabla A_{T,j} (\vecr) = -\int d^3 r  \phi (\vecr) \nabla \left[\sum_{j=1}^3 \partial_j A_{T,j} (\vecr)\right]  =  0
$$
we express
\beq
\vP= \sum_{a=1}^N (m_a\vv_a+q_a \vA_T(\vecr_a))
+ \epsilon_0 \int d^3 r \,\sum_{j=1}^3 E_{T,j}(\vecr) \nabla A_{T,j} (\vecr)
\eeq
In the absence of the charged matter (i.e. when all $q_a$'s are zero) the field part of this momentum becomes the momentum of the free radiation as we have already derived in  (\ref{radiation_mom}).

\subsection{More on the EM field angular momentum}

 \subsubsection{Relation to the classical expression}

 See Ref.\cite{Coh}, Complement $B_I$.

 \subsubsection{Spin 1 part of rotations of a vector field}

We can write the 2nd term in Eq. (\ref{rotating_vec_field}) as
 $$
  [\delta\valpha \times \vA]_j = \epsilon_{jkl} \delta\alpha_k A_l = -\frac{i}{\hbar} \delta\alpha_ks^k_{jl}A_l = -\frac{i}{\hbar} [\delta\valpha \cdot \vs]_{jl} A_l
  $$
where the matrices
$$
s^k_{jl} = i\hbar \epsilon_{jkl}
$$
are  spin 1 matrices written in cartesian components basis $x_1=x, x_2=y, x_3=z$ rather than in the more familiar spherical components basis  ($x_m,\; m=\pm1,0$ )
$$
x_{+1}  = -\frac{1}{\sqrt{2}} (x + iy)  \;\;\;, \;\;\; x_{-1}  = \frac{1}{\sqrt{2}} (x - iy) \;\;\;, \;\;\; x_{0}  = z
$$
 i.e. $x_m\sim  rY_{1m} (\theta,\phi)$.

One can easily verify that the commutators indeed have the correct form
\beq
[s^i, s^j] = i\hbar \epsilon_{ijn} s^n
\eeq
 For this must  prove that
$$
[s^i, s^j]_{kl} = -\hbar^2[ \epsilon_{kim}\epsilon_{mjl} - \epsilon_{kjm}\epsilon_{mil} ]
$$
is equal to
$$
i\hbar \epsilon_{ijn} s^n_{kl} = i\hbar \epsilon_{ijn} i\hbar \epsilon_{knl} = -\hbar^2  \epsilon_{ijn} \epsilon_{knl}
$$
Have
$$
\epsilon_{kim}\epsilon_{mjl} - \epsilon_{kjm}\epsilon_{mil}  = (\delta_{kj}\delta_{il} - \delta_{kl}\delta_{ij}) -( \delta_{ki}\delta_{jl} -
\delta_{kl} \delta_{ji}) = \delta_{kj}\delta_{il} - \delta_{ki}\delta_{jl}
$$
which indeed is equal to
$$
\epsilon_{ijn}  \epsilon_{knl} = \epsilon_{ijn}  \epsilon_{lkn} = \delta_{il} \delta_{jk} - \delta_{ik} \delta_{jl}
$$

\chapter{Photon-Matter Interactions}

This chapter is the  continuation of the chapter "Quantum Mechanics of  Electromagnetic Field".  We will use the quantum description of the EM field  discussed there to provide several simple examples of how photons are emitted and absorbed by quantum matter systems.

\section{Interaction Hamiltonian}
\subsection{Separating the interaction terms}
As was shown in the chapter "Quantum Mechanics of  Electromagnetic Field"  the Hamiltonian operator of the EM field interacting with (non relativistic) matter is
\beq \label{EM_matter_Ham}
\hat{H} = \sum_{a=1}^N \frac{1}{2m_a} [\hat{\vp}_a -q_a \hat{\vA}_T(\vecr_a)]^2 + V_{Coul} +
 \frac{ \epsilon_0}{2} \int\left[\hat{\vE}_T(\vecr)^2 + c^2 (\nabla \times \hat{\vA}_T(\vecr))^2\right] d^3r 
 \eeq
with
\eqna
 \hat{\vA}_T(\vecr)&=&\sum_{\vk \alpha}\left(\frac{\hbar }{2\epsilon_0\omega_k\Omega}\right)^{1/2}\left[\ha_{\vk\alpha}\vlambda_{\vk\alpha}e^{i\vk\cdot\vecr}+\ha^{\dagger}_{\vk\alpha}\vlambda_{\vk\alpha}e^{-i\vk\cdot\vecr} \right]  \\ 
 \hat{\vE}_T(\vecr) &=&\sum_{\vk \alpha}i\left(\frac{\hbar\omega_k }{2\epsilon_0\Omega}\right)^{1/2}\left[\ha_{\vk\alpha}\vlambda_{\vk\alpha}e^{i\vk\cdot\vecr}-\ha^{\dagger}_{\vk\alpha}\vlambda_{\vk\alpha}e^{-i\vk\cdot\vecr} \right] 
 \eqne
and
$$ 
V_{Coul} = \frac{1}{8\pi\epsilon_0} \sum_{a\ne b}^N \frac{q_a q_b}{|\vecr_a-\vecr_b|}
$$ 
 This expression can be written as
 \beq
 \hat{H}= \hat{H}_{matter}  + \hat{H}_r + \hat{H}_{matter-radiation\; interaction}
 \eeq
 with
 \eqna \label{parts_of_H}
  && \hat{H}_{matter}   =  \sum_{a=1}^N \frac{\hat{\vp}^2_a }{2m_a}  + V_{Coul} (\vecr_1, ..., \vecr_n) \nonumber \\
  && \hat{H}_r =  \frac{ \epsilon_0}{2} \int\left[\hat{\vE}_T(\vecr)^2 + c^2 (\nabla \times \hat{\vA}_T(\vecr))^2\right] d^3r   \\
   && \hat{H}_{matter-radiation\; interaction} 	= \hat{H}_{I1} +\hat{H}_{I2} \nonumber
   \eqne
   and
   \eqna \label{interact_Ham1}
   \hat{H}_{I1} &=& -\sum_{a=1}^N \frac{q_a}{2m_a} \left[ \hat{\vp}_a\cdot \hat{\vA}_T(\vecr_a) +  \hat{\vA}_T(\vecr_a)\cdot \hat{\vp}_a\right] \\
   \hat{H}_{I2} &=& \sum_{a=1}^N \frac{q_a^2}{2m_a} [\hat{\vA}_T(\vecr_a)]^2 \label{interact_Ham2}
   \eqne
   The expressions for  $\hat{H}_{I1}$ and $\hat{H}_{I2}$ depend on the coordinates and momenta of the particles and on the "coordinates"   $\hat{\vA}_T(\vecr)$ of the field. It is worth noting that the transversality of $\vA_T$ means that $\vp_a$ and $\hat{\vA}_T(\vecr_a)$ commute
   $$
 \sum_{i=1}^3 \left[ \hat{p}_{a,i} ,\hat{A}_{T,i}(\vecr_a)\right] = -i\hbar\nabla_a\cdot\hat{\vA}_T(\vecr_a) = 0
  $$
   so that the interaction  $\hat{H}_{I1} $ can be written as one term
   \beq
      \hat{H}_{I1} = -\sum_{a=1}^N \frac{q_a}{m_a} \hat{\vA}_T(\vecr_a)\cdot \hat{\vp}_a
   \eeq

  \subsection{Adding spin and external fields}
  When matter particles have spins one must add spin degrees of freedom $\vs_a$ to the particle coordinates $\vecr_a$. As a rule spinning particles have non zero magnetic moment $\vmu_a$\footnote{This is obvious for charged particles but in fact also neutral particles with spin, e.g. molecules, atoms, neutrons, etc,  may have non zero $\vmu$ due to the "spinning" charges inside the overall neutral system. Charged quarks in a neutron is an obvious example.} which is parallel to the spin and follows its dynamics. The proportionality relation between the corresponding operators is conventionally written
  \beq
  \hat{\vmu}_a = g_a\frac{q_a}{2m_a} \hat{\vs_a}
  \eeq
  where $g_a$ is the so called Lande factor or g-factor (see e.g. the appropriate section in the chapter "Motion in External Electromagnetic Field").

  Particle magnetic moments interact with the magnetic field so one must add a new term to the interaction Hamiltonian $\hat{H}_{matter-radiation\; interaction}$,
  \beq \label{interact_Ham3}
  \hat{H}_{I3} = -\sum_{a=1}^N \hat{\vmu}_a\cdot \hat{\vB}(\vecr_a)
  \eeq
  with the operator of the magnetic field (cf., the chapter "Quantum Mechanics of  Electromagnetic Field")
  \beq
  \hat{\vB}(\vecr) =\sum_{\vk \alpha}i\left(\frac{\hbar }{2\epsilon_0\omega_k\Omega}\right)^{1/2}\left[\ha_{\vk\alpha}(\vk\times\vlambda_{\vk\alpha})e^{i\vk\cdot\vecr}-\ha^{\dagger}_{\vk\alpha}(\vk\times\vlambda_{\vk\alpha})e^{-i\vk\cdot\vecr} \right]  
  \eeq

  We have up to now considered a closed matter-EM field system. One often encounters a situation in which in addition there are  external  fields acting on the matter particles.  Examples are  Coulomb potential of a heavy nucleus acting on atomic electrons or external magnetic field acting on electrons in Landau levels. Such external fields are to a good approximation classical with prescribed space and time dependence. In their presence the Hamiltonian (\ref{EM_matter_Ham}) should be modified by adding external classical vector potential, external scalar potential and external magnetic field. The full Hamiltonian will then have the form\,\footnote{Note that external fields influence the radiation only via matter. There is no direct effect on the dynamics of the radiation.   This is a consequence of the linearity of the Maxwell equations. }

 \eqna \label{Ham_with_spin_and_ext}
\hat{H} &=& \sum_{a=1}^N \frac{1}{2m_a} \left[\hat{\vp}_a -q_a\vA^{external}(\vecr_a, t) - q_a \hat{\vA}_T(\vecr_a\right]^2  +  V_{Coul} + \nonumber \\ && + \sum_{a=1}^N U^{external} (\vecr_a,t)   -\sum_{a=1}^N \hat{\vmu}_a\cdot \vB^{external}(\vecr_a, t)  - \\
&& -\sum_{a=1}^N \hat{\vmu}_a\cdot \hat{\vB}(\vecr_a) +
 \frac{ \epsilon_0}{2} \int\left[\hat{\vE}_T(\vecr)^2 + c^2 \nabla \times \hat{\vA}_T(\vecr)\right] d^3r \nonumber  \\
 V_{Coul} &=&\frac{1}{8\pi\epsilon_0} \sum_{a\ne b}^N \frac{q_a q_b}{|\vecr_a-\vecr_b|} \nonumber
 \eqne
 where we have also added the spin degrees of freedom interacting with external magnetic fields via the particles magnetic moments.

 \subsection{Disentangling radiation from the matter degrees of freedom}

The objects like $\hat{\vA}_T(\vecr_a)$ and $\hat{\vB}(\vecr_a)$ in the expressions  (\ref{interact_Ham1}), (\ref{interact_Ham2}) and (\ref{interact_Ham3}) are operator valued functions (fields) of operators (particle coordinates). It is easy and convenient to disentangle this complicated dependence using the identities
$$
\hat{\vA}_T(\vecr_a) = \int \delta(\vecr-\vecr_a) \hat{\vA}_T(\vecr) \, d^3r \; , \;
 \hat{\vA}^2_T(\vecr_a) = \int \delta(\vecr-\vecr_a) \hat{\vA}^2_T(\vecr)  \,d^3 r
 $$
 and
 $$
\hat{\vB}(\vecr_a) = \int \delta(\vecr - \vecr_a)\hat{\vB}(\vecr) \, d^3 r
$$

Using these  one can write the interactions (\ref{interact_Ham1}), (\ref{interact_Ham2}) and (\ref{interact_Ham3}) as
   \eqna  \label{Int_Ham_1}
   \hat{H}_{I1} &=& -\int d^3 r \sum_{a=1}^N \frac{q_a}{2m_a}\left[ \hat{\vp}_a\delta(\vecr - \vecr_a) +  \delta(\vecr - \vecr_a)\hat{\vp}_a\right] \cdot \hat{\vA}_T(\vecr)   \\
  \label{Int_Ham_2} \hat{H}_{I2} &=&\int d^3 r \sum_{a=1}^N \frac{q_a^2}{2m_a} \delta(\vecr-\vecr_a)[\hat{\vA}_T(\vecr)]^2   \\
 \label{Int_Ham_3}    \hat{H}_{I3} &=& -\int d^3 r \sum_{a=1}^N \hat{\vmu}_a\delta(\vecr-\vecr_a) \cdot \hat{\vB}(\vecr)
\eqne

The 1st and the 3rd of these expressions have a simple form
\beq \label{eq:HI1_interact}
 \hat{H}_{I1} = -\int  \hat{\vj} (\vecr) \cdot  \hat{\vA}_T(\vecr)  \, d^3 r
 \eeq
 and
\beq \label{eq:HI3_interact}
\hat{H}_{I3} = -\int  \hat{\vm}(\vecr) \cdot \hat{\vB}(\vecr) \, d^3 r
\eeq
with current operator
 $$
  \hat{\vj} (\vecr) =  \frac{1}{2}\sum_{a=1}^N \frac{q_a}{m_a}\left[ \hat{\vp}_a\delta(\vecr - \vecr_a)) +  \delta(\vecr - \vecr_a)\hat{\vp}_a\right]
 $$
and magnetization operator
$$
 \hat{\vm}(\vecr) = \sum_{a=1}^N \hat{\vmu}_a\delta(\vecr-\vecr_a)
 $$
 The second term $\hat{H}_{I2}$ simplifies when all the charges and masses of the particles are equal  $q_1= q_2 = ... =q_N=q$,
 $m_1=m_2 = ... = m_N=m$. Then
 \beq
    \hat{H}_{I2} =\frac{q}{m} \int \hat{\rho}(\vecr)[\hat{\vA}_T(\vecr)]^2 \, d^3 r
   \eeq
   with charge density operator
   $$
   \hat{\rho}(\vecr) =  \sum_{a=1}^N q\delta(\vecr-\vecr_a)
   $$

 \subsection{Resulting insights}
  \subsubsection{Matter creates, annihilates, scatters photons}
Qualitative insights into the nature of the interaction terms is gained if the expressions for the fields $\hat{\vA}_T(\vecr)$ and $\hat{\vB}(\vecr)$ in terms of the photon creation and annihilation operators written in the form\footnote{To simplify expressions we assume here and in the following that the polarization vectors for $\vk$ and $-\vk$ modes are chosen to be the same $\vlambda_{\vk\alpha} = \vlambda_{-\vk\alpha}$.   }
\eqna
\hat{\vA}_T(\vecr) &=&  \sum_{\vk \alpha}\left(\frac{\hbar }{2\epsilon_0\omega_k\Omega}\right)^{1/2}
\vlambda_{\vk\alpha}e^{i\vk\cdot\vecr}  (\hat{a}_{\vk\alpha}+\hat{a}^{\dagger}_{-\vk\alpha})   \\
 \hat{\vB}(\vecr)\; &=&\sum_{\vk \alpha}i\left(\frac{\hbar}{2\epsilon_0\omega_k \Omega}\right)^{1/2}(\vk\times\vlambda_{\vk\alpha}) e^{i\vk\cdot\vecr}  \left(\hat{a}_{\vk\alpha}+\hat{a}^{\dagger}_{-\vk\alpha} \right)
\eqne
are inserted in Eqs. (\ref{Int_Ham_1} - \ref{Int_Ham_3}).
The interaction term $\hat{H}_{I1}$ takes the form
\beq \label{first_intercat_H}
   \hat{H}_{I1}  = -  \sum_{\vk \alpha}\left(\frac{\hbar }{2\epsilon_0\omega_k\Omega}\right)^{1/2}
   ( \,\hat{\vj}_{-\vk} \cdot \vlambda_{\vk\alpha} ) (\hat{a}_{\vk\alpha}+\hat{a}^{\dagger}_{-\vk\alpha})
 \eeq
with
\beq \label{eq:Fc_of_vj}
\hat{\vj}_{\vk} = \int  \hat{\vj}(\vecr) e^{-i\vk\cdot\vecr} \, d^3 r  =    \sum_{a=1}^N \frac{q_a}{2m_a}\left[ \hat{\vp}_a e^{-i\vk\cdot\vecr_a}  +  e^{-i\vk\cdot\vecr_a} \hat{\vp}_a\right]
\eeq
It is seen that to 1st order\,\footnote{By  "to 1st order" here and in the following we mean that the interaction acts one time on a wave function. Note that in solving the Schr\"odinger equation the Hamiltonian acts "infinitely many  times" so to speak. This can be seen by viewing the time evolution
$$
\psi(t)= \exp[-(i/\hbar) \hat{H}(t-t_0)]\psi(t_0)= [1+(-i/\hbar) \hat{H}(t-t_0) +(-i/\hbar)^2 \hat{H}^2(t-t_0)^2 + ... ]\psi(t_0)
$$}  
this interaction acts by creating or annihilating single photons with (not surprising but worth noting) opposite signs of the momentum $\hbar\vk$.  It is also important to note that (as will become clearer later and especially in the chapter on Second Quantization) the expression
$$
\vlambda_{\vk\alpha}e^{i\vk\cdot\vecr}
$$
can often be regarded as a photon wave function having definite momentum $\vp = \hbar \vk$ and polarization $\vlambda_\alpha$.

Inserting the expression for $\hat{\vB}(\vecr)$ into the interaction  $\hat{H}_{I3}$, Eq.(\ref{Int_Ham_3}), one obtains
\beq \label{eq:H_I3_expanded}
    \hat{H}_{I3} = -\sum_{\vk \alpha}i\left(\frac{\hbar }{2\epsilon_0\omega_k\Omega}\right)^{1/2} [ \hat{\vm}_{-\vk} \cdot (\vk\times\vlambda_{\vk\alpha})
] \, (\hat{a}_{\vk\alpha}+\hat{a}^{\dagger}_{-\vk\alpha})
 \eeq
 with
 \beq
\hat{\vm}_{\vk} = \int  \hat{\vm} (\vecr) e^{-i\vk\cdot\vecr} \, d^3 r  = \sum_{a=1}^N \hat{\vmu}_a e^{-i\vk\cdot\vecr_a}
 \eeq
 One observes that this interaction term also creates or annihilates one photon in 1st order. The difference with  $\hat{H}_{I1}$ is that in the former case the photon creation or annihilation was "accompanied" with the "action" on the matter variables of the corresponding component $\hat{\vj}_{-\vk}$ of the current operator projected on the photon polarization $\vlambda_{\vk\alpha}$.  In $\hat{H}_{I3}$ this action is replaced with $ \hat{\vm}_{-\vk}$ component of the magnetization density $\hat{\vm}(\vecr)$ projected on $\vk\times \vlambda_{\vk\alpha}$.

 Turning now to the $\hat{H}_{I2}$, Eq. (\ref{Int_Ham_2}), we note that the presence of the square $[\hat{\vA}_T(\vecr)]^2$ means that the creation and annihilation operators will appear in this expression in the products
 $$
 \hat{a}_{\vk\alpha} \hat{a}_{\vk'\alpha'}  \;\;\;,\;\;\;  \hat{a}^{\dagger}_{\vk\alpha} \hat{a}^{\dagger}_{\vk'\alpha'} \;\;\;,\;\;\; \hat{a}^{\dagger}_{\vk\alpha} \hat{a}_{\vk'\alpha'} \;\; \;,\;\;\; \hat{a}_{\vk\alpha} \hat{a}^{\dagger}_{\vk'\alpha'}
 $$
which  shows that these interaction terms in 1sr order either create or destroy two photons or simultaneously create and destroy a photon with different momentum and polarization.

 \subsubsection{Matter "shifts", "mixes" the radiation oscillators}
   Let us recall that $\hat{\vA}_T(\vecr)$ is written in terms of the running plane waves as
$$
\vA_T(\vecr) = \frac{1}{\sqrt{\Omega\epsilon_0}}\sum_{\vk}\left(\vQ_{\vk} \cos (\vk\cdot\vecr)- \frac{1}{\omega_k} \vP_{\vk} \sin (\vk\cdot \vecr)  \right)
$$
where $\vlambda_{\vk\alpha}$'s are fixed polarization vectors orthogonal to $\vk$.
Recalling also that $\hat{H}_r$ is the sum of the normal modes oscillators
\beq \label{radiation_H}
\hat{H}_r =  \frac{1}{2}\sum_{\vk \alpha}\left(\hat{P}_{\vk\alpha}^2 +\omega_k^2 \hat{Q}_{\vk\alpha}^2\right)
\eeq
we find that in terms of  $\hat{P}_{\vk,\alpha}$'s and $\hat{Q}_{\vk \alpha}$ the Hamiltonian is written
\beq
 \hat{H} = \hat{H}_{matter}  + \frac{1}{2}\sum_{\vk \alpha}\left(\hat{P}_{\vk\alpha}^2 +\omega_k^2 \hat{Q}_{\vk\alpha}^2\right) + \sum_{\vk\alpha}\left(\hat{S}_{\vk\alpha} \hat{Q}_{\vk \alpha}+\hat{C}_{\vk\alpha}\hat{P}_{\vk\alpha}\right)  + \hat{H}_{I2}+\hat{H}_{I3}
\eeq
with
\eqna
\hat{S}_{\vk\alpha} &=& -\frac{1}{\sqrt{\Omega\epsilon_0}} \int \vlambda_{\vk\alpha}\cdot  \hat{\vj}(\vecr) \cos(\vk\cdot \vecr) \, d^3 r \\
\hat{C}_{\vk\alpha} &=&  \frac{1}{\sqrt{\Omega\epsilon_0 \omega^2}}\int \vlambda_{\vk\alpha}\cdot \hat{\vj} (\vecr)  \sin (\vk\cdot\vecr)  d^3 r   \nonumber
\eqne
Schematically one can say that via the $\hat{H}_{I1}$ interaction the matter causes shifts of the  oscillators of the radiation normal modes. The shift is in both the coordinates $Q_{\vk\alpha}$ and momenta $P_{\vk\alpha}$.  For fixed classical $S_{\vk\alpha}$ and $C_{\vk\alpha}$  each oscillator gets shifted
\beq \label{eq:shifted_ho_mode}
 \frac{1}{2}\left(\hat{P}_{\vk\alpha}^2 +\omega^2 \hat{Q}_{\vk\alpha}^2\right)  \to \frac{1}{2}\left[\left(\hat{P}_{\vk\alpha}-P_{\vk\alpha}^{(0)}\right) ^2 +\omega_k^2 \left(\hat{Q}_{\vk\alpha} - Q_{\vk\alpha}^{(0)}\right)^2\right] + E_{\vk\alpha}^{(0)}
\eeq
with $P_{\vk\alpha}^{(0)}$ , $Q_{\vk\alpha}^{(0)}$ and $E_{\vk\alpha}^{(0)}$ determined by  $S_{\vk\alpha}$ and $C_{\vk\alpha}$  in an obvious way. Of course in a real situation $S_{\vk\alpha}$ and $C_{\vk\alpha}$ are dynamical and quantized.

  Let us also note that the interaction term $\hat{H}_{I3}$ may schematically be viewed in a similar way as we outlined above for $\hat{H}_{I1}$ since it is  linear in $\hat{\vB}(\vecr) = \nabla \times \hat{\vA}_T(\vecr)$ and therefore in  $Q_{\vk\alpha}$ and $P_{\vk\alpha}$ variables.  
  
  The interaction term $\hat{H}_{I2}$ on the other hand is quadratic in   $\hat{\vA}_T(\vecr)$. Its dependence on the field normal modes variables is therefore quadratic depending on products   $Q_{\vk\alpha}Q_{\vk'\alpha'}$,
  $P_{\vk\alpha}P_{\vk'\alpha'}$ and $Q_{\vk\alpha}P_{\vk'\alpha'}$ mixing the normal modes $\vk\alpha$'s already in the 1st order.

\subsubsection{Generation of  coherent states. Schematic model of a laser}
Let us recall the properties of the coherent states which were discussed in the Section 6.1.3. of the Quantum Mechanics of  Electromagnetic Field chapter.  It was shown there that such states can be viewed as ground states of a shifted harmonic oscillator. Turning to the expression (\ref{eq:shifted_ho_mode}) we notice that if just one photon mode $\vk\alpha$ is selected and the current which "feeds" this mode is external, constant in time  and classical then the lowest eigenstate of the corresponding Hamiltonian will be a coherent state.   

Such a Hamiltonian can actually be used as a simplest schematic model to begin understanding  the quantum mechanics of the light emitted by a laser. Selecting a single mode is modeling (in the simplest way) of the laser resonator. The classical external current is (a very much simplified description of ) the source of excitations of the electric charges which de-excite by emitting photons into the resonator mode.  This shifted harmonic oscillator model obviously is extremely schematic and misses many important laser features and details. It nevertheless correctly indicates that a simple reasonable approximation to the state of light which (one mode) laser emits is a coherent state.

 \section{Emission and Absorption of Photons}
 In this section we discuss the details of quantum mechanical description of photon emission and absorption. We will do this treating the radiation-matter interaction using the perturbation theory and will limit ourselves to the leading 1st order terms. As should be clear from our discussion above the relevant terms for such 1st order processes are $\hat{H}_{I1}$ and  $\hat{H}_{I3}$.  We will begin by considering only the effect of  
 $\hat{H}_{I1}$ i.e. photon emission and absorption resulting from the change of the state of the electric current of the matter system. Classically this would correspond to emission of radiation by an alternating current (like e.g. in a simple antenna). The treatment of the photon emission by changing the spin states of matter, i.e. the effect of the  $\hat{H}_{I3}$ interaction term will fit naturally in the discussion of these processes in relation to the changes of the current magnetic moment, cf., Section \ref{sec:mag_dipole}.

Following this introduction we will begin by considering the Hamiltonian
\beq
\hat{H}=\hat{H}_0+  \hat{H}_{I1}
\eeq
where the unperturbed part is
\beq \label{unpert_H}
\hat{H}_0 =  \hat{H}_{matter} + \sum_{\vk\alpha}\hbar\omega_k \,\hat{a}^{\dagger}_{\vk\alpha}\hat{a}_{\vk\alpha}
\eeq
and where we dropped the constant vacuum energy term $E_{vacuum}=(1/2) \sum_{\vk\alpha}\hbar\omega_k$.

 \subsection{Paradigm of spontaneous emission of radiation - discrete matter level coupled to a photon continuum}
 
 \subsubsection{Unperturbed energies} 
 We assume that we know how to solve the matter Hamiltonian i.e. that we know its eigenstates and the corresponding eigenvalues
 \beq
 \hat{H}_{matter}|n\rangle = E_n | n \rangle
 \eeq
 We therefore know the eigenstates of the unperturbed $\hat{H}_0$, Eq. (\ref{unpert_H}) ,
 \beq \label{unpert_En}
 |n\rangle |\{N_{\vk,\alpha}\}\rangle  \;\;\; {\rm with \;\; eigenenergies} \;\;\;
 E(n,\{N_{\vk\alpha}\}) =   E_n + \sum_{\vk,\alpha} N_{\vk\alpha}\; \hbar\omega_k
 \eeq
We assume that (as is typical for atomic, molecular or nuclear systems) the low lying matter eigenenergies  in (\ref{unpert_En})
form discrete system of levels following by higher lying continuum states (like e.g. simplest hydrogen atom at rest\footnote{We ignore at the moment the center of mass motion of the emitting.  Its effects will be discussed below}).  Let us consider the sector of unperturbed levels with zero photons
$$E_n + 0\; {\rm photons}
$$
 and compare to the  corresponding levels in a one photon sector
 $$
 E_n + 1 \;{\rm photon} = E_n +\hbar\omega_k
 $$
 It is important to note that
 $$
 \hbar\omega_k=\hbar ck
 $$
 form a continuum of levels because of essentially continuum values of $k$ (for large quantisation volume).

\begin{figure}[H]
\centering \includegraphics[width=1.0\textwidth]{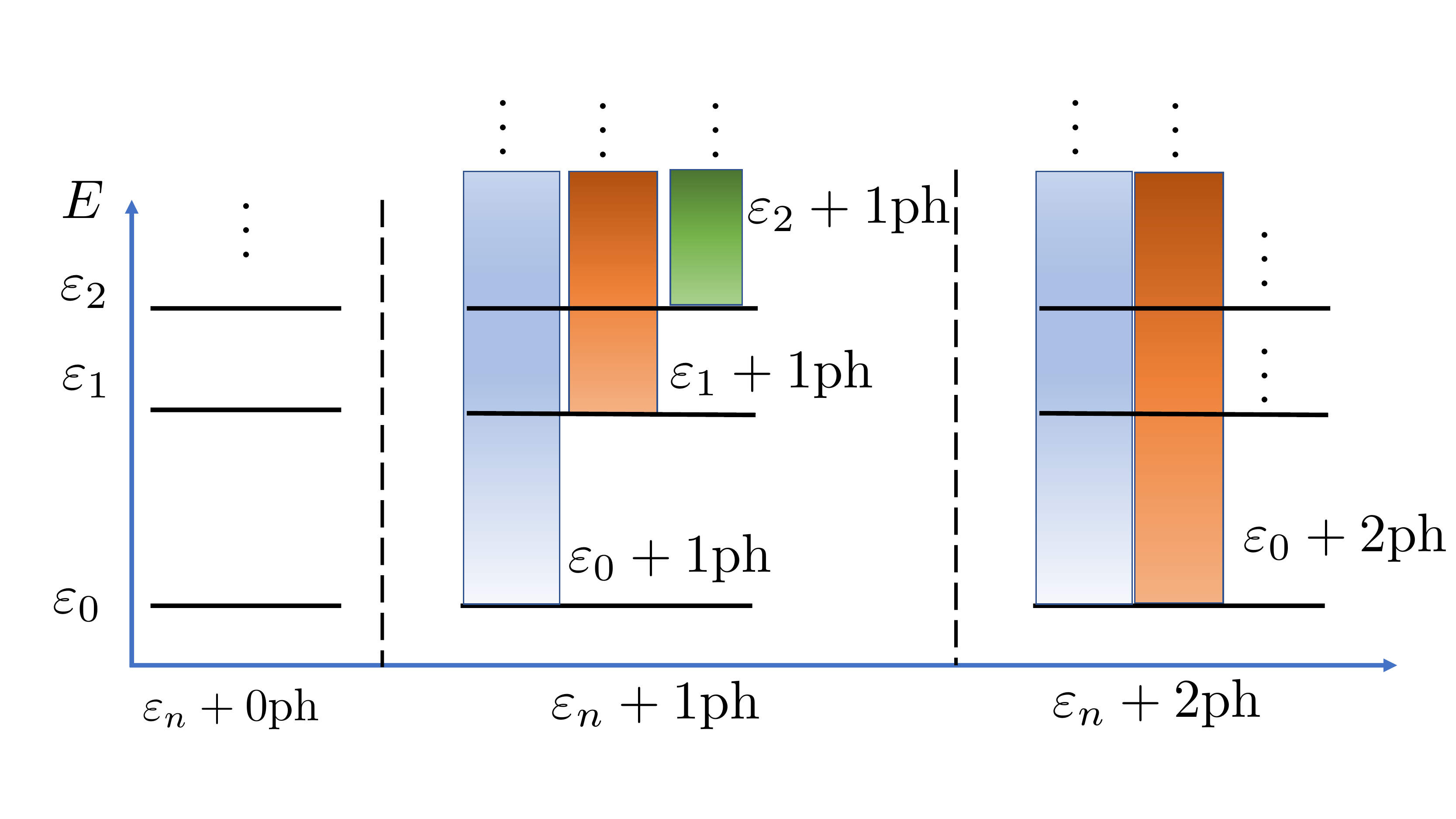}
\caption{The presence of the continuum of the photonic levels i.e. of photons in the continuum of the EM field modes (represented schematically as  colored bands in the figure) means that the discrete excited levels of  matter are embedded in this continuum. As explained in this section the coupling of the matter to the EM field means in turn that the discrete matter levels get "smeared" over the nearby continuum of photonic levels. The result is that their energy position gets shifted and they acquire a width becoming somewhat analogous to classical resonances.  }
 \label{fig:cont_photons}
\end{figure}

  Plotting these energies, cf. Fig. \ref{fig:cont_photons}, one can see discrete levels of the matter without photons "embedded" in the continuum of matter + one or more  photon levels.  The simplest is e.g. the first excited matter level with no photons
  $$
  |n=1\rangle |\{0_{\vk\alpha}\}\rangle  \;\;\;\; {\rm with} \;\;\;   E(1, \{0_{\vk\alpha}\})  =E_1 + 0 \;{\rm photons}
  $$
  vs the ground state $E_0$ plus one photon
  $$
  |n=0\rangle |1_{\vk\alpha}, \{0_{\vk'\alpha'}\}\rangle    \;\;\;\; {\rm with} \;\;\;   E(0, 1_{\vk\alpha}, \{0_{\vk'\alpha'}\}) =E_0 + \hbar\omega_k
  $$
   continuum of levels. 

   \subsubsection{ Coupling to the continuum - time domain. Exponential decay \label{sec:WW_model}}

We now turn to the discussion of what will the perturbation $\hat{H}_{I1}$ which has matrix elements connecting such levels  cause.  We will do this in the framework of  a simple model - a single discrete state coupled to a continuum of states. This is known as Weisskopf-Wigner model. We present here the main results for this  model. Details are found in the Appendix of this chapter, as well as in Ch.I-C3 and Complement $C_I$ of Ref.\cite{Cohen}.

We will use simplified notations.  Consider a quantum state with energy $\mathcal{E}_0$ and wave function $\psi_0$ imbedded into a broad continuum of levels with energies $\cE_\nu$ and wavefunctions $\psi_\nu$. In the notation of the previous section $\cE_0$ stands for $E(1, \{0_{\vk\alpha}\})$ while $\cE_\nu$ for $E(0 ; 1_{\vk\alpha}, \{0_{\vk'\alpha'}\})$ with the corresponding wave functions.

 Let $V$ be the interaction between the levels with matrix elements 
$$
V_{0\nu} = V_{\nu 0}^* \;\;\;, \;\;\; V_{\nu\mu}=V_{\mu\nu}^*
$$ 
We want to consider how the system develops in time if it was initially (say at $t=0$ ) prepared in the discrete state $\psi_0$. Formally we need to solve the \Sch equation of this system 
\beq \label{eq:Sch_eq_for_lvl_cpld_2cont}
i\hbar\frac{\partial\Psi(t)}{\partial t} = (\hat{H}_0 +\hat{V}) \Psi (t)
\eeq
with the initial condition 
\beq \label{eq:in_cond_lvl_cont}
\Psi(t=0) = \psi_0
\eeq
Let us write $\Psi(t)$ as an expansion in the basis of the unperturbed states $\{\psi_0, \psi_\nu\}$ 
\beq \label{eq:wwexpand}
\Psi(t)= c_0(t)\psi_0 e^{-i\cE_0 t/\hbar} +\int  c_\nu(t)\psi_\nu e^{-i\cE_\nu t/\hbar}\, d\nu
\eeq
where  for convenience we "pull out" the factors $e^{-i\cE_0 t/\hbar}$  and $e^{-i\cE_\nu t/\hbar}$ from the (yet undetermined) time dependent coefficients $c_0(t)$ and $c_\nu(t)$.  We note that the coefficient $c_0(t) e^{-i\cE_0 t/\hbar}$ determines the  time dependence of the "persistence  amplitude" of the initial state $\psi_0$ 
\beq \label{eq:perst_amp}
 \langle\psi_0|\Psi(t)\rangle =c_0(t) e^{-i\cE_0t/\hbar} 
\eeq
while the amplitudes $c_\nu(t)e^{-i\cE_\nu t/\hbar}$ provide the time dependence of the spreading of the initial discrete state over the continuum states.

Inserting the expansion (\ref{eq:wwexpand}) into the \Sch equation, using 
$$
\hat{H}_0\psi_0=\cE_0\psi_0 \;\;\; , \;\;\; \hat{H}_0\psi_\nu = \cE_\nu\psi_\nu
$$
and projecting on $\psi_0$ and $\psi_\mu$ we obtain coupled equations for the coefficients
\eqna \label{eq:eqs_for_c_0_mu}
i\hbar \frac{d c_0}{ dt} &=&\int V_{0\mu} c_\mu e^{-i\omega_{\mu 0} t} \, d\mu  \nonumber \\
i\hbar \frac{d c_\mu}{dt} &=&V_{\mu0} c_0 e^{-i\omega_{0\mu} t}  + \int V_{\mu\nu} c_\mu e^{-i\omega_{\nu\mu} t} \, d\nu   
\eqne
with the notation
$$
\omega_{\nu\mu }=(\cE_\nu-\cE_\mu)/\hbar  
$$
and initial conditions
\beq
c_0(0)=1 \;\;\;, \;\;\; c_\nu(0)=0
\eeq
The crucial step/approximation in the Weisskopf-Wigner approach is to neglect the coupling between the continuum levels, i.e. to set
\beq \label{eq:WW_approx1}
V_{\mu\nu}=0
\eeq
in the equations (\ref{eq:eqs_for_c_0_mu}). This approximation allows to integrate the second equation (recall that $c_0 (0)=0$)
\beq 
c_\mu(t) = \frac{1}{i\hbar} \int_0^t V_{\mu 0} e^{-i\omega_{0\mu} t'}  \, c_0(t') \, dt'
\eeq 
Inserting this into the first equation we obtain a single integro-differential equation for $c_0(t)$
\beq \label{eq:WW_approx2}
\frac{d c_0}{dt}  = \int_0^t K(t-t') c_0(t') dt'
\eeq
where we introduced notation for the kernel $K(t-t')$
\beq \label{kernel_of_eq}
K(t) = -\frac{1}{\hbar^2} \int |V_{0\mu}|^2 e^{i\omega_{0\mu}t}\, d\mu
\eeq
Equations of this type are called equations with memory (for obvious reason). The memory time is finite if the kernel $K(t)$ has finite "range"  $T$, i.e. vanishes for $t$ much larger than some finite time interval $T$.

Let  us make an important observation here - $K(t)$  is proportional to  the time correlation  of $\hat{V}(t)$ in the initial state $\psi_0$ 
\beq \label{eq:time_correlators}
\int d\mu \langle \psi_0 |\hat{V}|\psi_\mu \rangle \langle \psi_\mu\ |\hat{V}|\psi_0 \rangle e^{i(\cE_0 - \cE_\mu)t/\hbar} =
\langle \psi_0 |\hat{V}(t)\hat{V}(0)|\psi_0 \rangle
\eeq
where 
$$
\hat{V}(t)= e^{i\hat{H}_0 t/\hbar}\hat{V} e^{-i\hat{H}_0 t/\hbar}
$$
is the interaction $\hat{V}$ in the so called interaction representation.  In order to understand what this means for the spontaneous photon emission let us recall what are the unperturbed energies 
and the corresponding wave functions in that problem, cf,. Eq. (\ref{unpert_En})  and the following discussion.  
Let us also recall  the explicit form of the interactions, 
\beq
\hat{H}_{I1} = -\int  \hat{\vj} (\vecr) \cdot  \hat{\vA}_T(\vecr)  \, d^3 r\;\;\; , \;\;\;
\hat{H}_{I3} = -\int  \hat{\vm}(\vecr) \cdot \hat{\vB}(\vecr) \, d^3 r
\eeq
cf., Eqs.\,(\ref{eq:HI1_interact},\ref{eq:HI3_interact}). Using this in the correlator $\langle \psi_0 |\hat{V}(t)\hat{V}(0)|\psi_0 \rangle$ we observe that in this case it is a product of the matter part involving correlators of the current $\hat{\vj} (\vecr)$ or magnetization $ \hat{\vm}(\vecr) $ in the initial matter state  and the correlations 
$$
\langle vacuum |  \hat{A}_{T, a}(\vecr,t)  \hat{A}_{T, b}(\vecr,0)  | vacuum\rangle \;\; {\rm and} \;  \;\;\; \langle vacuum |  \hat{B}_{a}(\vecr,t)  \hat{B}_{b}(\vecr,0)  | vacuum\rangle 
$$
 of the components of the EM field in the vacuum. These correlators measure the vacuum fluctuations of the field which drive the matter (say an atom)  in an excited state to spontaneously  emit a
photon and decay to a lower state.   

Returning Eq.\,(\ref{eq:WW_approx2}) we note that it can be formally solved by Laplace transform. To invert the transform however one must use approximations.  In Appendix we discuss a different method of solving  Eq.\,(\ref{eq:WW_approx2}) using the Maslov approximation. 
To state the results it is useful to rewrite the integral $\int d\mu$ over the continuum states $\psi_\mu$ in Eq.\,(\ref{kernel_of_eq})  by splitting it into the integral over the states with a fixed energy $\cE_\mu=\cE$ following by the integral over $\cE$. This can be done using  
\beq \label{eq:sep_int_E}
\int d\mu ... =\int d\cE \int d\mu \, \delta (\cE-\cE_\mu) ...
\eeq
The kernel $K(t)$ is then 
\beq \label{eq:K_of_t}
\bes
K(t) =& -\frac{1}{\hbar^2}\int d\cE  \overline{|V_{0\mu}|^2}\big|_{_{\cE_\mu=\cE}}  e^{i(\cE_0 - \cE)t/\hbar} \;\;{ \rm with} \\
 & \;\;\;\;\;\;\;\;\;\; \overline{|V_{0\mu}|^2}\big|_{_{\cE_\mu=\cE}} =  \int d\mu \;\delta(\cE_0 -\cE_\mu) |V_{0\mu}|^2
\end{split}
\eeq
Using this we show in the Appendix that  using the Maslov approximation approach to solve Eq.\,(\ref{eq:WW_approx2}) one finds that the  time dependence of the "persistence  amplitude" Eq.\,(\ref{eq:perst_amp}) of the initial state $\psi_0$  in the long time limit  (cf., 
Eq.\,(\ref{eq:cond_on_Gamma_etc})) is given by the exponential 
\beq \label{eq:init_decay}
 \langle\psi_0|\Psi(t)\rangle \equiv c_0(t)e^{-i\cE_0t/\hbar} = b_0e^{-\Gamma t/2} e^{-i (\cE_0+\Delta\cE) t/\hbar}
\eeq
where $b_0$ is a constant which depends on the short times behavior of $c_0(t)$ and  
 \beq \label{eq:width_shift}
 \bes
 \Gamma = & \frac{2\pi}{\hbar} \overline{|V_{0\mu}|^2}\Big|_{_{\cE_\mu=\cE_0}} 
 = \frac{2\pi}{\hbar}  \int d\mu \;\delta(\cE_0 -\cE_\mu) |V_{0\mu}|^2 \\
 \Delta \cE &= \cP \int d\cE  \,\overline{|V_{0\mu}|^2}\Big|_{_{\cE_\mu=\cE}} \;\;\frac{1}{\cE_0-\cE}  
 \end{split}
\eeq
Here $\cP$ denotes the "principle value" of the integral, cf.,
$$
  \cP \int_a^b \frac{f(x)}{x} dx \equiv \lim_{\epsilon\to 0} \left[\int_{a}^{-\epsilon}dx + \int_{\epsilon}^b dx\right] \frac{f(x)}{x} \;\;, {\rm for} \; a<0\;, b>0
  $$
 We see that the "survival probability" of the initial state asymptotically 
decays exponentially  with $\Gamma$ controlling the decay rate
\beq
w_0(t) \equiv  |\langle\psi_0|\Psi(t)\rangle|^2 =  |b_0|^2 e^{-\Gamma t} 
\eeq
The inverse ratio $1/\Gamma$ is often called the lifetime of the level.  

We observe that $\Gamma$ is a sum
$$
 \Gamma = \int d\mu \Gamma_{0\to\mu}
 $$
 of partial $\Gamma_{0\;\to\;\mu} $'s given by 
 $$
 \Gamma_{0\;\rightarrow\;\mu} =  \frac{2\pi}{\hbar}  |V_{0\mu}|^2 \delta(\cE -\cE_\mu)
 $$
which are just the golden rule probabilities per unit time of transitions into particular continuous state $\psi_\mu$.

The irreversible dynamics of a discrete state decaying into a continuum may serve as a simple example of how irreversibility appears
 in a formally reversible theoretical framework. In this respect it is instructive to follow a chain of considerations which starts by replacing the continuum of levels by just one level, then a few, 
 then many but still discrete  and finally by  the continuum.  It should be clear that in the few levels case there will
  be finite times that  the system will "visit" back the initial level.  These "return times"  are growing with the number of levels and turning to infinite (i.e. to a decay) in the continuum case.

\subsubsection{Coupling to the continuum - energy domain. Line shape, shift and width }
 
 The quantity $\Delta \cE$ in Eq.\,(\ref{eq:init_decay})  is the energy shift of the unperturbed energy $\cE_0$ caused by the coupling via $V_{0\mu}$ of $\psi_0$ to the continuum of $\psi_\mu$'s.  To understand this statement better we would like to present now the "stationary" version of the above discussion, i.e. to determine how the discrete state $\psi_0$ of the unperturbed Hamiltonian $\hat{H}_0$ gets "smeared", i.e. becomes distributed over the exact states of the problem with the coupling to the continuum states. 
 
 We note that the solution  (\ref{eq:wwexpand})      
 which we found to the time dependent \Sch equation (\ref{eq:Sch_eq_for_lvl_cpld_2cont}) can be formally expanded in terms of the eigenfunctions $\Psi_\chi$ of the "full" Hamiltonian   $(\hat{H}_0+\hat{V}) \Psi_\chi = E_\chi \Psi_\chi$
 \beq
 \Psi(t) = \int d\chi \cA_\chi \Psi_\chi e^{-i\Omega_\chi t} \;\; , \;\;\; \Omega_\chi = E_\chi/\hbar
 \eeq
 with the expansion coefficients $\cA_\chi$ determined by the initial condition (\ref{eq:in_cond_lvl_cont})\footnote{We assume that the continuum eigenfunctions $ \Psi_\chi$ are normalized to the delta function $\langle  \Psi_\chi |  \Psi_{\chi'} \rangle =\delta(\chi-\chi')$}
 \beq
 \cA_\chi = \langle\Psi_\chi | \psi_0\rangle
 \eeq
 The amplitude (\ref{eq:init_decay}) can then be written as 
 \beq
 \langle \psi_0 |\Psi(t)\rangle = \int d\chi |\langle \psi_0 |\Psi_\chi \rangle |^2 e^{-i\Omega_\chi t}= \int dE  \overline{ |\langle \psi_0 |\Psi_\chi \rangle |^2}\big|_{_{E_\chi =E}} e^{-iEt/\hbar}
 \eeq
 where for the integral $\int d\chi$ we used the identity Eq.\,(\ref{eq:sep_int_E}) with the notation similar to Eq.\,(\ref{eq:K_of_t})  for 
 $\overline{ |\langle \psi_0 |\Psi_\chi \rangle |^2}\big|_{_{E_\chi =E}}$. We obtain
 \beq
 \overline{ |\langle \psi_0 |\Psi_\chi \rangle |^2}\big|_{_{E_\chi =E}} =  \frac{1}{2\pi} \int_{-\infty} ^\infty dt \langle \psi_0 |\Psi(t)\rangle e^{iE t/\hbar} 
 \eeq
 The left hand side is what we are interested in - the distribution of probabilities of the unperturbed discrete state $\psi_0$ among the exact stationary states of the problem.
 
  To evaluate the integral in the r.h.s. we need to extend the solution (\ref{eq:init_decay}) to negative times $t < 0$.   
 This is simply done by noting that all the elements of the solution going from  $c_\mu(t)$ to  $c_0(t)$ keep their formal expressions. The only difference is found in the discussion of the long time limit $t_0 \to \infty$ in Eq. (\ref{eq:t_0_to_infty_lim}) of the Appendix 
which must be replaced by $t_0 \to - \infty$. To calculate such  limit we will again use the shift in the energy integration contour but this time we will need to do this into the positive $Im \, \cE$ half plane.  It is easy to see that this will lead to the same result as for positive $t$  but with the sign change of $\Gamma$.  The integral above therefore consists of two parts 
$$
\frac{1}{2\pi}\left[ \int_{-\infty} ^0 dt e^{-i (\cE_0+\Delta\cE +i\hbar \Gamma /2 - E) t/\hbar} + 
\int_0^\infty dt e^{-i (\cE_0+\Delta\cE -i\hbar \Gamma 2-E) t/\hbar}\right] 
$$
which are easily evaluated with the result 
\beq \label{eq:BW_shape}
 \overline{ |\langle \psi_0 |\Psi_\chi \rangle |^2}\Big|_{_{E_\chi =E}} = \frac{1}{\pi} \frac{  \hbar \Gamma/2} { (E -\cE_0-\Delta\cE)^2 + (\hbar \Gamma/2)^2}
 \eeq
 This shows that the unperturbed discrete state with a fixed energy $\cE_0$ gets "smeared" over the exact states in the energy range $\hbar\Gamma$ shifted by $\Delta\cE$ relative to $\cE_0$. The function in the r.h.s. of the above equality is a Lorentzian (also known as Breit-Wigner distribution). One often says that the discrete state with a sharp position in energy "acquires"  a line shape with  a width and a shift. 
 
 One can also describe the result Eq.\,(\ref{eq:BW_shape}) as a discrete state turning into a resonance.  This due to the analogy with what happens to a classical harmonic oscillator with an oscillation frequency  $\omega_0$ under an  influence of the dissipative force $- \gamma v$. The oscillator motion (for unit  mass)
 $$
 q(t) = q_0 e^{-\gamma t/2} \sin (\omega t +\phi_0) \;\; \; , \; \;\; \omega = \sqrt{\omega_0^2 - \gamma^2/4}
 $$
 is damped oscillations with a shifted frequency and the amplitude exponentially decaying with time.

 \subsection{Photon emission rate} 
 
Following the discussion in the previous section  we will now use the Fermi golden rule
 \beq
 \Gamma_{i \to f} = \frac{2\pi}{\hbar} |\langle f|V|i\rangle|^2 \delta (E_n-E_0-\hbar\omega_k)
 \eeq
  to calculate the rate (probability per unit time) of spontaneous photon emission in which the state of matter changes from higher to lower energy state. Denoting these matter states as $|n\rangle$ and $|0\rangle$ we have in this case
  \beq \label{spont_emis_states}
  |i\rangle = |n\rangle |\{0_{\vk\alpha}\}\rangle \;\;\;\; |f\rangle = |0\rangle |1_{\vk\alpha}, \{0_{\vk'\alpha'}\}\rangle
  \eeq
where $\vk'\alpha'$ denote all photon states except $\vk\alpha$. We have already inserted the corresponding  initial and final energies $E_n$ and $E_0+\hbar\omega_k$ in the $\delta$ function above.

Using $\hat{H}_{I1}$ of Eq. (\ref{first_intercat_H}) with these initial and final states we calculate
\eqna
\langle f |\hat{H}_{I1} | i \rangle&=& \langle 0| \, \langle 1_{\vk\alpha}, \{0_{\vk'\alpha'}\}|\hat{H}_{I1}  |\{0_{\vk\alpha}\}\rangle \, |n\rangle  =   \nonumber \\
&=& -  \sum_{\vk'' \alpha''}\left(\frac{\hbar }{2\epsilon_0\omega_k''\Omega}\right)^{1/2}
   \langle 0 | \hat{\vj}_{-\vk''} \cdot \vlambda_{\vk''\alpha''} | n \rangle  
    \langle 1_{\vk\alpha}, \{0_{\vk'\alpha'}\}| \hat{a}_{\vk''\alpha''}+\hat{a}^{\dagger}_{-\vk''\alpha''} |\{0_{\vk\alpha}\}\rangle \nonumber
\eqne
We have a sum of products of the matter and the radiation matrix elements. The latter are trivial to calculate
\beq \label{mat_el_of_radiation}
  \langle 1_{\vk\alpha}, \{0_{\vk'\alpha'}\}| \hat{a}_{\vk''\alpha''}+\hat{a}^{\dagger}_{-\vk''\alpha''} |\{0_{\vk\alpha}\}\rangle = \delta_{-\vk'' , \,\vk}\delta_{\alpha'' \alpha}
\eeq
which means that only one term is not zero in the sum over the modes. 

So we obtain
\beq \label{eq:mat_el_ph}
\langle f |\hat{H}_{I1} | i \rangle = - \left(\frac{\hbar }{2\epsilon_0\omega_k\Omega}\right)^{1/2}
  \langle 0 | \hat{\vj}_{\vk} \cdot \vlambda_{\vk\alpha}  |n\rangle
\eeq
The appearance of the inverse quantization volume $1/\Omega$ in the square of  this expression is easy to understand. We are calculating the probability rate to find the emitted photon in a given $\vk$ momentum state.  For a macroscopically large $\Omega$ this probability is very small $\sim 1/\Omega$ but the values of $\vk$ are very dense. In fact their density is $\sim \Omega$ which will cancel the $1/\Omega$ in
 the probability rate. We will now consider an example showing this. 

\subsubsection{What a typical detector measures} 

Let us consider a practical situation in which the  emitted photons are detected by a detector placed sufficiently far from the emitting system and measuring all photons emitted in a small sold angle $d\gamma$ around $\vk$.
\begin{figure}[H]
\centering \includegraphics[width=0.5\textwidth]{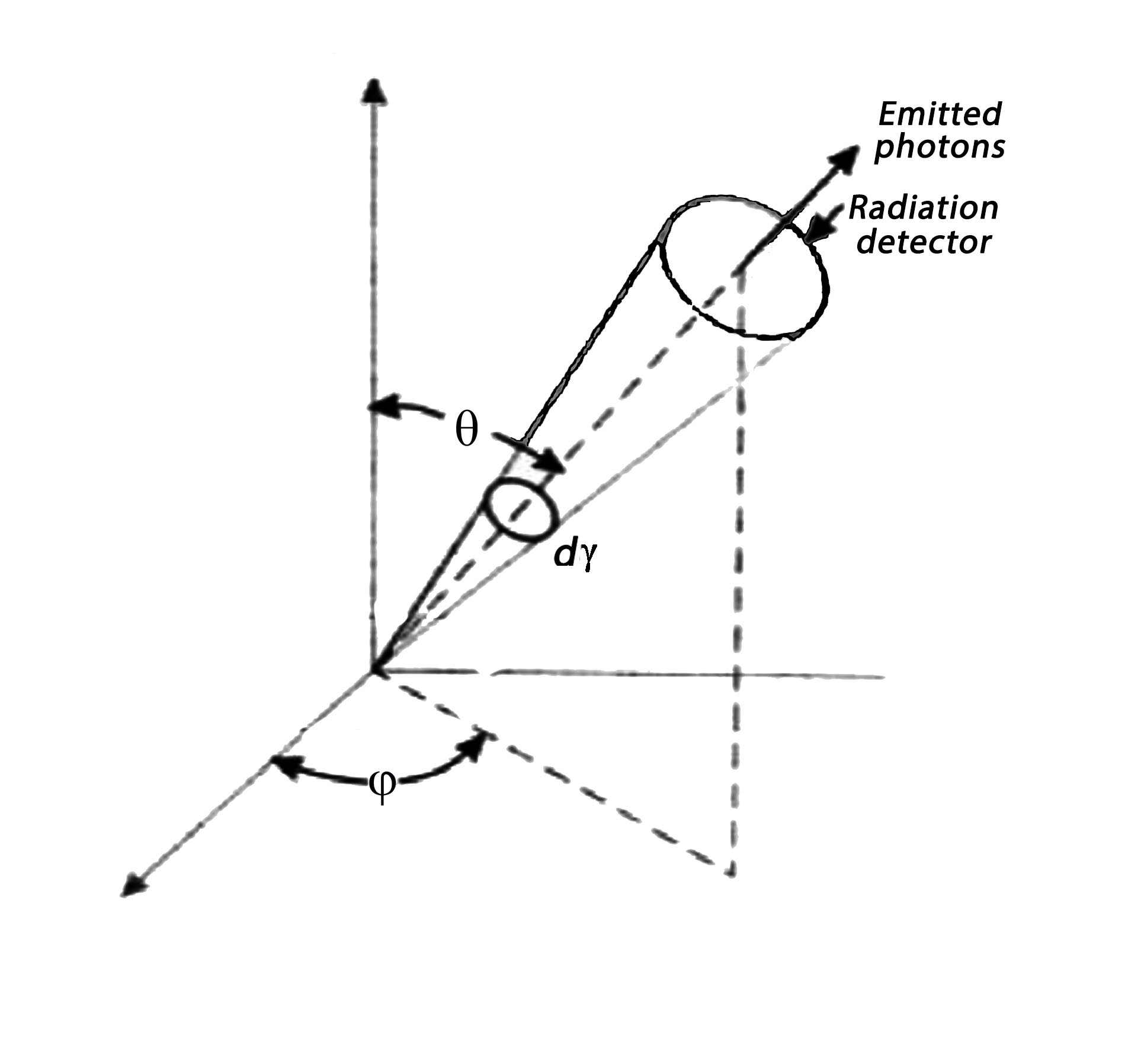}
\caption{Schematic geometry of the detection of emitted photons. The detector opening spans $d\gamma$ solid angle centered at the direction $\theta, \phi$  along which the emission rate is detected. }
 \label{fig:radiation_detect}
\end{figure} 
To calculate what the detector measures we note that the probability per unit time to measure a photon with a given polarization $\alpha$ and  momentum $\hbar \vk$ in a small "volume" $\Delta^3 k$ around a given $\vk$ is given by
$$
dw_{\vk\alpha} = \sum_{\vk' \;in\; \Delta^3 k} \Gamma_{i\to \vk' \alpha }\approx \Gamma_{i\to \vk \alpha }\times \left({\rm number\;of 
\; \vk's \; in \; \Delta^3 k} \right)\to \Gamma_{i\to \vk \alpha}   \frac{\Omega d^3k}{(2\pi)^3}
$$
where we denoted schematically by $\Gamma_{i\to \vk' \alpha }$ the rate of the photon emission into  $\vk' \alpha$ state and assumed that  $\Delta^3 k$ is small enough to have this rate changing little in the above sum. We 
have also conventionally switched to the differential $d^3k$ in our notations and used the expression $\Omega d^3 k/(2\pi)^3$ for the number of $\vk$'s in $d^3k$.

To continue with what we assumed this detector measures we should adjust the above expression to account for all the  $\vk$'s in the solid angle $d\gamma$. For this we express $d^3k = k^2 dk d\gamma$, keep $d\gamma$ fixed and integrate over $dk$.  Using the explicit expression for $\Gamma_{i\to \vk' \alpha }$ with matrix element (\ref{eq:mat_el_ph}) and changing to $k=\omega/c$ we have that the probability or more practically the relative number of photons per unit time measured  by the detector in repeated experiments is given by
\beq
dN_{\vk\alpha} = d\gamma \int  \frac{2\pi}{\hbar} \left(\frac{\hbar }{2\epsilon_0\omega\Omega}\right)
  | \langle 0 | \hat{\vj}_{\vk} \cdot \vlambda_{\vk\alpha}  |n\rangle|^2 \delta (E_n-E_0-\hbar\omega) \; \frac{\Omega \omega^2 d\omega}{(2\pi c)^3}
 \eeq
We note that $\Omega$ cancels out.  Using the $\delta$ function to do the integral we find 
\beq \label{rate_of_photon_emiss}
\frac{dN_{\vk\alpha}}{d\gamma} =\frac{\omega}{ 8 \pi^2 c^3 \epsilon_0 \hbar}   | \langle 0 | \hat{\vj}_{\vk} \cdot \vlambda_{\vk\alpha}  |n\rangle|^2
\eeq
where we must remember that $\omega$ and the magnitude of $\vk$ are fixed by the energy conservation
\beq \label{eq:omega_en_dif}
\hbar\omega=ck=E_n-E_0
\eeq

The above expression for the emission rate is the main result of this section. It shows all one needs in order to find the deexcitation rate with photon emitted in the small angle in the  direction $\vk$ with the polarization vector  $ \vlambda_{\vk\alpha} $.
One should be able to calculate the matrix element
 $$
  \langle 0 | \hat{\vj}_{\vk} |n\rangle
  $$
  of the $\vk$-th Fourier component of the matter current, then project it on the polarization  $\vlambda_{\vk\alpha}$, square the result and multiply by the coefficient  in front of (\ref{rate_of_photon_emiss}).
  
  Let us indicate that working in spherical coordinates 
\beq \label{eq:wave_vec_angles}
\vk = k(\sin\theta \cos\phi, \sin\theta \sin\phi, \cos\theta)
\eeq
a convenient choice  of linear polarization vectors for the photon emission problem is 
\beq \label{eq:pol_vec_conv}
\bes
\vlambda_1&=(\cos\theta\cos\phi, \cos\theta \sin\phi, -\sin\theta) \;\;\; , \;\;\; \lambda_2 = (-\sin\phi, \cos\phi, 0)   \\
& \vk\cdot\vlambda_{1,2} = \vlambda_1 \cdot \vlambda_2 =0
\end{split}
\eeq
This choice corresponds to $\vlambda_1$ lying in the $\vk, \ve_z$ plane, i.e. parallel to $\ve_\theta$ while $\vlambda_2$ is perpendicular to it, i.e. parallel to $\ve_\phi$. 

If the detector does not distinguish  between the photon polarizations (as is often the case) one must sum
  \beq
  \frac{dN_{\vk}}{d\gamma} = \sum_{\alpha=1,2} \frac{dN_{\vk\alpha}}{d\gamma}
  \eeq

 \subsubsection{The classical limit} 
 In the following sections we will discuss various properties and simplifications of the current matrix element in (\ref{rate_of_photon_emiss}).  Before that let us compare this expression with the corresponding classical result.  For this let us it by the photon energy $\hbar \omega$.  In this way we will find the power emitted by the system
 \beq \label{eq:power_emit}
\frac{d \mathcal{P}_{\vk\alpha}}{d\gamma} =\frac{\omega^2}{ 8 \pi^2 c^3 \epsilon_0 }   | \langle 0 | \hat{\vj}_{\vk} |n\rangle| \cdot \vlambda_{\vk\alpha} |^2
\eeq
Remarkably there is no explicit $\hbar$ dependence in this expression and the quantum mechanics manifests itself in the presence of the matrix element of the current.  

Comparing this expression with the classical result (cf., Ref.\cite{Baym}, p.279) one finds that the expressions are formally identical\footnote{One should remember the extra $1/ 4\pi\epsilon_0$ factor when passing from CGS to SI  of the square of electric charge}  
provided one identifies the matrix element of the current operator $ \hat{\vj}(\vecr)$ in quantum mechanical expression with the Fourier component  with  the frequency (\ref{eq:omega_en_dif})  of the classical  current $\vj(\vecr, t)$. 

This correspondence fits the semiclassical rule (cf., Sec.48 in Ref. \cite{LandL3})  that the matrix elements $f_{mn}$ in the classical limit approach the components   $f_{m-n}$ of the Fourier expansion of the classical function $f(t)$.  This rule was originally guessed by Heisenberg in his matrix quantum mechanics approach.

\subsubsection{Momentum conservation and recoil energy}
Let us consider the common case that the initial and final states of the photon emitting matter system are momentum eigenstates with total momentum $\vP_i$ and $\vP_f$ respectively. Isolated atoms, molecules, nuclei will be in such states. The initial and the final states in such systems will then be 
 $$
  |i\rangle = |n, \vP_i\rangle |\{0_{\vk\alpha}\}\rangle \;\;\;\; |f\rangle = |0, \vP_f\rangle |1_{\vk\alpha}, \{0_{\vk'\alpha'}\}\rangle
 $$
 with  the transition matrix element (\ref{eq:mat_el_ph}) 
 \beq \
|\langle f|V|i\rangle|^2 = \left(\frac{\hbar }{2\epsilon_0\omega_k\Omega}\right)
  | \langle 0, \vP_f | \hat{\vj}_{\vk}  |n, \vP_i\rangle  \cdot \vlambda_{\vk\alpha} |^2
\eeq

The operator $\hat{\vj}_{\vk}$ has the property that when acting on a matter state having a given total momentum $\vP$ it transforms this state  into a state with $\vP- \hbar\vk$.   
To show this  let us use the momentum operator  $\hat{\vP}=  \sum_{a=1}^N \hat{\vp}_a$ and calculate the action of its components $\hat{P}_m$ on the state which $\hat{\vj}_{\vk}$ generates acting on $ |n, \vP_i\rangle  $
\beq \label{eq:Pm_act_on_j_k}
\hat{P}_m \, \left( \hat{\vj}_{\vk}  |n, \vP_i\rangle\right) = \left[\hat{P}_m  \,,  \,\hat{\vj}_{\vk} \right]  |n, \vP_i\rangle + P^{(i)} _m \hat{\vj}_{\vk}  |n, \vP_i\rangle 
\eeq
where to avoid confusion we denoted by $ P^{(i)} _m $ the m-th component of the $\vP_i$ vector and used 
$\hat{P}_m  |n, \vP_i\rangle =   P^{(i)} _m |n, \vP_i\rangle $  in the second term on the r.h.s.  Let us now  calculate the commutator in the first term
using the explicit expression (\ref{eq:Fc_of_vj}) for $\hat{\vj}_{\vk}$ and the following relation for the components ${\hat\vp}_a$ of $\hat{\vP}$
$$
\left[{\hat\vp}_a \, , \,e^{-i\vk\cdot\vecr_b} \right] = -\delta_{ab} \, i\hbar \nabla_ae^{-i\vk\cdot\vecr_b} = -\delta_{ab}\,\hbar \vk e^{-i\vk\cdot\vecr_b}  
\; \;\; \to \;\;\; \left[\hat{P}_m  \,,  \,\hat{\vj}_{\vk} \right]  = -\hbar k_m   \,\hat{\vj}_{\vk} $$
This gives for Eq.\,(\ref{eq:Pm_act_on_j_k}) 
\beq
\hat{P}_m \left( \hat{\vj}_{\vk}  |n, \vP_i\rangle\right) = (\vP_i- \hbar\vk)_m \, \left( \hat{\vj}_{\vk}  |n, \vP_i\rangle\right)
\eeq
showing that indeed the state $\hat{\vj}_{\vk}  |n, \vP_i\rangle$ had a definite value of the momentum $\vP = \vP_i- \hbar\vk$.  Since states with different momenta are orthogonal this property means the transitions matrix elements $ \langle 0, \vP_f | \hat{\vj}_{\vk}  |n, \vP_i\rangle $ is non vanishing only for 
$$
\vP_i=\vP_f+\hbar\vk
$$
i.e. the emitted photons conserve the total momentum.

 The above discussion concerned the change of the momentum of matter systems emitting photons. But this recoil momentum $\Delta\vP\equiv \vP_f - \vP_1$ implies that there is also a corresponding recoil energy. This energy should in principle be included in the energy conservation relation 
  Eq.(\ref{eq:omega_en_dif}). However one can show that for typical photon momenta the recoil energy can to a very good approximation be neglected and the matter system assumed to remain at rest in its c.m. frame.  
  
  Indeed for the photon energy $\epsilon_{\rm photon} =\hbar \omega$ the momentum transferred  to the recoiling matter  is
   $p_{\rm recoil} = \hbar k = \hbar \omega /c$. Thus the matter recoil  kinetic energy 
 $\epsilon_{\rm recoil} = p_{\rm recoil}^2 /2 M  =(\hbar \omega)^2/2 Mc^2 $ where we assumed that the matter is non relativistic and work in a reference frame in which it was initially at rest.

The ratio of the recoil energy to the photon energy  is therefore 
 $$
 \frac{\epsilon_{\rm recoil}}{\epsilon_{\rm photon} } \sim \frac{\hbar\omega}{Mc^2}
 $$
which for typical emitting matter systems (molecules, atoms, nuclei) is
 $$
  \frac{ 1 eV \div 10 \, MeV}{(1\div 100) \, GeV} \ll 1
 $$
 so that the recoil energy is indeed negligible for such system.  
 
 Let us note that the dimensionless recoil velocity is given by the same expression
 $$
 \frac{v}{c} = \frac{p}{M c} = \frac{\hbar \omega}{Mc^2}
 $$ 
 For e.g. hydrogen atom this gives
 $$
 \frac{v_{\rm recoil}}{c} \sim \frac{10 eV}{10^9 eV}  \;\; \;\; \to \;\; \;\; v_{\rm recoil} \sim 10^{-8} c = 3 m/s
 $$

 \subsection{Long wavelength approximation}
 Consider two typical photon emitting quantum systems - atoms and nuclei and examine the relation between their sizes and the wavelengths of emitted photons. The latter are related to the photon energies as
 $$
  \lambda  = \frac{2\pi}{k}=  \frac{2\pi c}{ck}= \frac{2\pi\hbar c}{\hbar\omega}\approx \frac {6.28 \times 197 \; eV\cdot nm}{\hbar \omega} \approx \frac{1200 \; eV\cdot nm}{\hbar\omega}
 $$
 The typical atomic sizes are  $\sim 0.1 \div 0.2 \; nm$ while typical emission energies of atomic photons are
$ \sim 1 \div 10^3 \; eV$. This means that the emitted photon wavelengths are
$$
\lambda \sim (1.2 \div 1200) \; nm \gg   0.1 \div 0.2\; nm {\rm \; atomic \; sizes}
$$
 Similar result holds for nuclei for which the sizes are $5\div 10 fm$ while typical 
 emission energies are $(1\div 10) \; MeV$. So
 $$
 \lambda \approx \frac{1200 \; MeV\cdot fm}{\hbar\omega} \sim
120 \div 1200 \; fm \gg  5 \div 10 \; fm {\rm \; nuclear\; sizes}
 $$
 Similar estimates hold for solid state emission systems (there the typical size is the crystal unit cell, etc) and small molecules.

These estimates have important consequence for the evaluation and magnitude of the current matrix element in the emission rate expression Eq.\,(\ref{rate_of_photon_emiss}).  Writing it out explicitly
 \beq
   \langle 0 | \hat{\vj}_{\vk} |n\rangle = \int d^3 r \; e^{-i\vk\cdot\vecr}   \langle 0 | \hat{\vj}(\vecr)|n\rangle
  \eeq
 we see that the range of the integration where the integrand is not vanishing is determined by the matrix element of the current. So this range
 must be $|\vecr| \le a$ where $a \sim$ size of the emitting system. As we have seen above for the majority of the matter systems of interest 
 this range will be $a\ll\lambda$ - the wave lengths of the emitted photons.  This gives the condition $ka \ll1$ under which one can expand the exponent in the above integral
 \beq \label{LWA_expan}
    \langle 0 | \hat{\vj}_{\vk} |n\rangle = \int d^3 r (1-i\vk\cdot\vecr + ...)   \langle 0 | \hat{\vj}(\vecr)|n\rangle = 
     \langle 0 | \hat{\vj}_0|n\rangle -i  \int d^3r\, (\vk\cdot\vecr)  \langle 0 | \hat{\vj}(\vecr)|n\rangle  + ...
 \eeq
and keep only the lowest non-vanishing term.

This is the basis of the important element of the photon emission (and as we will see below photon absorption) treatment -- the Long Wavelength Approximation (LWA).

 \subsection{Electric dipole emission}
 Let us discuss the photon emission rate which one should expect retaining only the lowest term in the LWA expansion (\ref{LWA_expan}). We use
 \beq
  \hat{\vj}_0 = \int d^3 r  \sum_{a=1}^N\frac{q_a}{2m_a} [\hat{\vp}_a\delta(\vecr-\vecr_a) +  \delta(\vecr-\vecr_a) \hat{\vp}_a ] =
  \sum_{a=1}^N \frac{q_a}{m_a} \hat{\vp}_a 
  \eeq
  To evaluate matrix elements of this operator between matter eigenenergy states  it is convenient to use the commutation relation 
  $$
  [\vecr_a, \hat{H}_{matter}]  = [\vecr_a, \sum_{b=1}^N \frac{\hat{\vp}_b^2}{2m_b} ] =   
 i\hbar \frac{\hat{\vp}_a}{m_b}
$$
 Therefore
 \beq  
  \hat{\vj}_0  = \sum_{a=1}^N \frac{q_a}{m_a} \hat{\vp}_a = \frac{1}{i\hbar} [\hat{\vd}, \hat{H}_{matter}] \;\;\; {\rm with} \;\;\;\hat{\vd}=\sum_{a=1}^N q_a\vecr_a
 \eeq
 where $\hat{\vd}$ is the operator of the dipole moment of the matter system.

For the matrix element in the first term of (\ref{LWA_expan}) we therefore  have
\beq \label{eq:j_to_d}
 \langle 0 | \hat{\vj}_0|n\rangle  =  \frac{1}{i\hbar} \langle 0| [\hat{\vd}, \hat{H}_{matter}] | n \rangle = \frac{E_n-E_0}{i\hbar}  \langle 0| \hat{\vd} | n \rangle 
 \eeq
 where we used that $|0\rangle$ and $|n\rangle$ are eigenstates of $\hat{H}_{matter}$.
 Thus to lowest order in $ka$ 
 $$
  \langle 0 | \hat{\vj}_{\vk} |n\rangle  \approx -i\omega  \langle 0 | \hat{\vd}|n\rangle  
  $$
  Transitions described by these matrix elements are called \emph{electric dipole transitions}.
  Using this in the expression (\ref{rate_of_photon_emiss})  for the photon emission rate we obtain
  \beq \label{dipole_emission}
  \frac{dN_{\vk\alpha}}{d\gamma} =\frac{\omega^3}{ 8 \pi^2 c^3 \epsilon_0 \hbar}  
   | \langle 0 | \hat{\vd}|n\rangle \cdot \vlambda_{\vk\alpha} |^2
\eeq
Radiation described by this formula is called \emph{electric dipole} radiation. 

The above expression depends on three factors - the vector of the matrix elements of the dipole operator between the matter eigenstates, 
$$
 \vd_{on}\equiv \langle 0 | \hat{\vd}|n\rangle\;,
$$ 
the energy difference $\hbar\omega$ between these states and the polarization $\vlambda_{\vk\alpha}$ of the emitted photon. 

As we will see below the fact that the dipole operator $ \hat{\vd} $ is a vector allows to make many  general statements concerning the resulting  vector $\vd_{on}$ of matrix elements.   For the situation in which the initial and final states $|n\rangle$ and $|0\rangle$ are eigenstates of the angular momentum of the matter system it will be possible to determine when $\vd_{on}$ is non vanishing and to derive general relations between the components of  $\vd_{on}$, i.e. to find its direction. 

\begin{figure}[H]
\centering \includegraphics[width=0.5\textwidth]{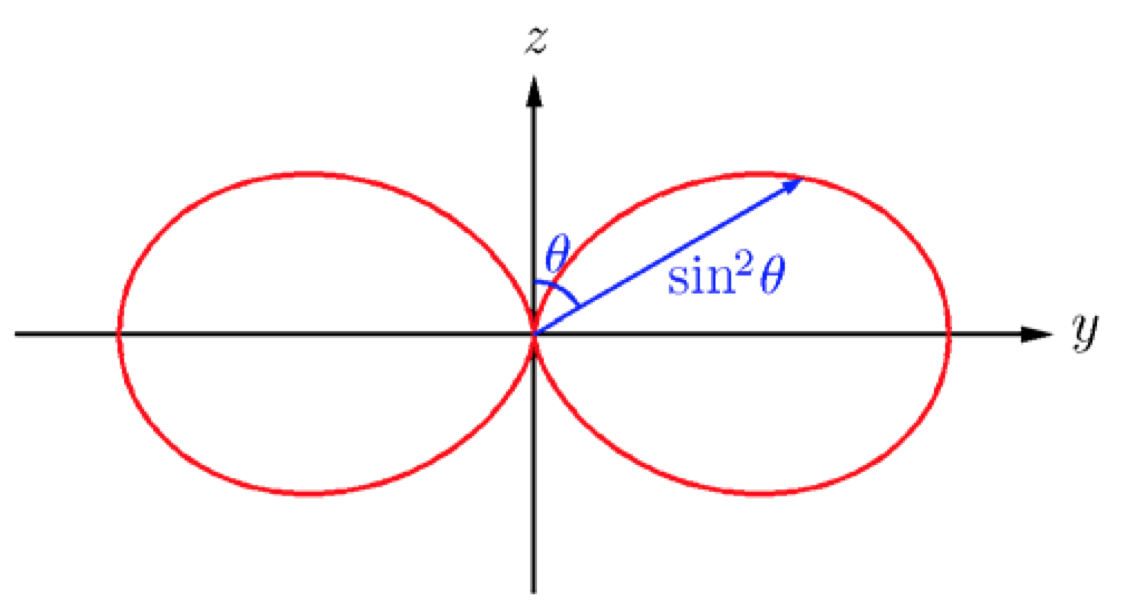}
 \includegraphics[width=0.3\textwidth]{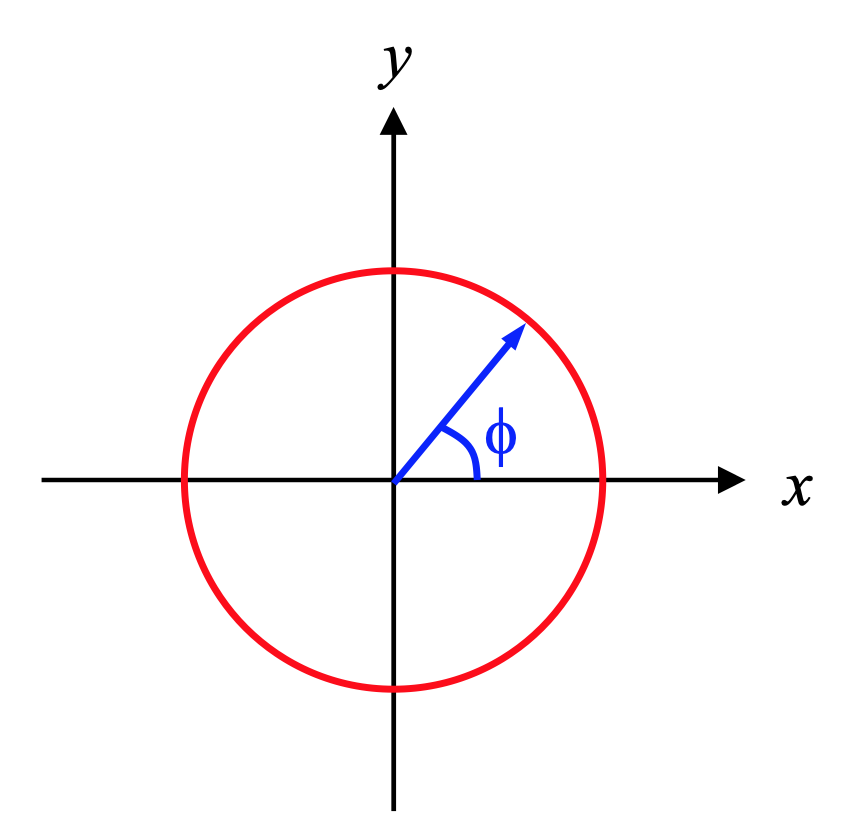}
\caption{Angular distribution of the electric dipole radiation vs the spherical angles $\theta$ and $\phi$ of the emitted photon wave vector  $\vk$ with the dipole matrix element vector $\vd_{on}$ chosen to lie along the z-axis. }
 \label{fig:dipole_emit}
\end{figure}

The direction $\vk$ of the photon emission enters the dipole emission rate Eq.\.(\ref{dipole_emission}) via its 
dependence on the polarization vectors $\vlambda_{\vk\alpha}$ which are perpendicular to $\vk$. It is obvious that  the angular distribution of the emitted photons is symmetric around the direction of the vector $\vd_{on}$.  Moreover  since $\vlambda_{\vk\alpha}$ are perpendicular to $\vk$, the emission rate is zero along the line of the direction of $\vd_{on}$.  
For an arbitrary direction of $\vk$ it is convenient to work with   
polarization vectors $\vlambda_{\vk 1}$ and $\vlambda_{\vk 2}$ which are respectively parallel  and perpendicular to the plane defined by $\vd_{on}$ and $\vk$.   Choosing the coordinate system  with $\vd_{on}$ along its  z-axis and denoting by $\theta$ and $\phi$ the spherical angles of $\vk$  it is easy to see that such a choice corresponds to Eq.\,(\ref{eq:pol_vec_conv}).  Then 
\beq \label{ang_dip}
|\vd_{on}\cdot \vlambda_{\vk 1}|^2  =|\vd_{on}|^2 \sin^2\theta
\eeq
We plot the resulting pattern in Fig.\,\ref{fig:dipole_emit}.  Clearly the emission rate with the polarization $\vlambda_{\vk 2}$  is identically zero for all the directions of such emission
$$
|\vd_{on}\cdot \vlambda_{\vk 2}|^2  = 0
$$

Multiplying Eq.\,(\ref{dipole_emission})  by $\hbar\omega$ gives the power emitted in the dipole radiation 
\beq
  \frac{d\mathcal{P}_{\vk\alpha}}{d\gamma} =\frac{\omega^4}{ 8 \pi^2 c^3 \epsilon_0}  
   | \langle 0 | \hat{\vd}|n\rangle \cdot \vlambda_{\vk\alpha} |^2
\eeq
As in our discussion following 
Eq.\,(\ref{eq:power_emit}) we note that there is no explicit $\hbar$ dependence in this expression 
and that the quantum mechanics manifests itself "only" in the matrix element of the dipole operator. Once again comparing with the 
classical expression, cf.\footnote{ https://farside.ph.utexas.edu/teaching/em/lectures/node95.html} one sees that this matrix element in the classical limit becomes the Fourier component of the classical dipole moment $\vd(t)$ with  frequency (\ref{eq:omega_en_dif}). 

Here is a pictorial representation of the classical electric dipole radiation.
\begin{verbatim}
https://www.youtube.com/watch?v=UOVwjKi4B6Y
\end{verbatim}

 \subsection{Angular momentum and parity selection rules}
 
 In discussing  photon emission by individual molecules, atoms, nuclei and subnuclear particles one deals with
  rotationally invariant matter Hamiltonians with eigenstates which are also eigenstates  of the total angular momentum (including the spin)
 $$
  \hat{\vJ} = \hat{\vL} + \hat{\vS} = \sum_{a=1}^N \vecr_a\times\hat{\vp}_a +\sum_{a=1}^N \hat{\vs}_a
  $$ 
  or more precisely of its square $\hat{\vJ}^2$ and one of its projections, commonly chosen as $\hat{J}_z$. The sum here is over the components of the molecule, atom, etc, which is under  consideration.  
  So in these (very common) cases the dipole matrix elements to be considered are 
  \beq \label{matrix_el_of_d}
   \langle 0 | \hat{\vd}|n\rangle  \to  \langle \nu_2 J_2 M_2 | \hat{\vd}|\nu_1 J_1 M_1 \rangle 
   \eeq
 where we indicated explicitly the angular momentum quantum numbers and denoted by $\nu$ all the remaining ones needed to completely specify the states of the matter system. For example levels of a particle with spin  $1/2$  moving in a spherically symmetric potential and with spin-orbit coupling have 4 quantum numbers 
   $$
   n_r\; , \; l\; ,\; j\; ,\; m
   $$ 
   so $\nu$ will stand in this case for $n_r, l$ numbers. 

   \subsubsection{Dipole moment is an $\ell=1$ object. Spherical components of vectors} 
   
   When the dipole operator $\hat{\vd}$ in the matrix element Eq.\,(\ref{matrix_el_of_d}) acts on the initial state $|\nu_1 J_1 M_1 \rangle $  it creates a state which doesn't have the same angular momentum and is in general expected to be a superposition of states with definite $J$'s and $M$'s. The vectorial character of $\hat{\vd}$ allows to determine the range of possibles values of these quantum numbers and to a certain extent also the coefficients in the resulting linear combination.  To demonstrates this it is useful to transform the vector $\hat{\vd}$ from cartesian to the so called spherical components. 
   
The general expressions for such components of any vector $\vv$ are by definition 
   \beq
   v_{\mu=1} = -\frac{v_x+iv_y}{\sqrt{2}}  \;\;, \;\; v_{\mu=0}=v_z \;\;, \;\;    v_{\mu=-1} = \frac{v_x-iv_y}{\sqrt{2}} = -v_{\mu=1}^*
   \eeq
   The scalar product of vectors in spherical components is expressed as 
   \beq
   \va\cdot\vb = \sum_{\mu=-1,0,1} (-1)^\mu a_\mu b_{-\mu} = \sum_{\mu=-1,0,1} a_\mu b^*_\mu
   \eeq
   The usefulness of forming the spherical components' combinations can be seen especially clear in terms of the spherical coordinates\footnote{Note that the subscript $v$ in the angles here indicates that they are not necessarily the same as of the real space coordinate vector $\{x,y,z\}$.}
   \beq
   v_x = v\sin\theta_v \cos\phi_v \;\;\;, \;\;\; v_y = v \sin\theta_v \sin \phi_v \;\;\;, \;\;\; v_z = v\cos\theta_v 
   \eeq
   so that
\beq \label{eq:vec_vs_Y1m}
v_0= v\cos\theta_v \;\;, \;\; v_{\pm 1} =  \mp \frac{1}{\sqrt{2}} v \sin \theta_v e^{\pm i\phi_v}    \;\;\; \to \;\;\; 
   v_{\mu}  = \sqrt{\frac{4\pi}{3}}\; v\,Y_{1\mu}(\Omega_v)  \;\;\; \mu=1,0,-1
   \eeq
   emphasizing that three components of a vector behave under rotations as $Y_{1\mu}$.  In the group theoretical terminology  one says that vectors transform as $j=1$ representation of the group $O(3)$ of rotations.  
   
   This of course holds true also for the vector $\hat{\vd}$ of the electric dipole moment. Let us now explore the consequences of this insight. 

\subsubsection{Dipole angular momentum selection rules - hydrogen atom first}

Let us start with a simplest case of electric dipole transitions in a hydrogen atom.  With its single electron the dipole operator and its spherical components  in this simple  system are just 
$$
\hat{\vd} =e\vecr \;\;  \to   \;\; \hat{d}_\mu =  e\sqrt{\frac{4\pi}{3}}\;  r \,Y_{1\mu}(\theta,\phi)
$$ 
We can ignore the spin and consider dipole transitions between the orbital eigenstates of the hydrogen atom 
$$
| n, l, m \rangle \to | n', l', m' \rangle
$$
 with the coordinate representation of these states having the familiar form 
$$
\langle \vecr | n, l, m \rangle = R_{nl}(r) Y_{lm}(\theta, \phi)
$$
In this representation electric dipole operator acting on the initial state  $| n, l, m \rangle$ results in a state $\hat{d}_\mu | n, l, m \rangle $ which in the coordinate representation is 
\beq \label{eq:d_act_onY}
\langle \vecr |\hat{d}_\mu | n, l, m \rangle =  \sqrt{\frac{4\pi}{3}} e r R_{nl}(r)  Y_{1\mu}(\theta,\phi)  Y_{lm}(\theta, \phi)
\eeq
Let us use the intuition from the quantum angular momentum algebra and view the product of the two spherical harmonics  $Y_{1\mu} Y_{lm}$ as an eigenfunction of the (quantum) sum of two angular momenta  $\ell_1 =1$ and $\ell_2=l$.  
As we know the resulting angular momentum $\ell$ has possible values given by
$$
l-1\; , \; l \; , \;  l+1 \;\;\; {\rm with \;the \; projection} \;\;  \mu+m
$$ 
Continuing with this understanding we expect that the state $\hat{d}_\mu | n, l, m \rangle$ is a linear combination of states with the above values of $\ell$ and its projection.   

Forming the dipole matrix element $\langle n',l',m' | \hat{d}_\mu |n, l, m \rangle$ means that the final state 
$|n',l',m' \rangle$ is projected on this linear combination.  The resulting overlap should be  zero unless the final angular momentum $l' m'$ is equal to one of the above values, i.e. satisfy the familiar triangular rule of adding angular momenta
\beq \label{eq:dipole_ang_mom_rule}
|l -1| \le l' \le l+1  \;\; , \;\; m' = m+\mu
\eeq
Formally these considerations are supported and extended by using the known expansion of the product of two spherical harmonics $Y_{1_1 m_1}(\theta,\phi) Y_{l_2 m_2}(\theta,\phi)$ viewed as a function of the angles $\theta , \phi$ in terms of the complete set $\{Y_{LM}(\theta,\phi)\}$ 
\beq \label{eq:exp_ofY}
Y_{l_1 m_1}(\theta,\phi) Y_{l_2 m_2}(\theta,\phi) = \sum_{L=0}^\infty \sum_{M=-L}^L   G_{L\, l_1\, l_2}^{M m_1 m_2}  \, Y_{LM}(\theta,\phi)   
\eeq
where $G_{L\, l_1\, l_2}^{M m_1 m_2} $ are the so called Gaunt coefficients which are proportional to the respective Clebsh-Gordan (CG) coefficients, cf. Ref.\cite{sph_h}, p.57
\beq \label{eq:Gnt_via_CG}
G_{L\, l_1\, l_2}^{M m_1 m_2} = a(l_1, l_2, L) \langle LM | l_1 m_1, l_2 m_2\rangle
\eeq
Here the proportionality factor $a(l_1, l_2, L)$ doesn't depend on the projections $m_1, m_2, M$. The  CG coefficient is zero unless
$$
|l_1 - l_2| \le L \le l_1+l_2\;\;\; {\rm and} \;\;\; M=m_1 +m_2
$$
which constraints the sum over $L$ in the expansion (\ref{eq:exp_ofY}) and  removes the sum over M. When applied to our case, Eq. (\ref{eq:d_act_onY}),  with the product $Y_{1\mu} Y_{lm}$ one recovers what we have guessed using qualitative arguments, i.e. the rules (\ref{eq:dipole_ang_mom_rule}). These are called electric dipole angular momentum selection rules. In words they state that only transition with at most one unit change in the angular momentum are allowed, i.e  $\Delta l =0, \pm 1$. Below we will complete the discussion of these rules by examining also the consequences of the parity conservation. 

Let us further observe that the dependence of the Gaunt coefficients on the angular momentum projection quantum numbers $M, m_1, m_2$ 
enter only via the CG coefficient. To see what this means for the electric dipole transitions let us sketch schematically the calculation of the dipole matrix element $\langle n',l', m' |\hat{d}_\mu |n, l, m \rangle$.  We will need to calculate 
\beq \label{eq:di_me_in_H}
\langle n',l',m' | \hat{d}_\mu |n, l, m \rangle =\int (\rm radial \; part) \int (\rm angular \; part)
\eeq
where 
\beq
\label{eq:rad_ang_int}
\bes
 \int (\rm radial \; part) &= e\sqrt{\frac{4\pi}{3}} \int_0^\infty  r^2 dr R^*_{n'l'}(r) r R_{nl}(r)  \\
\int (\rm angular \; part) &= \int  \; Y^*_{l'm'}(\theta, \phi) Y_{1\mu}(\theta,\phi)  Y_{lm}(\theta, \phi)\;\sin\theta d\theta d\phi
\end{split}
\eeq
Expansion (\ref{eq:exp_ofY}) shows that the angular integral equals the appropriate Gaunt coefficient 
$G_{l'\, 1\, l}^{m' \mu m}$. Using Eq.\,(\ref{eq:Gnt_via_CG}) we see that
\beq \label{eq:mat_vs_redm}
\langle n',l', m' | \hat{d}_\mu |n, l, m \rangle = \langle l m, 1 \mu| l'm'\rangle \langle n'l' || \hat{d} || n l \rangle
\eeq
where we introduced the common notation $\langle n'l' || \hat{d} || n l \rangle$ called reduced matrix element for the part of the full matrix element which is independent of $m, m'$ and $\mu$. In the present case it is the product of the radial part in (\ref{eq:rad_ang_int}) and the factor $a(l, 1 ,l')$ in the relation (\ref{eq:Gnt_via_CG}). 

Expression  (\ref{eq:mat_vs_redm}) is a particular case of a more general relation known as the Wigner-Eckart theorem which will be discussed in the next Section. It shows that the dipole matrix element dependence on $m, ,m'$ and $\mu$ is entirely determined by known (tabulated) CG coefficients, cf. Ref.\cite{sph_h}.  

From this it follows that if for given $nl$ and $n'l'$  quantum numbers one needs to find 
all the matrix elements $\langle n',l',m' | \hat{d}_\mu |n, l, m \rangle$ it is be enough to determine just one of them, say,  with $m=m', \mu=0$.   Using its 
value one can calculate the reduced matrix element $\langle n'l' || \hat{d} || n l \rangle$  and then all the $(2l+1)\times 3$ via the relation Eq.\,(\ref{eq:mat_vs_redm}) with appropriate CG coefficients.

We also note that  for given initial and final states the selection rules Eq.\,(\ref{eq:dipole_ang_mom_rule}) show that only one spherical component of the vector\footnote{We use the term "vector"  for complex valued  matrix elements of the dipole operator $\hat{d}$ for the brevity of presentation. It is the relative size of its three components that will be of our interest} $\langle n',l',m' | \hat{d}_\mu |n, l, m \rangle$ is non zero, that with $\mu = m'-m$. Let us recall that in the present context  this vector is what  was denoted $\langle 0 | \hat{\vd}|n\rangle$ in the expression (\ref{dipole_emission}) for the electric dipole emission rate.  We then conclude that the scalar product $\langle 0 | \hat{\vd}|n\rangle\cdot \vlambda$ in that expression has correspondingly only one term $\langle 0 | \hat{d}_\mu |n\rangle \lambda_\mu^{*}$ with that $\mu$ and the angular distribution is given by the angular dependence of $|\lambda_\mu|^2$.

Let us consider as an example the case of transitions between states with equal $m=m'$ for which the only non zero matrix element is $\langle n',l',m | \hat{d}_0 |n, l, m \rangle$. This is $\langle 0 |d_z |n\rangle$ in the notation of Eq.\,(\ref{dipole_emission}) and correspondingly the angular distribution of the emitted photons is given by $\lambda_z^2$ which for the choice (\ref{eq:pol_vec_conv})  of $\vlambda_1$ is given by Eq.\,(\ref{ang_dip}) and zero for $\vlambda_2$.  More examples and details will be considered in tutorials and homework.

\subsubsection{Dipole parity selection rule - hydrogen atom first}

\begin{figure}[H]
\centering \includegraphics[width=0.8\textwidth]{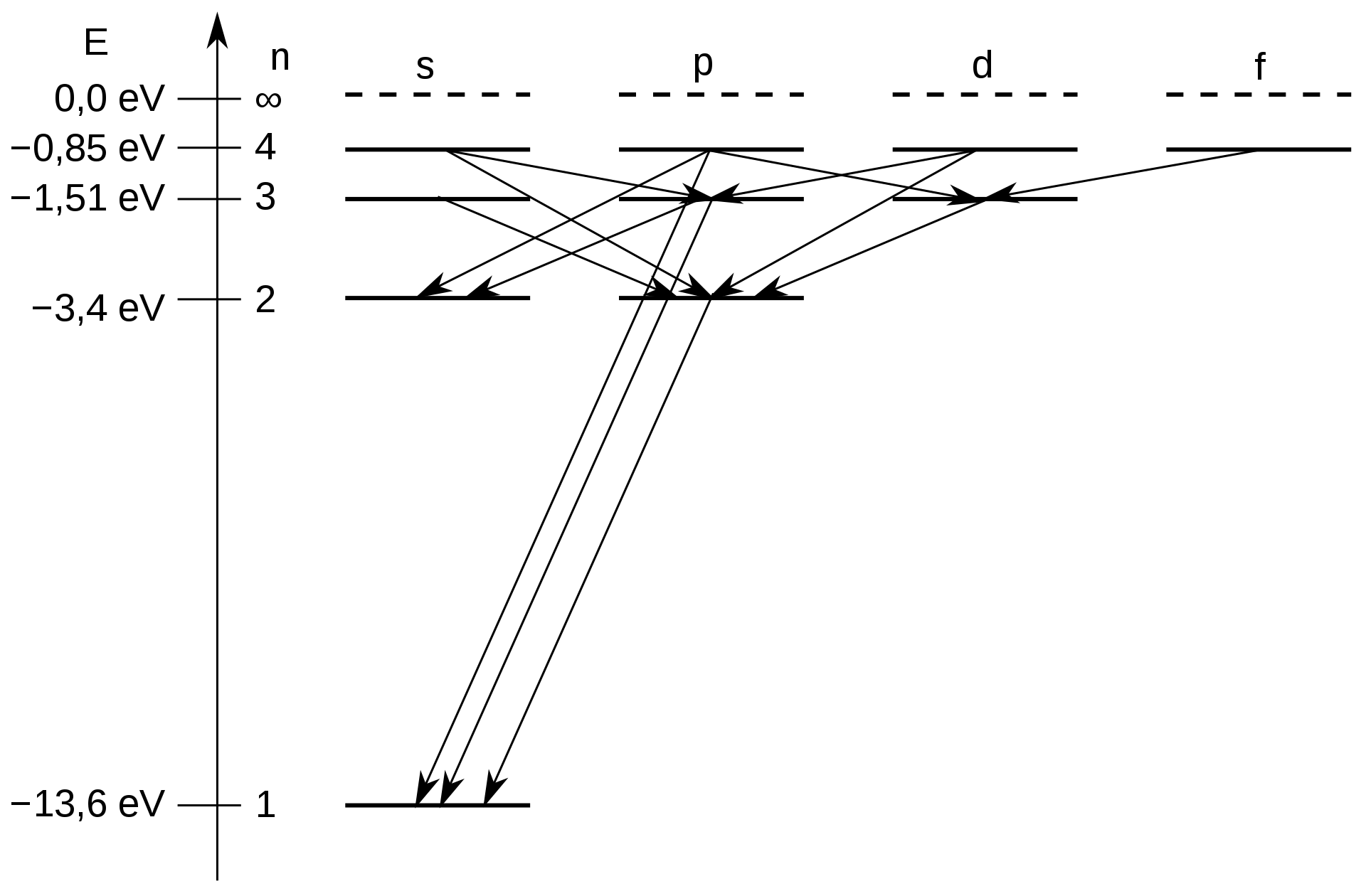}
\caption{Radiative transitions in hydrogen. Only dipole transitions between adjacent angular momentum columns are allowed, as per combined angular momentum and parity selection rule $\Delta l = \pm 1$}
 \label{fig:dip_trans_inH}
\end{figure}

Let us now examine limitations which parity conservation imposes  on the possible final states of electric dipole transitions from a given initial state.  We start by noticing that under the parity transformation $\vecr \to -\vecr$, i.e. under mirror reflection  
\beq \label{eq:mirror_parity_trans}
x, y, z \;\; \to \;\; -x, -y, -z
\eeq 
of the coordinate system the electric dipole operator changes sign $\hat{\vd} \to - \hat{\vd}$.  Let us make this coordinate change in the integral   Eq.\,(\ref{eq:di_me_in_H}).  In spherical coordinates this change is 
\[
r ,  \theta, \phi \;\; \to \;\; r, \pi - \theta, \phi + \pi
\]
so that the radial part doesn't change while  the spherical harmonics transform as\footnote{To see this start with the easy $Y_{ll}\sim \sin^l\theta e^{il\phi}$ and then use $Y_{lm} \sim \hat{L}_{-} Y_{l m+1}$ together with $\hat{L}_{-} $ being even under $\vecr \to -\vecr$ to show that all $Y_{lm}$ transform as $Y_{ll}$. } 
 \beq \label{eq:Ylm_under_parity}
 Y_{lm}(\theta, \phi) \to  Y_{lm}(\pi - \theta, \phi + \pi) = (-1)^l  \,Y_{lm}(\theta, \phi)
 \eeq
 The result is that the entire integral on the r.h.s. of Eq. (\ref{eq:di_me_in_H}) is equal to itself multiplied by $-(-1)^l (-1)^{l'}$.   This of course means that it is zero and together with it the matrix element 
 $\langle n',l',m' | \hat{d}_\mu |n, l, m \rangle$ is zero unless 
 \beq
 (-1)^l (-1)^{l'} = -1
 \eeq
 i.e. $l'$ and $l$ are of opposite parity (i.e. odd vs even or even vs odd). This is called parity selection rule. 
 Taken together with the  angular momentum we find that electric dipole selection  rules can be formulated as 
 \beq
 l' = l \pm 1
 \eeq 
 In Fig.\ref{fig:dip_trans_inH} we show examples of electric dipole transitions 
 
\subsubsection{Angular momentum selection rules - general view. The Wigner-Eckart theorem}
  
  In this section we will extend and formalize our discussion of the angular momentum selection rules from the simplest case of 
  radiative transitions in a hydrogen atom to a general case of any physical system (e.g. multi-electron atoms, nuclei, molecules) the Hamiltonian of which is invariant under rotations.  We will show that the main relation, Eq.(\ref{eq:mat_vs_redm}) holds for such systems  with all its consequences.
  
The general structure of the eigenstates  in  systems with rotationally invariant Hamiltonian is $|\nu  J M \rangle$,  cf., Eq.\,(\ref{matrix_el_of_d}),  with $\nu$ denoting all the quantum numbers needed to specify this state apart of the angular momentum $J$ and its projection $M$.   What this structure means is that under $O(3)$ rotations these states  transform as 
 \beq
 \bes
 \hat{U}(\alpha \vn)|\nu J M \rangle& \equiv e^{i\alpha \vn\cdot \hat{\vJ}} |\nu J M \rangle = \sum_{M'=-J}^J D_{M M'}^J (\alpha \vn)  |\nu J M' \rangle \\
 D_{M M'}^J (\alpha \vn) &=  \langle \nu J M'| e^{i\alpha \vn\cdot \hat{\vJ}} | \nu J M \rangle 
\end{split}
 \eeq
 i.e. the multiplets of states with different $J$'s do not mix.  Here we denoted by $\alpha$ the angle of rotation and by the unit vector $\vn$ the direction of the rotation axis. 
 
In our discussions of the hydrogen atom case we have seen that the vector character of the  dipole operator, i.e. its behavior under rotations played a very important part.  We will now generalize this discussion.  Let us recall  that under any unitary transformation which transforms wavefunctions as $|\psi\rangle \to U|\psi \rangle$  the  operators transform as $U \hat{f} U^{-1}$.  This is trivially seen by considering how the states obtained by acting with $\hat{f}$ transform
$$
\hat{f} |\psi\rangle \to U \hat{f} |\psi\rangle = U \hat{f} U^{-1} U |\psi\rangle
$$
which demonstrates that indeed $U \hat{f} U^{-1}$ acting  on transformed wavefunctions $U |\psi\rangle$ produces the correctly transformed result. 

Following this understanding one defines spherical tensor operators $\hat{T}_{j\mu}$ as a set of $2j+1$ operators  which transform among themselves under $O(3)$ rotations
$$
 \hat{U}^{-1} (\alpha \vn)\hat{T}_{j\mu} \hat{U} (\alpha \vn) = \sum_{\mu'} D_{\mu \mu'}^j  (\alpha \vn) \hat{T}_{j\mu'}
$$
Obviously the electric dipole operator $\hat{d}_\mu$ is an example of the spherical tensor  $\hat{T}_{j\mu}$  with rank $j=1$.  In the following section we will 
encounter examples of electric and magnetic multipole operators which will correspond to spherical tensors $\hat{T}_{j\mu}$ with higher rank $j$.  One also encounters similar expansions of  physical operators in terms of spherical tensor operators  $\hat{T}_{j\mu}$ in other fields of physics, e.g. in the context of atomic and nuclear shell models.  

To understand the properties of the spherical tensor operators let us examine how the state which is obtained  when $\hat{T}_{j\mu}$ acts on $|\nu J M\rangle$ behaves under rotations 
$$
U \hat{T}_{j\mu} |\nu J M\rangle = U \hat{T}_{j\mu}U^{-1} U |\nu J M\rangle = \sum_{\mu'} \sum_{M'}  D_{\mu \mu'}^j D_{M M'}^J    \hat{T}_{j\mu'}|\nu J M' \rangle
$$
The product of the two D matrices appearing here is identical to what would be obtained when rotating the direct product of states with angular momentum $j,\mu$ and $J,M$.  This suggests that $\hat{T}_{j\mu}$ acting on $ |\nu J M\rangle$  generates a state having total angular momentum equal (quantum mechanically) to the sum of  $j, \mu$ and $J, M$. This would mean that in the matrix 
\beq \label{eq:me_of_Tjm}
\langle \nu' J' M' | \hat{T}_{j\mu} | \nu J M\rangle
\eeq
only matrix  elements satisfying the quantum mechanical rules of summing the angular momenta  
\beq \label{eq:ang_mom_sr_gen}
 |J-j| \le J' \le J_1+j \;\; , \;\; 
M'=M+m
\eeq
can be non zero. 

These intuitive expectations find rigorous proof in the classic Wigner-Eckart theorem. It  generalizes the equality 
Eq.\,(\ref{eq:mat_vs_redm}) to matrix elements (\ref{eq:me_of_Tjm}), i.e. to the most general spherical tensor operators and eigenstates of  any physical system with spherical symmetry  
\beq \label{WE_theorem}
 \langle \nu' J' M' | \hat{T}_{j \mu} | \nu J M \rangle = \langle J' M' | j \mu, J M \rangle  \langle \nu' J' ||  \hat{T}_j  ||\nu J \rangle
 \eeq
 Here $\langle J' M' |j m, J M \rangle$ are the Clebsh-Gordan coefficients and  the notation  $\langle \nu' J'  || \hat{T}_j ||\nu J \rangle$ called  \emph{reduced matrix elements} stands for the  parts of the full matrix elements which are independent of the  projections $M, M' $ and  $m$. This dependence is fully incorporated in the  CG coefficients which also carry the information about the angular momentum selection rules, Eq.\,(\ref{eq:ang_mom_sr_gen}).
  
 As in the hydrogen atom case  the  reduced matrix elements represent the orientation independent context of the  original matrix elements, Eq.(\ref{eq:me_of_Tjm}).  To find them it is enough to calculate $\langle \nu' J' M'| \hat{T}_{jm}|\nu J M \rangle$ for  one particular set  of values of $M, m, M'=M+m$ and divide  the result by the corresponding CG coefficient.  For fixed $\nu J$ and $\nu' J'$ this amounts to just one calculation  to  determine all the $(2J+1)\times (2j + 1)$ matrix elements   in the left hand side of the relation (\ref{WE_theorem}) via the (known, tabulated) CG coefficients.

Finally let us note that the formal proof of Eq.\,(\ref{WE_theorem}) can be found in many references, e.g. p.\,252 in Ref.\,\cite{Sac}.

\subsubsection{An aside - review of the parity symmetry}

In our discussion of the parity selection rules in hydrogen atom they looked like a special case depending on the behavior  of the  spherical harmonics $Y_{lm}(\theta, \phi)$ under the transformation of the angles, Eq.\,(\ref{eq:Ylm_under_parity}). We now wish to generalized these considerations to photon radiation in more complicated systems. 

Parity transformation is an inversion transformation of a coordinate system in which all of its axes change signs, e.g. Eq.\,(\ref{eq:mirror_parity_trans}). Let us note that in two dimensions this transformation can be accomplished by a $\pi$ rotation of the axis.  This 
is not so in three dimensions where the coordinate system changes from right-handed to left-handed. This is the reason the parity transformation probes additional features in three dimensional physical systems. 

\begin{figure}[H]
\centering \includegraphics[width=0.8\textwidth]{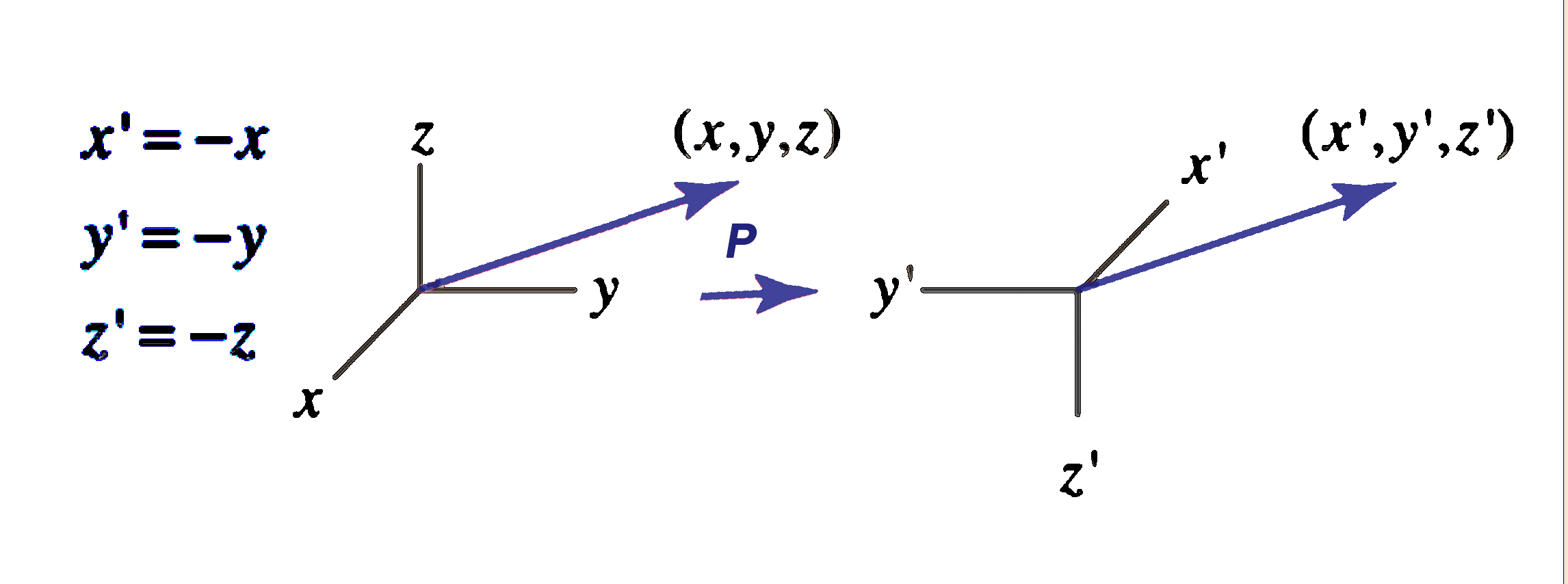}
\caption{Parity transformation - the same physics (e.g. the same particle position, momentum, etc) is seen in the inverted coordinate axes  system with  $\vecr \to -\vecr$, $\vp \to -\vp$ etc}
 \label{fig:parity_trans}
\end{figure}

Let us note that technically the coordinate inversion can be achieved by  a reflection in any plane, followed by a $\pi$ rotation about an axis normal to this plane.  We also note that under coordinate inversion vectors are expected to change signs, cf., Fig.\ref{fig:parity_trans}. However as we will see below there exist a category of vectors which do not change signs under parity transformation.  Such vectors are called axial vectors or pseudo-vectors to distinguish from the real vectors also called polar vectors.

Is the nature invariant with respect to the parity transformation? Historically this was a very important question and the brief answer is that  physical systems interacting via gravity, electromagnetic and strong interactions are invariant but the weak interactions violate this.  It is beyond the scope of  these lectures to go into the details of this statement, cf., Ref.\,\cite{Grif}.  Rather let us remain in the framework of what we study and examine this issue starting with the Hamiltonian given by Eq.\.(\ref{EM_matter_Ham}).  We observe that this Hamiltonian remains invariant if we change  
\beq
\vecr_a \to -\vecr_a,  \;\;\;, \;\;\; \vp_a \to - \vp_a \;\;\; \vA(\vecr)  \to -\vA (-\vecr) \;\;, \;\;  \vE(\vecr)  \to -\vE (-\vecr) 
\eeq
which is obviously the parity transformation. The extension to the remaining part $\hat{H}_{I3}$, Eq.\.(\ref{interact_Ham3}) of the (non relativistic) matter-EM field Hamiltonian is discussed in the Appendix \ref{app:App_H3} where it is shown that magnetic field $\vB(\vecr)$ and particles' angular momenta $\vl_a$ and spins $\vs_a$ are axial vectors, i.e. they do not change under the coordinate inversion.  

Let us now consider what does the invariance of the Hamiltonian under the parity transformation imply. It will be sufficient for our goals to limit the discussion to the matter part of the Hamiltonian $\hat{H}_{matter}$ in Eq. (\ref{parts_of_H}).    We  introduce the parity operator by defining its action on the wavefunctions of the matter particles
\beq
 \hat{P} \psi(\vecr_1, \sigma_1; \vecr_2, \sigma_2;  ... ;\vecr_N, \sigma_N)= \psi(-\vecr_1,\sigma_1 ; -\vecr_2, \sigma_2 ; ...; -\vecr_N, \sigma_N)
\eeq
or formally
$$
\langle \vecr_1, \sigma_1 ;...; \vecr_N, \sigma_N |\hat{P}|\psi\rangle = \langle -\vecr_1,\sigma_1; ...;-\vecr_N, \sigma_N|\psi\rangle 
$$
Here $\sigma$'s denote the particle spin variables (e.g. for spin 1/2 they are $\sigma = \pm 1/2$)  and we used the axial vector nature of the spins.

Clearly 
 $$
 \hat{P}^2 \equiv \hat{P} \hat{P} = 1 
 $$
 which means that
 \beq \label{Parity_op_1}
 \hat{P}=\hat{P}^{-1}
 \eeq
  As usual with symmetries the invariance of the matter Hamiltonian under the parity transformation means that to transform the result 
  of $\hat{H}$ acting on any $|\psi\rangle$ will produce  the same result as of $\hat{H}$ acting on the transformed $|\psi\rangle$
 \beq
 \hat{P}\left(\hat{H}_{matter} | \psi\rangle\right) = \hat{H}_{matter} \hat{P}|\psi\rangle
 \eeq
 Formally this means
 \beq \label{Parity_op_2}
 \hat{P}\hat{H}_{matter} = \hat{H}_{matter}\hat{P}\;\;\; \to  \;\;\; [\hat{H}_{matter}\;, \; \hat{P}] =0
 \eeq
 or using  (\ref{Parity_op_1})
 $$
 \hat{P}\hat{H}_{matter}\hat{P}=\hat{H}_{matter}
 $$
 Eigenstates of such Hamiltonian are or can be chosen to be eigenstates of $\hat{P}$. Indeed, acting with $\hat{P}$  
 $$
 \hat{H}_{matter} | n \rangle = E_n | n\rangle \;\; \to \;\; \hat{P}  \hat{H}_{matter} | n \rangle = E_n \hat{P}| n\rangle
 \;\; \to \;\;  \hat{H}_{matter} \hat{P} | n \rangle = E_n \hat{P}| n\rangle
 $$
  we see that if $| n \rangle$ is an eigenstate of $\hat{H}_{matter}$ so is $\hat{P}| n \rangle $ with the same eigenenergy $E_n$. 
 This implies one of the two possibilities - either $\hat{P}| n \rangle$ is proportional  to $| n \rangle$
 $$
 \hat{P}| n \rangle = {\rm const} | n \rangle
 $$
 or it is a different state.   In the former case we find
 $$
 \hat{P}^2| n \rangle ={\rm const} \hat{P}| n \rangle = {\rm const}^2 | n \rangle
 $$
 and since $\hat{P}^2=1$ have $\rm const^2 = 1\;\;\; \to \;\;\; {\rm const} = \pm 1$.
 
 When $\hat{P}| n \rangle$ is a different state from $| n \rangle$ we have degeneracy and can form linear combinations of these states 
 $$
 | n \rangle_{\pm} \equiv \frac{1}{2}(1\pm \hat{P})| n \rangle 
 $$
 which are eigenstates of $\hat{P}$
 \beq
 \hat{P}| n \rangle_{\pm} = \hat{P}  \frac{1}{2}(1\pm \hat{P}) | n \rangle =  \frac{1}{2}(\hat{P}\pm \hat{P}^2)|n\rangle  =  \pm \frac{1}{2}(1\pm \hat{P})|n \rangle =\pm | n \rangle_{\pm}
 \eeq
 exactly as in the non degenerate case.
 
 \subsubsection{Parity selection rule - general view \label{sec:parity_select}}
 We now want to learn what limitations the parity symmetry imposes on the dipole matrix elements 
 in the expression (\ref{dipole_emission}). Following the discussion in the previous section we can assume that the states $|0|\rangle$ and $| n \rangle$ have well defined parity which we denote respectively by $P_f$ and $P_i$. Then using
  $$
   \hat{P}^2 =1 \;\;\;{\rm and} \;\; \; \hat{P}\hat{\vd} |n\rangle = -\hat{\vd}  \hat{P} |n\rangle
  $$
   we can write
 \beq \label{dipole_parity_formal}
 \langle 0| \hat{\vd}| n \rangle = \langle 0| \hat{P}^2 \hat{\vd} | n \rangle =\left(\langle 0|\hat{P}\right) \left(\hat{P} \hat{\vd}| n \rangle\right) =  - (-1)^{P_f} (-1)^{P_i}   \langle 0| \hat{\vd}| n \rangle
  \eeq
  This means that must have  
  \beq \label{eq:dipole_parity_sr}
  (-1)^{P_f} (-1)^{P_i} = -1
  \eeq
  in order to have non zero dipole matrix element $ \langle 0| \hat{\vd}| n \rangle$. 
  
  The above relation for the parities of the initial and final states of the transition is called \emph{dipole parity selection rule}. Together with the dipole angular momentum selection rule they impose fairly stringent limitations  on the allowed pairs of matter states which can be "connected" by non zero radiative dipole transitions.
  
To conclude this section we note that formal manipulations  in (\ref{dipole_parity_formal}) actually take a very simple form if we write  explicitly the dipole matrix elements as integrals
$$ 
\langle 0| \hat{\vd}| n \rangle = \sum_{\sigma_1,...,\sigma_N}\int \psi_0^*(\vecr_1\sigma_1,...,\vecr_n \sigma_n) \left[\sum_{a=1}^N q_a \vecr_a \right]  \psi_n(\vecr_1\sigma_1,...,\vecr_n\sigma_n) \;  d^3 r_1.... d^3 r_n
$$
where $\sigma$'s denote the spin variables. Changing the integration variables $\vecr_a=-\vecr_a'$ and using 
$$
 \psi_n(-\vecr'_1\sigma_1,...,-\vecr'_n\sigma_n) = (-1)^{P_n}  \psi_n(\vecr'_1\sigma_1,...,\vecr'_n\sigma_n)  
 $$
reproduces the formal arguments of Eq.\,(\ref{dipole_parity_formal}).

In the following sections the above discussion will help to derive the parity selection rules for higher terms in the long wavelength expansion Eq.\,(\ref{LWA_expan}).
 
 \subsection{"Forbidden" (higher multipole)  transitions} 
 When the dipole matrix element between a pair of states $| n_i \rangle$ and $| n_f \rangle$ vanishes because of the selection rules  the radiative transitions between such states are traditionally called \emph{forbidden}.  But of course there is a possibility that the transitions  still occur via higher order terms in the long-wavelength expansion (\ref{LWA_expan}) of $ \langle 0 | \hat{\vj}_{\vk} \cdot \vlambda_{\vk\alpha}  |n\rangle$. 
 
 In this section we examine the next order term after the dipole in this expansion. This term is 
 \beq \label{2nd_order_LWA}
 -i \langle 0 |\int d^3r\, (\vk\cdot\vecr)  (\hat{\vj}(\vecr) \cdot \vlambda_{\vk\alpha}) |n\rangle
  \eeq
  It is useful to  transform the integrand (omitting the subscript of $\vlambda$ and using the summation convention)
  \beq \label{sym_and_antisym}
  (\vk\cdot\vecr)  (\hat{\vj} \cdot \vlambda) = k_l r_l \; \hat{j}_s  \lambda_s =  \frac{1}{2}k_l\lambda_s[(r_l \hat{j}_s+r_s \hat{j}_l) + (r_l \hat{j}_s - r_s \hat{j}_l)]
  \eeq
  We shall consider the two parts of this expression separately.
  
   \subsubsection{Electric quadrupole transitions}
 We start by considering the symmetric term in (\ref{sym_and_antisym}).  This term contributes 
 \beq \label{antisym_trans}
 -\frac{i}{2} k_l \lambda_s \int d^3r \langle 0 | r_l \hat{j}_s(\vecr)+ r_s \hat{j}_l(\vecr) |n\rangle
 \eeq
 in the transition matrix element  (\ref{2nd_order_LWA}). We will transform this expression using the continuity equation for the operators $\hat{\rho}$ and $\hat{\vj}$
 $$
 \frac{\partial \hat{\rho}(\vecr,t) }{\partial t} = -\nabla \cdot \hat{\vj}(\vecr,t) \equiv-\frac{\partial\hat{ j}_m}{\partial r_m}  \;\;\; ({\rm with\;summation\; over \; repeated \; indices})
 $$
  Let us consider
 $$
 \int d^3 r \, r_s r_l \frac{\partial \hat{\rho} }{\partial t} = -\int d^3 r \, r_s r_l \frac{\partial\hat{ j}_m}{\partial r_m} =
- \int d^3r [\delta_{ms}r_l + r_s \delta_{ml}] j_m =  \int d^3r [r_l j_s + r_s  j_l ]
$$
where we used the continuity equation followed by integration by parts. This resulting relation allows to express the matrix element of the symmetric term (\ref{antisym_trans}) as
$$
 -\frac{i}{2} k_l \lambda_s \int d^3r  \, r_s r_l  \langle 0 | \frac{\partial \hat{\rho} }{\partial t} |n\rangle
$$
Using the Heisenberg equation for $ \hat{\rho}$ we can write
$$
  \langle 0 | \frac{\partial \hat{\rho} }{\partial t} |n\rangle = \frac{1}{i\hbar}   \langle 0 | [\rho, H_{matter} ]  |n\rangle =\frac{E_n-E_0}{i\hbar}    \langle 0 | \rho |n\rangle = -i\omega \langle 0 | \rho |n\rangle
  $$
With this the symmetric term becomes 
\beq \label{eq:exp_sym_trm}
-\frac{\omega}{2}  k_l \lambda_s \int d^3r  \, r_l r_s  \langle 0 | \hat{\rho}|n\rangle
\eeq
Using  $\vk\cdot\vlambda \equiv k_l\lambda_l =0$ this can be written
\beq \label{eq:el_quad_init}
-\frac{\omega}{2}  k_l \lambda_s \int d^3r  \,( r_l r_s  - \frac{1}{3}\delta_{ls}r^2)  \langle 0 | \hat{\rho}|n\rangle = -\frac{\omega}{6}  k_l \lambda_s  \langle 0 | \hat{Q}_{ls}|n\rangle 
\eeq
where $ \hat{Q}_{ls}$ are components of the operator of the electric quadrupole tensor of the radiation emitting matter
\beq \label{quad_mom}
 \hat{Q}_{ls}=  \int d^3r  \,(3 r_l r_s  - \delta_{ls}r^2) \hat{\rho}(\vecr) = \sum_{a=1}^N q_a (3 r_{a,l} r_{a,s} - \delta_{ls}r^2_a)
 \eeq
The emitted photon parameters enter via the  factor $(\omega/6)  k_l \lambda_s $ in Eq.(\ref{eq:el_quad_init}) while the matter is represented by the electric quadrupole  moment operator.  Radiative transitions arising through this term are called \emph{electric quadrupole transitions}. 

 \subsubsection{Electric quadrupole moment is an $\ell=2$ object. Selection rules}
 
 Electric quadrupole moment is a symmetric traceless tensor. This means that it has 5 independent components which transform between themselves under rotations. This is similar to the 5 components of the second order spherical harmonic $Y_{2\mu}$ or in a more formal language to the 5 components of the $\ell=2$ representation (multiplet) of the group of rotations.  
 
 To see the relation it is useful to step back to Eq.\,(\ref{eq:exp_sym_trm}), write the product  $r_l r_s$  in  spherical components 
 $$
 r_m r_{m'} = \frac{4\pi}{3} r^2 Y_{1, m}(\Omega)  Y_{1, m'}(\Omega) \; \; {\rm with}  \;\; m,m'=-1, 0,1
 $$
 and use the relations Eqs.\,(\ref{eq:exp_ofY},\ref{eq:Gnt_via_CG}) for $l_1=l_2=1$ 
 $$
Y_{1, m}(\Omega)  Y_{1, m'}(\Omega)  = 
 \sum_{l=0,1,2}  a(1, 1, l)  \langle l\mu | 1, m; 1,m' \rangle Y_{l \mu} (\Omega) \;\;, \;\; \mu=m+m'
$$
where in this case
$$
 a(1, 1, l)  =\frac{3}{\sqrt{4\pi(2l+1)}}  \langle l0 | 1, 0; 1,0 \rangle 
$$
Here the allowed values of $l=0,1,2$ in the sum   correspond to adding  two units of angular momenta and are formally dictated by the CG coefficients. The  $l=1$ term in the sum vanishes since $\langle 10 | 1, 0; 1,0 \rangle =0$ reflecting the vanishing of the antisymmetric (vector) product of two identical vectors $Y_{1m}$, cf, Ref.\,\cite{sph_h}.  In the $l=0$ term (the scalar product) the corresponding CG coefficient $\langle 00 | 1, 0; 1,0 \rangle \sim \delta_{m,-m'}$ and since $r_m r_{m'}$ enter Eq.\,(\ref{eq:exp_sym_trm}) when written in spherical components as  
$$
\sum_{m, m'=-1,0,1} k_m^* \lambda_{m'}^*    \, r_m r_{m'}
$$
 it vanishes (as with such terms earlier) due to orthogonality $\vk\cdot\vlambda=0$.
  
We are thus left with only $l=2$ term which shows that in spherical components the cartesian tensor of the quadrupole moment  becomes (a linear combination of)  the five components of the  spherical representation of this tensor\footnote{Explicit expressions relating the spherical components $Q_{2\mu}$  to the Cartesian $Q_{ls}$, Eq.\,(\ref{quad_mom}) are 
$
Q_{20} = -(1/2)Q_{zz} \;, \; Q_{2,\pm1} = \pm (1/\sqrt{6}) (Q_{xz} \pm i Q_{yz} ) \;, \; Q_{2,\pm2} = -(1/2\sqrt{6}) ( Q_{xx} - Q_{yy} + 2i Q_{xy})
$ } 
 \beq
\hat{Q}_{2\mu} = \sqrt{\frac{4\pi}{5}}\int r^2 Y_{2\mu}(\theta, \phi) \hat{\rho}(\vecr)\, d^3 r = \sqrt{\frac{4\pi}{5}}\sum_{a=1}^N q_a r_a^2  Y_{2\mu}(\theta_a, \phi_a) 
\eeq
Summarizing the above and using $\langle 20 | 1, 0; 1,0 \rangle =\sqrt{2/3}$ we have for Eq.\,(\ref{eq:exp_sym_trm}) 
 \beq \label{eq:quad_trans_Q}
 \bes
  -\frac{\omega}{2} & k_l \lambda_s \int d^3r  \, r_l r_s  \langle 0 | \hat{\rho}|n\rangle =   \\
&=  -\frac{\omega}{\sqrt{6}}\sum_\mu \Phi_\mu(\Omega_k, \vlambda)  \langle 0 | \hat{Q}_{2\mu} | n \rangle  
 \end{split}
   \eeq
  where we denoted
 \beq 
 \bes
\Phi_\mu(\Omega_k, \vlambda) &= \sum_{m, m'=-1,0,1} \langle 2\mu | 1, m; 1,m' \rangle k_m^* \lambda_{m'}^* = \\ 
&= \sqrt{\frac{4\pi}{3}} \sum_{m, m'=-1,0,1} \langle 2\mu | 1, m; 1,m' \rangle\, k \,
Y^*_{1\mu}(\Omega_k) \lambda_{m'}^* 
\end{split}
\eeq 
The above expression is useful for finding the  selection rules for electric quadrupole transitions in physical  systems with eigenstates having definite angular momentum values
\beq \label{eq:Q_for_JM}
 \langle 0 | \hat{Q}_{2\mu} | n \rangle  \to  \langle \nu' J' M' |\hat{Q}_{2\mu}| \nu J M \rangle 
 \eeq
Using the Wigner-Eckart theorem, Eq.\,(\ref{WE_theorem}), for the operator of the electric quadrupole moment one has 
$$
 \langle \nu' J' M' | \hat{Q}_{2\mu} | \nu J M \rangle = \langle J' M' | 2 \mu, J M \rangle  \langle \nu' J' ||  \hat{Q}_2 ||\nu J \rangle
$$
which shows that the angular momentum selection rules for transitions with this operator, i.e. for electric quadrupole transitions  are
\beq  \label{eq:el_quad_sr_am}
J'=|J-2|, .... , J+2 \;\;\;, \;\;\; M'=M + \mu 
\eeq
Applying the parity transformation $\vecr_a \to -\vecr_a \; , \; a=1, ..., N$ to $\hat{Q}_{ml}$ , Eq. (\ref{quad_mom}) we see that it  does not change. Thus the parity selection rule for electric quadrupole transitions is
\beq
P_f=P_i
\eeq
This rule which is "opposite" to the dipole selection rule, Eq.(\ref{eq:dipole_parity_sr}), is the main reason why the electric quadrupole and magnetic dipole transitions explained below are the leading mechanisms for the transition, which are forbidden by the dipole selection rules. Since related to  higher order terms in the long wavelength expansion, Eq.\,(\ref{LWA_expan}) such transitions have order of magnitude smaller transition rates in the small $ka$ parameter than the allowed dipole transitions. 

This is of course for the levels which satisfy the angular momentum selection rules, Eq.\,(\ref{eq:el_quad_sr_am}). Transitions between levels with larger angular momentum differences  are controlled by higher terms in the LWA expansion, which are correspondingly weaker, cf., our discussion below and Ref.\,\cite{Baym}. In this respect an interesting situation arises when a matter system has a low lying high angular momentum state. If all the levels below such state  have  low angular momenta this state will have a long radiative lifetime. Such states are called isomeric and are metastable if probabilities of non radiative transitions (e.g. via collisions in gases or  phonon emission in solids) are small too.    

\subsubsection{Angular distribution of electric quadrupole radiation} 

\begin{figure}[H]
\centering \includegraphics[width=0.4 \textwidth]{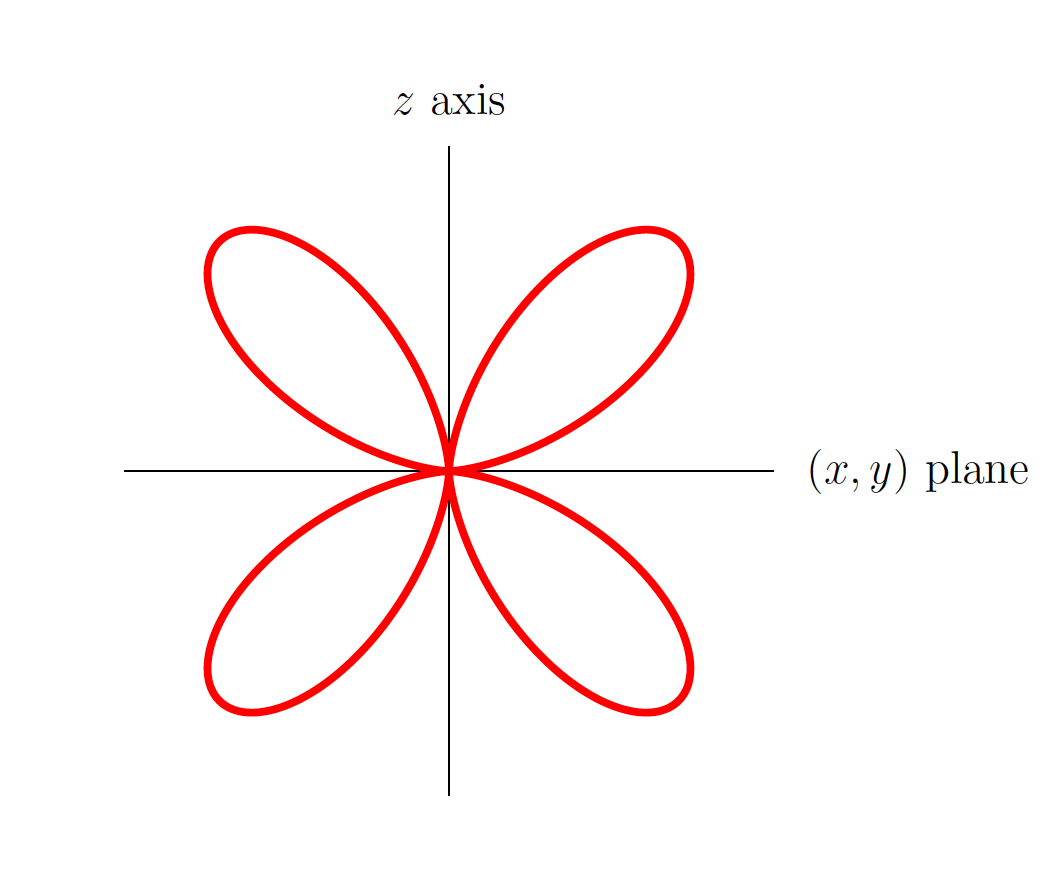}
\caption{Angular distribution of electric quadrupole radiation with $\vlambda_1$ polarization, cf., Eq.(\ref{eq:el_quad_dist}) plotted in a similar way as in Fig.\,\ref{fig:dipole_emit}.  Note that the independence of the azimuthal angle $\phi$ means that the 3D version of this figure is obtained by rotating it around the z-axis. }
 \label{fig:el_quad}
\end{figure}

Let us now briefly discuss the angular distribution of photons emitted in electric quadrupole  transitions in the very common case when the relation (\ref{eq:Q_for_JM}) is valid.  Since in this case $\mu$ is fixed, $\mu=M'-M$ only one term will remain in the sum in  Eq.\,(\ref{eq:quad_trans_Q}).  The angular distribution is obviously given by the corresponding function  $\Phi_\mu(\Omega_k, \vlambda)$ with the dependence on the angles of the polarization vectors
as given e.g. by the relations Eq.\,(\ref{eq:pol_vec_conv}).   The spherical components of the latter are
 \beq
 \bes
(\vlambda_1)_{\mu=\pm1} &= \mp \frac{\cos\theta e^{\pm i\phi}}{\sqrt{2} } \;\; \; , \;\; \;(\vlambda_1)_{\mu=0} = - \sin\theta  \nonumber \\
(\vlambda_2)_{\mu=\pm1} &= -\frac{i e^{\pm i\phi} }{\sqrt{2} } \;\;\;\;\; \; \;\;\; , \; \;\;  (\vlambda_2)_{\mu=0} = 0
\end{split}
\eeq
As an example let us consider transitions with $M=M'$,  $\mu=0$. Using 
$$
\langle 2\mu | 1, m; 1,-m\rangle = (-1)^{1-m}\frac{3m^2-1}{\sqrt{6}}
$$
one finds that the sum in (\ref{eq:sum_over_k_lam}) is $-3\sin\theta \cos\theta$ for the $\vlambda_1$ photon polarization while it vanishes for $\vlambda_2$.  Accordingly the corresponding angular distributions in electric quadrupole transitions are
\beq  \label{eq:el_quad_dist}
\frac{dN_{\vk1}}{d\gamma} \sim \sin^2\theta \cos^2\theta \;\;\;, \;\;\;   \frac{dN_{\vk2}}{d\gamma} =0
\eeq
cf., Fig.\,\ref{fig:el_quad}.

\subsubsection{Magnetic dipole transitions \label{sec:mag_dipole}} 
  The antisymmetric part  of (\ref{sym_and_antisym}) is conveniently expressed via vector products
  \beq
  \frac{1}{2}k_l\lambda_m(r_l \hat{j}_m- r_m \hat{j}_l) =\frac{1}{2}k_l\lambda_m \epsilon_{lmn}(\vecr\times\hat{\vj} )_n = 
  \frac{1}{2}(\vk\times\vlambda)\cdot(\vecr\times\hat{\vj} )
  \eeq
  which will contribute in Eq.\,(\ref{2nd_order_LWA}) as
 \beq \label{eq:orb_mag_mom}
 -i (\vk\times\vlambda_{\vk\alpha})\cdot \langle 0 | \frac{1}{2}\int d^3 r (\vecr\times\hat{\vj} (\vecr))| n\rangle
 \eeq
 The emitted photon parameters enter via the  factor $(\vk\times\vlambda_{\vk\alpha})$ while the matter is represented by the magnetic dipole moment  of the current generated by the orbital motion of the matter constituents, 
 $$
   \frac{1}{2}\int d^3 r (\vecr\times\hat{\vj}(\vecr) )= \frac{1}{2} \sum_{a=1}^N \frac{q_a}{2m_a}[-\hat{\vp}_a \times \vecr_a + \vecr_a\times \hat{\vp}_a ] = \sum_{a=1}^N \frac{q_a}{2m_a}(\vecr_a\times\hat{\vp}_a) .
 $$
 At this stage it is important to recall the interaction term $H_{I3}$, Eq.\,(\ref{eq:H_I3_expanded}) which we have not treated so far.  This term  also depends on the magnetic moments of the matter constituents.  However not  the ones generated by the orbital motion but rather by their intrinsic motion, i.e. their spins.   
 
 The contribution of $ \hat{H}_{I3}$ to the transition matrix element is straightforward to derive just following the same steps which led us to the expression (\ref{eq:mat_el_ph}) for $\langle f|\hat{H}_{I1}|i\rangle$ with the result
\beq \label{eq:mat_mag_ph}
\langle f|\hat{H}_{I3}|i\rangle = -i \left(\frac{\hbar }{2\epsilon_0\omega_k\Omega}\right)^{1/2}
   (\vk\times\vlambda_{\vk\alpha}) \cdot \langle 0 | \hat{\vm}_{-\vk} |n\rangle 
\eeq
where
$$
\hat{\vm}_{\vk} = \int  \hat{\vm} (\vecr) e^{-i\vk\cdot\vecr} \, d^3 r  = \sum_{a=1}^N \hat{\vmu}_a e^{-i\vk\cdot\vecr_a}
$$
In the long wavelength limit 
$$ 
\langle 0 | \hat{\vm}_{-\vk} |n\rangle \approx \langle 0 | \hat{\vm}_0|n\rangle =  \langle 0 | \int d^3\vecr \, \hat{\vm}(\vecr) \;  |n\rangle
=  \langle 0 |  \sum_{a=1}^N \hat{\vmu}_a | n \rangle
$$
Taking this term into account the expression (\ref{eq:orb_mag_mom}) becomes
\beq \label{eq:tot_mag_mom}
\bes
  -i (\vk\times\vlambda_{\vk\alpha})\cdot  \langle 0 | \hat{\vM}  | n\rangle \;\; , & \\
 {\rm with} \qquad \qquad \qquad \qquad \qquad \qquad  \qquad \qquad & \\
 \hat{\vM} = \int d^3 r  \left[ \frac{1}{2}\left( \vecr\times\hat{\vj} (\vecr) \right) +\hat{\vm}(\vecr) \right] &=  \sum_{a=1}^N \left[\frac{q_a}{2m_a}(\vecr_a\times\hat{\vp}_a) + \hat{\vmu}_a \right]
 \end{split}
 \eeq
 Radiative transitions arising through this term are called \emph{magnetic dipole transitions}. 

Electrons in atoms have equal charge to mass ratio $e/m$ so the orbital part of this expression reduces to 
 \beq
    \frac{e}{2m}\sum_{a=1}^N (\vecr_a\times\hat{\vp}_a) = \frac{e}{2m} \hat{\vL}
  \eeq
 while the spin part 
 \beq \label{eq:mag_m_spin}
  \sum_{a=1}^N \hat{\vmu}_a = g  \frac{e}{2m}  \sum_{a=1}^N \hat{\vs}_a = g  \frac{e}{2m} \hat{\vS}
  \eeq
  where $\vL$ and $\vS$ are respectively the total orbital angular momentum and the spin of the emitting matter system. In the spin part expression, Eq.\,(\ref{eq:mag_m_spin}) we used the
gyromagnetic ratio $g (e/2m)$ to relate intrinsic magnetic moments to the spins. The part $e/2m$ denotes the classical value while $g$ - known as the g-factor - is the dimensionless quantity to account for deviations from the classical $g=1$ value. Dirac relativistic equation  for spin 1/2 particles  predicts the value $g=2$ while field theoretical corrections change it slightly to $g=2(1+\alpha/2\pi + \cdots) \simeq 2.002 319...$ where $\alpha = 1/137$.
 
The expression (\ref{eq:tot_mag_mom})  for electrons is therefore  
 \beq \label{eq:mag_dp_atoms}
 -\frac{ie}{2m} (\vk\times\vlambda_{\vk\alpha})\cdot \langle 0 | \hat{\vL} + g \hat{\vS} | n \rangle
 \eeq
 i.e. it is proportional to the matrix element of the combination $ \hat{\vL} + g\hat{\vS} \approx  \hat{\vL} + 2 \hat{\vS}$ of the electronic angular momentum and spin.
 
\subsubsection{Selection rules and angular distribution of magnetic dipole transitions}
 
Let us now focus as in the electric dipole and quadrupole transitions on emitting systems with eigenstates having defined angular momentum values
\beq \label{eq:mag_dp_ang_mom}
 \langle 0 | \hat{\vM}  | n\rangle \to  \langle \nu', J', M' | \hat{M}_\mu  |\nu, J, M\rangle \;\;, \;\; \mu=-1, 0, 1
 \eeq
where we also introduced the spherical components of the vector $\hat{\vM}$. The angular momentum selection rules are obviously
 \beq
 J'=|J -1|, ..., J + 1 \;\;\;, \;\;\; M'=M + \mu 
 \eeq
 as in the electric dipole case. 
 
 However the parity selection rule is different.  Indeed the magnetic dipole is a pseudo vector as it does not change under the parity transformation $\vecr \to -\vecr \; , \vp \to -\vp$. Therefore  the parity will not change in the transitions i.e. the \emph{magnetic dipole} parity selection rule is  
 $$
 P_f=P_i
 $$
 
 Let us also note the following. One can rewrite the operator  $\hat{\vL} + g \hat{\vS}$ in Eq.\,(\ref{eq:mag_dp_atoms}) as $\hat{\vL} + \hat{\vS} +
 (g-1)\hat{\vS} = \hat{\vJ} + (g-1) \hat{\vS}$ where $\hat{\vJ}$ is the total angular momentum. In emitting systems with eigenstates as in Eq.\,(\ref{eq:mag_dp_ang_mom}) the operator $\hat{\vJ}$ can not cause transitions so that the magnetic dipole emission must go via "spin-flips", i.e. (in conventional language) via the change of the spin projection $S_z$. That in turn means that spin must not be a conserved quantity in the eigenstates of Eq.\,(\ref{eq:mag_dp_ang_mom}). Which implies that there must be a spin-orbit interaction in the matter Hamiltonian of the emitting system.  Thus magnetic dipole emission is the measure of such an interaction. 
 
 Finally let us address the angular distribution of photons emitted in magnetic dipole transitions. This is determined by the angular dependence of the components of  the vector $(\vk\times\vlambda_{\vk\alpha})$ in Eq.\,(\ref{eq:tot_mag_mom}) which are weighted by the vector of the matrix elements $\langle 0 | \hat{\vM}  | n\rangle$. Noting that our favorite choice Eq.\,(\ref{eq:pol_vec_conv}) of  polarizations form right handed system of unit vectors with the direction of $\vk$ we deduce that  the vectors $(\vk\times\vlambda_{\vk\alpha})$  with $\alpha=1$ and $\alpha=2$ are respectively proportional to $\vlambda_{\vk2}$ and $\vlambda_{\vk1}$. Therefore the angular distribution of the expression (\ref{eq:tot_mag_mom}) is identical to the electric dipole one with  $ \langle 0 | \hat{\vd}|n\rangle$ replaced by $\langle 0 | \hat{\vM}  | n\rangle$ and appropriate adjustment of the polarization vectors.

\subsubsection{General multipole expansion}
 
 What we have seen so far in our discussions of the electric dipole and quadrupole and magnetic dipole emissions is essentially a transformation of the terms in the Taylor expansion (\ref{LWA_expan}) to the expansion in terms of "angular" multipoles. The reason the latter is more appropriate is that the small parameter of the long wavelength expansion $ka\ll1$ concerns the "radial size" $a$ of the system, $|\vecr| \le a$ with obviously no  limitation on the angles. Perhaps the simplest familiar example of this is a "move" from the Taylor expansion of  the Coulomb potential 
 $$
  \phi(\vecr) = \frac{1}{4\pi\epsilon_0} \int _{r'\le a} \frac{\rho(\vecr')}{|\vecr - \vecr'|} d^3r'
  $$
  '"outside" of a charge distribution, $r > a$ to the multiple expansion  
 \beq 
  \bes \label{eq:Coul_pot_mult}
  {\rm  Taylor \; expansion \; - \;\; } \phi(\vecr) &=  
 \frac{1}{4\pi\epsilon_0 r}  \int \rho(\vecr')(1 + \vecr\cdot \vecr'/r^2 +. . .) d^3r' \; .  \\
  {\rm Multipole \; expansion \; - \;\; }  
\phi(\vecr) &= \frac{1}{4\pi\epsilon_0} \sum_{l=0}^\infty \sum_{m=-l}^{l} \frac{M^E_{lm}}{R^{l+1}} \; Y_{lm}(\Omega_r)
\end{split}
 \eeq
 with
 $$
 M^E_{lm} = \frac{4\pi}{2l+1} \int r^l \rho(\vecr)Y_{lm}^*(\Omega_{r}) d^3 r \;\;  - \;\;{\rm multipole\; electric \; moments}
 $$
 In deriving the latter expression one uses the known expansion of $1/|\vecr - \vecr'|$ into a sum of products 
  $Y_{lm}(\Omega_r) Y_{lm}^*(\Omega_{r'})$ which allows to factorize  the outside $r>a$ and the inside $r'\le a$ regions. The result is the Coulomb potential represented as a sum of multipole potentials which the electric multipole moments $M^E_{lm}$ generate. 
  
  Returning to our problem we want to find a similar multipole expansion for the expressions Eqs.\,(\ref{eq:mat_el_ph}) and 
  (\ref{eq:mat_mag_ph}).  Let's concentrate on the former and consider 
  \beq
    \hat{\vj}_{\vk}\cdot \vlambda = \int d^3 r \; \hat{\vj}(\vecr)\cdot \vlambda  \; e^{-i\vk\cdot\vecr} 
 \eeq
 Rather than expanding $\exp(-i \vk\cdot\vecr)$ in Taylor series as we have done in Eq.\,(\ref{LWA_expan}) we shall use 
 $$
 \exp(-i \vk\cdot\vecr) =4\pi \sum_{l=0}^\infty \sum_{m=-l}^{l} (-i)^l g_l(kr)  Y_{lm}(\Omega_k)  Y^*_{lm}(\Omega_r)
 $$
 where the spherical Bessel functions are $g_l(kr) = \sqrt{\pi/2kr} J_{l+1/2}(kr)$.  With this expansion of the exponent 
 \beq \label{eq:phot_mult_exp_start}
  \hat{\vj}_{\vk}\cdot \vlambda =4\pi\sum_{l=0}^\infty \sum_{m=-l}^{l}  \sum_{q=-1}^{1} (-i)^l   Y_{lm}(\Omega_k) \lambda_q \int d^3 r \;g_l(kr) \hat{j}^*_q(\vecr)Y^*_{lm}(\Omega_r)
 \eeq
 where we used the spherical components of the vectors  $\vlambda$ and $\hat{\vj}$.
 
 The resulting expression (\ref{eq:phot_mult_exp_start}) for  $\hat{\vj}_{\vk}\cdot \vlambda$  has similar features with 
 the multiple expansion Eq.\,(\ref{eq:Coul_pot_mult}) of the Coulomb potential.  It is a sum of terms each factorized in the product of components depending on the  photon variables $\vk$ and $\vlambda$ and the matter variables $\vj(\vecr)$. One still has the matter component depending on $k$ via $g_l(kr)$ but this will decouple in the long wavelength approximation (LWA)  $kr\ll1$ for which  $g_l(kr) \sim k^lr^l$ when the matrix element  $\langle 0| \hat{\vj}_{\vk}|n\rangle\cdot \vlambda$ is considered. 
 
 The remaining problem in the expansion Eq.\,(\ref{eq:phot_mult_exp_start}) is  that  photon and matter components in each term do not have  definite multipolarities.  This is most obvious in the photon related parts which transform under rotations as a product of $Y_{lm}(\Omega_k) $  and  $\lambda_q$, i.e. as a sum of representations $l-1, l$ and $l+1$.  The technical reason for this is trivially obvious - the photon related $Y_{lm}(\Omega_k)$ is coupled to the matter related $Y^*_{lm}(\Omega_r)$ and $\lambda_q$  to $\hat{j}^*_q(\vecr)$.  What one needs is to "recouple" the products  into the photon and the matter groups. This can be done using the Clebsch-Gordan completeness relation, cf., Ref.\,\cite{Baym}, p.\,338,
 $$
 \sum_{LM} \langle lm, 1q | l1 LM \rangle \langle l1 LM | lm' 1 q' \rangle =\delta_{mm'} \delta_{qq'} 
 $$
 Inserting it into Eq.\,(\ref{eq:phot_mult_exp_start})   one obtains
 \beq \label{eq:mult_exp_final}
  \hat{\vj}_{\vk}\cdot \vlambda =4\pi\sum_{L,l,M} (-i)^l    \Phi_{LM,l}(\Omega_k, \vlambda)  \hat{\mathcal{M}}_{LM,l}
  \eeq
  where with $g_l(kr) \approx (kr)^l/(2l+1)!! $  in the long wavelength limit have
  \beq \label{eq:photon_LM_ang_dist}
   \Phi_{LM,l}(\Omega_k,\vlambda) = \sum_{m,q} \langle lm, 1q | l1 LM \rangle k^lY_{lm}(\Omega_k) \lambda_q \;\; \; \; ,\; \; \; \;  l=L, L\pm 1\; ,
   \eeq
   and
   \beq
   \hat{\mathcal{M}}_{LM,l} = \frac{1}{(2l+1)!! } \sum_{m',q'} \int d^3 r \; r^l \langle l1 LM | lm' 1 q' \rangle
  \hat{j}^*_{q'}(\vecr)Y^*_{lm'}(\Omega_r)  \;\; \; , \; \;  l=L, L\pm 1\; ,
  \eeq
  The values of $l=L, L\pm 1$ in both expression correspond of course to the vector addition of a unit angular momentum
   of the vectors $\vlambda$ and $\vj$ to the $l$ of $Y_{lm}(\Omega_k)$  and $Y_{lm}(\Omega_r)$ respectively. 
  
  The general expansion (\ref{eq:mult_exp_final}) has the structure we were looking for. Both   $\Phi_{LM,l}(\Omega_k,\vlambda)$ and $ \hat{\mathcal{M}}_{LM,l}$ transform as the $M$ components of the $L$-th representation of rotations. While it is obvious for  the photon components  a bit more work is needed to show this for the integral representing  $\hat{\mathcal{M}}_{LM,l}$.  This is left as an exercise.  
  
 It is useful to consider a few simple cases.
 \newpage
{\bf Monopole  emission}
\\
\\
Starting with the  $L=0$ term it is easy to show that it vanishes.  Indeed for $L=0$ have  $M=0 \to m=-q$ and only $l=1$ as a possible value. So $k^lY_{lm}(\Omega_k) \to kY_{1m}(\Omega_k) \sim k_m$ and  
  $$
  \Phi_{00,1}\sim \sum_m (-1)^m k_m \lambda_{-m} =\vk\cdot \vlambda =0
  $$
  So - no monopole photon emission.  One can intuitively relate this to the fact that photons have spin 1 - one can't emit a photon without changing the emission system angular momentum by at least one unit.\\
  \\
  {\bf Dipole emission}

  The terms with $L=1$ have $l= 0$ and $l=1$. For $l=0$ have 
  $$
  \Phi_{1M,l=0} \sim \lambda_M \;\;, \;\; \hat{\mathcal{M}}_{1M,l=0} \sim \int d^3 r \hat{j}^*_{M}(\vecr)
  $$
  recovering the electric dipole case, cf. Eq.(\ref{eq:j_to_d}).
  
  For  the $l=1$ value   one has 
  \beq
  \bes
  \Phi_{1M,l=1} &= k \sum_q  \langle 1\; M-q, 1q | 11 1M \rangle Y_{1m}(\Omega_k) \lambda_{M-q} \sim (\vk\times\vlambda)_M \nonumber \\
 \hat{\mathcal{M}}_{1M,l=1} &= \frac{1}{3} \sum_{q} \int d^3 r \; r \langle 11 1M | 1 \; M-q 1 q \rangle \hat{j}^*_{q}(\vecr)Y^*_{1M-q}(\Omega_r) \sim
   \int d^3 (\vecr \times \hat{\vj}(\vecr))_M
   \end{split}
   \eeq
   i.e. the magnetic dipole emission. \\
\\   
{\bf Higher multipoles}   

Discussions of the higher values of $L$ cf., Ref.\cite{Baym}, p.376, confirms this pattern - the $l=L$ terms correspond 
   to magnetic multipoles while the $l=L \pm 1$ terms are electric radiation terms. So the sum over $L$ in Eq.(\ref{eq:mult_exp_final}) is the sum over different multipoles of the matter "vibrations" (quantum mechanical transition matrix elements) causing the photon  emission. 
  
  \subsubsection{Angular distribution, selection rules of the general multipole terms}
The expression (\ref{eq:photon_LM_ang_dist}) for $\Phi_{LM,l}(\Omega_k,\vlambda)$ can be interpreted as the probability
  amplitude of the photon emitted by the $l,L$ matter multipole into the solid angle $\Omega_k$ with polarization $\vlambda$. It reflects the expectation that the total angular momentum of a photon is a sum of its orbital angular momentum (encoded in  $ Y_{1m}(\Omega_k)$) and
  its unit spin (associated with the polarization vector). 
 
Angular momentum selection rules for the terms in the expansion Eq.\,(\ref{eq:mult_exp_final}) follow by applying the Wigner-Eckart theorem to matrix elements of the multipole moments operators between matter eigenstates  with defined angular momentum values
\beq  \label{eq:mat_el_mult}
\langle \nu', j', m' |  \hat{\mathcal{M}}_{LM,l} |\nu, j, m\rangle = \langle j'm' | L M, j m \rangle \langle \nu', j' ||\hat{\mathcal{M}}_{L,l} || \nu, j \rangle
 \eeq
 From this we have the angular momentum selection rules
 \beq
 |L-j| \le j' \le L+j\;\; , \;\; m'= m+M
 \eeq
 Perhaps not surprisingly they do not involve the $l$ index which distinguishes between electric and magnetic multipoles.  This index is important however in the parity selection rules.  Perhaps the fastest way to see this is to observe that $ \Phi_{LM,l}(\Omega_k,\vlambda)$ changes under the parity $\vecr \to -\vecr$ transformation as $(-1)^{l+1}$ where $l$ comes from the orbital $Y_{lm}(\Omega_k)$ while the extra minus from the polar vector of the polarization. So
 $$
 (-1)^{P_f}=(-1)^{l+1}(-1)^{P_i}
 $$ 
 As a final remark we note that the above arguments  based on the parity properties of the photons amplitudes could be made more formal and rigorous by examining how the matrix elements Eq.\,(\ref{eq:mat_el_mult}) behave under the parity transformation, cf., Ref. \cite{Baym}, p. 379. 
 
 \subsection{Induced photon emission}
 
 In our discussion above of the photon emission by an excited state of quantum matter (atom, solid, nucleus, molecule) we have assumed that prior to the emission (i.e. in the initial state) there were no photons present in the radiation mode $\vk\alpha$ into which the matter system emits the photon, cf. Eq. (\ref{spont_emis_states}). Such an emission is called {\em{spontaneous}}. 
 
  Let us now consider what happens if the initial state already contained N photons before the emission, i.e. have 
$$
  |i\rangle = |n\rangle | N_{\vk\alpha},  \{0_{\vk'\alpha'}\} \rangle \;\;\;\; |f\rangle = |0\rangle |(N_{\vk\alpha}+1, \{0_{\vk'\alpha'}\}\rangle
$$
With this change the calculation of the field matrix element in Eq. (\ref{mat_el_of_radiation}) becomes
$$
  \langle N_{\vk\alpha} +1, \{0_{\vk'\alpha'}\}| \hat{a}_{\vk''\alpha''}+\hat{a}^{\dagger}_{-\vk''\alpha''} |N_{\vk\alpha}, \{0_{\vk\alpha}\}\rangle = \delta_{-\vk'' , \,\vk}\delta_{\alpha'' \alpha} \sqrt{N_{\vk \alpha} +1} 
$$
because of the basic matrix element of the harmonic oscillator creation operator
$$
\langle N +1 |\hat{a}^\dagger |N\rangle = \sqrt{N+1}
$$
This produces the following result in the absolute values square of the interaction  
\beq
|\langle f|\hat{H}_{I1}|i\rangle|^2 =  \left(\frac{\hbar }{2\epsilon_0\omega_k\Omega}\right)
  | \langle 0 | \hat{\vj}_{\vk} \cdot \vlambda_{\vk\alpha}  |n\rangle|^2 \,  (N_{\vk\alpha} +1)
\eeq
which means that the emission rate $\Gamma_{n\to 0,\vk\alpha}$ is  $N_{\vk\alpha} +1$ times larger than in the spontaneous emission case.   
So just the initial presence of $N_{\vk\alpha}$ photons in the radiation modes into which the emission occurs leads to this increase of the emission rate. This effect is called {\em{induced or stimulated emission}}.  It is often interpreted as a quantum mechanical effect of "bosons like to stick together", i.e. to be in the same state  and is the key to the idea of lasers.  

Very schematically this idea can be outlined as following.  Assume a large number of identical  "emitters" (e.g. atoms, molecules, etc) which can be "continuously" excited to a certain energy level and then de-excite to low lying levels via photon emission. As we learned earlier the angular distribution and  polarization of the  emitted photons will depend on the angular momentum projections $M$ and $M'$ of the initial and final states but if only the initial energy is specified the $M$ values will be random and so will be the emitted photons directions and polarizations.    This is as long as only the spontaneous emission is considered. 

If  some particular  photon modes $\vk\alpha$ contain a (large) number of (pre emitted) photons then  high probability ($\sim\,N_{\vk \alpha}$)  induced emission, i.e. "lasing"  will occur into these particular modes. Schematically the needed accumulation of  photons in controlled modes is achieved e.g. by placing the emitters in a resonator.  This selects resonating modes in which photons "bounce back and forth" before escaping.  

All this is very sketchy of course. More detailed explanations are found in appropriate quantum optics literature. 

Let us note that historically it is common to write the expression for the emission rate as a sum of the term containing the $N_{\vk\alpha}$ and the term containing $1$ from the sum $N_{\vk\alpha} +1$
$$
\Gamma = \Gamma^{\rm induced} + \Gamma^{\rm spontaneous}
$$
This  the expression for the spontaneous photon emission rate (\ref{rate_of_photon_emiss}) is changed to 
\beq
\frac{dN_{\vk\alpha}}{d\gamma} = \left(\frac{dN_{\vk\alpha}}{d\gamma} \right) ^{\rm induced}  +   \left(\frac{dN_{\vk\alpha}}{d\gamma}\right) ^{\rm sponteneous}
\eeq
with 
\beq
 \left(\frac{dN_{\vk\alpha}}{d\gamma} \right)^{\rm induced}   = N_{\vk\alpha} \left( \frac{dN_{\vk\alpha}}{d\gamma} \right) ^{\rm sponteneous}
 \eeq
 
\subsection{Photon absorption}
Consider now the process of the photon absorption. We have
\beq
|i\rangle = |0\rangle |N_{\vk\alpha}, . . . \rangle \;\;\;, \;\;\; | f \rangle = |n \rangle | N_{\vk\alpha} -1 , . . . \rangle
\eeq
The matrix element of $\hat{H}_{I1}$ between these states gives 
\beq
|\langle f|\hat{H}_{I1}|i\rangle|^2 =  \left(\frac{\hbar }{2\epsilon_0\omega_k\Omega}\right)
  | \langle n | \hat{\vj}_{-\vk} \cdot \vlambda_{\vk\alpha}  | 0 \rangle|^2 \,  N_{\vk\alpha}
\eeq
Since
$$
\langle n | \hat{\vj}_{-\vk} \cdot \vlambda_{\vk\alpha}  | 0 \rangle = \langle 0| \hat{\vj}_{\vk} \cdot \vlambda_{\vk\alpha}  | n \rangle^* 
$$
we find equality relation 
\beq
\Gamma_{0\to n} ^{\rm absorption} = \Gamma_{n\to 0} ^{\rm induced \;  emission}
\eeq
for absorption and induced emission rates of photons with the same $\vk$ and $\vlambda$.   This relation is crucial for  laser
 physics.  Indeed it shows that having $N_{\vk\alpha}$ incident photons (per unit time) of energy $\hbar\omega_k$  a photon  has an equal probability of being absorbed by a ground-state atom or being duplicated (amplified!) via an induced emission  by an excited-state atom. To favor  emission over absorption, there need to be more excited-state atoms than ground-state atoms. This of course doesn't happen in thermally equilibrated systems. A non equilibrium situation must be created by adding energy via a process known as ``pumping'' in order to raise enough atoms to the upper level.  The result called ``population inversion'' leads to light amplification. Pumping may be electrical, optical or chemical.

\section{Appendix}
\subsection{Discrete level coupled to continuum}
Here we present details of a simple non perturbative approach to deal with the Weisskopf-Wigner model as defined in Section (\ref{sec:WW_model}). A more general treatment of this problem is reviewed in e.g. Ref.\cite{Gae}.  

\subsubsection{Neglecting coupling between continuum levels}

As was described following Eq.\,(\ref{eq:eqs_for_c_0_mu}) the crucial step/approximation in the Weisskopf-Wigner approach is to neglect the coupling between the continuum levels, i.e. to set $V_{\mu\nu}=0$.  This means that the Hamiltonian matrix in the basis $\{\psi_0, \psi_\nu\}$  has the "bordered" form
$$
H= 
\left(
\begin{array}{ccccc}
 E_0 & V_{01}  & ... & V_{0\nu} & ...   \\
 V_{10} & E_1  & ... & 0  & ....  \\
  ...& ...  & ... & ... & ...   \\
  V_{\nu 0} & 0 & ... & E_\nu & ... \\
  ... & ... & ... & ... & ... 
\end{array}
\right)
$$
Here we tacitly assumed discrete values of the $\nu$ index. Such matrices are easy to diagonalize especially when simplifying assumptions about $E_\nu$'s and $V_{0\nu}$ are made. This is described in e.g. Ref.\,\cite{Cohen}.

\subsubsection{Markov approximation}
Examining the integral expression Eq.\,(\ref{eq:K_of_t}) for the kernel $K(t)$ which for convenience we rewrite here 
$$
K(t) = -\frac{1}{\hbar^2}\int d\cE  \overline{|V_{0\mu}|^2}\big|_{_{\cE_\mu=\cE}}  e^{i(\cE_0 - \cE)t/\hbar}
$$
we observe that the integrand  is a product of in general a smooth function of $\cE$ and an exponential which oscillates in $\cE$ with the period 
$\sim \hbar/t$.  Denoting by $\Delta$ the scale over which $ \overline{|V_{0\mu}|^2}\big|_{_{\cE_\mu=\cE}}$ changes it is clear that the integral will tend to zero for long times $t\gg \hbar/\Delta$.  Under this condition the kernel $K(t)$ has the "range" 
$$
T \sim \frac{\hbar}{\Delta}
$$
Let us change the variable $t'$  in the integral in (\ref{eq:WW_approx2})  to $\tau=t-t'$ 
\beq
\frac{d c_0}{dt}= \int_0^t K(t-t') c_0(t') dt' = \int_0^t K(\tau)c_0(t-\tau) d\tau 
\eeq
For a given $t$ only the values of $c_0(t-\tau)$ within "memory times"  $\tau\le T$ of $K(\tau)$ contribute in the integral.  To simplify further we next assume that $c_0(t)$ changes little over the time $T$. We will address below the meaning of this assumption. When it is valid we can approximate under the integral
$$
c_0(t-\tau) \approx c_0(t)
 $$
 and write 
 \beq \label{full_Markov_approx}
 \frac{d c_0}{dt}=  c_0(t) \int_0^t K(\tau)d\tau 
 \eeq
 This approximation is called the Markov approximation - dynamics of $c(t)$, i.e. how it changes at the time $t$  depends only on its value at the time t and not on earlier times  $t'< t$,  i.e., it  has no memory of the past. 
 
 The integral on the right hand side of the above equation is a known function of $t$ so the equation can be integrated but let us first make one more simplification. We will be interested in the long time behaviour of $c_0(t)$ for $t \gg T$. Since by assumption $K(\tau)$ is small for $\tau\gg T$ we can approximate 
  $$
  \int_0^{t} K(\tau) d\tau \approx  \int_0^{\infty} K(\tau) d\tau 
  $$
Let us introduce the following notation for the real and imaginary part of this integral
\beq
 {\rm Im} \int_0^\infty K(\tau) d\tau  = -\frac{\Delta \cE}{\hbar} \;\;\;, \;\;\; Re\int_0^\infty K(\tau) d\tau =  \frac{\Gamma}{ 2} 
 \eeq
With this we have for the time dependence of the "persistence  amplitude" of the 
 initial state $\psi_0$ 
 \beq
 \langle\psi_0|\Psi(t)\rangle =c_0(t) e^{-i\cE_0t/\hbar}|_{t\gg T} = c(0) e^{-i(\cE_0+\Delta \cE)t/\hbar}e^{-\Gamma t/2}
  \eeq
 We will see below that $\Gamma$ is positive so this amplitude decays exponentially with the decay rate $\Gamma$. Its phase acquires  energy shift $\Delta \cE$. 
 
 We will  discuss the explicit form of $\Delta \cE$  and $\Gamma$ in the next subsection. Here we note that the time scale over which $c_0(t)$ 
 changes is $\sim 1/\Gamma$ or $\sim \hbar/\Delta\cE$.  This our assumption of $c_0(t)$ changing slowly in the interval $T=\hbar/\Delta$ means that must have
 \beq \label{eq:cond_on_Gamma_etc}
\hbar \Gamma \ll \Delta \;\;\;, \;\;\; \Delta \cE \ll \Delta
\eeq

 \subsubsection{Decay rate (width)  and the energy shift of a decaying state}
 
\begin{figure}[H]
\centering \includegraphics[width=0.7\textwidth]{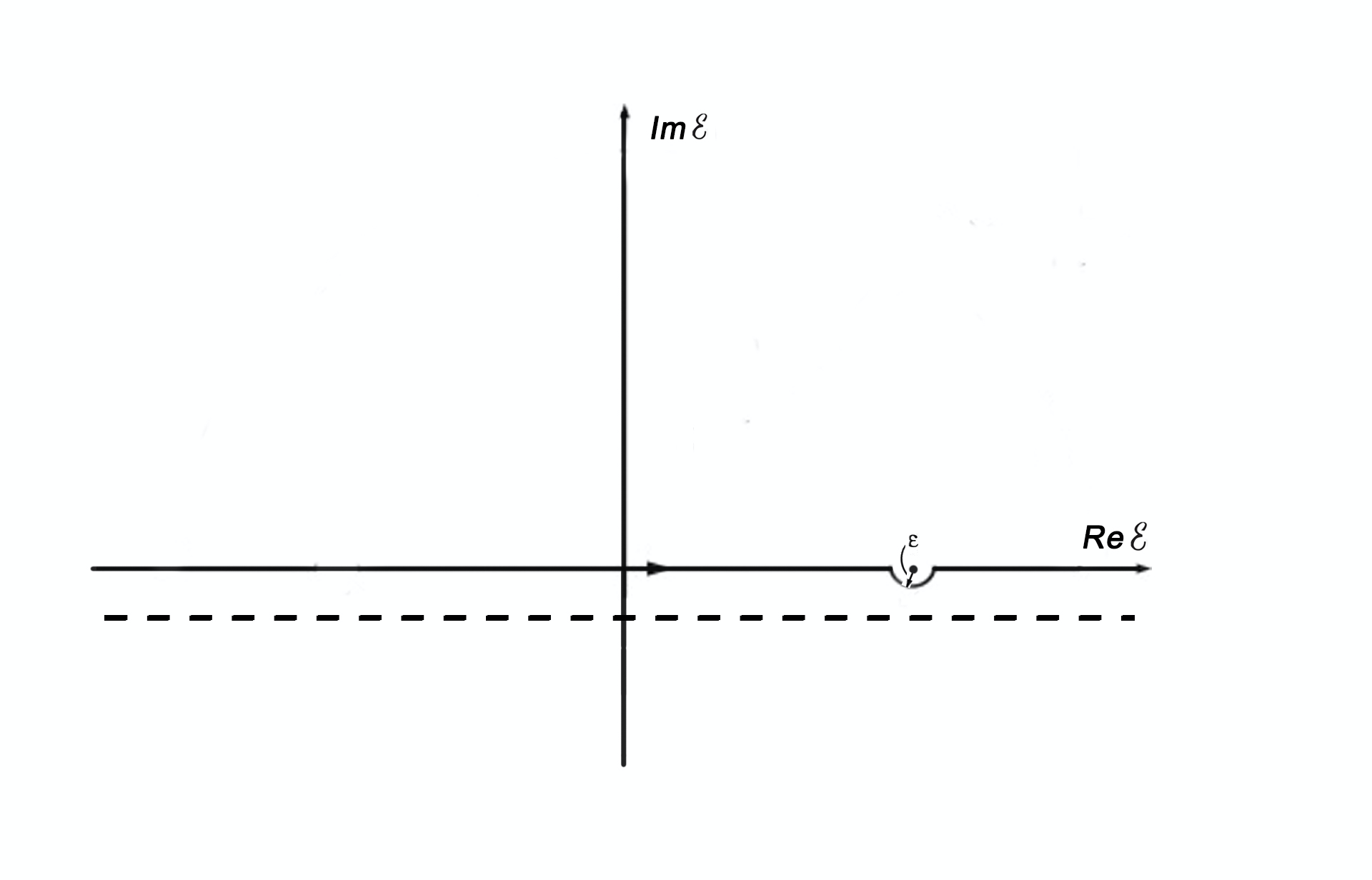}
\caption{The resulting integration contour in the $i\epsilon$ prescription. It was obtained first shifting the integration 
contour along the real axis  in Eq.(\ref{eq:t_0_to_infty_lim}) to the lower $Im\cE < 0$ half-plane, then letting  $t_0\rightarrow \infty$ for which the 1st term in the integrand vanishes and then bringing the contour back to the real axis with a small semicircle around the singularity point at $\cE=\cE_0$. Note that choosing the semicircular shape is a matter of convenience allowing to obtain easily the conventional result (\ref{eq:i_eps_formula}) as explained in the text. }
 \label{fig:contour}
\end{figure}

We now provide explicit expressions for $\Gamma$ and $\Delta\cE$.  Consider the integral   
\beq \label{eq:t_0_to_infty_lim}
\hbar \lim_{t_0\to \infty} \int_0^{t_0} K(\tau) d\tau =
i\lim_{t_0\to \infty} \int d\cE \, \overline{|V_{0\mu}|^2}\big|_{_{\cE_\mu=\cE}}  \;\;\frac{e^{i(\cE_0 -\cE)t_0/\hbar} -1}{\cE_0-\cE}
\eeq
where we used Eq.\,(\ref{eq:K_of_t})  for $K(t)$. To calculate the $t_0\to \infty$ limit we will use the following device (cf., Ref.\,\cite{LandL3}, Sec. 43). Let us shift the integration contour over $\cE$ slightly into the lower imaginary half plane (with $\rm Im \,\cE < 0$), cf. dashed line in Fig.\ref{fig:contour} 
 
This can be done without changing the value of the integral since the integrand has no singularities on the real axis\footnote{To be precise one should write the definite energy integral with its limits $\int_{\cE_{min}}^{\cE_{max}} d\cE ... $ and assume that the contour end points (which are fixed and can't be moved to the complex plane) give negligible contribution.}.  The integral above can then be separated into a sum of two 
$$
\int d\cE \, \overline{|V_{0\mu}|^2}\big|_{_{\cE_\mu=\cE}}\;\;\frac{e^{i(\cE_0 -\cE)t_0/\hbar}}{\cE_0-\cE} \;\;\; \;\;
 {\rm and} \;\;\;\; -\int d\cE \, \overline{|V_{0\mu}|^2}\big|_{_{\cE_\mu=\cE}} \;\;\frac{1}{\cE_0-\cE}
$$
This was not possible when the integration was over the real $\cE$ axis because each term separately is singular at $\cE=\cE_0$. 

 In the limit $t_0\rightarrow \infty$ the first integral tends to zero (due to the presence of the  $e^{Im \cE \,t_0}$ factor in its integrand) and we are left with the second integral. There we can bring the integration over $\cE$ back to the real axis taking care that it doesn't cross the pole at $\cE=\cE_0$,  cf., Fig.\ref{fig:contour}
 \beq\label{eq:integ_cont}
 \hbar \int_0^\infty K(\tau)d\tau = i  \int_{\rm contour \; in \; Fig. \ref{fig:contour}} d\cE \, \overline{|V_{0\mu}|^2}\big|_{_{\cE_\mu=\cE}}\;\;\frac{1}{\cE-\cE_0}  
 \eeq 
 It is convenient and conventional to view the resulting integration contour in the above integral in a following way.  Let us first add a small positive imaginary quantity $i\epsilon$ to $\cE_0$, then shift the contour to the real axis (it goes under the pole so there is no problem) and at the end consider the limit  of $\epsilon \to 0$ deforming the contour to prevent the pole crossing it.  
 
 This procedure is often called the $i\epsilon$ prescription and using it we write the expression (\ref{eq:integ_cont}) as
 \beq \label{eq:WWinteg}
 \hbar \int_0^\infty K(\tau)d\tau = i\lim_{\epsilon\to 0}\int d\cE \,\overline{|V_{0\mu}|^2} \;\;\frac{1}{\cE-\cE_0-i\epsilon}
 \eeq
 The integral here can be transformed using the formula
 \beq \label{eq:i_eps_formula}
 \lim_{\epsilon\to 0} \int_a^b \frac{f(x)}{x \pm i\epsilon} dx = \cP \int_a^b \frac{f(x)}{x} dx \mp i\pi f(0) 
 \eeq
 valid for $a<0$ and $b>0$. Here $\cP$ denotes the principle value of the integral
 \beq \label{eq:prin_val}
  \cP \int_a^b \frac{f(x)}{x} dx = \lim_{\epsilon\to 0} \left[\int_{a}^{-\epsilon}dx + \int_{\epsilon}^b dx\right] \frac{f(x)}{x} 
  \eeq
 The two terms in (\ref{eq:i_eps_formula}) correspond to the integral along the real axis with the excluded interval $-\epsilon < x <\epsilon$ and the integral along the semicircle of radius $\epsilon$ around the singularity at $x=0$, cf., Fig.\ref{fig:contour}.  The value of the second term is just half of the $\mp 2\pi i f(0)$ from the application of the  Cauchy's residue theorem to the full circle (or just calculating the integral using polar coordinates $ Re\, x = r\cos\phi, Im \, x = r\sin\phi$ in the complex plane). 
 Using this for the integral in (\ref{eq:WWinteg}) we arrive at the expressions (\ref{eq:width_shift}) for the energy shift $\Delta \cE$ and the width $\Gamma$.

\subsection{The  $\hat{H}_{I3}$ part of the Hamiltonian and the parity transformation \label{app:App_H3}}
The invariance under parity transformation of the part $\hat{H}_{I3}$, Eq.(\ref{interact_Ham3}) of the (non relativistic) matter-EM field Hamiltonian follows since   both $\vB$ and the particle spins $\hat{\vs}_a$ do not change their signs under the parity transformation. They are  axial (or pseudo) vectors. For the magnetic field this is  already seen in the Lorenz force
$$
\vF = q(\vE + \vv\times \vB)
$$
Since the force $\vF = md\vv/dt$ it must be a polar vector so must be $\vE$.  In the second term since $\vv$ is polar $\vB$ must be axial. 
This is also seen in the Maxwell equations 
$$
\nabla\times\vE =-\pd \vB/\pd t \;\;\;\; , \;\;\;\; c^2 \nabla\times\vB= \pd \vE/\pd t +\vj/\epsilon_0 
$$
 as well as in the relation  $\vB = \nabla \times \vA$.
 
Spin vectors $\vs$ are axial as they are part of the total angular momentum $\vj = \vl +\vs$ with the orbital part $\vl =\vecr \times \vp$ which obviously doesn't change under the parity transformation. It is not difficult to gain intuition about this peculiar property of $\vl$. Indeed the orbital angular momentum reflects/measures the magnitude and the direction of the "rotational" component with respect to the origin $\vecr=0$ in a (chosen) coordinate system of a particle motion at a position $\vecr$ moving  with the momentum $\vp$.
Changing the particle position $\vecr \to -\vecr$ and the momentum $\vp \to -\vp$ leave the direction and the magnitude of the rotational component of the motion the same.

\chapter{Quantization of the Schr\"odinger Field - The Second Quantization}

\section{Introduction}
Let us consider  the \Sch equation for a free particle
\beq \label{free_Sch_fld}
i\hbar\frac{\pd\psi(\arg)}{\pd t}=-\frac{\hbar^2}{2 m} \nabla^2 \psi(\arg)
\eeq
and regard it as equation for a classical field $\psi(\vr,t)$ just like  we regarded the Maxwell equations for the electromagnetic field. 
To remind - the Maxwell field after the quantization describes free quanta - photons - which behave like quantum  particles. Their energy-momentum relation $\epsilon=c |\vp|$ is determined by the classical dispersion relation $\omega= c|\vk|$ of the free EM waves supplemented with the basic QM particle-wave relations $\epsilon = \hbar\omega$ and $\vp=\hbar \vk$.  

For the free \Sch field the dispersion relation is read off the equation (\ref{free_Sch_fld}) as
$$
\hbar \omega = \frac{\hbar^2 |\vk|^2}{2m}
$$
which suggests that the quantization of this field will lead to the description of free quanta with the energy-momentum relation 
$$
\epsilon=  \frac{\vp^2}{2m}
$$ 
i.e. that of non relativistic particles.  This procedure  is called  {\em second quantization} for obvious reasons. 

We will start by confirming this picture and then extending  it to describe particles moving in an external potential and also interacting between themselves.  In the process of doing this we will  discover that the formalism describes identical particles obeying boson statistics. We will then understand how to extend the formalism  to describe particle  obeying fermion statistics. 

\section{Free \Sch field. Quantization}

The \Sch  field $\psi(\vr,t)$ is a scalar field and in that it is simpler than the vector EM field. It is however complex valued
unlike the real valued EM field. The last feature means that actually the  equation (\ref{free_Sch_fld}) should be considered as a pair
of equations for real and imaginary parts or equivalently for $\psi(\vr)$ and  its complex conjugate $\psi^*(\vr)$
\beq \label{eq:freeScheqanditscc}
i\hbar\frac{\pd\psi(\arg)}{\pd t}=-\frac{\hbar^2}{2 m} \nabla^2 \psi(\arg) \;\;\; ; \;\;\; -i\hbar\frac{\pd\psi^*(\arg)}{\pd t}=-\frac{\hbar^2}{2 m} \nabla^2 \psi^*(\arg)\; .
\eeq
Following the motivation outlined in the Introduction we consider the quantization of this field. We follow the standard quantization procedure and start by identifying  the hamiltonian structure and the canonical conjugate pairs of the \Sch field.

We note that the pair of equations (\ref{eq:freeScheqanditscc})
can be regarded as the Hamiltonian pair of equations with the Hamiltonian
\beq\label{eq:Hamiltonian_Sch_field}
H=\frac{\hbar^2}{2 m}\int d^3 r |\nabla \psi(\vr)|^2
\eeq
Indeed, the variation of this expression gives
\eqna
\delta H&=&\frac{\hbar^2}{2 m}\int d^3 r \left[ \nabla\psi^*(\vr)\cdot\nabla \delta \psi(\vr)+
 \nabla\delta\psi^*(\vr)\cdot\nabla  \psi(\vr)\right] = \nonumber \\
 &=& - \frac{\hbar^2}{2 m}\int d^3 r \left\{[\nabla^2\psi^*(\vr)]\delta \psi(\vr)+[\nabla^2\psi(\vr)]\,\delta \psi^*(\vr)\right\}
\eqne
Now regarding $\delta\psi(\vr)$ and $\delta\psi^*(\vr)$ as \textbf{independent} we read off that
\beq
\frac{\delta H}{\delta\psi(\vr)}=-\frac{\hbar^2}{2 m}\nabla^2\psi^*(\vr)\;\;\;\; ; \;\;\;\;
\frac{\delta H}{\delta\psi^*(\vr)}=-\frac{\hbar^2}{2 m}\nabla^2\psi(\vr)
\eeq
This shows  that the pair (\ref{eq:freeScheqanditscc}) is indeed the Hamiltonian pair provided one considers $\psi(\vr)$ and $i\hbar \psi^*(\vr)$ (i.e. their values at every space point $\vr$) as respectively \textbf{canonically conjugate coordinates and momenta},
\beq \label{eq:ScheqinHamiltform}
\frac{\pd\psi(\vr,t)}{\pd t}= \frac{\delta H}{\delta(i\hbar \psi^*(\vr,t))}\;\;\;\; , \;\;\;\;\frac{ \pd(i\hbar \psi^*(\vr,t))}{\pd t}= -\frac{\delta H}{\delta\psi(\vr,t)}
\eeq

\subsection{Separating the real and imaginary parts}
The complex valuedness of $\psi$ and $i\hbar\psi^*$ presents a slight problem in  applying the standard rules of the canonical quantization procedure. A possible way to avoid this problem is to transforms to the real and imaginary parts
\beq \label{eq:realandimag}
\psi={\rm Re}\psi+i {\rm Im}\psi\;\;\;;\;\;\; \psi^*={\rm Re}\psi- i {\rm Im} \psi
\eeq
We will proceed with this for a little while and use it to learn how to quantize using the original $\psi$ and $i\hbar\psi^*$. As we will see it will be a more convenient (and conventional) option.

 One must take care that the transformation (\ref{eq:realandimag}) is canonical to make sure that the transformed variables also form a canonical pair. This is achieved by
\beq  \label{eq:realandimag1}
\psi(\vr)=\frac{1}{\sqrt{2\hbar}}[\phi(\vr)+i\pi(\vr)]\;\;\;;\;\;\; \psi^*(\vr)=\frac{1}{\sqrt{2\hbar}}[\phi (\vr)- i\pi(\vr)]
\eeq
with real $\phi$ and $\pi$. To verify that $\phi$ and $\pi$ are canonical let us use the known property\footnote{cf, L. Landau and E. Lifshitz, Mechanics, Sec.45, Elsevier Ltd. 1976} that time independent canonical transformation
from a set $p_k$, $q_k$ to $P_k$, $Q_k$ obeys
 $$
 \sum_k p_k d q_k=\sum_k P_k d Q_k+dF
 $$
  where $dF$ is a total differential. In our case the sum over k is the integral over $\vr$ so that
\eqna
&&\int  i\hbar\psi^*(\vr)d\psi(\vr)d\vr =\int  i\hbar \frac{1}{2\hbar} \left[\phi(\vr)d\phi(\vr)+i\phi(\vr)d\pi(\vr)
-i\pi(\vr)d\phi(\vr)+ \right. \nonumber \\
&&+ \left.\pi(\vr)d\pi(\vr)\right] d\vr = \int \pi(\vr)d\phi(\vr) d\vr +d\int \frac{i}{4}\left[\phi^2(\vr)+
\pi^2(\vr)+2i\phi(\vr)\pi(\vr)\right]d\vr \nonumber
\eqne
showing that indeed $\phi$ and $\pi$ are canonical i.e. difference of the symplectic forms in the old and the new canonical variables is a complete differential.

Inserting (\ref{eq:realandimag1}) into the equations (\ref{eq:freeScheqanditscc}) we obtain
\beq \label{eq:freeScheqinreal}
\frac{\pd \phi(\arg)}{\pd t}=-\frac{\hbar}{2 m} \nabla^2 \pi(\arg) \;\;\; ; \;\;\; \frac{\pd\pi(\arg)}{\pd t}=\frac{\hbar}{2 m} \nabla^2 \phi(\arg)\; .
\eeq
The Hamiltonian becomes
\beq \label{eq:Haminreal}
H=\frac{\hbar^2}{2 m}\int d^3 r |\nabla \psi(\vr)|^2=\frac{\hbar}{4 m}\int d^3 r
\left[( \nabla \pi(\vr))^2 + (\nabla \phi(\vr))^2 \right]
\eeq
Its variation
\eqna
\delta H&=& \frac{\hbar}{2 m}\int d^3 r
\left[ \nabla \pi(\vr)\cdot\nabla \delta\pi(\vr)  +   \nabla \phi(\vr)\cdot \nabla\delta \phi(\vr)\right] = \nonumber \\
 &=& -\frac{\hbar}{2 m}\int d^3 r
\left[\nabla^2 \pi(\vr)\delta\pi(\vr)  +   \nabla^2 \phi(\vr) \delta \phi(\vr)\right]
\eqne
so that
\beq
\frac{\delta H}{\delta\phi(\vr)}=-\frac{\hbar}{2 m}\nabla^2\phi(\vr)\;\;\; , \;\;\;
\frac{\delta H}{\delta\pi(\vr)}=-\frac{\hbar}{2 m}\nabla^2\pi(\vr)
\eeq
Thus we see that Eqs. (\ref{eq:freeScheqinreal}) indeed are in the Hamiltonian form
\beq
\frac{\pd\phi(\arg)}{\pd t}=\frac{\delta H}{\delta\pi(\vr,t)} \;\;\; , \;\;\;
\frac{\pd\pi(\arg)}{\pd t}=-\frac{\delta H}{\delta\phi(\vr,t)} \; .
\eeq
with $\phi(\vr)$ as coordinates and $\pi(\vr)$ as momenta. These are real and we can  quantize the theory in the usual way by introducing wave functionals $\Psi[\phi(\vr)]$ and operators which act on them
\beq
\hat{\phi}(\vr) \Psi[\phi(\vr)] = \phi (\vr) \Psi[\phi(\vr)] \;\; ;  \;\; \hat{\pi}(\vr)\Psi[\phi(\vr)]=-i\hbar \frac{\delta}{\delta \phi(\vr)}\Psi[\phi(\vr)] \; .
\eeq
with the \Sch equation
\beq
i\hbar \frac{\partial}{\partial t} \Psi[\phi(\vr),t]= H_{op} \Psi[\phi(\vr),t]
\eeq
where the Hamiltonian operator is given by the expression (\ref{eq:Haminreal}) with $\phi(\vr)$ and $\pi(\vr)$ replaced by the corresponding operators
\beq
H_{op}=\frac{\hbar}{4 m}\int d^3 r
\left[( \nabla \hat{\pi}(\vr))^2 + (\nabla\hat{\phi}(\vr))^2 \right]
\eeq
We note that the commutation relations for the operators $\hat{\phi}(\vr)$ and $\hat{\pi}(\vr)$ are
$$
[\hat{\phi}(\vr), \hat{\phi}(\vr')] =  [\hat{\pi}(\vr),\hat{\pi}(\vr')] = 0
$$
\beq \label{eq:commrel1}
 [\hat{\pi}(\vr), \hat{\phi}(\vr ')]  =
 \hat{\pi}(\vr)\,\hat{\phi}(\vr') -\hat{\phi}(\vr')\,\hat{\pi}(\vr) = -i\hbar \frac{\delta \phi(\vr')}{\delta \phi(\vr)} = -i\hbar\delta(\vr-\vr')
\eeq

\subsection{Back to the complex valued field}

 As was already stated it is more convenient to work with complex valued field. Using (\ref{eq:realandimag1}) we introduce non hermitian combinations of the operators $\hat{\phi(\vr)}$ and $\hat{\pi}(\vr))$
 \beq \label{eq:nonhermops}
\hat{\psi}(\vr)=\frac{1}{\sqrt{2\hbar}}[\hat{\phi}(\vr)+i\hat{\pi}(\vr)]\;\;\;;\;\;\; \hat{\psi}^{+}(\vr)=\frac{1}{\sqrt{2\hbar}}[\hat{\phi}(\vr)- i\hat{\pi}(\vr)]
\eeq
We note here a clear analogy with the familiar  operators $\hat{a}$ and $\hat{a}^{+}$
$$
\hat{a}=\frac{1}{\sqrt{2\hbar}}[\hat{x} + i\hat{p}] \;\;\;,\;\;\;  \hat{a}^{+}=\frac{1}{\sqrt{2\hbar}}[\hat{x} - i\hat{p}] \;,
$$
the non hermitian combinations of coordinate and momentum operators for a single degree of freedom.

 In our case   we have such combinations (\ref{eq:nonhermops}) for every $\vr$,  i.e. for $\infty^3$ coordinate-momentum pairs.

 From the commutation relations (\ref{eq:commrel1})  we have
 \beq \label{eq:commrel2}
[\hat{\psi}(\vr), \hat{\psi}^{+}(\vr')] =\delta(\vr-\vr') \;\;\;, \;\;\;
  [\hat{\psi}(\vr), \hat{\psi}(\vr')]=0=[\hat{\psi}^{+}(\vr) , \hat{\psi}^{+}(\vr')]
\eeq
Looking back at   (\ref{eq:ScheqinHamiltform}) we observe that had we postulated the usual commutation relations for the operators corresponding to the complex field canonical coordinates and momenta $\psi(\vr)$ and $i\hbar\psi(\vr)$
  \beq
  [\hat{\psi}(\vr), i \hbar \hat{\psi}^{+}(\vr')]=i\hbar\delta(\vr - \vr')
  \eeq
we would have arrived at the same result Eq.\,(\ref{eq:commrel2}).

As we will soon see the commutation relations  (\ref{eq:commrel2}) will be essentially all (well almost all) we will need to know about the operators $\hat{\psi}(\vr)$ and $\hat{\psi}^{+}(\vr)$  in order to understand how they act on any wave function.

\subsection{The Hamiltonian of the free \Sch field }
The Hamiltonian operator is directly obtained from Eq.\,(\ref{eq:Hamiltonian_Sch_field}) by replacing $\psi(\vr)$ and $\psi^*(\vr)$ with  the operators  $\hat{\psi}(\vr)$ and $\hat{\psi}^{+}(\vr)$. One usually finds two expressions in the literature
\beq
H_{op}= \frac{\hbar^2}{2m}\int d^3 r \nabla \hat{\psi}^{+}(\vr) \nabla \hat{\psi}(\vr)
\eeq
or
\beq \label{eq:Ham_op_Sch_fld}
H_{op} = \int d^3 r \hat{\psi}^{+}(\vr) \left(-\frac{ \hbar^2 \nabla^2}{2m}\right)\hat{\psi}(\vr) =
-\frac{\hbar^2}{2m} \int d^3 r \hat{\psi}^{+}(\vr) \nabla^2 \hat{\psi}(\vr)
\eeq
The difference is obviously just a  "surface term" at large distances. This is an operator valued term so disregarding this difference means the requirement that all the wave functions of the field will produce  zero when acted upon by such  "surface" operators.

Note also the order of the operators chosen in the above expression for  $H_{op} $ with $\hat{\psi}^{+}(\vr)$ acting after $\hat{\psi}(\vr)$. As will become clear below this order of operators (called "normal ordering") assures that the vacuum of the theory has zero energy.

\subsection{The eigenstates.  Field quanta are free non relativistic particles}

Having established the form of the  Hamiltonian operator of the theory we should proceed to solve the \Sch equation of the theory
\beq
i\hbar \frac{\partial }{\partial t} |\Psi(t)\rangle  = H_{op} |\Psi(t)\rangle
\eeq
Note - to solve the \Sch equation for the quantum \Sch field!

\subsubsection{The normal modes} 
Since $H_{op}$ is time independent we can solve the above equation in a standard way by first finding the eigenfunctions of the \Sch field Hamiltonian i.e. solutions of
\beq
H_{op}\Psi=E\Psi
\eeq
  To this end we will go to the normal modes of the field.  As we know from the systems with finite number of degrees of freedom quadratic Hamiltonians become sums of independent terms when the original degrees of freedom are transformed to the normal modes. 
  
To find the normal modes let us recall that they are special solutions of the classical equations in which all the degrees of freedom of the physical system oscillate  with the same frequency.  Classical equations in the present case are just the field equations (\ref{eq:freeScheqanditscc}).   Their solutions with a given frequency  $\psi(\vr,t) = u(\vr) \exp(-i\omega t)$ 
satisfy
\beq \label{eq:norm_modes}
\frac{\hbar^2}{2m} \nabla^2 u(\vr) +\hbar\omega \, u(\vr)=0 \;\;
\eeq
and can be chosen as  plane waves
\beq \label{eq:plane_wave_basis}
u_{\vk}(\vr)= \frac{1}{\sqrt{\Omega}} e^{i\vk\cdot\vr} \;\; {\rm with} \;\; \hbar\omega = \frac{\hbar^2k^2}{2m}
\eeq
where we assumed the usual periodic boundary condition in a very large volume $\Omega$$$
\vk = \frac{2\pi}{\Omega^{1/3}}(n_x, n_y. n_z) \;\;\; {\rm with}\;\; n_x,n_y,n_z = 0, \pm1, \pm2, ...
$$
We now expand the field operators using these normal modes
\eqna\label{eq:exp_fld_ops}
\hat{\psi}(\vr)&=&\sum_{\vk}  \hat{a}_{\vk} u_{\vk}(\vr)=
\frac{1}{\sqrt{\Omega}}\sum_{\vk}  \hat{a}_{\vk} e^{i\vk\cdot\vr}  \nonumber \\
\hat{\psi}^{+}(\vr)&=&\sum_{\vk}  \hat{a}^{+}_{\vk} u^{*}_{\vk}(\vr)=\frac{1}{\sqrt{\Omega}} \sum_{\vk} \hat{a}^{+}_{\vk} e^{-i\vk\cdot\vr}
\eqne
The coefficients $ \hat{a}_{\vk}$ and $ \hat{a}_{\vk}^{+}$ in the above expansions of the field operators  are obviously operators.  This expansion must be viewed as a transformation from  a canonical set of $2\times \infty^3$ operators $\hat{\psi}(\vr)\;,\; \hat{\psi}^{+}(\vr)$  to another canonical set of $2\times \infty^3$ operators $\hat{a}_{\vk}\,,\, \hat{a}_{\vk}^{+}$.   Using orthonormality of $u_{\vk}$'s  it is easy to invert  (\ref{eq:exp_fld_ops})
$$
  \hat{a}_{\vk}  = \frac{1}{\sqrt{\Omega}} \int d^3 r \, \hat{\psi}(\vr)  e^{-i\vk\cdot\vr} \; \;\;,  \;\;\; \hat{a}^{+}_{\vk}  = \frac{1}{\sqrt{\Omega}} \int d^3 r \, \hat{\psi}^{+}(\vr)  e^{i\vk\cdot\vr}
$$
Using the commutations (\ref{eq:commrel2}) one can then find the commutation relations between $\hat{a}_{\vk}$'s and $\hat{a}_{\vk'}^{+}$'s,
 \beq \label{eq:com_rel_ak}
[\hat{a}_{\vk}\,,\, \hat{a}_{\vk'}^{+}]=\delta_{\vk \vk'} \;\;\;, \;\;\;\; [\hat{a}_{\vk}\,,\, \hat{a}_{\vk'}] =0= [\hat{a}_{\vk}^{+}\,,\, \hat{a}_{\vk'}^{+}]
\eeq
which of course express the harmonic oscillator character of the normal modes for each $\vk$ and their independence for different $\vk$'s. 

\subsubsection{Diagonalizing the field Hamiltonian}

Inserting the expansions (\ref{eq:exp_fld_ops}) into the Hamiltonian Eq.\,(\ref{eq:Ham_op_Sch_fld}) we obtain
a sum of independent (commuting) oscillators 
\beq \label{eq:free_H}
H_{op} = \sum_{\vk} \epsilon_{\vk}  \hat{a}^{+}_{\vk}  \hat{a}_{\vk}
\eeq
where we have denoted the energies of the oscillator quanta 
\beq
\epsilon_{\vk}=\frac{\hbar^2 \vk^2}{2m}
\eeq
Based on this it is trivial to find the eigenfunctions and eigenenergies of each term.  Clearly the eigenstates of this $H_{op}$ are products of the familiar harmonic oscillator-like states (cf., Appendix, Eq.\,(\ref{eq:eigen_of_n_i}))
\beq \label{eq:eigstates_of_H0}
|\{n_{\vk}\}\rangle = \prod_{\vk}|n_{\vk}\rangle =\prod_{\vk}\frac{(\hat{a}_{\vk}^{+})^{n_{\vk}}}{\sqrt{n_{\vk}!}}|0\rangle
\eeq
with eigenvalues which are 
\beq
E_{\{n_{\vk}\}} =  \sum_{\vk}\epsilon_{\vk} \, n_{\vk}\;\;\;\; {\rm with \;\; each} \;\; n_{\vk}=0, 1, 2, \dots
\eeq
So the eigenenergies of the free \Sch field are sums over the modes $u_{\vk}(\vr)$ of integer numbers $n_{\vk}$ of quanta with energies $\epsilon_{\vk}$. To understand the physics of these quanta it is useful to ask/determine what are their momenta. For this one must find the corresponding operator. We deal with this in the next section.

We note that the ground state corresponds to all $n_{\vk}=0$, i.e. it is the vacuum state $|0\rangle$.  Its energy is equal to zero  which was assured by the normal ordered form of $H_{op}$, Eq.\,(\ref{eq:Ham_op_Sch_fld}),  which we have adopted. Let us also note that in this formulation the only properties we will ever need of the vacuum state are that it gives zero when acted upon with anyone of the operators $\hat{a}_{\vk}$  and that it is normalized  
\beq
\hat{a}_{\vk}|0\rangle = 0  \;\; , \;\;  \langle 0 | 0 \rangle = 1   
\eeq

Let us also note that the most general states of the theory are linear combinations of the eigenstates (\ref{eq:eigstates_of_H0})
\beq \label{eq:comb_of_eignstts}
 |\Psi\rangle = \sum_{\{n_{\vk}\}}   C_{\{n_{\vk}\}}|\{n_{\vk}\}\rangle
\eeq
They may appear e.g. as solutions of the time dependent \Sch equation of the field
\beq \label{eq:time_dep_Sch_eq}  
i\hbar \frac{\partial |\Psi(t)\rangle}{\partial t} =   H_{op}|\Psi(t)\rangle
\eeq
with coefficients depending on time via the usual
$$
C_{\{n_{\vk}\}}(t)=C_{\{n_{\vk}\}}(0)\exp(-iE_{\{n_{\vk}\}}t/\hbar)
$$
We note that the number of particles in the above expressions for $C_{\{n_{\vk}\}}$ is given by
$$
N = \sum_{\vk} n_{\vk}
$$
It is important to note that nowhere in the formalism there appears a requirement that $N$ is fixed, i.e. has the same value in the e.g. expression for the general wave function $|\Psi\rangle$. The formalism in principle allows to have states with coherent combinations of different particle numbers. We will address this issue in the last section.

\subsubsection{Degeneracy of the normal modes.  Spherical waves}
The normal modes Eq.\,(\ref{eq:plane_wave_basis}) are clearly infinitely degenerate having the same frequency $\omega$ for all $\vk$ with the same $k=|\vk|$. This of course follows from the degeneracy of the solutions of the (free \Sch)  equation (\ref{eq:norm_modes}).  This degeneracy means that other sets  can be chosen for a given $k$. The familiar spherical or cylindrical waves rather than the plane waves would supply examples of such sets. 

Let us consider the spherical waves set of solutions
$$
u_{klm}(\vr) = R_{kl}(r)Y_{lm}(\theta,\phi) 
$$ 
with $l,m$ the angular momentum and its projection (for a free particle) and $R_{kl}(r)$ and $Y_{lm}(\theta,\phi)$ the radial and angular parts\footnote{Recall the solutions of the stationary \Sch equation for a free particle in spherical coordinates, cf.,  Sakurai, Modern Quantum Mechanics, Sec.3.7, Addison-Wesley, 1994.}. We can expand the field operators using such normal modes\footnote{For convenience we assume that $k$ values are made discrete by imposing boundary condition in a large spherical box}
\eqna \label{eq:ang_mom_norm_modes}
\hat{\psi}(\vr) &=& \sum_{klm} \hat{a}_{klm}\, u_{klm}(r,\theta,\phi) 
\nonumber \\
\hat{\psi}^{+}(\vr) &=& \sum_{klm} \hat{a}^{+}_{klm}\, u_{klm}^{*}(r,\theta,\phi) 
\eqne
with the operators 
$$
\hat{a}_{klm} = \int d^3r \hat{\psi}(\vr) \, u_{klm}^{*}(r,\theta,\phi)  \;\; , \;\; \hat{a}^{+}_{klm} = \int d^3r \hat{\psi}^{+}(\vr) \, u_{klm}(r,\theta,\phi)
$$
and (as can be easily checked) the commutation relations equivalent to Eq.\,(\ref{eq:com_rel_ak}) with $\vk$ and $\vk'$ indices replaced by $klm$ and $k'l'm'$.   

Inserting the expansions (\ref{eq:ang_mom_norm_modes}) into the Hamiltonian Eq.\,(\ref{eq:Ham_op_Sch_fld}) we obtain
\beq
H_{op} = \sum_{klm} \epsilon_{klm} \hat{a}^{+}_{klm} \hat{a}_{klm}
\eeq
As with the plane waves it is a sum of independent (commuting) oscillators with quanta energies depending only on $k$ 
$$
\epsilon_{klm} = \frac{\hbar^2 k^2}{2m}
$$
i.e. equal to the energy of the plane wave quanta - reflecting the degeneracy of the normal modes.

\subsection{Momentum and angular momentum}
\subsubsection{Field momentum}
We now discuss the total (mechanical) momentum of the \Sch field. To find it expression we could go back to the classical fields and use the Noether theorem. We prefer to find it by considering the generator of the translations $\vr\to \vr+ \va$ with a constant vector $\va$. The  field operators change as
$\hat{\psi}(\vr)\to \hat{\psi}(\vr +\va)$ and $\hat{\psi}^{+}(\vr)\to \hat{\psi}^{+}(\vr +\va)$. So we are looking for the operator $\vP_{op}$ with which
\beq
e^{-i\va\cdot\vP_{op}/\hbar}\left\{\begin{array}{c}\hat{\psi}(\vr) \\ \hat{\psi}^{+}(\vr)\end{array}\right\}
 e^{i\va\cdot\vP_{op}/\hbar} = \left\{\begin{array}{c}\hat{\psi}(\vr +\va)
 \\ \hat{\psi}^{+}(\vr +\va) \end{array}\right\}
 \eeq
 For  infinitesimal $\va$ this is
 \[
 (1-i\va\cdot\vP_{op}/\hbar)\left\{\begin{array}{c}\hat{\psi}(\vr) \\ \hat{\psi}^{+}(\vr)\end{array}\right\}(1+i\va\cdot\vP_{op}/\hbar)  =  \left\{\begin{array}{c}(1+\va\cdot\nabla)\hat{\psi}(\vr) \\ (1+\va\cdot\nabla)\hat{\psi}^{+}(\vr) \end{array}\right\}
 \]
which means that must have the commutator
\[
\left[\vP_{op}\;, \left\{\begin{array}{c}\hat{\psi}(\vr) \\ \hat{\psi}^{+}(\vr)\end{array}\right\}\right]  =  \left\{\begin{array}{c}i\hbar\nabla\hat{\psi}(\vr) \\ i\hbar\nabla \hat{\psi}^{+}(\vr) \end{array}\right\} 
\]
This is achieved with the expression
\beq \label{eq:fld_momentum}
\vP_{op}=\int d^3 r' \hat{\psi}^{+}(\vr')(-i\hbar \nabla_{\vr'})\hat{\psi}(\vr')
\eeq
Indeed
\eqna
&&\left[\vP_{op}, \left\{\begin{array}{c}\hat{\psi}(\vr) \\ \hat{\psi}^{+}(\vr)\end{array}\right\}\right]  =\int d^3r'\left[
 \begin{array}{c} \hat{\psi}^{+}(\vr')(- i\hbar\nabla_{\vr'})\hat{\psi}(\vr')\; , \; \hat{\psi}(\vr) \\ \hat{\psi}^{+}(\vr')(- i\hbar\nabla_{\vr'})\hat{\psi}(\vr')\; , \; \hat{\psi}^{+}(\vr) \end{array}\right] = \nonumber \\
=&&\int d^3r'\left\{
 \begin{array}{c} \delta(\vr-\vr') i\hbar\nabla_{\vr'}\hat{\psi}(\vr') \\ \hat{\psi}^{+}(\vr')(- i\hbar\nabla_{\vr'})\delta(\vr-\vr')\end{array}\right\} = \left\{\begin{array}{c}  i\hbar\nabla_{\vr}\hat{\psi}(\vr) \\ i\hbar\nabla_{\vr}  \hat{\psi}^{+}(\vr) \end{array}\right\} \nonumber
\eqne
where in the second line of the last equality we used integration by parts.

\subsubsection{Field quanta are free nonrelativistic particles}

The momentum $\vP_{op}$ commutes with the Hamiltonian $H_{op}$, Eq.\,(\ref{eq:Ham_op_Sch_fld})
\beq
[H_{op} , \vP_{op}]=0
\eeq
 Verifying this  explicitly with $H_{op}$ and $\vP_{op}$ written in terms of the field operators $\hat{\psi}(\vr)$ and $\hat{\psi}^{+}(\vr)$ is a good exercise which is left to the reader. Physically this is the result of the invariance of $H_{op}$ under the translation.

 Let us write $\vP_{op}$ in terms of the normal modes operators $\hat{a}_{\vk}$'s and $\hat{a}_{\vk'}^{+}$'s. Using the expansions (\ref{eq:exp_fld_ops}) in (\ref{eq:fld_momentum}) we obtain
 \beq
 \vP_{op}=\sum_{\vk}\hbar\vk \; \hat{a}_{\vk'}^{+}\hat{a}_{\vk'}
 \eeq
 This expression compared to  Eq.\,(\ref{eq:free_H}) trivially shows that indeed $\vP_{op}$ commutes with $H_{op}$. It has the same eigenfunctions (\ref{eq:eigstates_of_H0}) and its eigenvalues
 are
 \beq
 \vP_{\{n_{\vk}\}}=\sum_{\vk} \hbar\vk \; n_{\,\vk}
 \eeq
 This shows that each field quantum with energy $\epsilon_{\vk}$ carry momentum $p_{\vk}=\hbar \vk$. The energy momentum relation $\epsilon(\vp)$  follows from the explicit dependence of $\epsilon_{\vk}=\hbar^2\vk^2/2m$ on $\vk$
 \beq
 \epsilon_{\vk}(\vp)=\frac{|\vp_{\vk}|^{ 2}}{2m}
 \eeq
 which is the familiar energy-momentum relation of non relativistic particles. This indicates that quanta of the free \Sch field behave like such particles.

 \subsubsection{Field angular momentum}
 In analogy with the field momentum one can find the expression for the operator of the field angular momentum by considering infinitesimal rotations $\vr\to \vr +\delta\phi\, \vn\times \vr$ with $\delta \phi$ -  angle of rotation and $\vn$ - unit vector along the rotation axis (with the usual "right hand rule" convention).   As with the momentum we should look for the operator $\vL_{op}$ for which 
 \beq
e^{-i\delta\phi \vn\cdot\vL_{op}/\hbar}\left\{\begin{array}{c}\hat{\psi}(\vr) \\ \hat{\psi}^{+}(\vr)\end{array}\right\}
 e^{i\delta\phi \vn\cdot\vL_{op}/\hbar} = \left\{\begin{array}{c}\hat{\psi}(\vr +\delta\phi \vn\times \vr)
 \\ \hat{\psi}^{+}(\vr +\delta\phi \vn\times \vr) \end{array}\right\}
 \eeq
 For infinitesimal $\delta\phi$ it is straightforward to conclude that  $\vL_{op}$ must satisfy 
 \[
\left[\vn\cdot\vL_{op}\;, \left\{\begin{array}{c}\hat{\psi}(\vr) \\ \hat{\psi}^{+}(\vr)\end{array}\right\}\right]  =  \left\{\begin{array}{c}i\hbar(\vn\times\vr)\cdot\nabla\hat{\psi}(\vr) \\ i\hbar(\vn\times\vr)\cdot\nabla\hat{\psi}^{+}(\vr) \end{array}\right\}
=  \left\{\begin{array}{c}i\hbar \vn\cdot(\vr\times \nabla) \hat{\psi}(\vr) \\ i\hbar \vn\cdot(\vr\times \nabla) \hat{\psi}^{+}(\vr) \nonumber\end{array}\right\}
\]
This is achieved with the expression
\beq \label{eq:field_ang_mom}
\vL_{op}=\int d^3 r' \hat{\psi}^{+}(\vr')[\vr' \times (-i\hbar \nabla_{\vr'})]\hat{\psi}(\vr')
\eeq
Indeed
\eqna
&&\left[\vL_{op}, \left\{\begin{array}{c}\hat{\psi}(\vr) \\ \hat{\psi}^{+}(\vr)\end{array}\right\}\right]  =\int d^3r'\left[
 \begin{array}{c} \hat{\psi}^{+}(\vr')[\vr' \times (- i\hbar\nabla_{\vr'})] \hat{\psi}(\vr')\; , \; \hat{\psi}(\vr) \\ \hat{\psi}^{+}(\vr')
 [\vr' \times (- i\hbar\nabla_{\vr'})] \hat{\psi}(\vr')\; , \; \hat{\psi}^{+}(\vr) \end{array}\right] = \nonumber \\
=&&\int d^3r'\left\{
 \begin{array}{c} \delta(\vr-\vr') i\hbar (\vr' \times \nabla_{\vr'}) \hat{\psi}(\vr') \\ \hat{\psi}^{+}(\vr')[- i\hbar
 (\vr' \times \nabla_{\vr'})] \delta(\vr-\vr')\end{array}\right\} = \left\{\begin{array}{c}  i\hbar(\vr\times \nabla_{\vr})\hat{\psi}(\vr) \\ i\hbar (\vr\times \nabla_{\vr})  \hat{\psi}^{+}(\vr) \end{array}\right\}
\eqne
where in the second line of the last equality we used integration by parts.

The angular momentum operator commutes with the free field Hamiltonian, Eq.\,(\ref{eq:Ham_op_Sch_fld})
\beq
[H_{op} , \vL_{op}]=0
\eeq
The reader is advised to carry out this calculation the result of which essentially follows from the commutativity of 
the "first quantized" $h_0 =-\hbar^2 \nabla^2/2m$ and $\vl=\vr\times (-i\hbar \nabla)$ which enter the expressions of these operators. Physically of course it reflects the invariance of the free field $H_{op}$  under rotations.  Another useful calculation for the reader  to work out is to verify the validity of the standard commutation relations for the components of $\vL_{op}$
\beq
 [ L_{op,i} \, , \, L_{op,j}] = i\hbar\,\sum_{n} \epsilon_{ijn} \,L_{op,n}
 \eeq
 Here again the corresponding commutators of $\vl_i=[\vr\times (-i\hbar \nabla)]_i$'s which enter the expressions of $L_{op,i}$'s are the "cause" of this result.  

Following the experience of transforming the field momentum operator $\vP_{op}$ to the plane wave basis  it is instructive to consider transforming the field operators in the field angular momentum operator $\vL_{op}$,  Eq.\,(\ref{eq:field_ang_mom}) to the spherical wave normal modes basis $u_{klm}(\vr)$ as given in Eq.\,(\ref{eq:ang_mom_norm_modes}). In contrast to  $\vP_{op}$  the non commutativity of different components of $\vL_{op}$ leads to different forms of the expressions  for different $L_{op,j}$'s.   The simplest is for $ L_{op,z}$
$$
 L_{op,z} = \sum_{klm} \hbar m \,\hat{a}^{+}_{klm}\hat{a}_{klm}
 $$
 The expressions for $L_{op,x}$ and $L_{op,y}$ will contain non diagonal $m\to m\pm1$ terms. We leave for the reader to work this out explicitly.

 \section{Adding external potential}
 So the conclusions at this stage are that the quantized free \Sch field describes a collection of quanta which behave like free moving non interacting non relativistic quantum particles. We also note that these particles are identical (see longer discussion of this aspect in the following sections). It is therefore natural to ask how to include interactions of the particles and how to account for their statistics?

 \subsection{The Hamiltonian}
  We begin by considering the \Sch field in the presence of an external potential. The field equation is the familiar
 \beq \label{eq:Sch_eq_with_int}
 i\hbar \frac{\partial \psi(\vr,t)}{\partial t} = \left(-\frac{\hbar^2}{2m}\nabla^2 + U(\vr)\right)\psi(\vr,t) \equiv h\psi(\vr,t)
  \eeq
  with $h$ defined as
   \beq \label{eq:sp_hamiltonian}
h=  -\frac{\hbar^2}{2m}\nabla^2 + U(\vr)
 \eeq

 Following what we did in the case  of the free field, cf., Eq.(\ref{eq:freeScheqanditscc}) we consider this equation and its complex conjugate as the pair of Hamilton equations with $\psi(\vr)$ and
 $i\hbar \psi^{*}(\vr)$ as canonical variables and the following Hamiltonian
 \beq
 H=\int d^3r \left[\frac{\hbar^2}{2m}|\nabla\psi(\vr)|^2 +U(\vr)|\psi(\vr)|^2\right]
 \eeq
 Indeed from
 \eqna
 \delta H&=&\int d^3 r \left[\frac{\hbar^2}{2m}\left(\nabla\psi^{*}(\vr)\nabla\delta\psi(\vr) +
   \nabla \delta\psi^{*}(\vr)\nabla\psi(\vr)\right)  + \right.    \nonumber \\
&& \left.+ U(\vr)\left(\psi^{*}(\vr)\delta\psi(\vr)
 + \delta\psi^{*}(\vr)\psi(\vr)\right) \right]  \nonumber
 \eqne
 we find that Hamilton equations for $i\hbar \psi^{*}(\vr)$ and $\psi(\vr)$ 
 \eqna
 \frac{\partial \psi(\vr)}{\partial t}=\frac{\delta H}{\delta[i\hbar \psi^{*}(\vr)]} &=&\frac{1}{i\hbar}
 \left[-\frac{\hbar^2}{2m}\nabla^2\psi(\vr) + U(\vr)\psi(\vr)\right] \nonumber \\ 
\frac{\partial[ i\hbar \psi^{*}(\vr)]}{\partial t}=-\frac{\delta H}{\delta \psi(\vr)} &=&
 -\left[-\frac{\hbar^2}{2m}\nabla^2\psi^{*}(\vr) + U(\vr)\psi^{*}(\vr)\right] \nonumber
 \eqne
reproduce correctly the field equation (\ref{eq:Sch_eq_with_int}) and its complex conjugate.

 On this basis we will quantize this field following the by now familiar pattern
 \beq
 \psi(\vr) \to \hat{\psi}(\vr)\;\;\;, \;\; \psi^{*}(\vr) \to \hat{\psi}^{+}(\vr)
 \eeq
 with commutation relations (\ref{eq:commrel2}) and the Hamiltonian operator
 $$
  H_{op} = \int d^3r \left[\frac{\hbar^2}{2m}\nabla\hat{\psi}^{+}(\vr)\cdot\nabla\hat{\psi}(\vr) +
  U(\vr)\hat{\psi}^{+}(\vr)\hat{\psi}(\vr)\right]
  $$
  or in an equivalent form (cf., the remark after Eq.\,(\ref{eq:Ham_op_Sch_fld}))
  \beq \label{eq:Ham_for_interact_fld}
   H_{op} = \int d^3r \; \hat{\psi}^{+}(\vr)\left[-\frac{\hbar^2}{2m}\nabla^2 +
  U(\vr)\right]\hat{\psi}(\vr) \equiv \int d^3r \; \hat{\psi}^{+}(\vr) h \hat{\psi}(\vr)
  \eeq
  with $h$ defined above in Eq. (\ref{eq:sp_hamiltonian}). As in the case of the free field the general goal of the theory is to solve the \Sch equation (\ref{eq:time_dep_Sch_eq}) but with the Hamiltonian operator given by (\ref{eq:Ham_for_interact_fld}). As always the  general method of doing this is to find the eigenfunctions of this operator.

  Before this let us  note that the Heisenberg equation for the field operators calculated with  the Hamiltonian $H_{op}$, Eq.\,(\ref{eq:Ham_for_interact_fld})  coincides in form (as they should) with the wave equation (\ref{eq:Sch_eq_with_int}) which we have quantized
 $$
 i\hbar\frac{\partial \hat{\psi}(\vr,t)}{\partial t} = \left[ \hat{\psi}(\vr,t), H_{op}\right]  = h \hat{\psi}(\vr,t)
 $$
 with $h$ defined in Eq. (\ref{eq:sp_hamiltonian}).
The calculation of the commutator in this equation can be efficiently done  by commuting the operator $\hat{\psi}(\vr,t)$ through the elements of the expression $\int d^3r' \; \hat{\psi}^{+}(\vr',t) h \hat{\psi}(\vr',t)$ for $H_{op}$. Since the only non zero commutator of $\hat{\psi}(\vr,t)$ is with $\hat{\psi}^{+}(\vr',t)$ we get
 $$
  \left[ \hat{\psi}(\vr,t), H_{op}\right]  = \int d^3r' \; \delta(\vr-\vr') h \hat{\psi}(\vr',t) = h\hat{\psi}(\vr,t)
  $$
  The Heisenberg equation for $\hat{\psi}^{+}(\vr,t)$
  coincides in form with complex conjugate of Eq.\,(\ref{eq:Sch_eq_with_int}). 

 \subsection{The eigenstates. Field quanta are particles in the external potential}

  To find the eigenfunctions of the above Hamiltonian
  $$
  H_{op} \Psi=E\Psi
  $$
  we use the experience with the free field and look for the basis $u_i(\vr)$  to expand the field operators
   $\hat{\psi}(\vr)$ in which $H_{op}$ will become a sum of  decoupled commuting terms like Eq.\,(\ref{eq:free_H}) for the free field Hamiltonian.  Before doing this let us briefly consider the general aspects of changing basis.   
   
 \subsubsection{Changing basis}
 The transformation from $\hat{\psi}(\vr)$ and $\hat{\psi}^{+}(\vr)$   to $\hat{a}_{\vk}$ and $\hat{a}_{\vk}^{+}$ can be viewed as a particular example of a more general operator transformation
\beq \label{eq:gen_expnsn_of_fld_ops}
\hat{\psi}(\vr) = \sum_i \hat{a}_i u_i(\vr) \;\;\;, \;\;\; \hat{\psi}^{+}(\vr) = \sum_i \hat{a}_i^{+} u_i^{*}(\vr)
\eeq
with $\{u_i(\vr)\}$  - any complete orthonormal basis, i.e.  set of functions which obey
\eqna \label{eq:orth_compl}
\int d^3r \, u^{*}_i(\vr) u_j(\vr) &=& \delta_{ij} \;\;\;\;\; {\rm orthonormality}   \\  \sum_i u_i(\vr)u^{*}_i(\vr') &=& \delta(\vr-\vr') \;\;\;{\rm completeness}  \nonumber
\eqne
Inverting the transformation
\beq \label{eq:a_via_psi}
\hat{a}_i=\int d^3r \,\hat{\psi}(\vr)u^{*}_i(\vr) \;\;\;\;, \;\;\;\; \hat{a}^{+}_i=\int d^3r \hat{\psi}^{+}(\vr)u_i(\vr)
\eeq
and using the commutation relations (\ref{eq:commrel2}) for the field operators $\hat{\psi}(\vr)$ and $\hat{\psi}^{+}(\vr)$ and orthogonality of the basis set $\{u_i(\vr)\}$ one finds that the commutations of the $\hat{a}_i, \hat{a}^{+}_i$ set remain canonical
\beq
[\hat{a}_i,\hat{a}^{+}_j]=\delta_{ij}\;\;\;, \;\;\; [\hat{a}_i,\hat{a}_j]=0=[\hat{a}^{+}_i,\hat{a}^{+}_j]
\eeq

Let us note a useful  view of the expansion (\ref{eq:gen_expnsn_of_fld_ops}) as transforming "vectors" of operators from one basis to another. E.g.  a vector $\hat{\psi}_{\vr}$ (i.e the set $\{\hat{\psi}_{\vr}\}$with $\vr$ regarded as an index) in the operator valued Hilbert space of functions of $\vr$ gets transformed to the vector $\{\hat{a}_i\}$ in this space with the use of the transformation matrix $\{u_{\vr,i} \}$ (with $\vr$ in $u_i(\vr)$ regarded as index). The orthogonality and completeness relations (\ref{eq:orth_compl})  of the set $\{u_i(\vr)\}$ are just  the expressions of the unitarity of the matrix $\{u_{\vr,i}\}$.   In Appendix we review the properties of the operators $\hat{a}_j$ and $\hat{a}^{+}_i$ for  a general basis set   $\{u_i(\vr)\}$ and the quantum states which they generate. 

Using the expansions Eq.\,(\ref{eq:gen_expnsn_of_fld_ops}) in the expression Eq.\,(\ref{eq:Ham_for_interact_fld}) for $H_{op}$, we obtain 
\beq \label{eq:non_inter_H_gen}
 H_{op}=  \sum_{ij}  h_{ij}\hat{a}^{+}_i \hat{a}_j
 \eeq
 where
 $$
 h_{ij} \equiv  \int d^3r \, u^{*}_i(\vr) h u_j(\vr)
 $$
  are matrix elements of $h$ in the basis ${u_i(\vr)}$.   
  
  \subsubsection{The normal modes} 
  We now choose $u_i(\vr)$'s  to be  solutions of 
  \beq \label{eq:sp_eq}
  h u_i(\vr)=\epsilon_i u_i(\vr)
  \eeq
 These solutions  are obviously the normal modes of the field described by the linear equation (\ref{eq:Sch_eq_with_int}). Indeed  in a trivial way the field configurations $\psi(\vr,t) = u_i(\vr)\exp(-i\epsilon_i t/\hbar)$  solve the (classical) field equation (\ref{eq:Sch_eq_with_int}), i.e. in each of these configurations all the field degrees of freedom (indexed by $\vr$) oscillate  with the same frequency $\epsilon_i/\hbar$.
 
We note that in the non interacting limit $U(\vr)=0$ the operator $h$ reduces to
 $$
 h_0=-\frac{\hbar^2}{2m}\nabla^2
 $$
 and $u_i(\vr)$'s become the plane waves $u_{\vk}(\vr)$, Eq.(\ref{eq:plane_wave_basis}).
 
 It is important to observe that  $h_0$ and $h$ appear as operators acting on functions of $\vr$. As such they are very different from the operator $H_{op}$ which acts on the states of the field $\hat{\psi}$, like e.g. the states Eq.\,(\ref{eq:comb_of_eignstts}).  As was already noted the field operators are on the one hand operators in the space of the states of the field (and in this role $\vr$ is just an index labelling these operators) and on the other hand they are functions of $\vr$ on which the operator $h$ acts. Perhaps a helpful analogy is the quantized EM field in which the components of $\vE_{op}(\vr)$ and $\vB_{op}(\vr)$ are both operators and functions of $\vr$.  In the present context for  reasons which will  become  clear in the sections below operators like $h_0$ and $h$ will often be called single particle operators and  the bases of functions like $u_{\vk}(\vr)$ or  $u_i(\vr)$ - single particle bases.
 
In the basis of the  eigenstates of $h$ we have
  $$
  h_{ij}=\epsilon_i\delta_{ij}
  $$
  so that as in the free field case $H_{op}$ is a sum of independent (commuting) oscillators 
 \beq \label{eq:non_inter_H}
   H_{op} =\sum_i \epsilon_i \hat{a}^{+}_i \hat{a}_i
   \eeq
 corresponding to the "vibrations" of amplitudes of the normal modes Eq.\,(\ref{eq:sp_eq}). The eigenfunctions  of $H_{op}$ are products of eigenstates $|n_i\rangle$ of these field oscillators, i.e eigenstates of the operators 
 \beq \label{eq:mode_part_num}
 \hat{n}_i = \hat{a}^{+}_i \hat{a}_i
 \eeq
  cf., Appendix, Eq.\,(\ref{eq:eigen_of_n_i}) while the eigenenergies are the corresponding sums  
 \beq \label{eq:egnstates_of_inter_fld}
 |\Psi_{\{n_i\}}\rangle\equiv |n_1, n_2, \dots, n_i, \dots\rangle = \prod_i |n_i\rangle  =\prod_i \frac{(\hat{a}^{+}_i)^{n_i}}{\sqrt{n_i!}}|0\rangle \;\;\; , \;\;\; E_{\{n_i\}}=\sum_i\epsilon_i n_i
 \eeq
To conclude, the quantization of the  \Sch field in the presence of an external potential, 
 Eq.(\ref{eq:Sch_eq_with_int}) describes collections of independent quanta of the normal modes given by the solutions of the equation (\ref{eq:sp_eq}). Since this equation is just a \Sch equation for a single particle in the potential $U(\vr)$ we therefore obtained a description of systems of such particles in this potential occupying its eigenstates $u_i(\vr)$. 
 
 \subsubsection{The particle number operator.  $U(1)$ symmetry \label{sec:part_num}}

 We note that the operators $\hat{n}_i$, Eq.\,(\ref{eq:mode_part_num}) 
 "count" the number of particles $n_i$ in each single particle state $u_i(\vr)$. We had similar operators $\hat{n}_{\vk}$ in the free field case, cf., Eq.\,(\ref{eq:free_H}).  It is useful and important to introduce the total number of particles operator
 \beq \label{eq:particle_num_op}
N_{op} = \sum_i \hat{n}_i 
\eeq
which "measures" the sum of all $n_i$'s
\beq
N_{op} |\Psi_{\{n_i\}}\rangle=N_{op}|n_1, n_2, \dots, n_i,\dots\rangle = N |n_1, n_2, \dots, n_i,\dots\rangle \;\; {\rm with} \;\; N=\sum_i n_i 
\eeq
This operator has the same form in any complete orthonormal basis 
\beq \label{eq:part_num_in_diff_bases}
N_{op} = \sum_i   \hat{a}_i^{+}\hat{a}_i =  \sum_{\vk}  \hat{a}_{\vk}^{+}\hat{a}_{\vk} = 
\int d^3 r \, \hat{\psi}^{+}(\vr) \hat{\psi}(\vr)
\eeq 
as can be verified by inserting the expansions (\ref{eq:gen_expnsn_of_fld_ops}) with different sets $u_i(\vr)$ in the last integral. 

The result that the eigenfunctions (\ref{eq:egnstates_of_inter_fld})  of the Hamiltonian  (\ref{eq:non_inter_H}) are also eigenfunctions of the number operator (\ref{eq:particle_num_op})  is linked to the fact that $N_{op}$ commutes with the Hamiltonian
 \beq \label{eq:comm_of_H_and_N}
 [H_{op}, N_{op}]=0
 \eeq
so that the particle number is a conserved  quantum number in this theory, not a fixed quantity prescribed from 
 "outside".  
 
 Let us note that $H_{op}$ commutes with the individual mode number operators $\hat{n}_i$, Eq.\,(\ref{eq:mode_part_num}). This however is only for the eigenmodes of the field, i.e. for the single particle states Eq.\,(\ref{eq:sp_eq}).  The conservation of $N_{op}$ is a much more general property independent of the basis, cf. Eq.\,(\ref{eq:part_num_in_diff_bases}). It is intuitively related to the manner in which the operators $\hat{a}_j$ and $\hat{a}_i^{+}$ enter the general Hamiltonian Eq.\,(\ref{eq:non_inter_H_gen}) and can be traced  to the way the field Hamiltonian Eq.\,(\ref{eq:Ham_for_interact_fld}) contains  the field operators $\hat{\psi}^+$ and $\hat{\psi}$.  Formally this is reflected in the invariance of the expression (\ref{eq:Ham_for_interact_fld}) under a global (coordinate independent) phase transformation 
 \beq \label{eq:glob_U(1)}
  \hat{\psi}(\vr) \; \to \; e^{i\alpha}  \hat{\psi}(\vr) \;\;\; , \;\;\; \hat{\psi}^+(\vr) \; \to \; e^{-i\alpha} \hat{\psi}^+(\vr)
 \eeq
 In a more general context such a transformation is called a global $U(1)$ gauge transformation and the operator $N_{op}$ is its generator.  This means that 
 \beq \label{eq:gauge_U(1)}
 e^{-i\alpha N_{op}} \hat{\psi}(\vr)  e^{i\alpha N_{op}} = e^{i\alpha}  \hat{\psi}(\vr) \;\; , \;\; e^{-i\alpha N_{op}} \hat{\psi}^{+}(\vr)  e^{i\alpha N_{op}} = e^{-i\alpha}  \hat{\psi}^{+}(\vr) 
 \eeq
 As usual to prove this it is sufficient to consider an infinitesimal $\alpha$. It is enough to do this  for $\hat{\psi}(\vr)$
since the relation for $\hat{\psi}^{+}(\vr)$ is just the hermitian conjugate.  We have
\beq \label{eq:inf_U1_bos}
 (1-i\alpha N_{op}) \hat{\psi}(\vr)  (1+ i\alpha N_{op})  = (1+ i\alpha)  \hat{\psi}(\vr)  \; \to \; [N_{op}, \hat{\psi}(\vr)] = -\hat{\psi}(\vr) 
 \eeq
 Simple calculation supplies the proof
 $$
  [N_{op}, \hat{\psi}(\vr)] = \int d^3 r' [ \hat{\psi}^{+}(\vr') \hat{\psi}(\vr') , \hat{\psi}(\vr)]  =
  -  \int d^3 r' \delta(\vr' - \vr) \hat{\psi}(\vr') =-\hat{\psi}(\vr)
  $$
  Using Eq.\,(\ref{eq:gauge_U(1)}), the invariance of $H_{op}$ under (\ref{eq:glob_U(1)}) and denoting
 $$
 U_{op}(\alpha)  \equiv e^{-i\alpha N_{op}}
 $$
 one has
 \eqna
  U_{op}(\alpha)  H_{op} U_{op}^{+}(\alpha) &=& U_{op}(\alpha) \left[ \int d^3r \; \hat{\psi}^{+}(\vr) h \hat{\psi}(\vr)\right]U^{+}_{op}(\alpha)  = \nonumber \\
  &=& \int d^3r \; U_{op}(\alpha)\hat{\psi}^{+}(\vr) U_{op}^{+}(\alpha)\; h \; U_{op}(\alpha)\hat{\psi}(\vr)U^{+}_{op}(\alpha) =H_{op} \nonumber  
  \eqne
  For infinitesimal $\alpha$ 
  \beq
  U_{op}(\alpha)  H_{op} U_{op}^{+}(\alpha) \to (1-i\alpha N_{op})   H_{op} (1+ i\alpha N_{op})  = ( H_{op} - i\alpha[N_{op} , H_{op} ]  ) 
  \eeq
 and to have it equal to $H_{op}$ must have Eq.\,(\ref{eq:comm_of_H_and_N}).
  
 Going back to the eigenfunctions and eigenvalues of  $H_{op}$ we note that the general solution of the \Sch equation (\ref{eq:time_dep_Sch_eq})  with this $H_{op}$ is a familiar linear combination
 \beq
 |\Psi(t)\rangle = \sum_{\{n_i\}} C_{\{n_i\}}   |\Psi_{\{n_i\}}\rangle e^{-E_{\{n_i\}} t/\hbar}
 \eeq
 with (as always) the coefficients $C_{\{n_i\}} $ determined by the initial condition for $|\Psi(t)\rangle $ at $t=0$.  And we note that the formalism in principle allows to have states with coherent combinations of different particle numbers $N=\sum_i n_i$.  The choice to have a fixed $N$, i.e. to have it the same for all components in the above solution is in the freedom of setting the appropriate initial condition supported (conserved in time) by the commutativity of $H_{op}$ with $N_{op}$.

 \subsection{Working with  the field operators}

 The last equality in the expressions (\ref{eq:part_num_in_diff_bases}) for $N_{op}$ in terms of the field operators  represents $N_{op}$ as a sum (integral) over particle number operators $d{\hat n}(\vr)=\hat{\psi}^{+}(\vr)\hat{\psi}(\vr) d^3 r$ in the infinitesimal volume $d^3r$ situated at $\vr$. This suggest that
 \beq
 \hat{\rho}(\vr) =   \hat{\psi}^{+}(\vr)\hat{\psi}(\vr)
 \eeq
 is the particle density operator. This also explains what is the physical meaning of the
  field operators $\hat{\psi}^{+}(\vr)$. Indeed let us consider  a state
 \beq \label{eq:1part_coord_st}
 |\vr'\rangle  \equiv const \,\hat{\psi}^{+}(\vr')|0\rangle
 \eeq
 where we  introduced a multiplicative constant for normalization, see below. 
  Let us act on this state with the operator $\hat{\rho}(\vr)$
 \beq
  \hat{\rho}(\vr) |\vr'\rangle = const \,\hat{\psi}^{+}(\vr)\hat{\psi}(\vr)\hat{\psi}^{+}(\vr')|0\rangle
   =const \,\delta(\vr-\vr')\hat{\psi}^{+}(\vr')|0\rangle = \delta(\vr-\vr')  |\vr'\rangle
  \eeq
where we commuted $\hat{\psi}(\vr)$ with $\hat{\psi}^{+}(\vr')$to its right and then used $\hat{\psi}(\vr) |0\rangle =0$.  The result shows that $\hat{\psi}^{+}(\vr)$ acting on the vacuum state creates a particle at the position $\vr$. More precisely it creates delta like particle density at this position.

 What happens if several $\hpsis$'s act on the vacuum?  E.g. consider the state
 \beq \label{eq:Npart_coord_st}
 |\vr_1, \dots, \vr_N\rangle = const_N \, \hpsis(\vr_1)\dots \hpsis(\vr_N)|0\rangle
 \eeq
  Let us act on this state with $\hat{\rho}(\vr)$.  As in the one particle case we find the result by first commuting $\hpsi(\vr)$ through  $\hpsis(\vr_a)$'s to its right all the way to the vacuum. This calculation will appear in several places below so we show it in details
 \eqna \label{eq:act_psi}
&&\hat{\psi}(\vr) \; \prod_{a=1}^N\hat{\psi}^{+}(\vr_a)| 0 \rangle  =[\delta(\vr-\vr_1) +\hat{\psi}^{+}(\vr_1)\hat{\psi}(\vr)] \prod_{a\ne1}^N\hat{\psi}^{+}(\vr_a)| 0 \rangle =  \nonumber \\
&&=\delta(\vr-\vr_1)\prod_{a\ne1}^N\hat{\psi}^{+}(\vr_a)  +\hat{\psi}^{+}(\vr_1)[ \delta(\vr-\vr_2)+\hat{\psi}^{+}(\vr_2)\hat{\psi}(\vr)] \prod_{a=3}^N\hat{\psi}^{+}(\vr_a) ]|0 \rangle   = \nonumber \\
&& = \delta(\vr-\vr_1)\prod_{a\ne1}^N \hat{\psi}^{+}(\vr_a) +
\delta(\vr-\vr_2)\prod_{a\ne2}^N\hat{\psi}^{+}(\vr_a) +  \dots +  \nonumber \\
&& + \prod_{a=1}^{N-1}\hat{\psi}^{+}(\vr_a)[\delta(\vr-\vr_N) + \hat{\psi}^{+}(\vr_N)\hat{\psi}(\vr)]|0\rangle
=  \nonumber \\
&& \;\;\;\;\;\;\;\;\;\;\;\;\;\;\;\;\;\;\;\;\; = \left[\sum_{b=1}^N \delta(\vr-\vr_b) \prod_{a\ne b}^N\hat{\psi}^{+}(\vr_a)\right]|0\rangle
\eqne
Acting on this with $\hpsis(\vr)$ and using the delta function in each term to replace $\vr \to \vr_b$ in it we get
 $$
 \hat{\rho}(\vr)  |\vr_1, \dots, \vr_N\rangle = [\sum_{a=1}^N \delta(\vr-\vr_a) ] |\vr_1, \dots, \vr_N\rangle
 $$
 i.e. have $N$ particles (delta like particle densities) at  the positions $\vr_a, a=1, \dots, N$.
 In the same manner one can show that
 $\hat{\psi}^{+}(\vr)$ creates a particle at $\vr$ when it acts on any general state (discussed below). We also note that the result (\ref{eq:act_psi}) shows that $\hat{\psi}(\vr)$ destroys (annihilates)  a particle if its coordinates coincide with $\vr$.

 It is important to notice that $\hpsis$ and $\hpsi$ create and annihilate particles only when they  act to the right. Acting to the left they produce an opposite result - they correspondingly annihilate and create particles.
 For example the state $\langle\vr_1, \dots, \vr_N|$ is the hermitian conjugate of   $|\vr_1, \dots, \vr_N\rangle$ so
 \beq
 \langle \vr_1, \dots, \vr_N | = [const_N \ \hpsis(\vr_1)\dots \hpsis(\vr_N)|0\rangle]^{+} = \langle 0 | \hpsi(\vr_N)\dots \hpsi(\vr_1) (const_N)^{*}
 \eeq
 since $[\hpsis]^{+} = \hpsi$.  Thus the state $ \langle \vr_1, \dots, \vr_N |$ is the result of acting with $N$ $\hpsi$'s to the left on the vacuum $\langle 0|$.

 What is  the norm of  $|\vr_1, \dots, \vr_N\rangle$? Take as an example one particle state Eq.\,(\ref{eq:1part_coord_st})
 and calculate
 $$
  \langle \vr'|\vr\rangle = |const|^2\langle 0|\hpsi(\vr')\hpsis(\vr)|0\rangle = |const|^2\delta(\vr-\vr')
 $$
 The result shows that such a state is non normalizable.  This should not be surprising as one has a continuum of states labeled by $\vr$. 
 Just as with more familiar momentum states labeled by $\vp$.   Also  the momentum  states are non normalizable. The common regularization  is to make $\vp$ discrete by introducing very large but finite volume, i.e. to introduce an infrared cutoff.  In the same way one can make $\vr$ discrete by introducing a lattice of discrete $\vr$'s. If this is not done - then one can normalize as convenient. 
 
 As we will see in the next section the most common use of the states $|\vr\rangle$ or their  $N$ particle generalization $ |\vr_1, \dots, \vr_N\rangle$, Eq.\,(\ref{eq:Npart_coord_st}),  makes it  convenient to choose the normalization of these states as
$$
const_N = \frac{1}{\sqrt{N!}}
$$

 \section{Wave functions. Operators. Comparison with the first quantized description}

The quantization of the \Sch field is (for obvious reasons) called the second quantization. For the field governed by  Eq.(\ref{eq:Sch_eq_with_int}) this seems to result in an alternative description of quantum non interacting  particles in the external potential $U(\vr)$.

Here we want to understand if this description is indeed complete and how it is related to the standard quantum mechanical description of say $N$ particles with the wave function $\Phi(\vr_1, \vr_2, \dots, \vr_N, t)$ obeying the $N$ particle \Sch equation
 \beq \label{eq:sch_eq_1st_q}
 i\hbar \frac{\partial \Phi(\vr_1, ..., \vr_N,t)}{\partial t} =  \sum_{a=1}^N\left[ -\frac{\hbar^2}{2m}\nabla_a^2 + U(\vr_a) \right] \Phi(\vr_a,..., \vr_N, t)
 \eeq

 \subsection{Wave functions in the second quantization}

 \subsubsection{Coordinate representation. Second vs first quantization} 

The states $|\vr_1, \dots, \vr_N\rangle$ introduced in the previous section, cf., Eq. (\ref{eq:Npart_coord_st}), form a very convenient basis to represent a general N particles wave function in the second quantization
\beq  \label{eq:wf_in_sq_0}
|\Phi\rangle = \int \prod_{a=1}^N d^3r_a \;\Phi(\vr_1,\vr_2,\dots,\vr_N)|\vr_1, \dots, \vr_N\rangle
 \eeq
  The interpretation of this expression is quite clear - we have a linear combination of $N$ particles in different coordinate positions $\vr_1, \dots, \vr_N$   weighted each with the probability amplitude $\Phi(\vr_1,\vr_2,\dots,\vr_N)$.  These amplitudes form the wave function $|\Phi\rangle$ in the coordinate representation and clearly are equivalent to this wave function in the first quantization formalism.  We will see this equivalence even more explicitly  in  the discussions below of how physical operators of particle observables act on $|\Phi\rangle$.
  
As discussed in the Appendix in order to have both $ |\Phi\rangle$ and  $\Phi(\vr_1, \dots, \vr_N)$  normalized to unity, i.e. to have
\beq \label{eq:both_wf_norm_unity}
 \int d^3r_1 \dots d^3 r_N |\Phi(\vr_1,\vr_2,\dots,\vr_N)|^2 =1 \;\;\;\;\;
 {\rm and } \;\;\;\; \langle \Phi |\Phi\rangle = 1
 \eeq
one must choose the normalization $const_N =1/\sqrt{N !}$ in the definition (\ref{eq:Npart_coord_st}) of the states $|\vr_1, \dots, \vr_N\rangle$ as they appear in the relation (\ref{eq:wf_in_sq_0}) between $ |\Phi\rangle $ and   $ \Phi(\vr_1,\vr_2,\dots,\vr_N)$.  We thus have
 \beq \label{eq:wf_in_sq}
 |\Phi\rangle = \frac{1}{\sqrt{N!}}\int \prod_{a=1}^N d^3r_a \;\Phi(\vr_1,\vr_2,\dots,\vr_N)\hat{\psi}^{+}(\vr_1)\hat{\psi}^{+}(\vr_2)\dots \hat{\psi}^{+}(\vr_N)| 0 \rangle
 \eeq
  
\subsubsection{Permutation symmetry \label{sec:perm_sym_bos}}

 The commutativity properties of the field operators $\hat{\psi}^{+}(\vr)$   imply that the coordinate probability   amplitudes $\Phi(\vr_1,\vr_2,\dots,\vr_N)$ in Eq.\,(\ref{eq:wf_in_sq}) can not be arbitrary.  These functions must be symmetric under all possible permutations of the particles' coordinates. 
 
 Let us demonstrate this for the simplest case of two particles
 \beq \label{eq:Phi_of_two}
 |\Phi\rangle \equiv \frac{1}{\sqrt{2}} \int d^3r_1 d^3r_2 \Phi(\vr_1, \vr_2) \hat{\psi}^{+}(\vr_1)\hat{\psi}^{+}(\vr_2)|0\rangle
 \eeq
 Functions of two variables can belong to one of the two symmetry representations - symmetric or antisymmetric,
  $$
  \Phi_S (\vr_1, \vr_2) = \Phi_S (\vr_2, \vr_1) \;\; {\rm and} \;\;   \Phi_A (\vr_1, \vr_2) = -\Phi_A (\vr_2, \vr_1)
  $$ 
  and in general have
 $$
  \Phi(\vr_1, \vr_2) = \frac{1}{2}[ \Phi(\vr_1, \vr_2) +  \Phi(\vr_2, \vr_1)] +\frac{1}{2}[ \Phi(\vr_1, \vr_2) - \Phi(\vr_2, \vr_1)] \equiv \Phi_S (\vr_1, \vr_2) + \Phi_A (\vr_1, \vr_2)
  $$
It is straightforward to show that $|\Phi\rangle_A$ obtained with  $\Phi_A(\vr_1,\vr_2)$ in Eq. (\ref{eq:Phi_of_two}) vanishes identically. Have
 \eqna
 |\Phi\rangle_A=   \frac{1}{\sqrt{2}}\int d^3r_1 d^3r_2 \Phi_A(\vr_1, \vr_2) \hat{\psi}^{+}(\vr_1)\hat{\psi}^{+}(\vr_2)|0\rangle =
 \nonumber \\
 = -  \frac{1}{\sqrt{2}}\int d^3r_1 d^3r_2 \Phi_A(\vr_2, \vr_1) \hat{\psi}^{+}(\vr_1)\hat{\psi}^{+}(\vr_2)|0\rangle = \nonumber \\
= -  \frac{1}{\sqrt{2}}\int d^3r_1 d^3r_2 \Phi_A(\vr_2, \vr_1) \hat{\psi}^{+}(\vr_2)\hat{\psi}^{+}(\vr_1)|0\rangle  =  \nonumber \\
= -  \frac{1}{\sqrt{2}}\int d^3r_1 d^3r_2 \Phi_A(\vr_1, \vr_2) \hat{\psi}^{+}(\vr_1)\hat{\psi}^{+}(\vr_2)|0\rangle  = -|\Phi\rangle_A \nonumber
\eqne
where in the 3rd line we commuted $\hat{\psi}^{+}(\vr_1)\hat{\psi}^{+}(\vr_2) = \hat{\psi}^{+}(\vr_2)\hat{\psi}^{+}(\vr_1)$ and in the 4th line have interchanged the integration variables $\vr_1 \leftrightarrow \vr_2$. So we have proved that $|\Phi\rangle_A =-|\Phi\rangle_A $ which means that $|\Phi\rangle_A =0$.

The same proof obviously holds for any pair of coordinates in a general wave function $\Phi(\vr_1, \dots, \vr_N)$. Thus only  $\Phi(\vr_1, \dots, \vr_N)$'s  which are symmetric with respect to permutation of any two particles produce non zero result in Eq.\,(\ref{eq:wf_in_sq}).  This means that this is true also for $\Phi(\vr_1, \dots, \vr_N)$'s which are symmetric under permutations of any number of particles. Indeed (as is simple to understand\footnote{Cf, Messiah, Quantum Mechanics (Dover Books in Physics), Ch. XIV. Denote for example by $(1532476)$ a permutation $1\to5\to3\to2\to4\to7\to6\to1$. It can clearly be written as an ordered product $(15)(53)(32)(24)(47)(76)$ of transpositions (with right to left order)} and can be proved by induction) any such permutation can be decomposed into a product of permutations of two particles (transpositions).

The above symmetry under permutations of the wave functions is one of the most important features of the second quantization formalism. Together with the symmetry of the physical observables as represented by the operators as discussed below this property means that the quanta of the theory are bosons, i.e. identical particles obeying Bose statistics.  We will provide more details to this discussion in Section\,\ref{sec:part_statistics}.   

 \subsubsection{Occupation number representation}
 Expanding the field operators in Eq.\,(\ref{eq:wf_in_sq}) in an arbitrary complete and orthonormal single particle basis, cf., Eq.\,(\ref{eq:gen_expnsn_of_fld_ops}), we obtain
 \beq \label{eq:expan_in_prodN}
  |\Phi\rangle = \sum_{i_1,..., i_N} C_{i_1, ... , i_N} \hat{a}^{+}_{i_1} ... \hat{a}^{+}_{i_N}|0 \rangle
  \eeq
  with
  $$
 C_{i_1, ... , i_N} = \frac{1}{\sqrt{N!}}\int \prod_{a=1}^N d^3r_a \;\Phi(\vr_1,\vr_2,\dots,\vr_N)u ^{*}_{i_1}(\vr_1)\dots u^{*}_{i_N}(\vr_N)
 $$
 The coefficients $C_{i_1, ... , i_N} $ represent the function  $|\Phi\rangle$ in the basis of products of the single particle states   $u_i(\vr)$.   As we discussed in the previous Section the  functions  $\Phi(\vr_1,\vr_2,\dots,\vr_N)$ are symmetric with respect to permutations of the particle coordinates $\vr_a$'s.  One can use this to replace the products  of  $u_i(\vr)$'s in the above expression for $C_{i_1, ... , i_N}$'s by symmetrized products   
$$ 
u ^{*}_{i_1}(\vr_1)\dots u^{*}_{i_N}(\vr_N) \to const\,\sum_{P} u ^{*}_{i_1}(\vr_{p_1})\dots u^{*}_{i_N}(\vr_{p_N})
$$
with $P$ denoting the permutations of  particle coordinates $\vr_1, ... , \vr_N \to \vr_{p_1}, ... ,\vr_{p_N}$, the normalization constant 
$$
const = \sqrt{n_1! ... n_N! / N!}
$$
and appropriate adjustment of the expansion constants $C_{i_1, ... , i_N}$.  It is a useful exercise to work this out starting with the simple $N=2$ case, writing $u ^{*}_{i_1}(\vr_1)u ^{*}_{i_2}(\vr_2)$ as a sum of symmetric and antisymmetric products, with the antisymmetric part vanishing in the integral of its product and the symmetric $\Phi(\vr_1,\vr_2)$.

 It is useful and conventional to write the expansion (\ref{eq:expan_in_prodN})  using the notation of Eq.\,(\ref{eq:egnstates_of_inter_fld}) with the occupation numbers $n_i$ of the single particle states. In this representation the state $|\Phi\rangle$ will be written as
 \beq  \label{eq:wf_in_genb}
 |\Phi\rangle = \sum_{n_1, ..., n_i, ... ; {\rm with} \sum_i n_i=N} C_{n_1, ... , n_i, ...}|n_1, n_1, ..., n_i, ...\rangle
 \eeq
 with appropriate adjustment of the coefficients $C_{n_1, ... , n_i, ...}$. Such representation of the $N$ particles  wave functions is called  \textbf{occupation number representation}. It emphasises the fact that we are dealing with identical quanta (particles) so that all one needs is their numbers $n_i$ in each single particle  state. Note that in this representation one must "supply" infinite (actually $\infty^3$) set of (positive) integers $n_i$. But since they are subject to the constraint $\sum_i=N$ only $\le N$ of them are not zero.

\subsection{Operators in the second quantization}
In this section we want to establish how the operators of the physical observables act on  wave functions in the second quantization formalism.  In this way we will also understand much better the connection with the first quantization.

\subsubsection{The one body Hamiltonian}

We will start with the discussion of the action on  $|\Phi\rangle$  by the  Hamiltonian  (\ref{eq:Ham_for_interact_fld}).
Let us write it as a sum of two terms - kinetic and potential
\eqna
 H_{op} &=& K_{op} + U_{op}  \\
 K_{op}&=&\int d^3r \; \hat{\psi}^{+}(\vr)\left(-\frac{\hbar^2}{2m}\nabla^2 \right)\hat{\psi}(\vr) \;\;\; , \;\;\; U_{op} = \int d^3 r \hat{\psi}^{+}(\vr) U(\vr)\hat{\psi}(\vr)  \nonumber
 \eqne
 and let us consider first the action on $|\Phi\rangle$ of the potential part
 \beq
 U_{op}|\Phi\rangle=  \frac{1}{\sqrt{N!}}\int \prod_{a=1}^N d^3r_a \Phi(\vr_1,\vr_2,\dots,\vr_N)
 \int d^3 r \hat{\psi}^{+}(\vr) U(\vr)\hat{\psi}(\vr) \; \prod_{a=1}^N\hat{\psi}^{+}(\vr_a)| 0 \rangle
\eeq
Using the result Eq.\,(\ref{eq:act_psi}), multiplying it by  $U(\vr)$ and $\hat{\psi}^{+}(\vr)$ and doing the $d^3r$ integral
with the help of the $\delta$-functions we get
\beq  
 \int d^3 r \hat{\psi}^{+}(\vr) U(\vr)\hat{\psi}(\vr) \; \prod_{a=1}^N\hat{\psi}^{+}(\vr_a)| 0 \rangle  =
 \left[\sum_{b=1}^N U(\vr_b)\right] \prod_{a=1}^N\hat{\psi}^{+}(\vr_a)| 0 \rangle
\eeq
 so that
 \beq 
 U_{op}|\Phi\rangle=  \frac{1}{\sqrt{N!}}\int \prod_{a=1}^N d^3r_a \Phi'(\vr_1,\vr_2,\dots,\vr_N)
 \prod_{a=1}^N\hat{\psi}^{+}(\vr_a)| 0 \rangle
 \eeq
 with
 \beq \label{eq:pot_en_1st_quan}
  \Phi'(\vr_1,\vr_2,\dots,\vr_N)  =  \left[\sum_{a=1}^N U(\vr_a)\right]  \Phi(\vr_1,\vr_2,\dots,\vr_N)
  \eeq
We see that the action of the second quantized operator $U_{op}$ on $|\Phi\rangle$  is equivalent to the  action of the first quantized $\sum_a U(\vr_a)$ on $\Phi(\vr_1,\dots,\vr_N)$ i.e. on the first quantized partner of $|\Phi\rangle$.

To calculate the action of $K_{op}$ on $|\Phi\rangle$ is a bit more involved but straightforward.  The details are given in the Appendix \ref{sec:K_act_on_Phi} with the result 
\beq
K_{op}|\Phi\rangle= \frac{1}{\sqrt{N!}} \int \prod_{a=1}^N d^3r_a \Phi'(\vr_1,\vr_2,\dots,\vr_N) \prod_{a}^N\hat{\psi}^{+}(\vr_a)|0\rangle
\eeq
with
\beq \label{eq:kin_en_1st_quan}
 \Phi'(\vr_1,\vr_2,\dots,\vr_N)  = \left[\sum_{b=1}^N
\left(-\frac{\hbar^2}{2m}\nabla^2_{\vr_b}\right)\right] \Phi(\vr_1,\vr_2,\dots,\vr_N)
\eeq
As in the $U_{op}$ case we see that the action of $K_{op}$ on $|\Phi\rangle$ is  equivalent to the  action of the first quantized kinetic energy operator
 $$
 \sum_{a=1}^N \left(-\frac{\hbar^2}{2m}\nabla^2_{\vr_b}\right)
 $$
 on the first quantized partner $\Phi(\vr_1,\vr_2,\dots,\vr_N)$ of $|\Phi\rangle$

Combining these results we find that
\beq
H_{op}|\Phi\rangle =(K_{op} + U_{op})|\Phi\rangle =   \frac{1}{\sqrt{N!}} \int \prod_{a=1}^N d^3r_a \left[\sum_{a=1}^N h_a\right]\Phi(\vr_1,\vr_2,\dots,\vr_N) \prod_{a}^N\hat{\psi}^{+}(\vr_a)|0\rangle
\eeq
with the single particle hamiltonian $h$ given by Eq.\,(\ref{eq:sp_hamiltonian})

\subsubsection{Other one body operators}

The operators $K_{op}$, $U_{op}$ and $H_{op}$ discussed above are all of the type which in the first quantization formulation have the form
\beq \label{eq:one_body_op}
\hat{F}^{(1)}=\sum_{a=1}^N f ^{(1)} _a
\eeq
with each $f^{(1)} _a$ being a function of $\vr_a$ and $\vp_a=-i\hbar\nabla_a$. Such operators act on wave functions of $N$ particles but at one particle at a time. They are called one-body operators and the subscript which we attached to $\hat{F}^{(1)}$ and $f^{(1)}$ serves to make this distinction.

On the basis of our above discussion of the operators $K_{op}$, $U_{op}$ and $H_{op}$ we can make a general statement  that in the second quantization one body operators have the form
\beq \label{eq:one_body_op_sq}
F_{op}^{(1)}=\int d^3 r \hat{\psi}^{+}(\vr) f^{(1)} \hat{\psi}(\vr)
\eeq
where $f^{(1)}$ in the last expression is one (any) of the operators in the sum  (\ref{eq:one_body_op}) and it is acting on $\hat{\psi}(\vr)$ as a function of $\vr$.  E.g. angular momentum
\beq \label{ang_mom_1st_quant}
\vL = \sum_{a=1} ^N \vl_a \equiv \sum_{a=1}^N \vr_a \times (-i\hbar\nabla_{\vr_a})
\eeq
 becomes
\beq
\vL_{op} = \int d^3 r \hat{\psi}^{+}(\vr) [ \vr \times (-i\hbar\nabla_{\vr})] \hat{\psi}(\vr)
\eeq
in the second quantization formalism.

It is important to observe that the particle number $N$ which appears in the operators in the 1st quantization Eq.\,(\ref{eq:one_body_op}) is a part of their definition while the corresponding operators in the second quantization do not contain any information about $N$.  It is the wave functions on which these operators act,  like $|\Psi\rangle$ in the previous section which depend on  $N$. The second quantization $F_{op}^{(1)}$'s "are ready
 to act" on  $|\Psi\rangle$ with any value of $N$ including a linear combination with different $N$'s (see the section below on the general Fock space). At the same time these particular type of operators  do not change $N$ since they contain an  equal number of creation and annihilation operators - one of each type. But nothing intrinsically in the formalism prevents having operators which change $N$. In fact the elementary ones $\hat{\psi}(\vr)$  and $\hat{\psi}^{+}(\vr)$  do just that.

Formally the conserving $N$ property of the operators $F_{op}^{(1)}$ is expressed by  their commutativity with the particle number operator $N_{op}=\int d^3 r \hat{\psi}^{+}(\vr) \hat{\psi}(\vr) $ ,
 \beq
 [F_{op}^{(1)}, N_{op}]=0
 \eeq
which as in the case of $H_{op}$, Eq.\,(\ref{eq:comm_of_H_and_N}) follows from the invariance of Eq.\,(\ref{eq:one_body_op_sq}) with respect to the global $U(1)$ transformation Eq.\,(\ref{eq:glob_U(1)}). 

From our derivations in the previous section it should also be clear in details the "mechanics" of how the one body second quantized operators act on functions like $|\Psi\rangle$. Pictorially one can say that first the destruction operator $ \hat{\psi}(\vr) $ acts on $|\Psi\rangle$ "seeking out" all the particles at their position $\vr_a$,  $a=1,\dots,N$ and "annihilating" them one at a time. The result depends on the coordinates of the  particle positions. The operator $\hat{f}$ then acts on these coordinates and then the  operator $\hat{\psi}^{+}(\vr)$ puts the particles back ("creates" them) where they originally were.  All this gets weighted with the probability amplitude $\Phi(\vr_1, \vr_2,\dots,\vr_N)$ and integrated over all possible $\vr_a$'s. 

 \subsubsection{Particle interactions. Two body operators} 
 Let us now understand how do we write in the second quantization the operators which represent interactions between particles. The most common such operators are potential energy which is a sum of all pairwise interactions  (e.g. Coulomb interaction). Their form in the first quantization is
\beq \label{eq:two_body_int}
V=\frac{1}{2}\sum_{a,b=1, a\ne b}^N V(\vr_a - \vr_b)
\eeq
 As we see this is a sum of operators with each acting on two particles at a time. Such operators are called two body operators.

 Based on the experience of the previous sections it is not difficult to guess that the following corresponding expression holds in the second  quantization
 \beq \label{eq:22nd_q_interctn}
 V_{op} = \frac{1}{2}\int d^3 r d^3 r' \hat{\psi}^{+}(\vr)\hat{\psi}^{+}(\vr')V(\vr-\vr')\hat{\psi}(\vr')\hat{\psi}(\vr)
 \eeq
 To verify this guess let us do what we did with one body operators - let us act with this expression on the general N particle wave function in Eq.\,(\ref{eq:wf_in_sq}).
  \eqna
 V_{op}|\Phi\rangle&=&  \frac{1}{\sqrt{N!}}\int \prod_{a=1}^N d^3r_a \Phi(\vr_1,\vr_2,\dots,\vr_N)  \times  \\
&& \times  \frac{1}{2}\int d^3 r d^3 r' \hat{\psi}^{+}(\vr)\hat{\psi}^{+}(\vr')V(\vr-\vr')\hat{\psi}(\vr')\hat{\psi}(\vr) \prod_{a=1}^N\hat{\psi}^{+}(\vr_a)| 0 \rangle   \nonumber
\eqne
To evaluate this we use  the relation Eq.\,(\ref{eq:act_psi}), act on it with  $\hat{\psi}(\vr')$ and obtain
$$
\hat{\psi}(\vr')\hat{\psi}(\vr) \; \prod_{a=1}^N\hat{\psi}^{+}(\vr_a)| 0 \rangle  =
\left[\sum_{b\ne c}^N \delta(\vr'- \vr_c) \delta(\vr-\vr_b) \prod_{a\ne b,c}^N\hat{\psi}^{+}(\vr_a)\right]|0\rangle  \nonumber
$$
Using this we find
\eqna
 &&\frac{1}{2}\int d^3 r d^3 r' \hat{\psi}^{+}(\vr)\hat{\psi}^{+}(\vr')V(\vr-\vr')\hat{\psi}(\vr')\hat{\psi}(\vr) \prod_{a=1}^N\hat{\psi}^{+}(\vr_a)| 0 \rangle =  \nonumber  \\
&&\frac{1}{2}\int d^3 r d^3 r' \hat{\psi}^{+}(\vr)\hat{\psi}^{+}(\vr')V(\vr-\vr')\left[\sum_{b,c=1, b\ne c}^N \delta(\vr'- \vr_c) \delta(\vr-\vr_b) \prod_{a\ne b,c}^N\hat{\psi}^{+}(\vr_a)\right]|0\rangle  = \nonumber \\
 &&\;\;\;\;\;\;\;\;\;\;\;\;\;\;\;\;\;\;\; = \frac{1}{2}\sum_{b,c=1, b\ne c}^N V(\vr_b - \vr_c) \prod_{a=1}^N\hat{\psi}^{+}(\vr_a)|0\rangle   \nonumber
  \eqne
  and therefore
  \beq
  V_{op}|\Phi\rangle = \frac{1}{\sqrt{N!}} \int \prod_{a=1}^N d^3r_a\left[ \frac{1}{2}\sum_{b,c=1, b\ne c}^N V(\vr_b - \vr_c) \right]\Phi(\vr_1,\vr_2,\dots,\vr_N)  \prod_{a=1}^N\hat{\psi}^{+}(\vr_a)| 0 \rangle
  \eeq
  So indeed the action of $V_{op}$ on $|\Phi\rangle$ is equivalent to/results in the action of the first quantized $V$, Eq.\,(\ref{eq:two_body_int})  on $ \Phi(\vr_1,\vr_2,\dots,\vr_N) $

 The intuitive understanding of the expression (\ref{eq:22nd_q_interctn}) is similar to what we saw in the one body operators case - the operators $\hat{\psi}(\vr')\hat{\psi}(\vr)$ "search" to annihilate two particles  (as $\vr$ and $\vr'$ are integrated over) and thereby "reveal" their position. The function $V(\vr-\vr')$ weighs the result while the operators $\hat{\psi}^{+}(\vr)\hat{\psi}^{+}(\vr)$ put the particles back. All this is integrated over all possible positions $\vr$ and $\vr'$.

 The general form of the two body operator in the first quantization is
 \beq \label{1st_quan_2_body}
 \hat{F}^{(2)} = \frac{1}{2} \sum_{a,b=1, a\ne b}^N f^{(2)}_{ab}
 \eeq
with the second quantized counterpart
\beq \label{eq:two_body_op_sq}
F^{(2)}_{op}= \frac{1}{2}\int d^3 r d^3 r' \hat{\psi}^{+}(\vr)\hat{\psi}^{+}(\vr')f^{(2)}\hat{\psi}(\vr')\hat{\psi}(\vr)
 \eeq
 with  $f^{(2)}$ in general being a function of $\vr, \vr'$ and $\hat{\vp}=-i\hbar\nabla_{\vr}, \hat{\vp'}=-i\hbar\nabla_{\vr'}$.

 It is important to observe that also here as with the one body operators  the second quantized  operators do not contain any information about the number $N$ of the particles which is encoded in the wave functions on which these operators act.
 Also here the operators $F_{op}^{(2)}$ do not change the value of $N$ and commute with the particle number operator
 \beq
 [F_{op}^{(2)}, N_{op}]=0
 \eeq
 which is a consequence of the $U(1)$ unitary symmetry Eq.\,(\ref{eq:glob_U(1)}) which the operators $F_{op}^{(2)}$ posses.

 \subsubsection{Changing the single particle basis}
 The one and two body operators discussed above were expressed in terms of the basic field operators $\hat{\psi}(\vr)$ and $\hat{\psi}^{+}(\vr)$. It is easy and instructive to express them using the expansion  (\ref{eq:gen_expnsn_of_fld_ops}) of these operators in a general single particle basis $\{u_i(\vr)\}$. Inserting   (\ref{eq:gen_expnsn_of_fld_ops}) into the expressions (\ref{eq:one_body_op_sq}) and (\ref{eq:two_body_op_sq}) we obtain
 \eqna \label{eq:one_body_sq_1}
 F_{op}^{(1)} &=& \sum_{ij}\langle i|f^{(1)}|j\rangle \hat{a}^{+}_i \hat{a}_j  \\
 F_{op}^{(2)} &=& \sum_{ijkl}\langle ij|f^{(2)}|kl\rangle \hat{a}^{+}_i \hat{a}^{+}_j \hat{a}_l \hat{a}_k \label{eq:two_body_sq_1}
 \eqne
 where we used the notation for the matrix elements of elementary one and two body operators
 \eqna
\langle i|f^{(1)}|j\rangle &=& \int d^3r u^{*}_i(\vr)f^{(1)}u_j (\vr) \\
 \langle ij|f^{(2)}|kl\rangle &=&\int d^3r d^3 r' u^{*}_i(\vr)u^{*}_j(\vr')f^{(2)}u_k (\vr)u_l (\vr')
 \eqne
We draw attention to the "logic" of how the operators $F_{op}^{(1)}$ and $F_{op}^{(2)}$ in a general single particle basis  act on a wave function in this basis as written in the occupation number representation of  Eq.\,(\ref{eq:wf_in_genb}).  In the one body $F_{op}^{(1)}$ one starts with the operator $\hat{a}_j$ annihilating a particle in a (single particle basis) state $u_j$ reducing the corresponding $n_j$ occupation to $n_j-1$.  This is "weighted" with a corresponding amplitude to find this $n_j$ as encoded in the coefficients $C_{n_1, ... , n_j, ...}$ of the occupation number representation.  The following action of the operator $\hat{a}^{+}_i$ creates (puts back) a particle in the state $u_i$ and the result gets multiplied by the transition matrix element $\langle i|f^{(1)}|j\rangle$.  At the end one sums over all such transitions. The two body $F_{op}^{(2)}$ operates in a similar fashion but with two particles annihilation and creation and the sum over all two particle transitions.

 \subsection{Second quantization via commutators describes identical bosons \label{sec:part_statistics}}
The following important features of the above formalism  must be observed at this stage.  The first quantization operators which are counterparts of the operators in the second quantization are always \emph{symmetric sums} over all the particles or their pairs etc in the wave functions on which they act.  

The symmetry of these sums follows since all their terms are identical in acting on different particles. They have the same functional dependence on the coordinates and momenta with the same parameters - masses, charges, etc., e.g. same kinetic energy, same external potential, same inter-particle interactions, etc.  This is seen in the formal correspondence Eq.\,(\ref{eq:one_body_op}) $\,\to\,$ Eq.\,(\ref{eq:one_body_op_sq}) and Eq.\,(\ref{1st_quan_2_body}) $ \, \to \,$ Eq.\,(\ref{eq:two_body_op_sq})  and in the explicit examples in Eqs.\,(\ref{eq:pot_en_1st_quan}, \ref{eq:kin_en_1st_quan}, \ref{ang_mom_1st_quant}, \ref{eq:two_body_int}).   There is no possibility to have second quantized operators representing observables distinguishing  a particular particle, say $f^{(1)}_5$ or sets of particular particles, e.g $f^{(1)}_7  + f^{(1)}_{15}$.  This is a general feature of quantum systems of identical particles.

Let us also recall that as we discovered in Section \ref{sec:perm_sym_bos} the first quantization wave functions $\Phi(\vr_1, ..., \vr_N)$ which are counterparts of the second quantization $|\Phi\rangle$ are symmetric under the permutation of all the particles coordinates. This confirms the particle being identical and moreover obeying the spin-statistics theorem requirements for systems of  bosons.  

 Let us remind that the spin-statistics theorem, proved by Pauli, states  (in its first part) that the wave functions of a system of identical integer-spin particles must be symmetric under the exchange of the coordinates of any two particles. Such particles are bosons obeying the Bose-Einstein statistics. 
 
In our case of the  identical particles without spin the symmetry requirement dictated by the Pauli theorem is an additional rule which is imposed in the first quantization formalism on selecting the wave function solutions of the \Sch equation (\ref{eq:sch_eq_1st_q}).  As we have seen it is automatically fulfilled in the second quantization wave functions Eqs.\,(\ref{eq:wf_in_sq_0},\ref{eq:wf_in_sq}). 

 The second part of the spin-statistics theorem concerns wave functions of system of identical half-integer spin particles.  The theorem states that they  must be anti-symmetric under any pair of particle exchange. Such particles are fermions obeying the Fermi-Dirac statistics.   We will discuss in the next section how the second quantization allows for a simple and straightforward modification to be extended to the descriptions of fermions. 
 
Concerning the proof of the spin-statistics theorem - as Feynman states in his Lectures on Physics:  ''...An explanation has been worked out by Pauli from complicated arguments of QFT and relativity...but we haven't  found a way of reproducing his arguments on an elementary level..."\footnote{cf., I. Duck and E. Sudarshan, Towards an understanding of the spin-statistics theorem, Am. J. Phys., 66 (4) 1998}.

\subsection{Self interacting \Sch field.}

\subsubsection{Summing up. Interacting Hamiltonian}

To summarize we learned how to translate the wave functions and physical operators into the second quantization formalism. The $N$ particle wave function of $N$ bosons $\Phi(\vr_1, ..., \vr_N)$ should be symmetric and becomes the amplitude of the $|\vr_1, ..., \vr_N\rangle $ state in the expression (\ref{eq:wf_in_sq_0}) or the more explicit (\ref{eq:wf_in_sq}).

Dealing with operators one should first determine to which type they belong - one body, two body, etc.
Examples of one body operators are momentum, density, current, etc
\eqna
\vP &=& \sum_{a=1}^N \vp_a =\sum_{a=1}^N (-i\hbar\nabla_{\vr_a}) \;\;, \;\; \rho (\vr) = \sum_{a=1}^N \delta(\vr - \vr_a) \nonumber \\
 \vj (\vr) &=&  \frac{1}{2m}\sum_{a=1}^N[ \delta(\vr - \vr_a)(-i\hbar\nabla_{\vr_a})+ (-i\hbar\nabla_{\vr_a})\delta(\vr - \vr_a)] \nonumber
\eqne
Note that identical particles imply that any such operator is a sum of identical operators acting on each particle. So one takes one member of the sum and uses it in the expression (\ref{eq:one_body_op_sq}) to find the corresponding 2nd quantized operator. If one prefers a general basis $\{u_i\}$ rather than the coordinate representation  of the  field operators one needs to calculate  the matrix elements between all possible pairs of $u_i$'s and use them in the expression (\ref{eq:one_body_sq_1}). The number $N$ of particles appears  explicitly in the operators of the first quantized formalism but not in the 2nd quantization.

There are not too many examples of two body operators. Beside the two body interaction (\ref{eq:two_body_int}) there are various correlators like density-density or current-current, etc
$$
\rho_{op}(\vr)\rho_{op}(\vr') =  \sum_{a,b=1}^N \delta(\vr - \vr_a) \delta(\vr' - \vr_b) \;\;, {\rm etc}
$$
As in the one body case one must take one term in such a double sum and either use it in the expression (\ref{eq:two_body_op_sq}) with field operators or calculate all its  two particle matrix elements in a chosen basis of the single particle states $u_i$'s.  One should then form an expression (\ref{eq:two_body_sq_1}) using these matrix elements.

As a rule it is extremely rare to find 3 body operator but it is straightforwardly clear how to extend what we have learned to such cases.

Let us follow the above rules to write the full 2nd quantization Hamiltonian of a many body interacting  system. Consider its (most common) expression in the 1st quantization
\beq \label{eq:1st_quantized_H}
\hat{H} =    \sum_{a=1}^N \left[  \frac{\hat{\vp}_a^2}{2m} + U(\vr_a)\right] + \frac{1}{2}\sum_{a,b=1, a\ne b}^N V(\vr_a - \vr_b)  \;\;\; , \;\;\; \hat{\vp}_a = -i\hbar \nabla_a
\eeq
with externally fixed number $N$ of the particles. Assuming that the particles are bosons their particles statistics must be imposed "by hand"  allowing only symmetric wave functions.

The 2nd quantized version of the above Hamiltonian is
\beq \label{eq:interact_Ham_2nd_q}
H_{op} = \int d^3 r\hat{\psi}^{+}(\vr)\, h \, \hat{\psi}(\vr) + \frac{1}{2}\int d^3 r d^3 r' \hat{\psi}^{+}(\vr)\hat{\psi}^{+}(\vr')V(\vr-\vr')\hat{\psi}(\vr')\hat{\psi}(\vr)
\eeq
with
$$
h =   -\frac{\hbar^2}{2m}\nabla^2 + U(\vr)
$$
In a general single particle basis this Hamiltonian is written
\beq
H_{op} = \sum_{ij}\langle i|\, h \,|j\rangle \hat{a}^{+}_i \hat{a}_j +
\frac{1}{2}\sum_{ijkl}\langle ij|V|kl\rangle \hat{a}^{+}_i \hat{a}^{+}_j \hat{a}_l \hat{a}_k
\eeq
If one knows the solutions of the one body part, i.e. knows the eigenfunctions $u_i(\vr)$ and the eigenenergies $\epsilon_i$ of $h$, Eq.\,(\ref{eq:sp_eq}) one can "incorporate" this knowledge in the above expression for $H_{op}$. Using the set $\{u_i\}$ as the basis one has $\langle i|\, h \,|j\rangle = \epsilon_i\delta_{ij}$ and
\beq
H_{op} = \sum_{i} \epsilon_i \hat{a}^{+}_i \hat{a}_i+
\frac{1}{2}\sum_{ijkl}\langle ij|V|kl\rangle \hat{a}^{+}_i \hat{a}^{+}_j \hat{a}_l \hat{a}_k
\eeq
In the Mean Field Approximations chapter of this course we  shall discuss and give examples of even more optimal ways to choose the single particle basis which incorporate on the average the effect of the interaction term in many body systems.

 \subsubsection{Heisenberg equations. Classical limits - field vs particles}
 Let us consider the Heisenberg equations for the field operators $\hat{\psi}(\vr)$ and $\hat{\psi}^{+}(\vr)$ and the general interacting Hamiltonian (\ref{eq:interact_Ham_2nd_q}). We have
$$
 i\hbar\frac{\partial}{\partial t}
\left(
\begin{array}{c}
   \hat{\psi}(\vr,t)  \\
   \hat{\psi}^{+}(\vr,t) \\  
\end{array}
\right)
 = \left[\left(
\begin{array}{c}
   \hat{\psi}(\vr,t)  \\
   \hat{\psi}^{+}(\vr,t) \\  
\end{array}
\right)
, H_{op}\right]
$$
Straightforward calculations produce Hermitian conjugate equations
 \eqna \label{eq:non_linear_fld_eqs}
i\hbar\frac{\partial \hat{\psi}(\vr,t)}{\partial t} &=& \left[ -\frac{\hbar^2}{2m}\nabla^2 + U(\vr)\right]\hat{\psi}(\vr,t) +
\int V(\vr-\vr')  \hat{\psi}^{+}(\vr',t) \hat{\psi}(\vr',t) \, d^3 r' \;\hat{\psi}(\vr,t) \nonumber \\
\\
-i\hbar\frac{\partial \hat{\psi}^{+}(\vr,t)}{\partial t} &=& \left[ -\frac{\hbar^2}{2m}\nabla^2 + U(\vr)\right]\hat{\psi}^{+}(\vr,t) +
\int V(\vr-\vr')  \hat{\psi}^{+}(\vr',t) \hat{\psi}(\vr',t) \, d^3 r' \;  \hat{\psi}^{+}(\vr,t) \nonumber
\eqne
These equations find many uses in the theory of many-particle systems. The Green's functions method provides a good example\footnote{cf., Quantum Theory of Many-Particle Systems, A. Fetter and J. Walecka, Dover, 2003}.  

Here we want to point out a simple but conceptually important aspect - their classical limit. Like in other quantum systems this limit is intuitively obtained by replacing coordinate and momentum operators by the corresponding classical functions of time turning Heisenberg equations into classical Hamilton equations.  In the above equations (\ref{eq:non_linear_fld_eqs})  this means replacing $\hat{\psi}(\vr,t)$ and $\hat{\psi}^{+}(\vr,t)$ by the c-number (classical, commuting) functions $\psi(\vr,t)$ and $\psi^{*}(\vr,t)$.  In the non interacting limit $V=0$ the resulting wave equations bring us back to where we started, cf., the \Sch equation (\ref{eq:Sch_eq_with_int}). The classical limit of the fully interacting case leads to  a non linear \Sch equation with cubic non linear term controlled by the interaction. 

 Let us add two more remarks. 
 
 a) Our intuitive "derivation" of the classical limit of Eq.\,(\ref{eq:non_linear_fld_eqs}) requires formal justification which will be discussed in the chapter "Mean Field Approximations for Many Body Problems".  On the intuitive level  the classical limit of the quantum field corresponds to physical processes in which very large number of quanta (particles) are "condensed" in the same wave mode, i.e the same single particle state.  
 
 b) The classical limit referred to above is different from the common classical limit for the N particle Hamiltonian  Eq.\,(\ref{eq:1st_quantized_H}).  The latter is given by replacing the operators for the particle coordinates and momenta by the classical variables in the corresponding Heisenberg equations. It is easy to show that this results  in the classical Hamilton equations
\beq 
\frac{d\vr_a}{dt}= \frac{\partial H}{\partial \vp_a} \;\;\; , \;\;\; \frac{d\vp_a}{dt}= -\frac{\partial H}{\partial \vr_a} 
\eeq
We therefore have two classical limits - the "field" classical limit for the fields $\psi(\vr,t)$ and $\psi^{*}(\vr,t)$  vs the more familiar "particle" classical limit for the particle coordinates $\vr_a$'s and momenta $\vp_a$'s. The latter classical limit is the limit of $\hbar \to 0$ while the former is achieved for the large number  $N_0 \gg 1$ of condensed quanta (i.e. the boson particles) of the theory.

\section{Fermions -- another alternative of the second quantization}

 As we have learned so far the quantization of the \Sch field leads to a very efficient and elegant description of many particle bosonic systems in all their aspects. A natural question is if this treatment can be extended to systems of fermions.  
 
 \subsection{Quantization via anticommutators} 
 A clear hint towards a positive answer can be found in our discussions in Section \ref{sec:perm_sym_bos}. There we saw that the symmetry of the bosonic wave functions was assured by the most basic  property of the field operators ${\psi}^{+}(\vr_a)$'s creating the particles - their commutativity.  As we will now show there is a consistent way of  quantizing the \Sch field by postulating anticommutativity of the basic operators. This single change of the quantization postulate will lead to a description of many fermion systems similar to the second quantized formalism for many bosons.

Dealing with fermions one must introduce spin variable together with position coordinates in order to describe the particles of the theory. Accordingly we start with the classical field which is described by functions
 \beq \label{class_fermi_fld}
 \psi_\sigma(\vr) \;\;\;\;  {\rm and} \;\;\;    \psi^{*}_\sigma(\vr)
 \eeq
 with the spin projection index $\sigma=\pm 1/2$  (we assume spin $1/2$ fermions as by far the most common).  It is often useful to write/view these functions in the explicit spinor form as
 $$
\left(  \begin{array}{c}
   \psi_{1/2}(\vr)     \\
    \psi_{ -1/2}(\vr)
\end{array}\right)   \;\;\;\; {\rm and} \;\;\; \left(  \begin{array}{c}
   \psi^*_{1/2}(\vr)     \\
    \psi^*_{-1/2}(\vr)
\end{array}\right)
$$
 
 We quantize this field  by introducing two sets of operators
$$
 \psis(\vr) \;\; {\rm and} \;\; \hpsis_{\sigma}(\vr) 
$$
with $\vr$ and $\sigma=\pm 1/2$ labelling each set.   We need to define the space of states on which these $2\times \infty^3$ operators act and the results of their action. 
We have seen with the bosonic field $\psi(\vr)$ treated above that to achieve this it was sufficient to define an abstract vacuum state $|0\rangle$ and the commutation relations between the field operators.  Following this we could define the basis of the space of states on which the operators act and calculate any matrix element for any given operator.  

Following this experience we start  by defining the vacuum state $|0\rangle$ with the properties
\eqna \label{def_vac_ferm}
&a)& \;\;\; \langle 0 | 0 \rangle = 1  \\
&b)&   \;\;\; \psis(\vr)|0\rangle = 0 \;\;\; {\rm for \;\; all \;\;  values \;\; of } \;\; \vr \;\;{\rm and} \;\; \sigma  \nonumber
\eqne
This we supplement with imposing (posulating) the anticommutation relations as follows
 \eqna \label{eq:anti_com}
 && \psis(\vr)\hpsis_{\sigma'}(\vr') +  \hpsis_{\sigma'}(\vr') \psis(\vr) \equiv \{\psis(\vr), \hpsis_{\sigma'}(\vr') \} =   \delta_{\sigma \sigma'}\delta(\vr-\vr')  \nonumber  \\
 && \psis(\vr)\hpsi_{\sigma'}(\vr') + \hpsi_{\sigma'}(\vr') \psis(\vr) \equiv \{\psis(\vr), \hpsi_{\sigma'}(\vr')\}   = 0
   \\
 &&\hpsis_{\sigma}(\vr)\hpsis_{\sigma'}(\vr')  +  \hpsis_{\sigma'}(\vr')\hpsis_{\sigma}(\vr) =  \{\hpsis_{\sigma}(\vr)\, \hpsis_{\sigma'}(\vr') \}  = 0 \nonumber
 \eqne
 where the curly brackets $\{\;\;, \;\;\; \}$ define anticommutators.
 
 As we will demonstrate below these two definitions are sufficient to define a quantum mechanical fermion field with any dynamics.  We note that while the definition of the vacuum is the same as in the bosonic case the anticommutation relations define a new quantization "paradigm" which is different from the familiar canonical quantization via the commutators.    
\subsection{Fermions in external potential}
In order to understand the consequences of the new quantization scheme defined above we start by considering a simple example - particles in an external potential.

\subsubsection{The field equations and the Hamiltonian}

The dynamical equation for the field (\ref{class_fermi_fld}) in an external potential  is a generalization of the 
Eq.\,(\ref{eq:Sch_eq_with_int})  to include the spin
\beq \label{fld_eq_with_spin}
i\hbar \frac{\partial \psi_\sigma (\vr,t)} {\partial t} = \sum_{\sigma'} h_{\sigma \sigma'} \psi_{\sigma'}(\vr,t)
\eeq
As an example we consider the following $h_{\sigma \sigma'}$
\beq
h_{\sigma \sigma'}=\delta_{\sigma \sigma'}\left(- \frac{ \hbar^2}{2m}\right)\nabla^2 + U_{\sigma \sigma'}(\vr)
\eeq
We assumed a spin dependent external potential, like for instance the interaction of the spin with an inhomogeneous magnetic field (e.g. in the Stern-Gerlach experiment)
$$U_{\sigma \sigma'}(\vr) =-\gamma \vB(\vr)\cdot \vs_{\sigma \sigma'} 
$$
with a constant $\gamma$ and vector $\vs$ of spin $1/2$ matrices.

Using our experience with the spinless field and appropriately generalizing it we consider the equation (\ref{fld_eq_with_spin}) and its complex conjugate as the pair of Hamilton equations with $\psi_\sigma(\vr)$ and $i\hbar \psi^{*}_\sigma(\vr)$ as canonical variables and the following classical Hamiltonian function
 \beq
 H=\sum_{\sigma \sigma'} \int d^3r \left[\delta_{\sigma\sigma'}\frac{\hbar^2}{2m}|\nabla\psi_\sigma(\vr)|^2 +U_{\sigma\sigma'}(\vr)\psi^*_\sigma(\vr) \psi_{\sigma'}(\vr)\right]
 \eeq
 Indeed from
 \eqna
 \delta H&=& \sum_{\sigma \sigma'} \int d^3 r \left\{\delta_{\sigma\sigma'}\frac{\hbar^2}{2m}\left[\nabla\psi^{*}_\sigma (\vr)\nabla\delta\psi_{\sigma'}(\vr) +
   \nabla \delta\psi^{*}_\sigma (\vr)\nabla\psi_{\sigma'}(\vr)\right]  + \right.    \nonumber \\
&& \left.+ U_{\sigma\sigma'}(\vr)\left[\psi^{*}_\sigma(\vr) \delta\psi_{\sigma'}(\vr)
 + \delta\psi^{*}_\sigma(\vr)\psi_{\sigma'}(\vr)\right] \right\}  \nonumber
 \eqne
 we find
 $$
 \frac{\partial \psi_\sigma(\vr)}{\partial t}=\frac{\delta H}{\delta[i\hbar \psi^{*}_\sigma(\vr)]} =\frac{1}{i\hbar}
\sum_{\sigma\sigma'} \left[-\delta_{\sigma\sigma'}\frac{\hbar^2}{2m}\nabla^2\psi_{\sigma'}(\vr) + U_{\sigma\sigma'}(\vr) \psi_{\sigma'}(\vr)\right]
 $$
 $$
\frac{\partial[ i\hbar \psi^{*}_\sigma(\vr)]}{\partial t}=-\frac{\delta H}{\delta \psi_\sigma(\vr)} =
 -\sum_{\sigma\sigma'} \left[-\delta_{\sigma\sigma'}\frac{\hbar^2}{2m}\nabla^2\psi^{*}_{\sigma'}(\vr) + U_{\sigma\sigma'}(\vr) \psi^*_{\sigma'}(\vr)\right]
 $$
 which reproduce correctly the field equation (\ref{fld_eq_with_spin}) and its complex conjugate.

 On this basis we quantize this spinor field by replacing it with the field operators 
 \beq
 \psi_\sigma(\vr) \to \hat{\psi}_\sigma(\vr)\;\;\;, \;\; \psi^{*}_\sigma(\vr) \to \hat{\psi}^{+}_\sigma(\vr)
 \eeq
 with the anticommutation relations (\ref{eq:anti_com}) and the Hamiltonian operator
 \beq
  H_{op} = \sum_{\sigma \sigma'} \int d^3r \left[\delta_{\sigma\sigma'}\frac{\hbar^2}{2m}\nabla\hat{\psi}^{+}_\sigma(\vr)\cdot\nabla\hat{\psi}_{\sigma'}(\vr) +
  U_{\sigma\sigma'}(\vr)\hat{\psi}^{+}_\sigma(\vr)\hat{\psi}_{\sigma'}(\vr)\right]
  \eeq
  or in an equivalent form (cf., the remark after Eq.\,(\ref{eq:Ham_op_Sch_fld}))
\beq \label{eq:fermion_H}
   H_{op} =  \sum_{\sigma \sigma'}  \int d^3r \; \hat{\psi}^{+}_\sigma(\vr)\left[-\delta_{\sigma\sigma'}\frac{\hbar^2}{2m}\nabla^2 +
  U_{\sigma\sigma'}(\vr)\right]\hat{\psi}_{\sigma'}(\vr) = \sum_{\sigma \sigma'}\int d^3 r \, \hpsis_{\sigma'}(\vr) h_{\sigma' \sigma} \psis(\vr)
\eeq

\subsubsection{Transforming to the normal modes} 
Let us now solve the quantum mechanical problem defined by the Hamiltonian (\ref{eq:fermion_H}).
This is not hard since it is quadratic. We need to find its normal modes. Following a very similar route as in dealing with (\ref{eq:Ham_for_interact_fld}) we consider a single particle equation
\beq \label{sp_sch_eq_fermions}
\sum_{\sigma'} h_{\sigma \sigma'} u_i(\vr,\sigma)=\epsilon_i u_i(\vr,\sigma)
\eeq
The set $\{u_i(\vr,\sigma)\}$ is complete and orthonormal in the space of functions of $\vr, \sigma$
\beq
 \sum_{\sigma} \int d^3 u_i^{*} (\vr,\sigma) u_j(\vr,\sigma) = \delta_{ij}   \;\;\;,\; \;\;
 \sum_i   u_i(\vr,\sigma) u_i^{*} (\vr',\sigma')  = \delta_{\sigma \sigma'} \delta(\vr-\vr')
\eeq
  We expand the field operators using this set
\beq \label{eq:expand_fermi}
\psis(\vr)=\sum_i \hat{a}_i u_i(\vr,\sigma) \;\;, \;\;
\psiss(\vr) = \sum_i \hat{a}^{+}_i u^{*}_i(\vr,\sigma)
\eeq
The operators $\hat{a}_i$ and $\hat{a}^{+}_i$ can be expressed as
\beq \label{a_i_via_psi_fermi}
\hat{a}_i = \sum_\sigma \int d^3 r \, \psis(\vr)u^{*}_i(\vr,\sigma) \;\;, \;\; \hat{a}^{+}_i = \sum_\sigma \int d^3 r \, \psiss (\vr) u_i(\vr,\sigma)
\eeq
Using the anticommutators (\ref{eq:anti_com}) and the completeness of the set $\{u_i(\vr,\sigma)\}$ it is easy to see that $\hat{a}_i$'s and $\hat{a}^{+}_i$'s satisfy anticommutation relations too
\beq
\{\hat{a}_i , \hat{a}^{+}_j \}=\delta_{ij}  \;\;, \;\; \{\hat{a}_i , \hat{a}_j \}= 0 = \{\hat{a}^{+}_i , \hat{a}^{+}_j \}
\eeq
Inserting the expansions (\ref{eq:expand_fermi}) in the Hamiltonian (\ref{eq:fermion_H}) we obtain
\beq \label{eq:Hamil_fermi}
H_{op}  = \sum_{i} \epsilon_i  \hat{a}^{+}_i  \hat{a}_i
\eeq
exactly as in the bosonic case but with the operators obeying the anticommutation relations.

\subsubsection{ The eigenstates.  Working with anticommiting $\hat{a}$'s and $\hat{a}^{+}$'s }
The Hamiltonian  (\ref{eq:Hamil_fermi}) is a sum of commuting parts. Indeed as is easy to verify that
\beq
[\hat{n}_i,\hat{n_j}]= 0
\eeq
where we denoted
\beq
\hat{n}_i = \hat{a}^{+}_i  \hat{a}_i
\eeq
We need to find the eigenfunctions of $\hat{n}_i$'s. We follow the same construction as in the bosonic case, cf., Sec.\ref{sec:rev_of_bos}.  We note that from Eqs.(\ref{a_i_via_psi_fermi}) it follows that 
the vacuum state $|0\rangle$ defined in (\ref{def_vac_ferm}) is annihilated by all $\hat{a}_i$'s \beq
\hat{a}_i|0\rangle =0 \;\;\; {\rm for} \;\;\; {\rm all} \;\; i{\rm 's}
\eeq
Since it is also annihilated by all $\hat{n}_i$'s it is clearly an eigenstate of the Hamiltonian  (\ref{eq:Hamil_fermi}) with zero energy eigenvalue.

We now define one particle  states
$$ |1_i\rangle \equiv \hat{a}_i^{+}|0\rangle
$$
for any $i$. We note the following properties of such states
\eqna \label{eq:ferm_pwr_a}
\langle 1_i|1_i\rangle &=& \langle 0|\hat{a}_i \hat{a}_i^{+}|0\rangle = \langle 0|1 -\hat{a}^{+}_i \hat{a}_i|0\rangle =1 \nonumber \\
\langle 0|1_i\rangle &=& \langle 0|\hat{a}^{+}_i |0\rangle =0 \;\;\;\; , \;\;\;\; \hat{a}^{+}_i |1_i\rangle =
(\hat{a}^{+}_i )^2|0\rangle =0
\eqne
In the 1st equality we used the anticomutation relation $\{\hat{a}_i , \hat{a}^{+}_i\}= 1$ and
$$
\langle 1_i| \equiv [\hat{a}^{+}_i |0\rangle]^{+} =  \langle 0|[\hat{a}^{+}_i ]^{+}= \langle 0|\hat{a}_i
$$
In the 2nd equality we used
$$
\langle 0|\hat{a}^{+}_i = [\hat{a}_i |0\rangle]^{+}=0
$$
In the 3rd we used the anticommutator
$$
\{\hat{a}_i^{+},\hat{a}_i^{+}\}=2 [\hat{a}_i^{+}]^2 =0
$$
Remarkably this last relation is the expression of the Pauli exclusion principle that two (or more) identical fermions cannot occupy the same quantum state  - in this case the state $u_i$.

The most relevant for us property of the states $|1_i\rangle$ is that they are eigenstates of $\hat{n}_i$ with eigenvalue $n_i=1$
\beq
\hat{n}_i|1_i\rangle = \hat{a}^{+}_i\hat{a}_i \hat{a}^{+}_i|0\rangle = \hat{a}^{+}_i[1- \hat{a}^{+}_i\hat{a}_i]|0\rangle = \hat{a}^{+}_i|0\rangle =|1_i\rangle
\eeq
The last relation in Eq.\,(\ref{eq:ferm_pwr_a}) means that  there are only two eigenstates of each $\hat{n}_i$ - $|0\rangle$ and $|1_i\rangle$ with respective eigenvalues $n_i=0$ and  $n_i=1$.  

It follows then that the eigenfunctions of the Hamiltonian  (\ref{eq:Hamil_fermi}) are the products of all possible eigenstates of
 $\hat{n}_i$
$$
|\Psi_{\{n_i\}}\rangle = |n_1, n_2, \dots, n_i, \dots\rangle = \prod_i |n_i\rangle = \prod_i [\hat{a}^{+}_i]^{n_i}|0\rangle   \;\;\; {\rm with} \;\;\; n_i = 0 \;\;{\rm or}\;\; 1 
$$
and with the corresponding eigenenergies
$$
E_{\{n_i\}} = \sum_i \epsilon_i n_i  \;\;\; {\rm with} \;\;\; n_i = 0 \;\;{\rm or}\;\; 1 \;\;; \;\; N=\sum_i n_i
$$
The restriction of the occupations $n_i$ to 0 or 1 is of course another expression of the Pauli principle and is a direct result of the anti-commutation relations which we assumed in the process of the quantization. 

As is the bosonic case the total number of particles $N$ is an eigenvalue of the total particle number operator
\beq \label{tot_part_num_fermi}
N_{op} = \sum_i \hat{n}_i
\eeq
which commutes with the Hamiltonian $H_{op}$, Eq.(\ref{eq:Hamil_fermi}). We will expand on this below in Section \ref{sec:part_num_ferm}.

So to summarize - the solution of this problem amounts to solving the single particle \Sch equation (\ref{sp_sch_eq_fermions}) and then populating (filling in) the resulting single particle states $u_i$ with $N$ particles according to the Pauil principe. This solution is of course identical to what we would obtain in the 1st quantization formalism for $N$ fermions with the difference that there $N$ was a fixed, given parameter of the problem while it is a quantum number and can take any value in the 2nd quantization formalism. 

\subsubsection{Spin independent potential} 
Let us discuss an important  limiting case of the single particle hamiltonian  in Eq.(\ref{fld_eq_with_spin}) which is spin-independent, i.e. diagonal in spin indices
$$
h_{\sigma\sigma'} =\delta_{\sigma\sigma'} h \;\;\; {\rm with} \;\;\; h=-\frac{h^2}{2m}\nabla^2 + U(\vr) 
$$
and correspondingly 
$$
H_{op} =  \sum_{\sigma }\int d^3 r \, \hpsis_{\sigma}(\vr) h \psis(\vr)
$$
The normal modes are then products of space and spin parts
$$
u_i(\vr,\sigma)  = u_k(\vr) \chi_s(\sigma)
$$
with $u_k(\vr)$ solving
$$
hu_k(\vr) = \epsilon_k u_k(\vr) 
$$
and $\chi_s(\sigma), \; s=\pm 1/2$ being just two orthogonal space independent spinors, e.g.
$$
\chi_{1/2} = \left(  \begin{array}{c}  1   \\ 0 \end{array}\right) \;\;\;\;, \;\;\;\; \chi_{-1/2} = \left(  \begin{array}{c}  0  \\ 1 \end{array}\right)
$$
The single particle energies $\epsilon_k$ are now spin degenerate and the expansion in normal modes has the form
$$
\psis(\vr)=\sum_{k s} \hat{a}_{k s} u_k(\vr) \chi_s(\sigma) \;\;, \;\;  \psiss(\vr) = \sum_{ks} \hat{a}^{+}_{ks} u^{*}_k(\vr)\chi^*_s(\sigma)
$$
with the commutation relations
$$
\{\hat{a}_{ks} , \hat{a}^{+}_{k's'} \}=\delta_{kk'} \delta_{ss'}  \;\;, \;\; \{\hat{a}_{ks} , \hat{a}_{k's'} \}= 0 = \{\hat{a}^{+}_{ks} , \hat{a}^{+}_{k's'} \}
$$
The Hamiltonian is expressed as
$$
H_{op}  = \sum_{ks} \epsilon_k \hat{a}^{+}_{ks}  \hat{a}_{ks} = \sum_{ks} \epsilon_k \hat{n}_{ks}
$$
The  number operators $\hat{n}_{ks}$ commute and their eigenfunctions are easily found as before to be $|0\rangle$ and $|1_{ks}\rangle \equiv \hat{a}^+_{ks}|0\rangle$ with corresponding eigenvalies $n_{ks}=0  \; {\rm and} \; 1$.  The eigenfunctions of $H_{op}$ are then
$$
|\Psi_{\{n_{ks}\}}\rangle = |n_1, n_2, \dots, n_{ks}, \dots\rangle = \prod_{ks} |n_{ks}\rangle = \prod_{ks} [\hat{a}^{+}_{ks}]^{n_{ks}}|0\rangle \;\;\; {\rm with} \;\;\; n_{ks} = 0 \;\;{\rm or}\;\; 1 
$$
and with the corresponding eigenenergies
$$
E_{\{n_{ks}\}} = \sum_{ks} \epsilon_k n_{ks} \;\;\; {\rm with} \;\;\; n_{ks} = 0 \;\;{\rm or}\;\; 1 \;\;; \;\; N=\sum_{ks} n_{ks}
$$

\subsubsection{The particle number operator \label{sec:part_num_ferm}}

As in the bosonic case it is  useful to express the total particle number operator $N_{op}$ in terms of the field operators. Using (\ref{a_i_via_psi_fermi}) in the expression (\ref{tot_part_num_fermi}) we obtain
$$
N_{op} = \sum_\sigma \int d^3r \hpsi^+_\sigma(\vr)\hpsi_\sigma(\vr)
$$
As in the bosonic case this operator is the generator of the global $U(1)$ gauge transformation, the analogue of Eq.\,(\ref{eq:glob_U(1)} for the fermion field
\beq \label{eq:UI_ferm}
  \hat{\psi}_\sigma(\vr) \; \to \; e^{i\alpha}  \hat{\psi}_\sigma(\vr) \;\;\; , \;\;\; \hat{\psi}_\sigma^+(\vr) \; \to \; e^{-i\alpha} \hat{\psi}_\sigma^+(\vr)
 \eeq
It is indeed easy to check that "despite" the  anticommutation relations for the fermion field operators the relation Eq.\,(\ref{eq:inf_U1_bos}) holds for each spin component
$$
[N_{op}, \hat{\psi}_\sigma (\vr)] = -\hat{\psi}_\sigma(\vr) 
$$
and therefore so is the corresponding generalization of 
Eq.\,(\ref{eq:gauge_U(1)})
$$
 e^{-i\alpha N_{op}} \hat{\psi}_\sigma(\vr)  e^{i\alpha N_{op}} = e^{i\alpha}  \hat{\psi}_\sigma(\vr) \;\; , \;\; e^{-i\alpha N_{op}} \hat{\psi}_\sigma^{+}(\vr)  e^{i\alpha N_{op}} = e^{-i\alpha}  \hat{\psi}_\sigma^{+}(\vr) 
$$
Since the Hamiltonian  Eq.\,(\ref{eq:fermion_H}) is invariant under this transformation it commutes with $N_{op}$.

\subsubsection{Working with the fermion field operators}

The expression for the particle number operator shows 
that 
$$
\hat{\rho}_\sigma(\vr) = \hpsi^+_\sigma(\vr)\hpsi_\sigma(\vr)
$$
is the density operator of particles with the spin projection $\sigma$.
Let us consider  a state
 \beq
 |\vr,\sigma\rangle  \equiv \hat{\psi}^{+}_\sigma(\vr)|0\rangle
 \eeq
  and let us act on it with the operator $\hat{\rho}_{\sigma'}(\vr')$. Using the anticommutation relations (\ref{eq:anti_com}) to commute $\hat{\psi}_{\sigma'}(\vr')$ towards $|0\rangle$ and using Eq.\,({\ref{def_vac_ferm}) we find
 \beq
  \hat{\rho}_{\sigma'}(\vr') |\vr,\sigma\rangle = \hat{\psi}^{+}_{\sigma'}(\vr')\hat{\psi}_{\sigma'}(\vr')\hat{\psi}^{+}_\sigma(\vr)|0\rangle
   =\delta_{\sigma\sigma'}\delta(\vr-\vr')\hat{\psi}^{+}_{\sigma'}(\vr')|0\rangle = \delta_{\sigma\sigma'}\delta(\vr-\vr')  |\vr,\sigma\rangle
  \eeq
which shows that $\hat{\psi}^{+}_\sigma(\vr)$ creates a particle at the position $\vr$ with spin projection $\sigma$. More precisely it creates delta like particle density of particles with spin projection $\sigma$ at this position.

Continuing as we did in the boson case let us consider the state
 \beq \label{Npart_fermns_st}
 |\vr_1 \sigma_1, \dots, \vr_N \sigma_N\rangle = const_N \, \hpsis_{\sigma_1}(\vr_1)\dots \hpsis_{\sigma_N}(\vr_N)|0\rangle
 \eeq
 where we  introduced a multiplicative constant for normalization, see below.  Acting on this state  with the operator $\hat{\psi}_\sigma(\vr)$, commuting it towards $| 0\rangle$ and using Eq.\,({\ref{def_vac_ferm}) we get
  \beq \label{act_psi_on_Npart_ferms_st}
 \hat{\psi}_\sigma(\vr) |\vr_1\sigma_1, \dots, \vr_N\sigma_N\rangle = const_N \sum_{a=1}^N(-1)^{P_a} \delta_{\sigma\sigma_a} \delta(\vr-\vr_a) \prod_{b\ne a}^N \hpsis_{\sigma_b}(\vr_b)|0\rangle
 \eeq
where $P_a$ is the parity of the number of permutations one needs to make in order to move $ \hat{\psi}_\sigma(\vr)$ to the right of $\hpsis_{\sigma_a}(\vr_a)$. The result (\ref{act_psi_on_Npart_ferms_st})  means that $\hat{\psi}_\sigma(\vr)$ destroys (annihilates)  one particle if its coordinates coincide with $\vr$ and its spin projection with $\sigma$. In doing this it also changes the sign of the resulting part of the wave function  if the permutation number $P_a$ is odd.  In this way it's action is sensitive to the order of the destroyed particle in the wave function.

 We can use the above result to act on the state (\ref{Npart_fermns_st})  with the operator $\hat{\rho}_\sigma(\vr)$.  We obtain in the same manner as in the boson case (cf., Eq.(\ref{eq:act_psi}))
 $$
 \hat{\rho}_\sigma(\vr)  |\vr_1, \dots, \vr_N\rangle = [\sum_{a=1}^N \delta_{\sigma \sigma_a} \delta(\vr-\vr_a) ] |\vr_1, \dots, \vr_N\rangle
 $$
 showing that this state describes $N$ particles (delta like particle densities) with spin projections $\sigma_a$ at  the positions $\vr_a, \; a=1, \dots, N$.

\subsection{Relation to the first quantization}

In this Section we  follow a similar development as in the boson case but with the additional spin index in the field operators and anti-commutation instead of the commutation relations. 

\subsubsection{The wave functions}
 
Consider the fermionic version of the N particles wave function in the second quantization
 \beq \label{eq:wf_in_sq_f}
 |\Phi\rangle = \frac{1}{\sqrt{N!}} \sum_{\sigma_1, ... \sigma_N }\int \prod_{a=1}^N d^3r_a \;\Phi(\vr_1\sigma_1,\vr_2\sigma_2,\dots,\vr_N \sigma_N)\hat{\psi}_{\sigma_1}^{+}(\vr_1) \dots \hat{\psi}^{+}_{\sigma_N}(\vr_N)| 0 \rangle
 \eeq
The interpretation of this expression is quite clear - we have a linear combination of $N$ particles in positions $\vr_1, \dots, \vr_N$   with spin projections $\sigma_1, ... , \sigma_N$ weighted each by the probability amplitude $\Phi(\vr_1 \sigma_1,\vr_2 \sigma_2,\dots,\vr_N \sigma_N)$. The anticommutation of $\hat{\psi}^+_\sigma(\vr)$'s assures that this amplitude is antisymmetric with respect to the exchange of any pair of $(\vr,\sigma)$'s\footnote{As was already discussed in the bosonic case one  can prove that any permutation of N objects can be achieved by a an ordered "product" (sequence) of pairwise transpositions.}.  This amplitude is clearly the  first quantization partner of the wave function $|\Phi\rangle$ 

 As in the bosonic case the normalization of $|\Phi\rangle $ assures that it is normalized, i.e. $ \langle \Phi |\Phi\rangle = 1$ provided 
 the amplitude $\Phi(\vr_1 \sigma_1 ,\dots, \vr_N \sigma_N)$ is
$$
 \sum_{\sigma_1, ..., \sigma_N} \int \prod_{a=1}^N d^3r_a |\Phi(\vr_1 \sigma_1,\vr_2\sigma_2,\dots,\vr_N \sigma_N)|^2 =1 
 $$
In the arbitrary single particle basis $u_i(\vr,\sigma)$ the above wave function looks exactly as in the boson case
 \beq
  |\Phi\rangle = \sum_{i_1,..., i_N} C_{i_1, ... , i_N} \hat{a}^{+}_{i_1} ... \hat{a}^{+}_{i_N}|0 \rangle
  \eeq
  with the "only" difference that the operators $\hat{a}^+_i$'s are anticommuting.
  
As in the bosonic case it is useful and practical to work with the wave functions in the \textbf{occupation number representation}, cf., Eq\,(\ref{eq:wf_in_genb}),
\beq 
 |\Phi\rangle = \sum_{n_1, ..., n_i, ... ; {\rm with} \, n_i=0 \; {\rm or}\,1 \, , \, \sum_i n_i=N} C_{n_1, ... , n_i, ...}|n_1, n_1, ..., n_i, ...\rangle 
 \eeq
 with the "only" difference that the fermionic occupations $n_i$'s  are restricted  to be zero or one.

\subsubsection{The operators}

As in the bosonic case the operators in the 1st quantized formulation of fermions are classified as one-body, two-body, etc.

{\bf One body operators} \\
To remind - these operators act on wave functions of identical particles one particle at a time and have a general form given by the expression (\ref{eq:one_body_op}).  The difference in the present fermion case is  that 
each $f^{(1)} _a$ operator in addition to being a function of $\vr_a$, $\hat{\vp}_a=-i\hbar\nabla_a$ may also depend on the spin matrices  $\vs_a$.  This means that in general   $f^{(1)} _a$'s are $2 \times 2$ spinor matrices with matrix elements depending on $\vr_a$ and $\hat{\vp}_a$, cf., the example of $h_{\sigma\sigma'}$ in Eq.(\ref{fld_eq_with_spin}). 

In a very similar way as in the bosonic case one can show (cf., Appendix \ref{sec:calc_1b_ferm})  that in the second quantization one body operators have the form
\beq \label{eq:1body_op_fermi}
F_{op}^{(1)}=\sum_{\sigma\sigma'} \int d^3 r \hat{\psi}_{\sigma'}^{+}(\vr) f^{(1)}_{\sigma'\sigma} \hat{\psi}_{\sigma}(\vr)
\eeq
where $f^{(1)}_{\sigma\sigma'} $ is one (any) of the operators in the sum (\ref{eq:one_body_op}) generalized to include the spin dependence. It is acting on $\hat{\psi}_\sigma(\vr)$ as a spinor function of $\vr$.  The expression (\ref{eq:fermion_H}) for the Hamiltonian in an external potential provides a good example of such an operator.

{\bf Two body operators}

The two body operators for identical particles with spins in the 1st  quantization have the same form (\ref{1st_quan_2_body}) as in the bosonic case but with the elementary operators $f^{(2)}_{ab}$ in general depending in addition  to $\vr_a, \vr_b, \hat{\vp}_a$ and $ \hat{\vp}_b$ also on the spin matrices $\vs_a , \vs_b$.  An example is given by the so called spin exchange term in a (phenomenological) two particle interaction
$$
\frac{1}{2} \sum_{a,b=1; a\ne b}^N \left[ V(\vr_a - \vr_b) + W(\vr_a - \vr_b)\,( \hat{\vs}_a\cdot \hat{\vs}_b)\right]
$$
For simplicity we will consider  only  spin independent  $f^{(2)}_{ab}$. One can show that such two body operators in the fermionic 2nd quantization  have a form similar to the bosonic expression (\ref{eq:two_body_op_sq}) with the addition of the spin indices in the field operators 
$$
F^{(2)}_{op}= \frac{1}{2}\sum_{\sigma \sigma'} \int d^3 r d^3 r' \hat{\psi}^{+}_\sigma(\vr)\hat{\psi}^{+}_{\sigma'}(\vr')f^{(2)}
\hat{\psi}_{\sigma'}(\vr')\hat{\psi}_\sigma(\vr)
$$
 with $f^{(2)}$ being a function of $\vr, \vr'$ and $\hat{\vp}=-i\hbar\nabla_{\vr}, \hat{\vp'}=-i\hbar\nabla_{\vr'}$
 \footnote{The general spin dependent two body $F^{(2)}_{op}$ will have the pairwise $f^{(2)}$'s  depending in addition on the spin operators  
 $\hat{\vs}$,$\hat{\vs'}$ of the particles' pairs.  This means they will be four index matrices $f^{(2)}_{\sigma\sigma', \sigma" \sigma{'''}}$ 
 and the expression for  $F^{(2)}_{op}$  will be 
 $$  
 F^{(2)}_{op}= \frac{1}{2}\sum_{\sigma \sigma' \sigma{''}\sigma{'''}} \int d^3 r d^3 r' \hat{\psi}^{+}_\sigma(\vr)\hat{\psi}^{+}_{\sigma'}(\vr')
 f^{(2)}_{\sigma\sigma', \sigma{''} \sigma{'''}}
\hat{\psi}_{\sigma{'''}}(\vr')\hat{\psi}_{\sigma{''}}(\vr)
$$
with the corresponding generalization of the expression (\ref{eq:two_part_me}) 
 $$
 \langle ij|f^{(2)}|kl\rangle =\sum_{\sigma \sigma' \sigma{''}\sigma{'''}}\int d^3r d^3 r' u^{*}_i(\vr,\sigma)u^{*}_j(\vr',\sigma')
  f^{(2)}_{\sigma\sigma', \sigma{''} \sigma{'''}}
 u_k (\vr, \sigma{''})u_l (\vr',\sigma{'''}) 
 $$
}. 
Note the relative order of the field operators.  Since they anticommute it is important to  keep it.

{\bf General single particle basis}

To obtain the expression for $F_{op}^{(1)}$  and  $F_{op}^{(2)}$ in a general basis $u_i(\vr,\sigma)$ one just has to expand the field operators in their expressions in this basis, cf., Eq.(\ref{eq:expand_fermi}). The result has identical form to the bosonic expressions
(\ref{eq:one_body_sq_1}) and  (\ref{eq:two_body_sq_1}) but the matrix elements have spin summations in addition to space coordinates integrals
 \eqna \label{eq:two_part_me}
\langle i|f^{(1)}|j\rangle &=& \sum_{\sigma\sigma'} \int d^3r u^{*}_i(\vr,\sigma')f^{(1)}_{\sigma' \sigma}u_j (\vr,\sigma) \nonumber \\
 \langle ij|f^{(2)}|kl\rangle &=&\sum_{\sigma\sigma'}\int d^3r d^3 r' u^{*}_i(\vr,\sigma)u^{*}_j(\vr',\sigma')f^{(2)}u_k (\vr, \sigma)u_l (\vr',\sigma')
 \eqne
 where for the two-body operator we write  only for the simple (but very common)  case of the spin independent $f^{(2)}$.

\subsection{Interacting fermions}

\subsubsection{Hamiltonian}
The most common Hamiltonian  of interacting fermions has the form 
\eqna \label{interact_Ham_2ndq_fermi}
H_{op} &=& \sum_{\sigma\sigma'}\int d^3 r\hat{\psi}^{+}_{\sigma'}(\vr)\, h_{\sigma'\sigma} \, \hat{\psi}_\sigma(\vr) +  \\
& & \hspace{1cm} +\frac{1}{2} \sum_{\sigma\sigma'} \int d^3 r d^3 r' \hat{\psi}^{+}_\sigma(\vr)\hat{\psi}^{+}_{\sigma'}(\vr')V(\vr-\vr')\hat{\psi}_{\sigma'}(\vr')\hat{\psi}_\sigma(\vr) \nonumber 
\eqne
with
$$
h_{\sigma' \sigma} =   -\delta_{\sigma'\sigma} \frac{\hbar^2}{2m}\nabla^2 + U_{\sigma'\sigma}(\vr)
$$
and a spin independent two body interaction.  In a general single particle basis this Hamiltonian is  
\beq
H_{op} = \sum_{ij}\langle i|\, h \,|j\rangle \hat{a}^{+}_i \hat{a}_j +
\frac{1}{2}\sum_{ijkl}\langle ij|V|kl\rangle \hat{a}^{+}_i \hat{a}^{+}_j \hat{a}_l \hat{a}_k
\eeq
As in the boson case  if the solutions of the non interacting part are known, i.e. if one knows  the eigenfunctions of the single particle Hamiltonian $h$, cf. Eq. (\ref{sp_sch_eq_fermions}) (e.g. Coulomb wave functions in atoms) one can use the operators $\hat{a}^{+}_i$ ,  $\hat{a}_j$ in this basis.  The matrix $\langle i|\, h \,|j\rangle$ is then diagonal making the first term in $H_{op}$ trivial 
\beq
H_{op} = \sum_{i} \epsilon_i\hat{a}^{+}_i \hat{a}_i+
\frac{1}{2}\sum_{ijkl}\langle ij|V|kl\rangle \hat{a}^{+}_i \hat{a}^{+}_j \hat{a}_l \hat{a}_k
\eeq
and helping to "focus attention" on the particle interactions.

\subsubsection{Heisenberg equations. No classical limit}

Despite anticommutation relations  of the  fermion field operators $\hat{\psi}_\sigma(\vr)$ and $\hat{\psi}_\sigma^{+}(\vr)$ the Heisenberg equations for these operators 
$$ 
 i\hbar\frac{\partial}{\partial t}
\left(
\begin{array}{c}
   \hat{\psi}_\sigma(\vr,t)  \\
   \hat{\psi}_\sigma^{+}(\vr,t) \\  
\end{array}
\right)
 = \left[\left(
\begin{array}{c}
   \hat{\psi}_\sigma(\vr,t)  \\
   \hat{\psi}_\sigma^{+}(\vr,t) \\  
\end{array}
\right)
, H_{op}\right]
$$
for the general interacting Hamiltonian (\ref{interact_Ham_2ndq_fermi}) have the same formal appearance as for bosons apart of the presence of the spin indices. It is a useful exercise for the reader to work this out explicitly.  The equation for $\hat{\psi}_\sigma(\vr)$ is
\beq \label{eq:Heis_eq_ferm}
i\hbar\frac{\partial \hat{\psi}_\sigma(\vr,t)}{\partial t} = \sum_{\sigma'} h_{\sigma \sigma'} \hat{\psi}_{\sigma'}(\vr,t) +
 \int  V(\vr-\vr') \sum_{\sigma'} \hat{\psi}_{\sigma'}^{+}(\vr',t) \hat{\psi}_{\sigma'}(\vr',t) 
 d^3 r' \, \hat{\psi}_\sigma(\vr,t) 
 \eeq
 and the Hermitian conjugate of this equation for $\hat{\psi}_{\sigma}^{+}(\vr,t)$.  We note that unlike the boson case these equations do not have  classical limit. This for the obvious reason that Pauli principle and formally the anti commutation relations of the field operators prevent having more than one fermion in any given field mode\footnote{The so called anticommuting c-numbers (Grassman variables) are often related to the classical limit of fermionic second quantized operators.  In a very crude way they are obtained by setting to zero all the anticommutators in Eq.\,(\ref{eq:anti_com}),  
 $$
 \{\psis(\vr), \hpsis_{\sigma'}(\vr') \} =  \{\psis(\vr), \hpsi_{\sigma'}(\vr')\}   = \{\hpsis_{\sigma}(\vr)\, \hpsis_{\sigma'}(\vr') \}  = 0 
 $$
This is in (again a crude) analogy with the classical limit of the bosonic case in which all the canonical commutators vanish, cf., Berezin, F. A., “The Method of Second Quantization,” Academic Press, 1965.  

The Grassman variables are most often used in constructing functional integrals for femionic systems, cf., Negele, J. W., and Orland, H., “Quantum Many-Particle Systems,” Perseus Books Group, 1998, pp.25-37 }. 
 
\subsubsection{Mean field approximation}
Let us assume for simplicity the spin independent $U(\vr)$) 
and write the Heisenberg equation (\ref{eq:Heis_eq_ferm})  in the following form 
\beq
i\hbar\frac{\partial \hat{\psi}_\sigma(\vr,t)}{\partial t} = \left[ -\frac{\hbar^2}{2m}\nabla^2 + U(\vr) +
 \int  V(\vr-\vr') \hat{\rho}(\vr',t)  d^3 r' \right] \, \hat{\psi}_\sigma(\vr,t)
 \eeq
with
$$
\hat{\rho}(\vr,t) = \sum_{\sigma} \hat{\psi}_{\sigma}^{+}(\vr,t) \hat{\psi}_{\sigma}(\vr,t) 
$$
The potential $U(\vr)$ in these equations is formally modified by the last term which is the convolution of the two body interaction $V(\vr-\vr')$   and the operator of the particle density $\hat{\rho}(\vr',t)$.  For a classical particle density function $\rho(\vr,t)$ this term would have a natural meaning of the potential which the particles of the system induce\footnote{Cf., the footnote on the next page}.
In quantum mechanic context one can qualitatively think of $\hat{\rho}(\vr,t)$ as a random variable the probability amplitude distribution of which is determined by the wave function  $|\Phi\rangle$ of the many fermion system under consideration.  

Given $|\Phi\rangle$ one can write 
 $$
  \hat{\rho}(\vr) = \langle \Phi |\hat{\rho}(\vr) |\Phi \rangle +\delta  \hat{\rho}(\vr)
  $$ 
separating the average and the fluctuations of $\hat{\rho}(\vr,t)$.
It is natural to ask if neglecting the fluctuations would be a good approximation.  This would certainly greatly simplify the problem. 
It would also be in line with similar approximations known in other fields under the name "mean field approximation"\footnote{There is an important aspect which must be addressed first.  This is related to the fact that $\hat{\rho}(\vr,t)$ and therefore its average includes all the particles in the system  while the mean field potential acting on any given particle 
$$
\int  V(\vr-\vr') \rho(\vr',t)  d^3 r' 
$$
must exclude this particular particle. This problem is elegantly solved in  the Hartree-Fock method described in the Mean Field Approximations chapter}.  
In the many-fermion systems such mean field approximations were first introduced in atomic physics by Hartree and then supplemented by Fock to result in the Hatree-Fock method.  We will address these developments in a separate chapter. Mean mean approximation and its extensions play a very important role in theoretical treatment of such many fermion systems as atoms, nuclei and solids.  

\section{The Fock space.}
In the first quantization formalism we encountered the notion of the Hilbert space.  For N particles this was the space of all functions of $N$ variables 
\eqna
&& \Phi(x_1, x_2, ...., x_N)    \; \;\;\;\;\;  {\rm with} \; \nonumber \\ 
&& x_a = \vr_a \;, \; a = 1, ..., N, \;  {\rm symmetrized \;\; for \; spinless \;\; bosons} \; , \nonumber  \\
 && x_a=\vr_a , \sigma_a \; ,\;  a = 1, ..., N, \; {\rm antisymmetrized \;\; for \; fermions} \nonumber 
\eqne
The operators acting on such functions didn't change the particle number $N$.  The situation is different in the second quantization formulation. Here already the most elementary operators $\hat{\psi}(\vr)$, $\hat{\psi}^+(\vr)$, $\hat{a}_i$, $\hat{a}_i^+$, etc., change the particle number and the most general wave function should be a linear combination of functions like $\Phi_N$  with different $N$'s and including the vacuum
\eqna 
|\Phi\rangle &=& C^{(0)} |0\rangle + \sum_i C_i^{(1)} \hat{a}_i^+ |0 \rangle + \sum_{ij} C_{ij}^{(2)} \hat{a}_i^+ \hat{a}_j^+|0 \rangle + .... +  \nonumber \\
&  +  & \sum_{i_1i_2.. } C_{i_1 i_2...i_N }^{(2)} \hat{a}_{i_1}^+ \hat{a}_{i_2}^+... \hat{a}_{i_N}^+|0 \rangle + ......
\eqne
The Hilbert space of all such functions is called the Fock space and is a direct sum 
\eqna
\left( \rm vacuum \right)  \bigoplus \left( \rm 1 \; particle \; Hilbert \; space \right) \bigoplus \left ( \rm 2 \; particle \; Hilbert \; space \right) \bigoplus \; ...   \\
... \; \bigoplus \left ( \rm N \; particle \; Hilbert \; space \right) \bigoplus\;  ... \qquad \qquad \nonumber
\eqne

 \section{Appendix}
 \subsection{Bosons - reviewing the properties of $\hat{a}$' s and $\hat{a}^{+}$'s \label{sec:rev_of_bos}}
\subsubsection{The vacuum state}
Let us defined a special state denoted $|0\rangle$. We shall call this state a vacuum state. The only properties we will ever need of this state are that it gives zero when acted upon with anyone of the operators $\hat{a}_i$  and that it is normalised
\eqna
\hat{a}_i|0\rangle &=& 0 \;\;\; i=1,2,... \\
\langle 0 | 0 \rangle &=& 1    \nonumber
\eqne

\subsubsection{Single mode}
We start by considering the pair  $\hat{a}_i , \hat{a}^+_i$ of operators with a fixed index $i$. We will call them operators of a single mode $u_i(\vr)$.  We then define the state (following an analogy with the oscillator ladder operators)
\beq
| 1_i \rangle \equiv \hat{a}^{+}_i| 0\rangle
\eeq
As is easy to see this state is normalised. Indeed using the commutation relations and the properties of $|0\rangle$ find
$$
\langle1_i| 1_i\rangle = \langle 0|\hat{a}_i\hat{a}^{+}_i|0\rangle = \langle 0|1+\hat{a}^{+}_i\hat{a}_i|0\rangle = \langle 0|0\rangle = 1
$$
Also have orthogonality
$$
\langle 1_i|0\rangle  = \langle 0| \hat{a}_i|0\rangle = 0
$$
In the same way we define
\beq
|2_i\rangle = {\rm const}\; \hat{a}^{+}_i|1_i\rangle = \frac{1}{\sqrt{2}}\;\hat{a}^{+}_i|1_i\rangle
\eeq
The normalization constant  is found as $const=1/\sqrt{2}$ by calculating the norm
\eqna
\langle 2_i|2_i\rangle &=& |const|^2 \langle 1_i|\hat{a}_i\hat{a}^{+}_i|1_i\rangle=|const|^2 \langle 1_i|\hat{a}_i\hat{a}^{+}_i \hat{a}^{+}_i|0\rangle =  \nonumber \\
&=& |const|^2 \langle 1_i|(1+ \hat{a}^{+}_i\hat{a}_i)\hat{a}^{+}_i|0\rangle
= |const|^2 [\langle 1_i|\hat{a}^{+}_i|0\rangle +  \nonumber \\
&+& \langle 1_i|\hat{a}^{+}_i(1+\hat{a}_i^{+}\hat{a}_i)|0\rangle ] = 2|const|^2\langle 1_i|1_1\rangle =2|const|^2 \nonumber
\eqne
We have  orthogonality
$$
\langle 2_i | 1_i\rangle = \frac{1}{\sqrt{2}}\langle 1_i| \hat{a}_i |1_i\rangle = \frac{1}{\sqrt{2}} \langle 1_i| \hat{a}_i \hat{a}_i^{+}|0_i\rangle = \frac{1}{\sqrt{2}} \langle 1_i| 1+\hat{a}_i^{+} \hat{a}_i|0_i\rangle =\frac{1}{\sqrt{2}} \langle 1_i| 0_i\rangle =0
$$
and even more trivially
$$
\langle 2_i|0_i\rangle = \frac{1}{\sqrt{2}}\langle 1_i | \hat{a}_i|0\rangle =0
$$
By iterating we define
\beq \label{eq:num_part_one_mode}
|n_i\rangle=\frac{1}{\sqrt{n_i}}\hat{a}_i^{+}|n_i-1\rangle =\frac{1}{\sqrt{n_i(n_i-1)}}(\hat{a}_i^{+})^2|n_i-2\rangle = \dots =\frac{1}{\sqrt{n_i!}}(\hat{a}_i^{+})^{n_i}|0\rangle
\eeq
One can prove that the resulting states $|n_i\rangle$ form orthogonal set
$$
\langle m_i| n_i\rangle = 0  \;\;\;\; {\rm for} \;\;\; m_i\ne n_i
$$
Indeed writing 
$$
\langle m_i| n_i\rangle = \frac{1}{\sqrt{m_i n_i}}\langle 0 | (\hat{a}_i)^{m_i} (\hat{a}_i^{+})^{n_i} |0 \rangle
$$
and commuting each $\hat{a}_i$'s to the right all the way to $|0\rangle$ one proves this to vanish for $m_i\ne n_i$. 

We also have
\eqna
\hat{a}^{+}_i|n_i\rangle &=&  \hat{a}_i^{+}\frac{1}{\sqrt{n_i!}}(\hat{a}_i^{+})^{n_i}|0\rangle  =\sqrt{n_i+1}\frac{1}{\sqrt{(n_i+1)!}}(\hat{a}_i^{+})^{n_i+1}|0\rangle   \nonumber \\
\hat{a}_i|n_i\rangle &=&  \hat{a}_i\frac{1}{\sqrt{n_i!}}(\hat{a}_i^{+})^{n_i}|0\rangle  =
n_i\frac{1}{\sqrt{n_i!}}(\hat{a}_i^{+})^{n_i-1}|0\rangle = \sqrt{n_i} \frac{1}{\sqrt{(n_i-1)!}}(\hat{a}_i^{+})^{n_i-1}|0\rangle \nonumber
\eqne
where the factor $n_i$ in the second equality of the second line results from commuting $\hat{a}_i$ through $n$ operators in $(\hat{a}^{+}_i)^{n_i}$ to get it acting on $|0\rangle$. The above calculation shows that
\beq
\hat{a}_i^{+}|n_i\rangle = \sqrt{n_i+1}|n_i+1\rangle \;\;\;\; , \;\;\;\; \hat{a}_i|n_i\rangle = \sqrt{n_i}|n_i-1\rangle
\eeq
Note also that  by hermitian conjugation 
\beq
\langle n_i|\hat{a}_i  = \sqrt{n_i+1}\langle n_i+1 | \;\;\;\; , \;\;\;\; \langle n_i| \hat{a}_i^{+} = \sqrt{n_i} \langle n_i-1|
\eeq
The last two sets of  equalities define the action of the operators $\hat{a}_i$ and $\hat{a}_i^{+}$ on any state "belonging" to the mode $u_i(\vr)$. Indeed for any such state $|\xi_i\rangle$ we can determine the result of acting on it with $\hat{a}_i$ or $ \hat{a}_i^{+}$ by writing it as  a linear combination
$|\xi_i\rangle = \sum_{n_i} c_{n_i} |n_i\rangle$ of the basis states $|n_ i\rangle $. 

Let us now consider the operator $\hat{n}_i = \hat{a}_i^{+} \hat{a}_i$. The basis states $|n_i\rangle$, Eq.\,(\ref{eq:num_part_one_mode})  are its eigenstates
\beq \label{eq:eigen_of_n_i}
\hat{n}_i |n_i\rangle =  \hat{a}_i^{+} \hat{a}_i |n_i\rangle =\sqrt{n_i} \;\hat{a}_i^{+} |n_i-1\rangle = n_i|n_i\rangle
\eeq
This operator is the i-th mode number operator. 

\subsubsection{Many modes}
We now generalize the above single mode construction to all modes of the complete set $u_i(\vr)$. This is easily done mostly because pairs of $\hat{a}_i$ and $\hat{a}_i^{+}$ commute for different $i$'s.
 The general multimode analogue of the states $|n_i\rangle$ is
\beq \label{eq:occ_num_basis}
|\{n_i\}\rangle\equiv |n_1,n_2,\dots, n_k, \dots\rangle =\prod_i |n_i\rangle = \prod_i \frac{1}{\sqrt{n_i!}}(\hat{a}_i^{+})^{n_i}|0\rangle
\eeq
The operators $\hat{a}_i$ and $\hat{a}_i^{+}$ act on these states as
\eqna
\hat{a}_i^{+}|n_1, \dots , n_i , \dots \rangle &=& \sqrt{n_i+1}|n_1,\dots, n_i+1, \dots \rangle  \nonumber \\ \hat{a}_i|n_1, \dots , n_i , \dots \rangle &=&\;\;\; \;\; \sqrt{n_i}\; |n_1,\dots, n_i-1, \dots \rangle
\eqne
and have number operators for all modes
\beq
\hat{n}_i |n_1,\dots, n_i, \dots \rangle = \hat{a}_i^{+}\hat{a}_i|n_1,\dots, n_i, \dots \rangle = n_i |n_1,\dots, n_i, \dots \rangle
\eeq
It is useful and important to introduce the total particle number operator
\beq 
N_{op} = \sum_i \hat{n}_i = \sum_i   \hat{a}_i^{+}\hat{a}_i
\eeq
which "measures" the sum of all $n_i$'s
\beq
N_{op}|n_1, n_2, \dots, n_i,\dots\rangle = \left(\sum_i n_i\right) |n_1, n_2, \dots, n_i,\dots\rangle
\eeq

 \subsection{Bosons - wave function normalization \label{sec:wf_norm} }
 Let us consider the norm  of the wave function Eq.\,(\ref{eq:wf_in_sq_0}) 
 \eqna \label{eq:norm_of_Phi}
 \langle \Phi |\Phi\rangle &=&  \\ 
&=&  \int \int \prod_{a,b=1}^N d^3r_a  d^3r'_b \;\Phi^*(\vr'_1,\vr'_2,\dots,\vr'_N)\Phi(\vr_1,\vr_2,\dots,\vr_N)\langle \vr'_1, \dots, \vr'_N|\vr_1, \dots, \vr_N\rangle \nonumber
 \eqne
 We need to evaluate the overlap $\langle \vr'_1, \dots, \vr'_N |\vr_1, \dots, \vr_N\rangle$.  In a straightforward way by commuting $\hpsi(\vr'_a)$'s to the right all the way to the vacuum state  $|0\rangle$ we obtain
 \eqna
&&\langle \vr'_1, \dots, \vr'_N |\vr_1, \dots, \vr_N\rangle =|const_N|^2 \langle 0|\hpsi(\vr'_N), ... \hpsi(\vr'_1)  \hpsis(\vr_1)\dots \hpsis(\vr_N)|0\rangle  = \nonumber \\
 &&\;\;\;\;\;\;\;\; =|const_N|^2   \langle 0|\hpsi(\vr'_N), ... \hpsi(\vr'_2)\sum_{a=1}^N\delta(\vr'_1 -\vr_a)\prod_{b\ne a}^N   \hpsis(\vr_b) |0\rangle =       \nonumber \\
  &&=|const_N|^2   \langle 0|\hpsi(\vr'_N), ... \hpsi(\vr'_3)\sum_{a=1}^N\delta(\vr'_1 -\vr_a)\sum_{b=1, b\ne a}^N \delta(\vr'_2-\vr_b)\prod_{c=1, c\ne a,b}^N   \hpsis(\vr_c) |0\rangle =     \nonumber \\
  &&= ... = |const_N|^2 \sum_{a=1}^N\delta(\vr'_1 -\vr_a)\sum_{b=1, b\ne a}^N \delta(\vr'_2-\vr_b)\sum_{d=1, d\ne a,b,c}^N\delta(\vr'_3 -\vr_d).. \langle 0 | 0 \rangle = \nonumber \\
  &&= |const_N|^2 \sum_P \prod_{a=1}^N \delta(\vr'_a - \vr_{Pa})
 \eqne
where $P$ stands for permutations of the particle indices $a=1, 2, .., N$.  There are $N!$  permutations of $N$ indices and therefore $N!$ terms in the last sum.  

Using this result in Eq.\,(\ref{eq:norm_of_Phi}) one can use the delta functions to reduce the norm $\langle \Phi |\Phi\rangle$ to a sum of integrals
$$
\langle \Phi |\Phi\rangle =  |const_N|^2 \sum_P  \int \prod_{a=1}^N d^3r_a \Phi^*(\vr_{P1},\vr_{P2},\dots,\vr_{PN})\Phi(\vr_1,\vr_2,\dots,\vr_N)
$$
Since $\Phi(\vr_1,\vr_2,\dots,\vr_N)$ is symmetric with respect to the permutations of its arguments the above $N!$ integrals are identical 
$$
\langle \Phi |\Phi\rangle= N!  |const_N|^2  \int \prod_{a=1}^N d^3r_a |\Phi(\vr_1,\vr_2,\dots,\vr_N)|^2
$$
which leads to the consistent normalization conditions to unity of both first and second quantization wave functions   Eq.\,(\ref{eq:both_wf_norm_unity})  for the choice of the const as
$$
const_N = \frac{1}{\sqrt{N!}}
$$

\subsection{Bosons - calculating $K_{op}|\Phi\rangle$  \label{sec:K_act_on_Phi}}

As with $U_{op}|\Psi\rangle$ we start by considering
$$
\int d^3r \; \hat{\psi}^{+}(\vr)\left(-\frac{\hbar^2}{2m}\nabla^2_{\vr} \right)\hat{\psi}(\vr)  \prod_{a=1}^N\hat{\psi}^{+}(\vr_a)| 0 \rangle
$$
Using in this expression the result (\ref{eq:act_psi})
and
$$
\nabla^2_{\vr}\delta(\vr-\vr_b) = -\nabla_{\vr} \nabla_{\vr_b}\delta(\vr-\vr_b) = \nabla^2_{\vr_b}\delta(\vr-\vr_b)
$$
we get  it in the form
$$
\int d^3r \; \hat{\psi}^{+}(\vr)\left[\sum_{b=1}^N\left(-\frac{\hbar^2}{2m}\nabla^2_{\vr_b}\right)\delta(\vr-\vr_b) \prod_{a\ne b}^N\hat{\psi}^{+}(\vr_a)\right]|0\rangle
$$
Therefore
\eqna
K_{op}|\Phi\rangle=  \frac{1}{\sqrt{N!}}\int \prod_{a=1}^N d^3r_a \Phi(\vr_1,\vr_2,\dots,\vr_N)\int d^3r \; \hat{\psi}^{+}(\vr)\left(-\frac{\hbar^2}{2m}\nabla^2_{\vr} \right)\hat{\psi}(\vr)  \prod_{a=1}^N\hat{\psi}^{+}(\vr_a)| 0 \rangle =  \nonumber \\
=  \frac{1}{\sqrt{N!}}\int \prod_{a=1}^N d^3r_a \Phi(\vr_1,\vr_2,\dots,\vr_N)\int d^3r \; \hat{\psi}^{+}(\vr)\left[\sum_{b=1}^N\left(-\frac{\hbar^2}{2m}\nabla^2_{\vr_b}\right)\delta(\vr-\vr_b) \prod_{a\ne b}^N\hat{\psi}^{+}(\vr_a)\right]|0\rangle = \nonumber  \\
=  \frac{1}{\sqrt{N!}}\int d^3r \; \hat{\psi}^{+}(\vr)\int \prod_{a=1}^N d^3r_a\left[ \sum_{b=1}^N
\left(-\frac{\hbar^2}{2m}\nabla^2_{\vr_b}\right) \Phi(\vr_1,\vr_2,\dots,\vr_N)\delta(\vr-\vr_b) \prod_{a\ne b}^N\hat{\psi}^{+}(\vr_a)\right]|0\rangle \nonumber
\eqne
in the last line we changed the order of integration and then did integration by parts (twice) to free the delta functions and transfer $\nabla^2_{\vr_b}$ to act on $\Phi(\vr_1,\dots,\vr_N)$.

Changing the order of integrations back again and using the delta functions we obtain

\subsection{Fermions - calculating  $F_{op}|\Phi\rangle$ \label{sec:calc_1b_ferm}}

Deriving the action of the operator $F_{op}^{(1)}$, Eq.\,(\ref{eq:1body_op_fermi}) on the many fermion wave function Eq.\,(\ref{eq:wf_in_sq_f}) let us start by applying the part $\hat{\psi}_{\sigma'}^{+}(\vr) f^{(1)}_{\sigma'\sigma}$ of  $F_{op}^{(1)}$ to the expression (\ref{act_psi_on_Npart_ferms_st}), with the result
\beq \label{eq:part_res_ferm}
const_N \sum_{a=1}^N f^{(1)}_{\sigma'\sigma}(\vr)\delta_{\sigma\sigma_a} \delta(\vr-\vr_a) \prod_{b<a} \hpsis_{\sigma_b}(\vr_b)\hpsis_{\sigma'}(\vr) \prod_{b>a} \hpsis_{\sigma_b}(\vr_b)|0\rangle
\eeq
Here we for simplicity assumed that $f^{(1)}_{\sigma'\sigma}$ is a function of $\vr$ only so that we could bring $\hpsis_{\sigma'}(\vr)$ "through it" and commute to where $\hpsis_{\sigma_a}(\vr_a)$ was. This commuting generated additional the factor $(-1)^{P_a}$ giving overall unity when combined with the same factor in Eq.\,(\ref{act_psi_on_Npart_ferms_st}). We note that for $f^{(1)}_{\sigma'\sigma}$ depending on $-i\hbar \nabla_{\vr}$ one should use the intermediate integration by parts in analogy with what we did in the kinetic energy case with bosons, cf., Appendix \ref{sec:K_act_on_Phi}. 

To finish the calculation let us sum the result (\ref{eq:part_res_ferm}) over $\sigma$ and $\sigma'$ and integrate over $\vr$. Using $\delta_{\sigma\sigma_a}$ to perform the sum over $\sigma$ and 
$\delta(\vr-\vr_a)$ to do the integral we obtain
$$
F_{op}^{(1)}\prod_{a=1}^N \hpsis_{\sigma_a}(\vr_a) |0\rangle = \sum_{a=1}^N \sum_{\sigma'} f^{(1)}_{\sigma'\sigma_a}(\vr_a) \prod_{b<a} \hpsis_{\sigma_b}(\vr_b)\hpsis_{\sigma'}(\vr_a) \prod_{b>a} \hpsis_{\sigma_b}(\vr_b)|0\rangle
 $$
This gives
\eqna
&&F_{op}^{(1)}|\Phi\rangle = \frac{1}{\sqrt{N!}} \sum_{\sigma_1, ... \sigma_N }\int \prod_{a=1}^N d^3r_a \;\Phi(\vr_1\sigma_1,\vr_2\sigma_2,\dots,\vr_N \sigma_N) F_{op}^{(1)}\prod_{a=1}^N \hpsis_{\sigma_a}(\vr_a) |0\rangle = \nonumber \\
&&=\frac{1}{\sqrt{N!}} \sum_{\sigma_1, ... \sigma_N }\int \prod_{b=1}^N d^3r_b \;\Phi(\vr_1\sigma_1,\vr_2\sigma_2,\dots,\vr_a\sigma_a, \dots,\vr_N \sigma_N)  \times \nonumber \\
&& \;\;\;\;\;\;\;\;\;\;\;\;\;\;\;\;\; \times \sum_{a=1}^N \sum_{\sigma'} f^{(1)}_{\sigma'\sigma_a}(\vr_a)\prod_{b<a} \hpsis_{\sigma_b}(\vr_b)\hpsis_{\sigma'}(\vr_a) \prod_{b>a} \hpsis_{\sigma_b}(\vr_b)|0\rangle   = \nonumber \\
&& = \frac{1}{\sqrt{N!}} \sum_{\sigma_1, ... \sigma_N }\int \prod_{b=1}^N d^3r_b \;\sum_{a=1}^N \sum_{\sigma'} f^{(1)}_{\sigma_a\sigma'}(\vr_a)\Phi(\vr_1\sigma_1,\vr_2\sigma_2,\dots,\vr_a\sigma', \dots,\vr_N \sigma_N)  \times \nonumber \\
&& \;\;\;\;\;\;\;\;\;\;\;\;\;\;\;\;\;\;\;\;\;\;\;\;\;\;\;\;\; \times\prod_{b<a} \hpsis_{\sigma_b}(\vr_b)\hpsis_{\sigma_a}(\vr_a) \prod_{b>a} \hpsis_{\sigma_b}(\vr_b)|0\rangle   \nonumber
\nonumber
\eqne
where after the last equality sign we have used the presence of sums over both $\sigma_a$ and $\sigma'$ and interchanged notation of their summation variables  $\sigma_a \leftrightarrow\sigma'$.  This finally gives
$$
F_{op}^{(1)}|\Phi\rangle = \frac{1}{\sqrt{N!}} \sum_{\sigma_1, ... \sigma_N }\int \prod_{a=1}^N d^3r_a \;\Phi'(\vr_1\sigma_1,\vr_2\sigma_2,\dots,\vr_N \sigma_N)\prod_{a=1}^N \hpsis_{\sigma_a}(\vr_a) |0\rangle 
$$
with 
$$
\Phi'(\vr_1\sigma_1,\vr_2\sigma_2,\dots,\vr_N \sigma_N) = \left[\sum_{a=1}^N \sum_{\sigma'} f^{(1)}_{\sigma_a\sigma'}(\vr_a)\right]\Phi(\vr_1\sigma_1,\vr_2\sigma_2,\dots,\vr_a\sigma', \dots,\vr_N \sigma_N) 
$$

\end{document}